\documentclass[floats,floatfix,showpacs,amssymb,prd,twocolumn,superscriptaddress,nofootinbib,nolongbibliography,reprint]{revtex4-1}

\usepackage{amssymb,amsmath,verbatim,mathtools,needspace,enumitem,etoolbox,graphicx,physics,microtype,afterpage,xspace,tabularx,lmodern,multirow}
\usepackage{textcomp, gensymb}

\usepackage[normalem]{ulem}

\usepackage[normalem]{ulem}
\usepackage{placeins}
\usepackage{comment}
\usepackage[dvipsnames, usenames]{xcolor}
\definecolor{linkcolor}{rgb}{0.0,0.3,0.5}
\usepackage[unicode, colorlinks=true, linkcolor=linkcolor, citecolor=linkcolor, filecolor=linkcolor, urlcolor=linkcolor, linktocpage, breaklinks]{hyperref}
\usepackage[all]{hypcap}
\usepackage[T1]{fontenc}
\usepackage[utf8]{inputenc}
\usepackage{mathrsfs}
\usepackage[usenames,dvipsnames]{xcolor}
\hypersetup{colorlinks=true,citecolor=romared,linkcolor=romared,urlcolor=romared}

\definecolor{romared}{RGB}{142,0,28}

\newcommand{\nn}{\nonumber}

\newcommand{\be}{\begin{equation}}
\newcommand{\ee}{\end{equation}}

\def\be{\begin{equation}}
\def\ee{\end{equation}}
\newcommand{\beq}{\begin{eqnarray}}
\newcommand{\eeq}{\end{eqnarray}}

\usepackage{aas_macros}
\usepackage{makecell}
\usepackage{soul}
\usepackage[nolist,nohyperlinks]{acronym}
\usepackage{orcidlink}
\usepackage{lipsum}

\acrodef{LSC}[LSC]{LIGO Scientific Collaboration}
\acrodef{BH}{black hole}
\acrodef{NS}{neutron star}
\acrodef{PN}{Post-Newtonian}
\acrodef{BBH}{binary black-hole}
\acrodef{BNS}{binary neutron-star}
\acrodef{NSBH}{neutron-star black-hole}
\acrodef{NR}{numerical relativity}
\acrodef{GW}{gravitational wave}
\acrodef{PSD}{power spectral density}
\acrodef{aLIGO}{Advanced Laser interferometer Gravitational-Wave Observatory}
\acrodef{AZDHP}{aLIGO zero detuned high power density}
\acrodef{GR}{general relativity}
\acrodef{PE}{parameter estimation}
\acrodef{LAL}{LIGO algorithm library}
\acrodef{TPI}{tensor-product interpolant}
\acrodef{SVD}{singular value decomposition}
\acrodef{SNR}{signal-to-noise ratio}
\acrodef{ODE}{ordinary differential equation}
\acrodef{PDE}{partial differential equation}
\acrodef{ROM}{reduced order model}
\acrodef{QNM}{quasi-normal mode}
\acrodef{IMR}{inspiral-merger-ringdown}
\acrodef{LVK}{LIGO-Virgo-KAGRA}
\acrodef{SXS}{Simulating eXtreme Spacetimes}

\newcommand{\jhu}{\affiliation{William H. Miller III Department of Physics and Astronomy, Johns Hopkins University, 3400 North Charles Street, Baltimore, Maryland, 21218, USA}}
\newcommand{\ciera}{\affiliation{Center for Interdisciplinary Exploration and Research in Astrophysics (CIERA), 1800 Sherman Ave, Evanston, IL 60201, USA}}

\graphicspath{{../Plots/}}

\newcommand{\orcid}[1]{\href{https://orcid.org/#1}{\includegraphics[width=10pt]{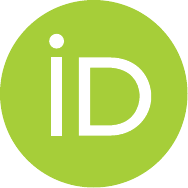}}}
\newcommand{\Mx}[1]{\ensuremath{Q_{#1}}}
\newcommand{\Mxr}[2]{\ensuremath{Q_{{#1}, {#2}}}}
\newcommand{\Mxrm}[2]{\ensuremath{Q^{(m)}_{{#1}, {#2}}}}
\newcommand{\Mr}[1]{\ensuremath{Q^{(f)}_{{#1}}}}
\newcommand{\Mrm}[1]{\ensuremath{Q^{(f,\,m)}_{{#1}}}}
\newcommand{\GIST}{Ref.~\cite{Giesler:2019uxc}}

\newcommand{\ben}{\begin{enumerate}}
\newcommand{\een}{\end{enumerate}}

\def\nn{\nonumber}
\def\be{\begin{equation}}
\def\ee{\end{equation}}
\def\beq{\begin{eqnarray}}
\def\eeq{\end{eqnarray}}
\def\f{\frac}

\def\nn{\nonumber}

\def\apjl{\rm{ApJ}}
\def\apjs{\rm{ApJS}}
\def\aap{\rm{A\&A}}

\def\nat{\rm{Nature}}

\graphicspath{{./Plots/}}

\allowdisplaybreaks

\begin{document}

\pagenumbering{arabic}

\title{Agnostic black hole spectroscopy: quasinormal mode content of numerical relativity waveforms and limits of validity of linear perturbation theory}

\author{Vishal Baibhav \orcid{0000-0002-2536-7752}}
\ciera
\author{Mark Ho-Yeuk Cheung \orcid{0000-0002-7767-3428}}
\jhu
\author{Emanuele Berti \orcid{0000-0003-0751-5130}}
\jhu
\author{Vitor Cardoso \orcid{0000-0003-0553-0433}}
\affiliation{CENTRA, Departamento de F\'{\i}sica, Instituto Superior T\'ecnico -- IST, Universidade de Lisboa -- UL,
Avenida Rovisco Pais 1, 1049-001 Lisboa, Portugal}
\affiliation{Niels Bohr International Academy, Niels Bohr Institute, Blegdamsvej 17, 2100 Copenhagen, Denmark}
\author{Gregorio Carullo \orcid{0000-0001-9090-1862}}
\affiliation{Niels Bohr International Academy, Niels Bohr Institute, Blegdamsvej 17, 2100 Copenhagen, Denmark}
\affiliation{Theoretisch-Physikalisches Institut, Friedrich-Schiller-Universit{\"a}t Jena, 07743, Jena, Germany}
\affiliation{Dipartimento di Fisica “Enrico Fermi”, Università di Pisa, Pisa I-56127, Italy}
\affiliation{INFN sezione di Pisa, Pisa I-56127, Italy}
\author{Roberto Cotesta \orcid{0000-0001-6568-6814}}
\jhu
\author{Walter Del Pozzo \orcid{0000-0003-3978-2030}}
\affiliation{Dipartimento di Fisica “Enrico Fermi”, Università di Pisa, Pisa I-56127, Italy}
\author{Francisco Duque \orcid{0000-0003-0743-6491}}
\affiliation{CENTRA, Departamento de F\'{\i}sica, Instituto Superior T\'ecnico -- IST, Universidade de Lisboa -- UL,
Avenida Rovisco Pais 1, 1049-001 Lisboa, Portugal}
\date{\today}

\begin{abstract}
Black hole spectroscopy is the program to measure the complex gravitational-wave frequencies of merger remnants, and to quantify their agreement with the characteristic frequencies of black holes computed at linear order in black hole perturbation theory. In a ``weaker'' (non-agnostic) version of this test, one {\em assumes} that the frequencies depend on the mass and spin of the final Kerr black hole as predicted in perturbation theory. Linear perturbation theory is expected to be a good approximation only at late times, when the remnant is close enough to a stationary Kerr black hole. However, it has been claimed that a superposition of overtones with frequencies fixed at their asymptotic values in linear perturbation theory can reproduce the waveform strain even at the peak. Is this overfitting, or are the overtones physically present in the signal? To answer this question, we fit toy models of increasing complexity, waveforms produced within linear perturbation theory, and full numerical relativity waveforms using both agnostic and non-agnostic ringdown models. We find that higher overtones are unphysical: their role is mainly to ``fit away'' features such as initial data effects, power-law tails, and (when present) nonlinearities. We then identify physical quasinormal modes by fitting numerical waveforms in the original, agnostic spirit of the no-hair test. We find that a physically meaningful ringdown model requires the inclusion of higher multipoles, quasinormal mode frequencies induced by spherical-spheroidal mode mixing, and nonlinear quasinormal modes. Even in this ``infinite signal-to-noise ratio'' version of the original spectroscopy test, there is convincing evidence for the first overtone of the dominant multipole only well after the peak of the radiation.
\clearpage
\end{abstract}

\maketitle

\tableofcontents

\clearpage

\section{Introduction}

A striking aspect of black hole (BH) perturbation theory is its formal analogy with quantum mechanics. This analogy follows from the fact that after separation of the angular variables using tensor spherical harmonics with angular indices $(\ell,\,m)$, the scattering of gravitational waves (GWs) off a Schwarzschild BH becomes formally equivalent to a Schr\"odinger-like equation with a potential barrier. This is true for both odd-parity~\cite{Regge:1957td} and even-parity~\cite{Zerilli:1970se,Zerilli:1970wzz} perturbations, which fully characterize the linear dynamics of the Schwarzschild spacetime.

Once the boundary conditions for the scattering problem were understood, Vishveshwara realized that the response of the BH to an incoming pulse of radiation is characterized by a superposition of damped exponentials with discrete frequencies and damping times, now commonly known as the ``ringdown'' by analogy with the dying tones of a vibrating bell~\cite{Vishveshwara:1970zz}. The damping occurs because (unlike many textbook problems in quantum mechanics) the BH scattering problem is not self-adjoint: BH spacetimes absorb gravitational radiation at the horizon and emit radiation at spatial infinity -- hence the name ``quasinormal'' modes (QNMs), as opposed to the ``normal'' modes of self-adjoint physical systems~\cite{Nollert:1999ji,Kokkotas:1999bd,Berti:2009kk,Konoplya:2011qq}.

The correspondence between BH spectra and atomic spectra was repeatedly used at the formal level in the development of QNM theory during the 1970s. Press~\cite{Press:1971wr} used the analogy to prove that the BH would not ``divest itself of the unwanted perturbations in a single large belch'', but rather radiate initial perturbations with multipole index $\ell$ ``only gradually, yielding a long and nearly sinusoidal wave train of gravitational radiation.''  The QNMs identified by Press were shown to play an important role in physical processes producing gravitational radiation -- for example, when particles fall radially into the BH~\cite{Davis:1971gg}.

Chandrasekhar and Detweiler~\cite{Chandrasekhar:1975zza} used again the quantum mechanical analogy to prove the isospectrality of even and odd perturbations, and to compute the so-called ``overtones'' of a Schwarzschild BH. For given angular indices $(\ell,\,m)$, there is a whole ``tower'' of QNM frequencies $\omega_{\ell m n}$ that can be sorted by the magnitude of their imaginary part. Typically $n=0$ denotes the fundamental mode, and increasing values of $n$ correspond to larger imaginary parts and shorter damping times.
Deeper connections between the quantum mechanical scattering problem and the gravitational scattering problem emerged in the work by Ferrari, Mashhoon, Schutz and Will~\cite{Ferrari:1984zz,Mashhoon:1985cya,BLOME1984231,Schutz:1985km}. After Teukolsky's breakthrough proof of the separability of the perturbation equations for rotating (Kerr) BHs~\cite{Teukolsky:1972my,Teukolsky:1973ha,Press:1973zz,Teukolsky:1974yv}, Detweiler pointed out that the fundamental QNM frequency of a Kerr BH (the one with the smallest imaginary part and longest damping time) depends only on its mass and spin~\cite{Detweiler:1977gy}, so -- at least conceptually -- the relation can be inverted to identify the Kerr BH parameters from a knowledge of the frequency and damping time.

While GW astronomy was a long time coming, the potential {\em observational} implications were clear to the first theorists studying the gravitational spectrum of Kerr BHs.
Detweiler's landmark calculation of the Kerr QNM spectrum ends with a remarkably prophetic statement: ``After the advent of GW astronomy, the observation of [the BH’s] resonant frequencies might finally provide direct evidence of BHs with the same certainty as, say, the 21 cm line identifies interstellar hydrogen''~\cite{Detweiler:1980gk}. An even deeper analogy between Kerr perturbations and the quantum mechanical treatment of the H$_2$ ion~\cite{Jaffe1934,1935PCPS...31..564B} was exploited by Leaver to compute the Kerr spectrum with high accuracy using continued fraction techniques~\cite{Leaver:1985ax}.

Practical attempts to implement the spectroscopy program in GW data analysis started later. Echeverria quantified the measurability of the frequency and damping time of the fundamental mode~\cite{Echeverria:1989hg} (see also~\cite{Finn:1992wt}).  The merger of two comparable-mass BHs was identified early on as one of the most promising LIGO-Virgo sources~\cite{1987thyg.book..330T}, but predicting which QNMs would be excited as a result of the merger was essentially a matter of guesswork before the first numerical BH merger simulations.

In the late 1990s, a conjectured correspondence between large-$n$ QNMs and BH area quantization~\cite{Hod:1998vk,Dreyer:2002vy} triggered several studies of overtones in various BH spacetimes. 
By the time Dreyer and collaborators introduced the term ``BH spectroscopy''~\cite{Dreyer:2003bv}, the idea had been explored by the GW community for decades (there was also a flourishing industry of research on ``GW asteroseismology,'' trying to infer the properties of ultradense matter from the analogous problem for compact stars: see e.g.~\cite{Kokkotas:1999bd,Kokkotas:1999mn}). The fact that overtones should generically contribute to the GW signal had been proven in many different contexts. For example, the classic study of Oppenheimer-Snyder collapse to a Schwarzschild BH by Cunningham, Price and Moncrief clearly identified  the first overtone (see Fig.~12 of Ref.~\cite{Cunningham:1978zfa}). Inclusion of one overtone was shown to better fit the waveforms from infalling particles with finite kinetic energy in the Schwarzschild case~\cite{Ferrari:1981dh}, as well as the waveforms from the first simulations of rotating collapse to a Kerr BH~\cite{Stark:1985da}.

The fact that multiple modes contribute to the ringdown offers the opportunity to characterize the remnant as a Kerr BH.
The idea of BH spectroscopy is quite simple (see~\cite{Kokkotas:1999bd,Berti:2009kk,Berti:2018vdi,Cardoso:2019rvt} for reviews).
In general relativity (GR), the two GW polarizations $h_{+,\,\times}$ can be decomposed as
\begin{equation}\label{eq:h_spherical}
h_+ - i h_\times \equiv \sum_{\ell m} h_{\ell m}(t) _{-2}Y_{\ell m}(\iota,\,\phi)\,,
\end{equation}
where the spin-weighted spherical harmonics $_{-2}Y_{\ell m} (\iota, \phi)$ depend on two angles that characterize the direction from the source to the observer.\footnote{As we discuss below, spin-weighted {\em spheroidal} harmonics $_{-2}S_{\ell m} (\iota, \phi)$ are more appropriate to study perturbations of Kerr BHs. This produces spherical-spheroidal mode mixing~\cite{Berti:2005gp,Berti:2014fga}.}

Within linearized BH perturbation theory, Leaver~\cite{Leaver:1986gd} proved that each multipolar component of the waveform at intermediate times -- after the ``prompt response'', and before the onset of power-law tails -- is described by a superposition of QNMs:
\begin{equation}\label{eq:model}
  h_{\ell m}(t) \equiv \sum_n A_{\ell m n} e^{-i\left[\omega_{\ell m n}(t - t_{\rm start}) + \phi_{\ell m n}\right]}\,.
\end{equation}
Here,
$t_{\rm start}$ is an arbitrary starting time.
In linearized GR, the complex Kerr QNM frequencies $\omega_{\ell m n}$
depend only on the remnant BH mass $M_f$ and dimensionless spin $\chi_f$, but not on the nature of the perturbation, and are known to very high accuracy~\cite{Berti:2005ys,RDwebsites}.
On the contrary, the QNM amplitudes $A_{\ell m n}$ and phases $\phi_{\ell m n}$ depend on the astrophysical process causing the perturbation.

Green's function techniques imply that the QNM amplitudes $A_{\ell m n}$ can be factorized as a product
of complex ``excitation factors'' $B_{\ell m n}$ that depend only on the remnant's mass and spin and complex-valued, initial-data dependent integrals $I_{\ell m n}$~\cite{Leaver:1986gd,Andersson:1995zk,Andersson:1996cm,Berti:2006wq,Zhang:2013ksa,Oshita:2021iyn,Lagos:2022otp}.
However, the excitation amplitudes of the overtones in a binary merger were unknown before the first numerical BH merger simulations.
Heuristic arguments suggested that comparable-mass BH mergers may have ringdown signal-to-noise ratio (SNR) roughly comparable to the inspiral SNR~\cite{Flanagan:1997sx}, while other astrophysical processes would be much less efficient at exciting QNMs~\cite{Berti:2006hb}. Early work on BH spectroscopy trying to quantify the {\em measurability} of different multipoles and different overtones had to rely on educated guesses~\cite{Berti:2005ys}.

Our understanding of ringdown excitation changed dramatically after the 2005 numerical relativity (NR) breakthrough~\cite{Pretorius:2005gq,Campanelli:2005dd,Baker:2005vv}. Fits of NR simulations revealed that the radiation from a BBH merger is dominated by the $\ell=|m|=2$ spherical harmonic multipole, while higher multipoles are subdominant~\cite{Buonanno:2006ui,Berti:2007fi}.
A superposition of overtones with frequencies fixed at the values predicted for the asymptotic Kerr remnant can fit the waveform even before the peak, but Ref.~\cite{Buonanno:2006ui} questioned the physical meaning of extending these fits to the peak of the radiation: ``The Kerr QNM frequencies and decay constants are computed assuming that the mass and angular momentum they carry away constitute a negligible perturbation on the system. This raises the question as to whether or not the radiated energy and angular momentum are affecting the QNM fits. This issue will, of course, become more significant as the fits are pushed to earlier times.''

As our understanding of QNM excitation has improved since 2005, so have GW data analysis techniques.  The first detection of GWs from the binary black hole (BBH) merger GW150914~\cite{LIGOScientific:2016aoc} marked the beginning of a new era in astronomy. Since then, the LIGO-Virgo-KAGRA (LVK) Collaboration~\cite{LIGOScientific:2014pky, VIRGO:2014yos,KAGRA:2020tym} has reported 90 events of probable astrophysical origin during the first three observing runs~\cite{LIGOScientific:2018mvr,LIGOScientific:2020ibl,LIGOScientific:2021usb,LIGOScientific:2021djp}.  These GW signals, in combination with those detected by independent groups~\cite{Nitz:2018imz,Nitz:2020oeq,Nitz:2021uxj,Venumadhav:2019lyq,Zackay:2019btq}, can be used for various tests of GR in the strong-field regime~\cite{LIGOScientific:2019fpa, LIGOScientific:2020tif, LIGOScientific:2021sio}.

While in principle a perfect knowledge of the dominant QNM frequency in a BH binary merger (which we now know to be the mode with $\ell=m=2$, $n=0$) could be used to infer both the mass and the spin of the remnant, in practice a single mode is not sufficient to get accurate and unbiased values for these quantities: mass and spin estimates can and should be improved by combining different multipolar components~\cite{Berti:2007zu} and by including overtones~\cite{Baibhav:2017jhs}.  Since multiple modes will always be excited to some extent, we must first understand which combination of modes will dominate the signal~\cite{Berti:2005ys}. Are we going to observe a combination of low-$n$ modes for different multipoles, or are higher overtones of the $\ell=m=2$ component dominant? Can we measure the frequencies and damping times accurately enough to resolve the modes? The answers to these questions depend on the properties of the merger remnant progenitors and on the sensitivity of the detectors~\cite{Berti:2007fi,Kamaretsos:2011um,Kamaretsos:2012bs,London:2014cma,Bhagwat:2016ntk,Baibhav:2017jhs,Thrane:2017lqn,London:2018gaq,Baibhav:2018rfk,Baibhav:2020tma,Cook:2020otn,JimenezForteza:2020cve,Ota:2021ypb,Li:2021wgz,MaganaZertuche:2021syq}.

Initial estimates of the detectability and resolvability of different modes used a Fisher matrix approximation, which is only valid for large SNRs~\cite{Echeverria:1989hg,Finn:1992wt,Berti:2005ys}. SNR estimates based on NR-calibrated amplitudes were instead presented in ~\cite{Berti:2007zu,Kamaretsos:2011um,Kamaretsos:2012bs}. In the first ``modern'' Bayesian treatment of BH spectroscopy~\cite{Gossan:2011ha}, the ringdown was parametrized in terms of mass, spin and a single deviation parameter, reducing the number of free parameters and related correlations. A search for beyond-Kerr signatures based on this model is more sensitive, but less generic than one where all the modes are measured independently (i.e., the ``classical'' formulation of BH spectroscopic tests: see the discussion in Sec.~\ref{sec:discussion} below).
The ``resolvability'' of the modes was quantified in terms of Bayes factors.
Ref.~\cite{Meidam:2014jpa} introduced an improved model for ringdown amplitudes and sources, focusing on the Einstein Telescope (ET) detector~\cite{Punturo:2010zz}.
After the first detections, updated forecasts of our ability to observe the fundamental modes with $\ell=m=3$ and $\ell=m=4$ were computed within a Bayesian framework. Reference~\cite{Bhagwat:2019bwv} concluded that subdominant fundamental modes with an amplitude of 0.1 (0.3) relative to the fundamental mode with $\ell=m=2$ could be detected with SNR of 30 (15) in the late-time ringdown without assuming NR constraints on the amplitudes. Relying on a model calibrated to NR~\cite{Borhanian:2019kxt}, these estimates were revised in Ref.~\cite{Cabero:2019zyt}. This work confirmed previous Fisher-based estimates~\cite{Berti:2016lat} (see also~\cite{Ota:2021ypb}), concluding that Voyager-class detectors would be necessary to have ``decisive'' Bayesian evidence for the presence of two modes in a single detection.

All of these studies focused on ringdown-only parameter estimation, ignoring the pre-ringdown signal.
The first study of ringdown signals from complete inspiral-merger-ringdown simulations windowed the signals at the ringdown start time to avoid spurious pre-ringdown contributions in the frequency domain, and showed that percent-level ``no-hair'' tests are possible by combining multiple loud sources detected by the LIGO-Virgo network~\cite{Carullo:2018sfu}.
The need to tune specific windowing parameters can be overcome by formulating the test directly in the time domain. The uncorrelated case was considered in Ref.~\cite{DelPozzo:2016kmd}, while the case of a nondiagonal autocovariance matrix was tackled in Ref.~\cite{Carullo:2019flw}, applying the time-domain method to search for multiple (fundamental) modes in GW150914 and constraining parametric deviations from the GR spectrum.
This formalism was extended by adopting a truncated likelihood formulation with a fixed ringdown starting time in Ref.~\cite{Isi:2019aib}, and applied to the search of the first ringdown overtone in GW150914 data.
These works led to the construction of a full ringdown pipeline built around the \texttt{pyRing} package~\cite{Carullo:2019flw,pyRing}, which was used by the LVK collaboration to produce a catalog of ringdown observations and to search for beyond-GR signatures~\cite{LIGOScientific:2020iuh,LIGOScientific:2020ufj,LIGOScientific:2020tif, LIGOScientific:2021sio}.
The searches employed models of increasing complexity, ranging from agnostic superpositions of damped sinusoids to templates calibrated against BBH simulations~\cite{London:2018gaq}.
The \texttt{pyRing} pipeline was later applied to search for BH charges~\cite{Carullo:2021oxn}, set bounds on the BH information emission mechanism~\cite{Carullo:2021yxh}, 
investigate heuristic models of ``area-quantized'' BHs~\cite{Laghi:2020rgl}, and constrain modified gravity corrections in a perturbative framework~\cite{Maselli:2019mjd,Carullo:2021dui}.
The truncated formulation has also been implemented in the \texttt{ringdown} package~\cite{Isi:2021iql}.

In data analysis, the omission of overtones may lead to biases in the remnant mass and spin estimates~\cite{Baibhav:2017jhs}. However, QNM tests often relied only on fundamental modes with different values of $(\ell,\,m)$ for two main reasons. The first reason is practical: overtones are short-lived and difficult to confidently identify in the data~\cite{London:2014cma}. The second reason (and the focus of this work) is conceptual: it is unclear whether multiple overtones have physical meaning, or they just happen to phenomenologically fit the nonlinear part of the merger signal~\cite{Buonanno:2006ui}.

Giesler et al.~\cite{Giesler:2019uxc} focused on the $\ell=m=2$ multipole of the radiation, and showed that including overtones up to $n=7$ in the ringdown model improves the mismatch with NR simulations for all times $t\geq t_{\mathrm{peak}}$. Here the ``peak time'' $t_{\mathrm{peak}}$ is defined as the time at which $|h_+^2 + h_\times^2|$ has a maximum. According to \GIST, the inclusion of higher overtones ``provides an
unbiased estimate of the true remnant parameters'' and the low mismatch ``implies that the spacetime is well described as a linearly perturbed BH with a fixed mass and spin as early as the peak.''
This {emphasis} on linearity {prompted} a sequence of additional investigations, both on the modeling and on the observational side~\cite{Ota:2019bzl, Bhagwat:2019dtm,JimenezForteza:2020cve,Bustillo:2020buq,Okounkova:2020vwu,Mourier:2020mwa,Cook:2020otn,Dhani:2020nik,Dhani:2021vac,Finch:2021iip,Finch:2021qph,Ota:2021ypb,Forteza:2022tgq}.
Some works extended the idea to counterrotating modes and higher multipoles~\cite{MaganaZertuche:2021syq}, included an even larger numbers of overtones~\cite{Forteza:2021wfq}, and proposed possible explanations for the apparent simplicity of the signal~\cite{Okounkova:2020vwu,Jaramillo:2022oqn}.

If higher overtones could be measured by starting at the peak, the larger ringdown SNR would open the door to more precise tests of GR.  This argument (that we will challenge below) motivated a reanalysis of GW150914~\cite{Isi:2019aib} where the post-peak waveform was fitted with a QNM superposition including overtones, claiming evidence for ``at least one overtone [...] with $3.6 \sigma$ confidence.''
This claim is at odds with Ref.~\cite{Bustillo:2020buq} and with the subsequent LVK analysis~\cite{LIGOScientific:2020tif}, both reporting weak evidence (with a log-Bayes factor of $\sim 0.6$) in favor of the model including two modes relative to the model including only one. A recent reanalysis of GW150914 found no evidence in favor of an overtone in data after the peak~\cite{Cotesta:2022pci}. Around the peak, the log-Bayes factor does not indicate the presence of an overtone, while the support for a nonzero amplitude is sensitive to changes in the starting time much smaller than the overtone damping time. GW150914-like injections in neighboring segments of the real detector noise suggest that noise can artificially enhance evidence for an overtone.
The matter was further debated in Refs.~\cite{Isi:2022mhy,Finch:2022ynt,Ma:2023vvr,Ma:2023cwe}, using different choices for the likelihood, noise estimation, sampling rate and analysis duration.

In this paper we set aside the controversial issue of identifying overtones in real data (but we discuss the implications of our findings on this debate in Sec.~\ref{sec:discussion}), and we {carefully} reanalyze the main conclusion of \GIST: can we reliably conclude from an analysis of NR simulations that the entire post-peak waveform is described by a superposition of overtones consistent with linear BH perturbation theory on a fixed Kerr background? To improve readability, we present our main conclusions in the following executive summary.

\subsection{Executive summary}

The main goal of the paper is to understand which physical modes can be extracted from GW signals and at which point in time, with a special (but non-exclusive) focus on overtones. To this end, we will often compare two broad classes of models: one in which all complex frequencies $\omega_{\ell m n}$ are fixed to match theoretical predictions from BH perturbation theory~\cite{Berti:2005ys,RDwebsites}, and one in which some (or all) of these frequencies are left free to vary.
The first class leads to a ``weak'' version of BH spectroscopy tests (often used in recent investigations), while the second corresponds to the original, ``agnostic'' spectroscopy proposal.

While the non-agnostic method leads to an easier extraction of the modes, it must be used with great care: by a-priori \textit{assuming} the presence of short-lived modes, it can often lead to false evidence for unphysical contributions, as in the case of higher overtones.
Generically, when looking for overtones, we will show that it is very important to take into account subdominant contributions to the waveform, such as tails in linear theory or spherical-spheroidal mode mixing in full NR simulations.

The notation used for our fitting models is introduced in Sec.~\ref{subsec:QNMmodels}, and it is summarized in Table~\ref{tab:models}.

\noindent
{\bf \em Extracting overtones in linearized theory.}
In Sec.~\ref{sec:linearovertones} we show that the ``original'' BH spectroscopy test, where one  agnostically searches for multiple QNMs in the data, is hard to carry out using overtones even within linear perturbation theory (and in the absence of noise). To understand why, we introduce three toy models of increasing complexity, such that linear perturbation theory is valid {\em by construction}: (i) a hypothetical ``pure ringdown'' (i.e., a pure superposition of damped exponentials, which is never expected to describe a real signal\footnote{In Appendix~\ref{app:Green} we use Green's function techniques to make two important conceptual points. First, a ``pure ringdown'' never exists: other components (including prompt emission and backscattering) will always affect the waveform and cause amplitudes to change in time. Second, even in the linear regime (and even for a Minkowski background) the ``ringdown starting time'' is ill-defined: it depends on which of the radiation properties (amplitude or energy) that we monitor, and there is no mathematical or physical basis to claim a well-defined instant as ``the'' ringdown starting time.}); (ii) a ``pure ringdown'' model to which we add a power-law tail; and (iii) a numerical waveform constructed by a replica of the original Vishveshwara scattering experiment~\cite{Vishveshwara:1970zz}.
These toy models illustrate that even within linear perturbation theory, it is not sensible to fit the waveform with $\sim 7$ overtones starting at the peak of the radiation.

We prove this point in two parts. We first ask: can we recover the known frequencies by fitting? As a test of the fitting algorithm, we show that this is possible in the ``pure ringdown'' model by using a ``bootstrap'' procedure: i.e., we first identify longer-lived modes in an agnostic way, and then we fix their complex frequencies when searching for shorter-lived ones.
In the ``ringdown+tail'' model, however, as we increase the number of fitting modes, the minimum of the mismatch ${\cal M}$ between the fitting model and the waveform keeps decreasing as we get closer to the waveform peak, {\em even when the free mode does not approach the expected overtone frequency.} 
Adding modes to the fitting model can reduce the mismatch even if the mode frequency is unphysical. Therefore a small mismatch is not sufficient evidence to claim the presence of an overtone.
The individual modes stop converging towards their expected values when their mismatch drops below the mismatch induced by the tail: at that point, the mode is effectively trying to fit the tail, and the mismatch saturates.
Even if the overtones are physically present in the waveform (as in this toy model), fitting with many overtones is not optimal from a data-analysis point of view, because the fit is not robust against small contaminations. This is an important limitation for agnostic tests of GR: tiny contaminations can hinder our ability to extract higher-overtone frequencies, even if the modes are physically present in the signal.
The toy models also allow us to better understand the behavior of the fitted QNM amplitudes.
For the (unrealistic) ``pure ringdown'' waveforms, all of the overtone amplitudes converge to constant values at late times.
When we add a tail, there is one remarkable difference: the fitted amplitudes blow up exponentially at some critical time -- more specifically, when the highest (fastest-decaying) overtone in the fitting model starts to pick up the contamination due to the tail.

The conclusion of this exercise is clear:
{\em {unless additional physics is taken into account} the original, agnostic BH spectroscopy test is unfeasible for all overtones (including $n=1$), and only possible at late times for the fundamental mode, even within linear perturbation theory.}
When we consider ringdown waveforms resulting from the scattering of a Gaussian wave packet in linearized gravity, an agnostic damped-sinusoid fit cannot recover the correct frequencies for any of the overtones. Even if we fix the frequencies to their known values, there is no convincing evidence that overtones with $n>2$ are present in the signal.
If we insist to use multiple overtones to test GR, we should start fitting the waveform at times significantly after the peak.
Many of the lessons learned in linearized theory carry over to the
full GR case.

\noindent
{\bf \em Are post-peak BBH waveforms linear?}
Reference~\cite{Giesler:2019uxc} claimed that the waveform resulting from the merger of two comparable-mass BHs can be (i) adequately described by linear perturbation theory starting from the peak of the strain, (ii) well modeled by a combination of QNM overtones, and (iii) used to test GR by identifying the overtones in the signal.
In Sec.~\ref{sec:postpeaklinearity} we revisit these claims. We present three different arguments against the validity of the linear approximation.

First, we show that a constant-amplitude overtone superposition does not work in the BH merger case, and the amplitudes of the overtones change significantly when we change the fitting window. This mode-amplitude evolution has two important implications: (i) overtone models with $N \geq 2$ are unphysical, because they try to overfit other features of the waveform; (ii) models with at most one overtone ($N\leq 1$) are physical, but they can only be used for meaningful spectroscopy tests at late times.

These conclusions are reinforced in three appendices. In Appendix~\ref{sec:NRerror} we demonstrate that using NR waveforms with different resolutions and different extrapolation orders makes almost no difference in the amplitude fits, and therefore NR errors cannot explain the time variation of the fitted amplitudes. In Appendix~\ref{sec:systematics} we investigate the effect of a spurious late-time constant observed in the SXS simulations.
We find that this spurious constant has only a small impact at very late times, and that it does not affect our conclusion that it is not possible to extract the first overtone at the peak. In Appendix~\ref{sec:NRtoy} we examine a complex-exponential toy model for SXS:BBH:0305 which includes an estimate of the numerical noise, confirming and strengthening the main conclusions of Sec.~\ref{sec:linearovertones}.

A second issue with a linear perturbation theory interpretation of post-peak ringdown concerns overfitting.
How many overtones are really necessary to minimize the mismatch between ringdown waveforms and the full inspiral-merger-ringdown waveform?  Which QNMs are most effective at minimizing the mismatch and reproducing the correct values of the remnant mass and spin?

{Recent studies claim that the inclusion of the fundamental mode and $7$ overtones provides a very accurate description of the ringdown up to the peak strain amplitude, and significantly reduces the uncertainty in the extracted remnant properties~\cite{Giesler:2019uxc}.} However, we show that the higher overtones lead to very small mismatches by merely overfitting the waveforms. Furthermore, we argue that these higher overtones try to fit other physics (such as time variation in the QNM amplitudes due to initial data, an evolving spacetime background, and nonlinearities) close to the merger. The addition of several overtones allows for better extraction of the fundamental mode and the first overtone, which carry most of the information about the remnant properties, because they effectively ``fit away'' poorly understood physics.

In Appendix~\ref{sec:frankenstein} we construct an unphysical post-peak BBH waveform, and show that the overtones can still fit it with similar accuracy. The fact that using the overtones allows us to improve the measurement of $\chi_f$ and $M_f$ even for this unphysical hybrid waveform supports the conclusion that overtones can match any early post-peak waveform portion, thus allowing the dominant mode to correctly fit the late post-peak waveform: it is the fundamental mode that really carries most of the information about the remnant BH properties.

In Sec.~\ref{sec:epsilon} we go beyond mismatches and ask, in the same vein: which overtones are necessary to correctly extract the remnant's properties?
We swap individual modes with random damped exponentials. If the ``fake'' random-frequency mode still fits the waveform with similar or better accuracy, or if it still extracts the remnant properties accurately, we can conclude that the originally ``swapped'' overtone was not really necessary. We use this argument to show that higher overtones do not play a significant role in extracting the remnant's properties either. Overtones with $n\geq 2$ do not significantly contribute to the extraction of the remnant's parameters, and therefore there is no motivation to include them in the modeling.

\noindent
{\bf \em Agnostic BH spectroscopy: extracting complex frequencies from the waveform.}
Since the inclusion of several overtones leads to overfitting, in Sec.~\ref{sec:howmany} we adopt a different strategy. Rather than imposing a priori that the known overtones associated to a given $(\ell,\,m)$ spin-weighted spheroidal harmonic are present in the $(\ell,\,m)$ spin-weighted spherical harmonic multipole of the NR data, we consider the complex frequencies as free fitting parameters. This is a much stronger test (in fact, it is the original BH spectroscopy proposal at infinite SNR), and therefore it should lead to more robust conclusions about which modes are truly present in the data.

We first focus on the $\ell=m=2$ spherical harmonic multipole and we ask: how many QNM frequencies can we extract without assuming any (no-hair theorem enforced) relation between them?
We show that
(i) in general, the mismatch between the fitting model and the waveform is lower when we keep the frequencies free;
(ii) many of the fitted damped exponentials robustly converge towards known QNM frequencies, naturally selecting the physical modes that contribute to the ringdown signal; and
(iii) it is essential to include spherical-spheroidal mode-mixing to identify the correct modes. In fact, including modes due to spherical-spheroidal mixing is essential to extract the first overtone (at late times) from the dominant $(2,\,2)$ multipole. 
At least three free modes are required to extract the first overtone, and it is easier to extract the long-lived fundamental modes than the fast decaying overtones, even if the latter have a much larger amplitude. 

We then study a variety of different fitting models. In some of these fitting models we include the dominant spherical-spheroidal mode at the expected frequency; in others, we do not. We adopt both a ``fully agnostic'' strategy, in which we include more and more free modes, and a ``bootstrap'' strategy in which we identify modes, fix them, and then search for an additional free frequency.

The study of these different fitting models supports an important conclusion: {\em the only identifiable physical modes in the $\ell=m=2$ multipole of the radiation are $(2,\,2,\,0)$, $(2,\,2,\,1)$ and $(3,\,2,\,0)$.} Higher overtones ($N>1$) cannot be robustly identified by free-frequency fitting.
Furthermore, once the ``physical'' ringdown modes (typically the fundamental mode, the first overtone, and the mode-mixing contribution) have been fitted for, additional free modes have a tendency to simply track the characteristic ``GW frequency chirp'' at early times. In fact, this observation has already been used in the gravitational waveform modeling community, where ``pseudo-QNMs'' were introduced in the context of the effective-one-body framework to model the rapid transition from the inspiral GW frequency to the post-merger QNM frequency ``plateau''~\cite{Pan:2011gk,Damour:2014yha,Brito:2018rfr}.

How do the frequencies inferred by fitting free damped exponentials translate into mass and spin estimates?
In Sec.~\ref{subsec:massspinest} we limit attention to the $\ell=m=2$ spin-weighted spherical multipole. We show that a free fit with two modes near the peak would give significantly biased mass and spin values (whether we include spherical-spheroidal mode mixing or not), and therefore it cannot be used for BH spectroscopy test. Even in the infinite-SNR limit,
including the first overtone never allows for percent-level estimates of the mass and spin: it may allow for $\sim 10\%$ estimates of the mass and spin at late times for loud enough signals, but only if we carefully take into account mode mixing and higher multipoles.

In fact, an analysis of the $\ell=m=4$ multipole (Sec.~\ref{sec:lm44}) shows that nonlinearites are also important: the nonlinear $(2,2,0)(2,2,0)$ QNM is easier to recover than the first linear overtone $(4,4,1)$ in a free-frequency fit. In Appendix~\ref{sec:harmonics} we study other subdominant multipoles.

\noindent
{\bf \em Can the first overtone be extracted in the presence of subdominant multipoles?}
In real-world data analysis problems, the waveform will in general be a superposition of several multipoles.
In Sec.~\ref{sec:sumharmonics} we ask: how many free modes would be necessary to extract the first overtone $(2,\,2,\,1)$ in this more realistic scenario?

We find that even in the relatively optimistic case of a face-on binary, when only the $(2,2)$, $(3,2)$ and $(4,2)$ multipoles significantly contribute to the strain, the extraction of the first overtone $(2,\,2,\,1)$ requires four free frequencies, and it relies on the successful extraction of the (long-lived) fundamental modes of subdominant multipoles. The extraction of overtones becomes significantly more difficult when the binary is not face-on.
Therefore, in practical data analysis applications it would be hard to extract the first overtone from GW150914-lke signals. Even when the overtone can be confidently identified, the fundamental modes of subdominant multipoles generally yield more reliable BH spectroscopy tests.

In Sec.~\ref{sec:discussion} we discuss the observational implications of our work; the possible role of nonlinearities and pseudospectral instabilities in destabilizing higher overtones; how parametrizing the QNM amplitudes may facilitate  BH spectroscopy, and the Occam penalties associated with the inclusion of several modes; the difference between ``weak'' and ``strong'' spectroscopy tests; and the role of beyond-Kerr parametrizations of the QNM spectrum as effective detection templates or tools to constrain modified gravity theories.

\begin{table*}
\setlength{\tabcolsep}{0.8em}
\def\arraystretch{1.3}
\caption{Fitting models used in this paper. All models include $N$ overtones (i.e., $N+1$ QNMs when mode mixing is not included, and $N+2$ QNMs when mode mixing is included). When present, the mode-mixing contribution has fixed frequency, but free amplitude and phase.}
\begin{center}
\begin{tabularx}{0.85\textwidth}{l  l  l   l   l}
\hline \hline  
Model& Description & Mode mixing & {Number of modes} & {Number of parameters} \\ \hline \hline
$\Mx{N}(t)$ & all frequencies fixed & no &  $N+1$ & $2(N+1)$\\
$\Mxr{N}{N_f}(t)$ & $N_f$ free frequencies & no & $N+1$ & $2(N+N_f+1)$\\
$\Mxrm{N}{N_f}(t)$ & $N_f$ free frequencies & yes & $N+2$ & $2(N+N_f+2)$\\
$\Mr{N}\equiv\Mxr{N}{N+1}$ & all frequencies free & no & $N+1$ & $4N+4$\\
$\Mrm{N}\equiv\Mxrm{N}{N+1}$ & all frequencies free & yes & $N+2$ & $4N+6$\\ \hline \hline
\end{tabularx}
\end{center}
\label{tab:models}
\end{table*}

\section{Fitting models and notation}
\label{subsec:QNMmodels}

Let us begin by introducing some notation for the fitting models adopted in the rest of this paper. 
We will often compare two broad classes of models: 
one in which the real and imaginary parts of the complex frequencies $\omega_{\ell m n}$ are fixed to match theoretical predictions from BH perturbation theory~\cite{Berti:2005ys,RDwebsites},
and one in which most (or all) of these frequencies are left free to vary.
The first scenario corresponds to the ``weak'' version of the no-hair test used in recent investigations, while the second reflects the original ``stronger'', i.e. more agnostic, proposal.

In the first class of models, we fit the waveform by $N+1$ damped exponentials with complex frequencies fixed to the QNMs $\omega_{\ell m n}$ of a remnant with known mass and spin:
\begin{align}\label{eq:Q_n_model}
 \Mx{N}(t)\equiv &\sum\limits_{n=0}^{N}  A_{\ell m n} e^{-i [\omega_{\ell m n} (t - t_{\rm peak}) +  \phi_{\ell m n} ] }\,,
\end{align}
where the only unknowns are $A_{\ell m n}$ and $\phi_{\ell m n}$. In this case, we assume that the complex frequencies $\omega_{\ell m n}$ of all modes present in the signal are known {\em a priori}. 
In this sum $n$ is the overtone number. Since $Q_{0}(t)$ corresponds to a model with only one mode -- the fundamental mode ($n = 0$), or ``$0$-th overtone'' --  the $Q_{N}(t)$ model contains $N$ overtones, and $N + 1$ modes in total.
This model has been employed, for example, in Refs.~\cite{Buonanno:2006ui,Giesler:2019uxc}.

In a more general category of models, we will assume that the complex frequencies of {\em some} modes are known {\em a priori}, while the complex frequencies of others are unknown. In this case we fit the waveform by $N+1$ damped exponentials where the complex frequencies $\omega_{n_f}$ of $N_f$ ``free'' modes are unknown, while the frequencies of the remaining $N-N_f+1$ modes are fixed to the QNMs $\omega_{\ell m n}$ of a remnant with known mass and spin:
\begin{align}
\Mxr{N}{N_f}(t)&\equiv \sum\limits_{n=0}^{N-N_f}  A_{\ell m n} e^{-i [\omega_{\ell m n} (t - t_{\rm peak}) +  \phi_{\ell m n}]}\nn\\&+  \sum\limits_{n_f=0}^{N_f-1} A_{n_f} e^{-i [\omega_{n_f} (t - t_{\rm peak}) +  \phi_{n_f}]}
\end{align}
This second class of models has $2(N+N_f+1)$ fitting parameters:  $2(N-N_f+1)$ real amplitudes and phases $A_{\ell m n},\, \phi_{\ell m n}$;  $2N_f$ real amplitudes and phases $A_{n_f},\, \phi_{n_f}$; and the $2N_f$ complex free frequencies $\omega_{n_f}$. The notation $\Mxr{N}{N_f}(t)$ is a reminder that we are fitting the waveform by {\em an $N$-overtone model, where $N_f$ modes are ``free''} (in the sense that their frequencies are not fixed).

In order to take into account spherical-spheroidal mixing, we will sometimes need to add another mode to $\Mxr{N}{N_f}(t)$:
\begin{align}
 \Mxrm{N}{N_f}(t)&=\sum\limits_{n=0}^{N-N_f}  A_{\ell m n} e^{-i [\omega_{\ell m n} (t - t_{\rm peak}) +  \phi_{\ell m n}] }\nn\\&+  \sum\limits_{n_f=0}^{N_f-1} A_{n_f} e^{-i [\omega_{n_f} (t - t_{\rm peak}) +  \phi_{n_f}] }\nn\\
 & +  A_{\ell'm 0} e^{-i [\omega_{\ell'm 0} (t - t_{\rm peak}) +  \phi_{\ell'm 0}]}\,.
\end{align} 
The subscript ``$(m)$'' in $\Mxrm{N}{N_f}(t)$ is a reminder that we also include the ``mode-mixing'' contribution $A_{\ell'm 0} e^{i \omega_{\ell'm 0} (t - t_{\rm peak}) + i \phi_{\ell'm 0}}$. For simplicity, we will only consider the dominant mode-mixing contribution. This comes from $\ell'=\ell+1$ when $\ell=m$, and from $\ell'=\ell-1$ when $\ell>m$ and $\ell>2$. The model now has $2(N+N_f+2)$ unknowns: besides $A_{\ell m n},\, \phi_{\ell m n}$, $A_{n_f},\, \phi_{n_f}$, and the complex frequencies $\omega_{n_f}$, we also have the amplitude and phase $A_{\ell' m 0},\, \phi_{\ell'm 0}$ of the mode-mixing contribution on the last line.
The $\Mxrm{N}{N_f}(t)$ model contains $N+2$ QNMs.

Two special subclasses of models will be of special interest below. In one subclass, {\em all} modes have free complex frequencies, and there are no fixed modes. In another subclass, we will only fix the complex frequency of the spherical-spheroidal mixing mode, while the rest of the complex frequencies are kept free. These cases correspond to setting $N_f=N+1$ in the $\Mxr{N}{N_f+1}(t)$ and $\Mxrm{N}{N_f+1}(t)$ models, respectively, and we will denote them with the following shorthand notation:

\be
\Mr{N}\equiv\Mxr{N}{N+1},\quad\quad \Mrm{N}\equiv\Mxrm{N}{N+1}.
\ee

For reference, the fitting models we will consider below are summarized in Table~\ref{tab:models}.

As a final note: while the post-merger waveforms computed in numerical relativity that will be the main focus of this paper are complex, the linear waveform model $\mathbf{\Psi_{\rm num}}$ that we will consider as a warm-up problem below is real, because we specify real initial conditions for the time evolution. When the waveform is real, we will simply consider the real part of the fitting models, but otherwise we will use the more general complex models. 

\section{Extracting overtones in linearized theory\label{sec:linearovertones}}

Before turning to BBH mergers in full nonlinear GR, in this section we investigate linear perturbations in GR, as well as toy models built to elucidate the main features observed in the linear regime.
This will allow us to build some understanding of what to expect in the full GR nonlinear case while working in a controlled setting, and highlight the limitations of waveform fits by a superposition of damped exponentials.
In fact, we will show that even when linear perturbation theory is valid {\em by construction}, it is not sensible to model the ringdown by fitting the waveform with $\sim 7$ overtones starting at the peak.
More generally, although fitting the waveform at the peak with more overtones yields smaller fit residues, the model fails to pass further basic consistency checks.

We will also see that extracting high-overtone frequencies at the peak of the waveform
to test GR does not yield robust results.
Even when the waveform is {\em by construction} a combination of QNMs with a small contamination from other components (such as power-law tails), the high-overtone frequencies estimated by fitting are easily biased by these subdominant contributions.
Many of the lessons learned in these simple settings will carry over to the full GR case.

\subsection{Preliminaries}

As we recall in Appendix~\ref{app:Green}, the starting time of the ringdown regime is an ill-defined quantity even within linear perturbation theory, because the Green's function always contains additional contributions (most notably, a prompt response, a tail due to backscattering of radiation, and effects coming from the build-up of initial data). 
What we want to understand now -- insisting on modeling the waveform as a superposition of damped exponentials -- is the precision to which one can recover a hypothetical ``pure ringdown'' waveform, and the physical grounds for claiming the presence or absence of overtones.

For a linearly perturbed Schwarzschild BH geometry, after separation of the angular variables and working in the {time} domain, the linearized Einstein field equations imply that odd-parity (or axial) perturbations are governed by the Regge-Wheeler equation~\cite{Regge:1957td}
\begin{equation}
    \frac{\partial^2 \mathbf{\Psi_{\rm num}}}{\partial r_*^2} - \frac{\partial^2 \mathbf{\Psi_{\rm num}}}{\partial t^2} - V_{\rm RW} \mathbf{\Psi_{\rm num}} = 0\,,\label{equation_regge_wheeler}
\end{equation}
where the Regge-Wheeler potential is
\begin{equation}
 V_{\rm RW}=\left(1-\frac{2M}{r}\right)\left[\frac{\ell(\ell+1)}{r^2}-\frac{6M}{r^3}\right]\,,
\end{equation}
and the tortoise coordinate $r_*$ is defined by the relation $dr/dr_*=1-2M/r$. 
For the purpose of this discussion we focus on the dominant, quadrupolar component of the radiation ($\ell=2$) and we denote the wavefunction as $\mathbf{\Psi_{\rm num}}$ to emphasize that it is computed {\em numerically} by solving the Regge-Wheeler equation within linear BH perturbation theory. Even-parity (or polar) perturbations, governed by the Zerilli equation~\cite{Zerilli:1970se,Zerilli:1970wzz}, are known to be isospectral to odd-parity perturbations and behave in a qualitatively similar way{~\cite{Glampedakis:2017rar}}.

We compute the waveform $\mathbf{\Psi_{\rm num}}(t,r)$ by imposing the following initial conditions for Eq.~\eqref{equation_regge_wheeler}:
\begin{equation}
\mathbf{\Psi_{\rm num}}(0,r)=0\,,\qquad \frac{\partial \mathbf{\Psi_{\rm num}}}{\partial t}(0,r)=\exp\left[ - \frac{(r_* - r_0)^2}{2\sigma^2} \right]\,, \label{eq:RW_initial_conditions}
\end{equation}
where $r_0 = 4 M$ and $\sigma = M/2$. We then perform a time evolution and extract the time-domain waveform $\mathbf{\Psi_{\rm num}}(t)$ at future null infinity, following Refs.~\cite{Zenginoglu:2011zz,Cardoso:2021vjq}. {We evolve the initial data using a two-step Lax-Wendroff method with second-order finite differences~\cite{Krivan:1997hc, Pazos-Avalos:2004uyd}, which has been extensively used in the past in the study of late-time tails in Kerr~\cite{Zenginoglu:2012us, Burko:2013bra} and extreme mass-ratio inspirals~\cite{Sundararajan:2007jg, Cardoso:2022whc}.}

Our goal is to analyze $\mathbf{\Psi_{\rm num}}(t)$ and see how well it can be fitted by a QNM model $Q_{N}(t)$ (here and below we will adopt the convention that the waveform we fit is denoted in boldface, while the fitting model is not).

\begin{figure*}
  \includegraphics[width=\textwidth]{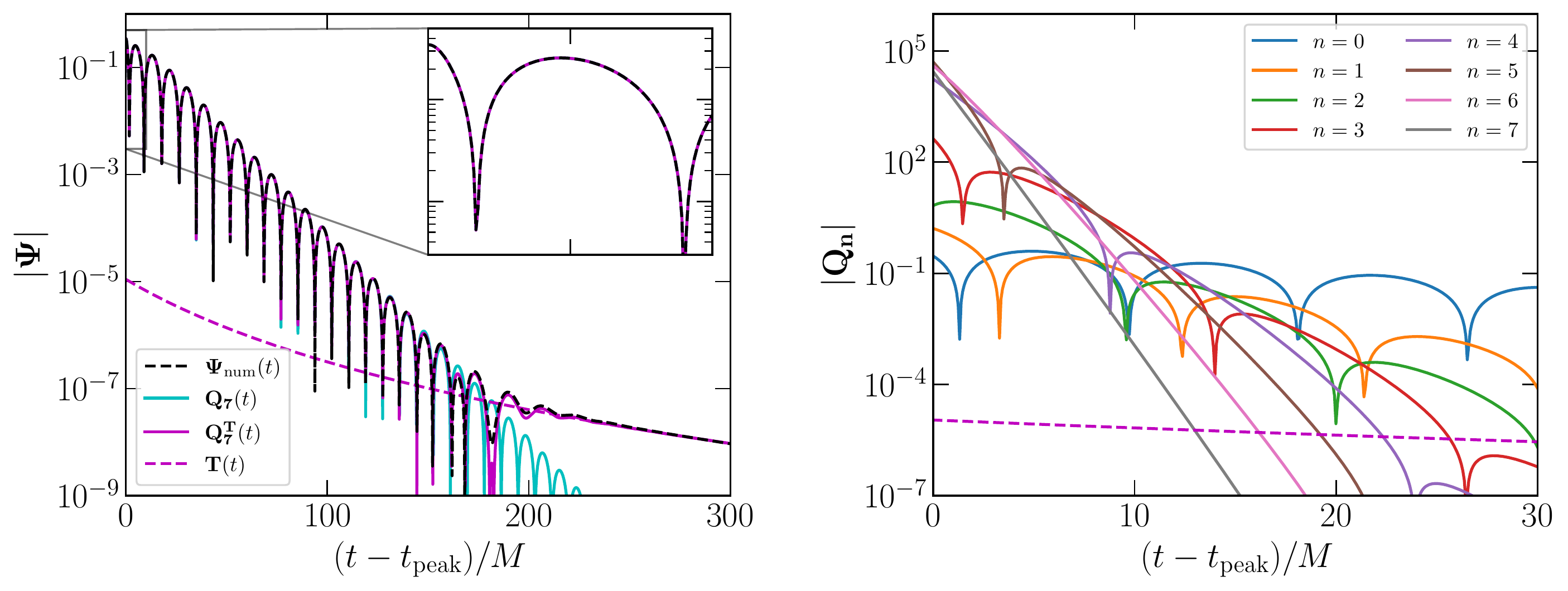}
  \caption{Left panel: the three waveforms considered in this section. 
  $\mathbf{\Psi_{\rm num}}(t)$ is the waveform extracted at future null infinity by solving the time-domain Regge-Wheeler equation \eqref{equation_regge_wheeler} with the initial conditions of Eq.~\eqref{eq:RW_initial_conditions}.
  $\mathbf{Q_7}(t)$ is a ``pure ringdown'' fit of $\mathbf{\Psi_{\rm num}}(t)$ starting at $t_0 = t_{\rm peak}$, with a fundamental mode and $7$ overtones fixed at the Schwarzschild frequencies computed in perturbation theory.
  $\mathbf{Q^T_7}(t)$ is the same as $\mathbf{Q_7}(t)$, but we also add a power-law tail to mimic the contamination due to backscattering of radiation. 
  The tail amplitude is obtained by fitting $\mathbf{\Psi_{\rm num}}(t)$ at times $t > 200$.
  The inset shows a zoomed-in view near the peak.
  Right panel:
  the individual modes (including the fundamental mode and $7$ overtones) that make up $\mathbf{Q_7}(t)$. 
  For reference, the extrapolated tail added to $\mathbf{Q^T_7}(t)$ (purple dashed line) is plotted in both panels.
  Note that in the right panel, the overtones with $n \geq 5$  drop below the extrapolated tail before completing even half of an oscillation cycle.
    }
  \label{fig:toyamplvsnoisetail}
\end{figure*}

In general, a full time evolution of the linear equation~\eqref{equation_regge_wheeler} will contain a ``prompt response'' component that depends on the initial data, as well as a late-time power law tail~\cite{Leaver:1986gd}. We will try to quantify the impact of these additional components below. For now, we just try to fit $\mathbf{\Psi_{\rm num}}(t,r)$ using a QNM model comprising a finite number of exponentially damped sinusoids.
Since for the moment we focus on $\ell = 2$ and on a nonrotating BH background, we can drop the index $\ell$ and set $m=0$ without loss of generality. For economy of notation, in this section we denote the amplitude $A_{\ell m n}$  of a generic mode by $A_n \equiv A_{20n}$, and similarly for the phases $\phi_n$ and the complex frequencies $\omega_{n}$. Therefore our fitting model is
\begin{equation}
    Q_{N}(t) \equiv \sum\limits_{n=0}^{N}  A_{n} e^{-i [\omega_{n} (t - t_{\rm peak}) +\phi_n]}, \quad t \in (t_0, t_{\rm end})\,. \label{eq:QNMmodel}
\end{equation}
The range of the fit is chosen to be $(t_0, t_{\rm end})$.
For all fits in this section we set $t_{\rm end} = 200 M$, but we have verified that the results would not change significantly if we used a larger value for $t_{\rm end}$.

Given a best-fit model $\psi$ to a waveform $\Psi$, we can quantify the goodness of fit by computing the mismatch
\begin{equation}
\mathcal{M} \equiv 1 - \dfrac{\langle \Psi | \psi \rangle}{\sqrt{\langle \Psi | \Psi \rangle \langle \psi | \psi \rangle}},
\end{equation}
where the scalar product is defined as
\begin{equation}
    \langle f | g \rangle = \int^{t_{\rm end}}_{t_0} f(t) g^*(t) dt,
    \label{eq:mismatch}
\end{equation}
and an asterisk denotes complex conjugation.
In this section both $\Psi=\mathbf{\Psi_{\rm num}}(t)$ and $\psi=Q_{N}(t)$ are real because we specified real initial data, but complex conjugation will be important later on, when we will consider complex waveforms from binary BH merger simulations in full GR.

In our fits we use a Levenberg-Marquardt nonlinear least-squares algorithm. We have cross-checked our results by comparing two different implementations, using either the \textsc{python} package \textsc{SciPy}~\cite{2020SciPy-NMeth} or the \textsc{NonlinearModelFit} function in \textsc{Mathematica}.

\subsection{Some considerations on overfitting}

By fitting Eq.~\eqref{eq:QNMmodel} to the solution of Eq.~\eqref{equation_regge_wheeler}, we found that the linear waveform $\mathbf{\Psi_{\rm num}}(t)$ can indeed be fitted ``well'' by a model with seven overtones $Q_7$ if the fitting range starts at the peak of $|\mathbf{\Psi_{\rm num}}|$, i.e. $t_0 = t_{\rm peak}$: the mismatch between the best fit waveform and the data is small, $\mathcal{M} \sim 10^{-8}$.
This result agrees with Ref.~\cite{Giesler:2019uxc}.

However, a problem immediately arises: by including more fitting parameters we can (in principle) decrease $\mathcal{M}$ indefinitely, even if the new parameters are not physically well-motivated and simply overfit the signal. 
Specifically, when adding more overtones to our fit model, we risk overfitting the early part of the ringdown, ``fitting away'' any contamination close to the peak of the waveform (due e.g. to the prompt response) with the rapidly decaying overtones.
Indeed, as will be clear from the following discussion, a small $\mathcal{M}$ is a {\em necessary but not sufficient} condition to conclude that the fitting model is consistent with the actual waveform or well-motivated.

To showcase overfitting issues, we consider a toy waveform $\mathbf{Q_7}$ consisting of a fundamental mode and 7 overtones, where the amplitude and phase of each mode is obtained by fitting $\mathbf{\Psi_{\rm num}}$ at $t_0 - t_{\mathrm{peak}} = 0 M$.
In other words, $\mathbf{Q_7}$ is a reconstruction of $\mathbf{\Psi_{\rm num}}$ with the $Q_7$ model  (recall that we use bold symbols for the waveforms to be fitted, and normal symbols for the fitting model: in this case  $\mathbf{Q_7}$ denotes the ``waveform,'' and $Q_7$ is the fitting model).

By comparing the black and {cyan} lines in the left panel of Fig.~\ref{fig:toyamplvsnoisetail} we see that the waveform is indeed well approximated by the $\mathbf{Q_7}$ model as early as the peak.
In the right panel we plot the fitted individual overtone modes from $n = 0$ to $n = 7$ of $\mathbf{Q_7}$.
The first few overtones ($n \lesssim 3$) consist, as expected, of sinusoidal oscillations modulated by an exponential decay. 
However, higher overtones ($3 \lesssim n \leq 7$) have a larger oscillation period and a shorter exponential decay time. If we focus on the early part of the waveform, the high-order overtones look like exponentials rather than damped oscillators because of their low quality factor.\footnote{The quality factor ${\mathcal Q}_{\ell m n}=-{{\rm Re}  (\omega_{\ell m n}) }/(2\ {{\rm Im}  (\omega_{\ell m n}) })$ is essentially the ratio between the decay time scale $\tau$ and the oscillation period.}

When we fit an exponentially decaying waveform with a ringdown model using a nonlinear least-squares algorithm, or when we compute the mismatch~\eqref{eq:mismatch}, the squared residue is dominated by the earlier part of the waveform. Then the higher overtones act effectively as ``bumps,'' removing early-time parts of the waveform that could not be well fitted by the lower overtones.

In linear perturbation theory, an important early-time contribution is the ``prompt response'' due to the initial wavepacket that propagates directly towards null infinity without scattering off the light ring. 
For the BBH merger waveforms considered in later sections, the early-time waveform includes the merger phase and any nonlinearities that may be present close to the peak.

The higher overtone phases can -- and often do -- (anti)align to produce destructive interference between modes: this (not neessarily physical) destructive interference reduces the mismatch, allowing ringdown models with many overtones to fit the early part of the waveform.
We see hints of this destructive interference in Fig.~\ref{fig:toyamplvsnoisetail}: note that the higher overtones can have amplitudes as high as $\sim 10^4$, much larger than the peak amplitude of the actual waveform (which is $\lesssim 1$). 

The bottom line is that fitting the ringdown with many overtones requires great care because of their exponentially decaying nature. A low mismatch with a multiple-overtone fitting model can easily mislead us into believing that the overtones are physically present when, in fact, they are just an unphysical artifact that produces good fits to other components of the signal. Therefore, a small mismatch is not sufficient to argue that the fitting model is a good representation of the waveform.
At the very least,
we should test whether the complex overtone frequencies assumed in our fitting model are actually those that best fit the waveform.
Moreover, the fits should not be very sensitive to the choice of $t_0$ and $t_{\rm end}$. The fitted amplitudes and phases should be consistent within about one period of oscillation of the corresponding QNMs, to exclude the possibility that overtones are just ``bumps'' fitting nonperiodic components of the waveform with their first half-a-period.

In fact, even in linear perturbation theory (where we neglect any possible nonlinearities associated to BBH mergers), the amplitudes of the QNMs are expected to vary close to the peak of the signal due to initial data effects (see Appendix~\ref{app:Green}).
The $Q_7$ model (a superposition of constant-amplitude damped sinusoids) is insufficient to consistently fit the whole ringdown even in this simple case, and we will observe amplitude modulations near the peak as we vary the starting time of the fit.

\subsection{Controlled fitting experiments}

Let us perform some controlled experiments to test whether model $Q_7$ (with fixed frequencies) is a good representation of the linear ringdown waveform found by solving Eq.~\eqref{equation_regge_wheeler}.
Here we build analytical toy-model waveforms to mimic the main features of the actual linear waveform, and to understand what our fitting procedure would return if the simulated waveform were {\em exactly} described by a combination of damped sinusoids.
We consider three such toy-models: (a) a ``pure'' ringdown waveform consisting only of damped sinusoids;
(b) a waveform consisting of damped sinusoids plus a Price power-law tail~\cite{Price:1971fb}, as expected in linear perturbation theory;
(c) the actual linear waveform found by solving Eq.~\eqref{equation_regge_wheeler}.

The details of the waveforms are as follows:

\begin{itemize}
\item[(a)] $\mathbf{Q_7}(t)$, {\em pure damped sinusoids}. This waveform is a particular realization of the (real-valued) fitting model $Q_7$, as specified in Eq.~\eqref{eq:QNMmodel}, whose mismatch with $\mathbf{\Psi_{\rm num}}(t)$ is $\sim 10^{-8}$:
\begin{equation}
    \mathbf{Q_{7}}(t) \equiv {\rm Re} \sum\limits_{n=0}^{7}  A_{n} e^{-i [\omega_{n} (t- t_{\rm peak})  + \phi_n ] }.
\end{equation}
The frequencies $\omega_n$ are the overtone frequencies of the $\ell = 2$ multipole of a Schwarzschild BH, i.e., $\omega_n \equiv \omega_{20n}$.
The amplitudes $A_{n}$ and phases $\phi_n$ are fixed by fitting $\mathbf{\Psi_{\rm num}}(t)$ with the $Q_7$ model at $t - t_{\rm peak} = 0 M$, as explained earlier and shown in the left panel of Fig.~\ref{fig:toyamplvsnoisetail}.
The toy waveform is denoted in bold fonts to distinguish it from the fitting model.

\item[(b)] $\mathbf{Q^T_7}(t)$, {\em damped sinusoids with contamination from a power-law tail}. The waveform in linear theory must contain contributions from a Price power-law tail due to backscattering of GWs, {which scales as $t^{-2\ell+3}$ at late times for perturbing fields of any spin, and for momentarily static initial data such as those used here~\cite{Price:1971fb}.
At late times, the power-law tail dominates the signal. To understand the dominant ($\ell=2$) mode, we}
model the tail of the numerical waveform $\mathbf{\Psi_{\rm num}}(t)$ by
\begin{equation}
     \mathbf{T}(t) \equiv A_{\rm tail} (t - t_{0,\rm tail})^{-5}.
\end{equation}
A fit of $\mathbf{\Psi_{\rm num}}(t)$ with this model at $t - t_{\rm peak} > 200 M$ yields $A_{\rm tail} \approx 9.26 \times 10^{4}$ and $t_{0,\rm tail} \approx -96.9$.
This backscattering is expected to affect also the earlier part of the ringdown, and we approximate its effects by extrapolating the fitted tail to early times $t \in (0 M, 200 M)$.
Then, we can construct a more realistic toy model by adding the extrapolated tail to $\mathbf{Q_7}(t)$:
\begin{equation}
    \mathbf{Q^T_7}(t) =\mathbf{Q_7}(t) + \mathbf{T}(t).
\end{equation}
This toy waveform is shown as the solid purple curve in the left panel of Fig.~\ref{fig:toyamplvsnoisetail}, while the extrapolated tail is shown by dashed purple lines in both the left and right panels.
Our approximation of the backscattering effects is admittedly crude.
Nonetheless, this toy waveform should be sufficient to understand whether backscattering can affect the fits.
In fact, we have found qualitatively similar results when we replace the power-law tail by a constant shift in amplitude, by a gaussian noise floor, or by the estimated numerical noise error intrinsic in $\mathbf{\Psi_{\rm num}}(t)$.
We decided to show results for a power-law tail (instead of numerical noise contributions) because the extrapolated tail has larger amplitude than the noise in the time range of interest.

\item[(c)] $\mathbf{\Psi_{\rm num}}(t)$, {\em linear perturbation theory waveform}.
This is the waveform found by solving Eq.~\eqref{equation_regge_wheeler} with the initial conditions Eq.~\eqref{eq:RW_initial_conditions}, extracted at null infinity.
\end{itemize}

\begin{figure*}
  \includegraphics[width=\textwidth]{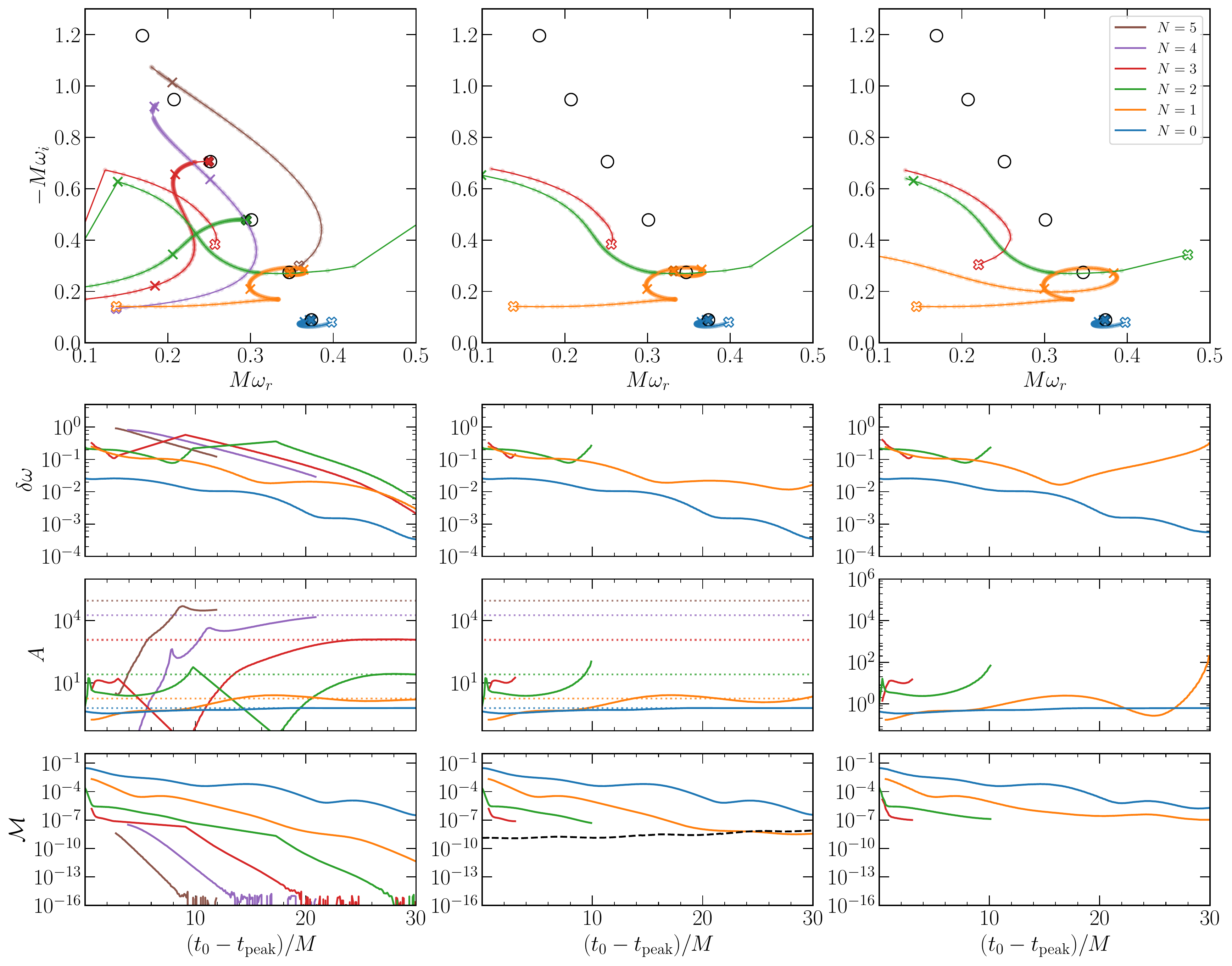}
  \caption{
    Frequency-agnostic fits of the $\mathbf{Q_7}(t)$ (left column), $\mathbf{Q^T_7}(t)$ (middle column) and $\mathbf{\Psi_{\rm num}}$ (right column) waveforms with the $Q_{0, 1}(t), Q_{1, 1}(t), \dots Q_{5, 1}(t)$ models, each consisting of $0, 1, \dots 5$ overtones, with the QNM frequency of the highest overtone in the model left as a free (complex) parameter for the fit.
  \textit{Top row}: recovered frequency of the free-frequency mode in each model. Each curve corresponds to a fitting model $Q_{N, 1}(t)$ with a different number of overtones.
  Each point on the curve corresponds to a fit starting at a different $t_0$, where $t_0$ is increased in $0.1 M$ steps. The empty thick cross corresponds to $t_0 - t_{\rm peak} = 0 M$, the two thin crosses correspond to $10 M$ and $20 M$, and the solid thick cross corresponds to $30 M$.
  The empty black circles are the expected overtone frequencies $\omega_0, \omega_1, \dots \omega_5$ (from bottom to top) of the $\ell m = 20$ harmonic for a Schwarzschild BH, used to construct the $\mathbf{Q_7}(t)$ and $\mathbf{Q^T_7}(t)$ toy waveforms.
  Spurious data points (e.g., points that move away from neighboring points by a disproportionate distance, or points that ``jump'' to a location outside of the plotting domain) are removed, hence some curves start later than $0 M$, terminate prematurely before $30 M$, or jump discontinuously at intermediate times, especially for the higher overtones.
  We also remove the same spurious points in the bottom three rows of this figure.
  We do not show the $N \geq 4$ curves for $\mathbf{Q^T_7}(t)$ and $\mathbf{\Psi_{\rm num}}$  (center and right column) because they do not approach any of the expected overtone frequencies meaningfully.
  \textit{Second row}: deviation of the free fitted frequency from the expected overtone frequency, as defined in Eq.~\eqref{eq:deltaomega}.
  \textit{Third row}: amplitude of the free-frequency mode.
  Although the starting time $t_0$ of the fit is varied, we define the amplitude of the mode to be the amplitude that we would measure by extrapolating back to $t_{\rm peak}$.
  In the left and central panels, we plot the injected amplitudes of each mode of the toy waveforms by horizontal dotted lines.
  \textit{Bottom row}: mismatch between the best-fit model and the actual waveform, as defined in Eq.~\eqref{eq:mismatch}.
  In the central panel, we also plot the mismatch between $\mathbf{Q_7}(t)$ and $\mathbf{Q^T_7}(t)$ (black dashed line), i.e., the mismatch induced by the tail.
}
  \label{fig:fitfreqtoys}
\end{figure*}

\subsection{Recovering QNM frequencies through an agnostic fit}

We now test whether we can recover the QNMs in the above toy waveforms in a {\em frequency-agnostic} manner: we consider a fitting model $Q_{N, N_f}$ in which some of the QNM frequencies are free parameters to be determined by the fit, instead of fixing all of the frequencies to the theoretically predicted overtone frequencies.
If the waveform is well described by a superposition of QNMs with certain overtone frequencies, we should be able (at least in principle) to recover these frequencies with our fitting procedure.

\subsubsection{Pure damped sinusoids}
Consider first the simplest toy model waveform $\mathbf{Q_7}(t)$, a superposition of damped sinusoids that mimic the full linear waveform $\mathbf{\Psi_{\rm num}}(t)$.
Recall that the frequencies of the QNMs used to construct $\mathbf{Q_7}(t)$ are those of the first $7$ overtones of a Schwarzschild BH.

We start by fitting the waveform with a single damped sinusoid, leaving the real and imaginary parts of the QNM frequency as two free parameters to be recovered by the fit. In our notation, then, the fitting model is $Q_{0, 1}(t)$, and the single free mode should converge to the fundamental mode when the fit starts at a sufficiently late time.
As shown by the blue curve in the top left panel of Fig.~\ref{fig:fitfreqtoys}, the fitted frequency converges to the theoretically predicted fundamental QNM frequency of Schwarzschild BHs. In the second row we plot the deviation
\begin{equation}
    \delta\omega=M|\omega_r+i\omega_i-\omega_{\rm ref}|, \label{eq:deltaomega}
\end{equation}
where $\omega_{\rm ref}$ denotes the reference value of the complex QNM frequency that we expect to find ($\omega_{200}$ in this case). The blue curve in the left panel of the second row shows that $\delta \omega \sim 10^{-4}$ when $t_0 - t_{\rm peak} \sim 30 M$.

We can now add one more QNM to our fitting model. The least demanding procedure to look for the first overtone is to fix the frequency of the fundamental QNM in the fit. Now our fitting model is $Q_{1, 1}(t)$: a model with one overtone (plus the fundamental mode, so two modes in total), in which one complex frequency (the frequency of the first overtone) is a free fitting parameter.
As shown by the orange curve in Fig.~\ref{fig:fitfreqtoys}, the free mode converges to the first overtone frequency as expected, this time with $\delta \omega \sim 10^{-3}$ when $t_0 - t_{\rm peak} \sim 30 M$.

We repeat this procedure by adding more modes to our model. Each time we fix the frequencies of the modes we have already recovered, leaving only one mode frequency free. When we use the fitting models $Q_{2, 1}(t), Q_{3, 1}(t) \dots$, we find that we can recover the frequencies to an accuracy $\delta \omega \lesssim 10^{-2}$ up to the third overtone, even when the actual waveform always contains $7$ overtones.
For overtone numbers $N \geq 4$, however, the fitted frequency becomes noisy and diverges before we reach $t_0 - t_{\rm peak} \sim 30 M$. The $|\delta\omega|$ curves are truncated at the point where the divergence occurs.\footnote{Here we are using a particular realization of the waveform $\mathbf{Q_7}(t)$. We could use other realizations by choosing a different set of amplitudes and phases. For example, if we fit the amplitudes and phases of $\mathbf{\Psi_{\rm num}}(t)$ at a time $t_0 - t_{\rm peak} \neq 0$, the agnostic frequency fit could recover the correct frequency up to $N = 5$ for some cases, and the ``amplitude constancy'' test explained later in this section could also work slightly better.}
For overtone numbers $N \geq 6$, the fits become very computational costly and they do not converge well, so we do not include the results in the plot.

Each additional mode in the fitting model increases the number of fitting parameters by two (the amplitude and the phase of each mode). This makes it difficult to locate the global minima in nonlinear least-squares fitting when the waveform is contaminated by unaccounted-for higher overtones (i.e., close to the peak) and when the overtones have decayed significantly (i.e., at late times).
At late times, the higher-overtone fits only fail once the mismatch between the fitting model and the actual waveform reaches machine precision.
Even when our toy waveform contains $7$ overtones (more than the number of overtones included in the fitting model), we can still recover the $N \leq 3$ overtones at sufficiently late times.

These results indicate that if the lowest ($N \leq 3$) overtones {are dominant} in full GR waveforms, our fitting procedure should return their correct frequencies.
For full GR BBH waveforms, we will find very weak evidence of the third overtone even under the most lenient requirement (i.e., when we impose the weak ``consistency test'' that the mode amplitude should be approximately constant for a brief period in time $\sim 2 M$). Therefore, including $N \gtrsim 4$ overtones in the fitting model is not useful anyway. We will confirm and reinforce these conclusions below.

\subsubsection{Damped sinusoids with tail}
Next, we test whether adding even the simplest subdominant contamination to the waveform would hinder our ability to recover the QNM frequencies.
In the middle panels of Fig.~\ref{fig:fitfreqtoys} we repeat the agnostic QNM fitting procedure using the same fitting model $Q_{N, N_f}(t)$, but on the toy waveform $\mathbf{Q^T_7}(t)$.
We find that the recovery of the fundamental mode and first overtone are almost as good as for the toy waveform $\mathbf{Q_7}(t)$, but the higher overtones ($n \geq 2$) are not recovered.

Perhaps the most instructive outcome of this experiment is shown in the bottom central panel of Fig.~\ref{fig:fitfreqtoys}. In that panel, the horizontal black dashed line corresponds to the mismatch between $\mathbf{Q_7}(t)$ and $\mathbf{Q^T_7}(t)$, i.e., the mismatch induced by the tail. It is clear that the individual modes stop converging towards their expected values around the time where their mismatch drops below the mismatch induced by the contamination. At that point, the mode is effectively trying to fit the contamination, and the mismatch saturates.
The results are qualitatively similar when we replace the tail with other types of injected subdominant contaminations (e.g., gaussian noise).
Clearly, if we increase the amplitude of the injected contamination, the recovery of the overtones becomes even worse.
In summary: this toy waveform illustrates that the presence of an expected subdominant contamination (beyond the ``pure ringdown'' signal) can prevent a robust extraction of the frequencies of the higher overtones.

What is worse, the free-mode frequency for fitting models with $N \geq 2$ does not converge to any particular value. This is an indication that the fit does not decisively ``pick up'' any QNM in the waveform.
However, as we increase $N$ the minimum mismatch ${\cal M}$ keeps decreasing and getting closer to the waveform peak, {\em even when the free mode does not approach the expected overtone frequency.}
This shows that adding modes to the fitting model can reduce the mismatch even if the mode frequency is unphysical. Therefore {\em a small mismatch is not sufficient evidence to claim the presence of an overtone}.
Numerical waveforms with high accuracy are necessary to confidently identify overtones in the data, and it is preferable to have a good model of all the sources of non-QNM contamination (including tails, noise, and nonlinearities).

On the other hand, the results for this toy waveform seem to imply that a failure to identify overtones with a frequency-agnostic search is not a proof that the overtones are absent, either. The overtones could be physically present (as they are in this toy model), but the fits could be failing simply because of small, subdominant contaminations.
However, even if the overtones were physically present, including many overtones in our ringdown waveform model might not be optimal  from a data-analysis point of view, because the model is not robust against even small contaminations. This is an important point if we want to test GR by extracting different frequencies in an agnostic manner: {\em any small contamination would hinder our ability to extract higher-overtone frequencies even if the modes are physically present in the signal.}

\subsubsection{Linear perturbation theory}

Finally, in the right panels of Fig.~\ref{fig:fitfreqtoys} we consider the more realistic case of fitting $\mathbf{\Psi_{\rm num}}(t)$, the time domain solution of Eq.~\eqref{equation_regge_wheeler} with initial conditions given by Eq.~\eqref{eq:RW_initial_conditions}, extracted at null infinity.
The left panel of Fig.~\ref{fig:toyamplvsnoisetail} shows that $\mathbf{\Psi_{\rm num}}(t)$ (black solid line) results from a combination of QNMs, contamination due to backscattering, and direct propagation of the initial wavepacket.
Moreover, for reasons explained with a toy model in Appendix~\ref{app:Green}, the initial data contamination means that the ``effective'' QNM amplitude should typically increase continuously until it approaches a constant.
In this more realistic model, extracting the overtone frequencies should be even harder than in the toy models examined previously.

This expectation is validated by our fits. While we can recover the fundamental mode ($\delta \omega \lesssim 10^{-3}$) with a free-frequency fit, {\em none of the overtone frequencies converge} to the theoretically expected values at late times.
In fact, only the first overtone (orange line) passes relatively close to the ``correct'' theoretical value ($\delta \omega \lesssim 10^{-1}$) for a short time interval $\sim 1 M$ around $t_0 - t_{\rm peak}\sim 16M$.
This test shows that, in the presence of unmodelled subdominant contributions, we can never recover the correct overtone frequencies by an agnostic fit to test GR.

{\em The original BH spectroscopy test is unfeasible for all overtones, and only possible at late times for the fundamental mode, even within linear perturbation theory}.

\begin{figure*}
  \includegraphics[width=\textwidth]{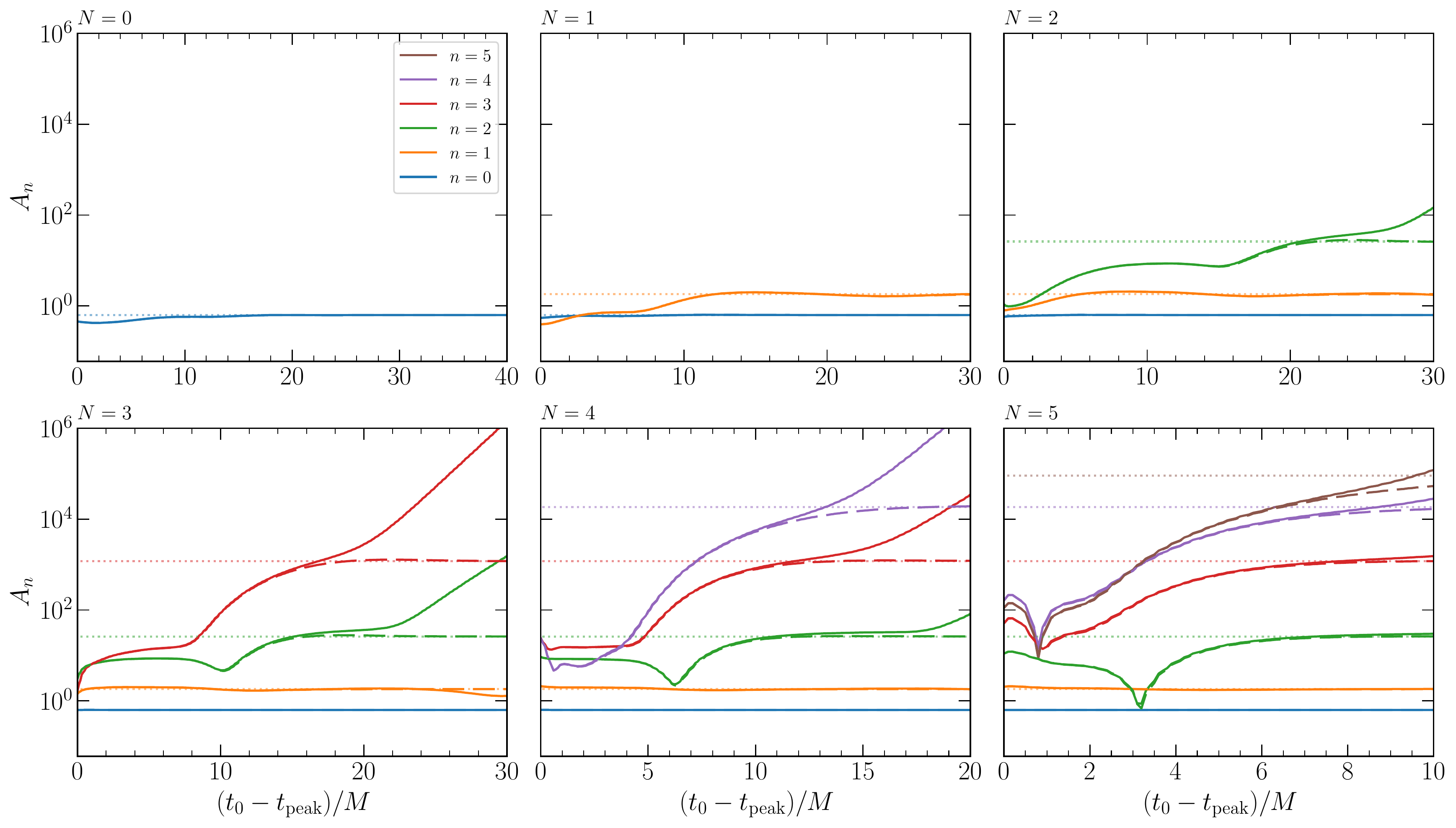}
  \caption{The amplitudes of QNMs fitted to the $\mathbf{Q_7}(t)$ (long dashed lines) and $\mathbf{Q^T_7}(t)$ (solid lines) toy waveforms, when the frequencies of all modes are fixed to the ``true'' frequencies of the corresponding overtones.
  Each panel corresponds to the fit results from a model $\Mx{N}(t)$ with a different number of overtones $N$.
  Different colors in each panel correspond to the fitted amplitudes of the different overtones $(n = 0, 1, \dots, N)$.
  While the starting time of the fit $t_0$ is varied, we always plot the value of the amplitudes extrapolated back to $t_{\rm peak}$.
  The injected values of the amplitude in the toy models $\mathbf{Q_7}(t)$ and $\mathbf{Q^T_7}(t)$ are shown as faint, horizontal dotted lines.
  Note the different time ranges in different panels.  
  }
  \label{fig:toysinusoidnoisetail}
\end{figure*}

\begin{figure*}
  \includegraphics[width=\textwidth]{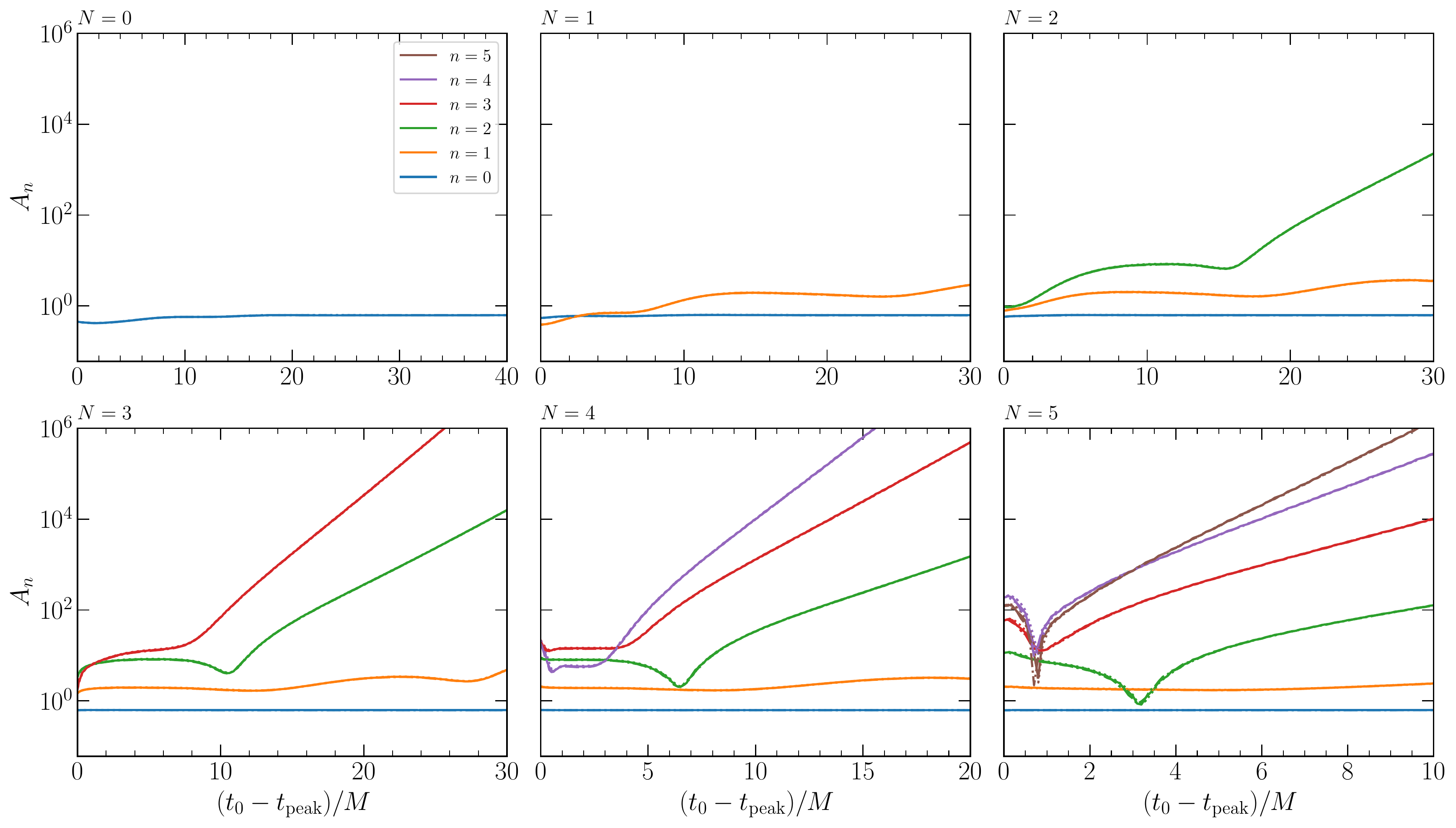}
  \caption{Same as Fig.~\ref{fig:toysinusoidnoisetail}, but for the linear waveform $\mathbf{\Psi_{\rm num}}(t)$.
  The dotted and dash-dot lines show the results for the simulations with grid spacing $dt = dr = 1.25\times 10^{-2} M$ and $6.25\times 10^{-3} M$, while solid lines show the results obtained from Richardson extrapolation.
  The three curves are similar, especially at late times, indicating that the variation of the amplitude and the exponential blow up at later times is not primarily driven by numerical error. 
  When compared to Fig.~\ref{fig:toysinusoidnoisetail}, the duration in which the amplitudes look flat is significantly shorter for the $N \leq 4$ fit models.
  For $N = 5$, only the $n = 0,1$ overtones show signs of constancy.
  Note the different time ranges in different panels. }
  \label{fig:fitamplRW}
\end{figure*}

\subsection{Recovering QNM amplitudes at fixed frequencies}

Since the original, agnostic spectroscopy test seems too ambitious even within linear perturbation theory, let us consider a more modest goal. In this ``weak'' version of the BH spectroscopy test, we assume the frequencies of the QNMs that should be present in the waveform to be known.
In other words, we fit our waveforms with the $Q_7(t)$ model (see Table~\ref{tab:models}): the complex frequencies in the model are fixed at the injected values for each QNM, and only the mode amplitudes and phases are free fitting parameters.

Our toy models will illustrate two main points:

\noindent
(i) If the waveform could really be described by a superposition of QNMs with the ``right'' frequencies, the fitted QNM amplitude should not change sensitively when we choose to fit the waveform at different starting times.

\noindent
(ii) The linear waveform $\mathbf{\Psi_{\rm num}}(t)$ gives inconclusive results when we consider $N \gtrsim 4$ (or arguably even fewer) overtones in the model. This is further evidence that we cannot reliably identify overtones with $N \gtrsim 4$, even within linear theory and in a ``weak'' formulation of BH spectroscopy tests. 

\subsubsection{Pure damped sinusoids}
In Fig.~\ref{fig:toysinusoidnoisetail} we plot QNM amplitudes fitted to various models when we change the starting time of the fit $t_0$.
As in Fig.~\ref{fig:fitfreqtoys}, for any $t_0$ we extrapolate the amplitude back to $t_{\rm peak}$ to ``unfold'' its exponential decay, so the plotted quantity is really the amplitude at $t_{\rm peak}$, as defined in Eq.~\eqref{eq:Q_n_model}.
In other words: if the plotted curve is a flat horizontal line, the amplitudes extracted at different values of $t_0$ are consistent with each other.

For the ``pure QNM'' toy waveform  $\mathbf{Q_7}(t)$ (long dashed lines in Fig.~\ref{fig:toysinusoidnoisetail}), we find that if we start the fit late enough, the recovered amplitudes ``flatten out'' to their injected values (shown as faint dotted lines), even when our fitting model contains fewer QNMs than the ``true'' pure ringdown waveform (i.e., when $N \leq 7$). 

The different panels show a clear trend: as we add more overtones, the lower overtones converge to a flat line earlier and earlier.
{\em Note that all of the overtone amplitudes (not only a subset) converge towards a constant at late times.}
This behavior is evidence that the model is a complete representation of the true waveform, and that the fit has good convergence properties.

\subsubsection{Damped sinusoids with tail}
Next, we apply the same procedure to the toy waveform $\mathbf{Q^T_7}(t)$.
The results are shown as solid lines in Fig.~\ref{fig:toysinusoidnoisetail}.
The fitted amplitudes are practically the same found in the pure damped sinusoid case for early starting times.

However, there is one remarkable difference: the amplitudes fitted to waveforms contaminated by power-law tails {\em blow up exponentially} at some critical time. 
When the injected contamination increases in amplitude, this ``critical blow up'' occurs at earlier times.
In fact, the exponential blow up occurs when the highest (fastest-decaying) overtone in the fitting model starts to pick up the contamination, and tries to fit the power-law tail with an exponential even at late starting times $t_0$, instead of following the expected exponential decay.
This results in an exponential blow-up in the figure, since we always plot the amplitude as defined at $t_{\rm peak}$.
We will observe an analogous behavior when fitting NR waveforms in the full, nonlinear theory.
Nonetheless, when the fitting model contains many overtones (e.g., $N = 5$), the lower overtone amplitudes (e.g., those with $n \leq 3$) are still roughly constant at late times.
From Fig.~\ref{fig:toysinusoidnoisetail}, we also conclude that a nonconstant amplitude at early times cannot be attributed to backscattering or some other subdominant contamination, because the early variation of the amplitude is similar with or without the contamination.

\subsubsection{Linear perturbation theory}
In Fig.~\ref{fig:fitamplRW} we consider $\mathbf{\Psi_{\rm num}}(t)$, the waveform computed within linear perturbation theory, and we fix the frequencies to the standard QNM frequencies of a Schwarzschild BH. While the amplitudes are flat within certain time ranges, their consistency is not as apparent as in the simpler toy models we considered earlier.

Consider e.g. the model with $N = 4$. For the toy waveform $\mathbf{Q^T_7}(t)$ in Fig.~\ref{fig:toysinusoidnoisetail}, the amplitude of the $n = 2$ overtone is consistent across a range of $\sim 10 M$ at $t_0 - t_{\rm peak} \gtrsim 7M$.
However, the corresponding amplitude when fitting $\mathbf{\Psi_{\rm num}}(t)$ is only constant at times $t_0 - t_{\rm peak} < 5M$, over a time range shorter than half the period of the fundamental mode ($P_{n = 0}/2 \sim 8.4 M$).
When we go to $N = 5$, none of the amplitudes is ever constant, except for the $n = 0$ and $n = 1$ modes.
This is different from the previous toy models, where all overtone amplitudes with $n \leq 3$ are stable at late times.
Also, for the previous toy models, the amplitudes stabilize earlier for fits with higher $N$, and across different $N$ the individual amplitudes all stabilize to the same value (faint dotted lines).
None of these behaviors is observed for $\mathbf{\Psi_{\rm num}}(t)$: compare e.g. the $N = 4$ and $N=5$ panels in Fig.~\ref{fig:fitamplRW}.
This implies that for $\mathbf{\Psi_{\rm num}}(t)$, the model breaks down at $N = 5$, and we should certainly not go further than that.
A cautious reader may even argue that we can only identify overtones with $n\leq 2$, because for $n=3$ and higher the amplitude is flat over times smaller than $P_{n = 0}/2$.\footnote{Moreover, a constant amplitude is necessary but not sufficient to infer that QNMs are present in a waveform. 
If in our fitting model we were to assume the presence of a QNM whose real frequency is very similar to a QNM that is actually present in the waveform, but whose decay time is significantly different, the fitted amplitudes could also turn out to be approximately flat for a brief period close to the peak.
This is why agnostic frequency fits represent the most robust method to identify an overtone.}

In fact, the $\mathbf{Q^T_7}(t)$ template considered above has been engineered to mimic $\mathbf{\Psi_{\rm num}}(t)$, so the flatness of the amplitude between these two waveforms should have been similar. Differences could arise either because the backscattering effect has been poorly modeled by extrapolating the power-law tail to early times in $\mathbf{Q^T_7}(t)$, or perhaps because there are additional contamination in the linear theory waveform.
For example, while the QNMs correspond to poles in the Green's function for the Regge-Wheeler potential, the ``prompt response'' (i.e., the direct propagation of the initial wave packet towards spatial infinity: see Appendix~\ref{app:Green}), is expected to contaminate the waveform.
Even without such contamination, the QNM amplitudes are varying close to the peak of the waveform: they build up continuously according to the shape of the initial data, so constant-amplitude QNMs should not be used to model the waveform too close to the peak (see Appendix~\ref{app:Green} and \cite{Lagos:2022otp}).
Pseudospectral instabilities may also cause alterations to the ringdown waveform, especially for higher overtones~\cite{Jaramillo:2020tuu,Jaramillo:2021tmt,Cheung:2021bol,Berti:2022xfj} (perturbations of the fundamental mode have been shown to affect the time-domain ringdown waveform only minimally~\cite{Berti:2022xfj}, but in principle the instability of the overtones could be observable~\cite{Jaramillo:2021tmt}).

We have estimated the noise in the numerical solution of the Regge-Wheeler equation $\mathbf{\Psi_{\rm num}}(t)$, and we have found it to be subdominant when compared to the extrapolated tail used for constructing $\mathbf{Q^T_7}(t)$.
Consistently with this finding, if we use a lower-resolution calculation of $\mathbf{\Psi_{\rm num}}$, the fitting results in Fig.~\ref{fig:fitamplRW} do not change significantly. 
In other words, the exponential blow-up is likely driven by physical effects (power-law tails) rather than by numerical noise.
This may not be true for comparable-mass BBH mergers.
As far as we know, a power-law tail has not yet been confidently
identified in BBH simulations in full GR.

\subsection{Take-home messages}
The lessons learned from fitting the three toy waveforms (Fig.~\ref{fig:fitfreqtoys}) can be summarized as follows:
\begin{itemize}
\item[(a)] If the waveform consisted of a pure superposition of damped QNMs, as in model $\mathbf{Q_7}(t)$, we should be able to recover the QNM frequencies by fitting the waveform up to at least $n = 3$, and the amplitude of each QNM would be consistent across different starting times of the fit. This applies to both the case where the frequencies are free, and to the case where they are fixed to their ``exact'' values.
\item[(b)] Contaminations that are subdominant close to the waveform peak, such as the power-law tails injected in $\mathbf{Q^T_7}(t)$, limit our ability to agnostically recover the overtone frequencies as required for testing GR, even if the overtones exist in the waveform and are dominant.
However, if we assume that overtones are present and we fix their frequencies in the fitting model, while the amplitudes will blow up exponentially at a later time, they should be approximately constant at intermediate times for overtones up to $n=3$. The constancy of the lower overtones should improve when we add more overtones to the model.
\item[(c)] An agnostic damped-sinusoid fit cannot recover the correct frequencies for any of the overtones when we consider ``true'' ringdown waveforms computed within linearized gravity $\mathbf{\Psi_{\rm num}}(t)$.
A fit of the linearized waveform with frequencies fixed to their known values does not show convincing evidence that overtones with $n \gtrsim 2$ are present in the signal.
In fact, the identification of the overtones is significantly more problematic than in the toy waveforms $\mathbf{Q_7}(t)$ and $\mathbf{Q^T_7}(t)$.
\end{itemize}

The conclusions of this exercise are quite clear.

First and foremost, using a small-mismatch criterion is not sufficient to conclude that overtones are present in the waveform.
Overtones are prone to overfit the early part of the waveform, because the rapidly decaying higher overtones, combined together, are just fine-tuned ``bumps'' that can fit away other sources of contamination. 

Even if the ringdown following a BBH merger were precisely described by linear theory, this would not imply that a superposition of multiple QNMs is sufficient for waveform modelling. Even the linearized waveforms are plagued by physical contamination from the prompt response and tails, and this makes it hard to conclude whether higher overtones are present, even if we assume the overtone frequencies to be known.

Moreover, the gaussian scattering example implies that it is difficult to use more than $n=2$ overtones to test GR.
In fact, Figure~\ref{fig:fitfreqtoys} shows that -- even in linear perturbation theory and for nonrotating BHs -- recovering the theoretically predicted QNM frequencies is difficult {\em even for the first overtone}, unless we fine-tune the starting time $t_0$ of the fit. A ``blind'' (percent-level) precision measurement of QNM frequencies is only feasible for the fundamental mode.

If we insist to use $0 < n \lesssim 2$ overtones to test GR, we should start fitting the waveform {\em at times significantly after the peak} (e.g. $\gtrsim 10 M$ after the peak for a model with one overtone). This is because the overtone amplitudes are roughly constant only at late times, whether or not the waveform is linear starting from the peak.\footnote{The time at which we should start the fit depends on the initial conditions used when solving for the linear waveform, and on the error tolerance we are willing to accept when we fit the frequency and amplitude. Note also that the Regge-Wheeler waveform is related to $\ddot{h}$, the second derivative of the GW strain, so the time delay needed for fitting $h$ in BBH merger waveforms might be significantly different. We will return to this topic below.}

While it is true that the fitting model we use for overtone extraction is incomplete (because a tail is clearly present in the numerical linear waveform),
the key point is that similar unmodeled linear {\em and nonlinear} contributions will certainly be present in full GR. Hence we can expect overtone recovery to be affected by similar issues in more realistic cases, in the absence of extremely accurate analytical models. (Incidentally, we have also tried to fit $\mathbf{\Psi_{\rm num}}$ with a power-law tail in addition to QNMs, but we could not confidently identify the portion of the waveform where {the tail starts being dominant}.)
Additional contributions -- including nonlinear effects~\cite{MaganaZertuche:2021syq,Sberna:2021eui} and nonlinear QNMs~\cite{Ma:2022wpv,Mitman:2022qdl,Cheung:2022rbm,Kehagias:2023ctr,Kehagias:2023mcl} -- have indeed been found in BBH merger waveforms in full GR. As a linear superposition of QNMs cannot fit a linear waveform in a self-consistent manner, we would expect the model to perform even worse when fitting the post-merger waveform of two comparable-mass BHs.
The next sections will confirm these expectations. In our analysis of the ringdown of BBH mergers simulated in full GR, the $n \geq 2$ overtones cannot be confidently identified, and the first overtone can only be identified at times $\approx 10M$ after the waveform peak.

In Appendix~\ref{sec:NRtoy} we repeat some of the present analysis on a complex-valued toy waveform constructed to mimic the NR post-merger waveform. The conclusions are qualitatively similar, if not stronger. When we consider a complex toy model consisting of 7 overtones, the frequency-agnostic fits work better than those presented here, and the fitted amplitudes are even more stable. 
In other words, a failure of these test for a complex NR waveform is even stronger indication that the waveform cannot be modeled by a superposition of overtones.

\section{Are post-peak BBH waveforms linear?}
\label{sec:postpeaklinearity}

Let us now turn to the real problem of interest: fitting waveforms in full GR. Fitting overtones in ringdown signals is a notoriously difficult problem even in the absence of instrumental noise (see, e.g.,~\cite{Dorband:2006gg,Buonanno:2006ui,Berti:2007dg,London:2014cma,Cook:2020otn,MaganaZertuche:2021syq}).
Besides the physical effects discussed so far, time variations in the inferred mode amplitudes in full GR can occur because the mass and spin of the remnant extracted from numerical simulations vary significantly close to the peak of the radiation~\cite{Buonanno:2006ui,Berti:2007fi,Baibhav:2017jhs,Sberna:2021eui, Cotesta:2022pci}.
Let us illustrate this point more concretely.

To facilitate comparison with previous work, we will follow \GIST \,and focus on the GW150914-like NR waveform SXS:BBH:0305 in the Simulating eXtreme Spacetimes (SXS) catalog~\cite{Mroue:2013xna,Boyle:2019kee}.
The waveform represents a BH binary with mass ratio of $1.22$, primary dimensionless spin $\chi_1=0.33$ aligned with the orbital angular momentum, and secondary dimensionless spin $\chi_2=-0.44$ antialigned with the orbital angular momentum. The merger remnant in this simulation has final mass $M_f=0.9520M$ and dimensionless spin $\chi_f=0.6921$.

Reference~\cite{Giesler:2019uxc} suggested that the addition of several overtones is necessary to appropriately model ringdown and to infer the final mass and spin. 
When starting to fit at $t\geq t_{\rm peak}$ , where $t_{\rm peak}$ is defined as the time where the $(\ell,\,m)=(2,\,2)$ component of the strain has a maximum,
overtones up to $n=7$ were included to obtain an unbiased estimate.
Earlier work had indeed found that using a single mode can lead to large systematic errors on the inferred mass and spin of the remnant~\cite{Buonanno:2006ui,Berti:2007zu,Baibhav:2017jhs}.

As pointed out in early systematic studies of ringdown from nonspinning BH merger simulations~\cite{Buonanno:2006ui,Berti:2007fi}, the fact that a linear superposition of damped exponentials can reproduce the merger waveform for $t\geq t_{\rm peak}$ does not necessarily imply that the time evolution of the background and nonlinearities can be ignored.
A significant fraction of the mass and angular momentum is being radiated away from the system post-merger, while the Kerr QNM frequencies are computed assuming a fixed background.

\begin{figure}[t]
   \includegraphics[width=0.48\textwidth]{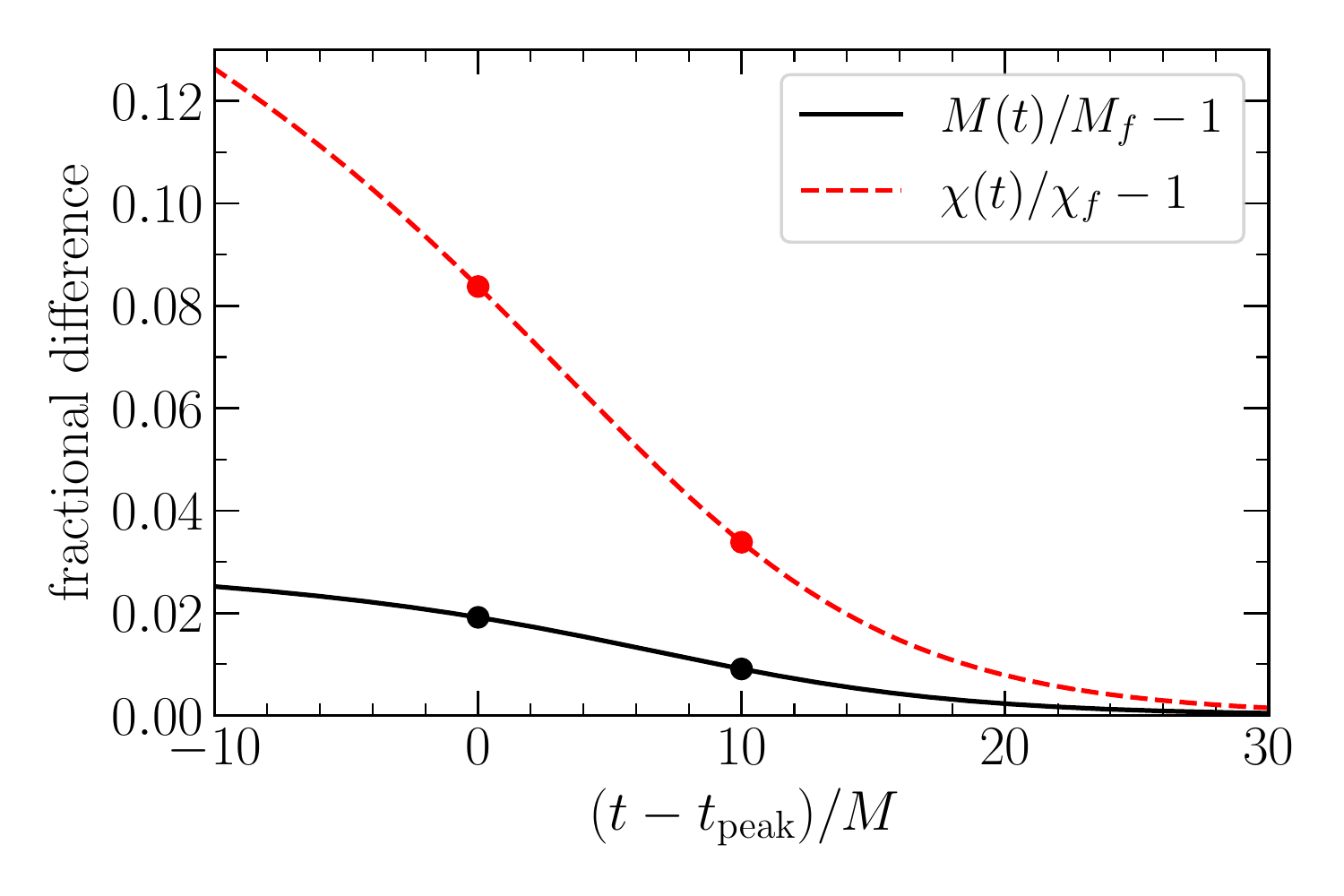}
    \caption{Evolution of the remnant mass and dimensionless spin for the GW150914-compatible simulation SXS:BBH:0305. The background is rapidly evolving at least until $t-t_{\rm peak}\simeq 20M$, where $t_{\rm peak}$ is defined as the time at which the strain $h$ has a maximum.}
    \label{fig:mchievol}
\end{figure}

In Fig.~\ref{fig:mchievol} we show the difference between the BH mass and angular momentum and their asymptotic values, computed for the SXS:BBH:0305 simulation following the procedure outlined in Ref.~\cite{Ruiz:2007yx}. We find that the remnant mass and dimensionless spin differ from their asymptotic value by $2\%$~($1\%$) and $8\%$~($4\%$) at $t = t_\mathrm{peak}$ ($t = t_\mathrm{peak} + 10M$), respectively.
Such large variability in the background spacetime can significantly complicate the analysis, and there is no reason a priori why the simple model of a linearly perturbed BH with a fixed mass and spin should work around the peak. In fact, several authors pointed out that modeling waveforms close to the peak of the radiation by a linearly perturbed BH with a fixed mass and spin leads to conceptual issues~\cite{Bhagwat:2017tkm,Bhagwat:2019dtm,JimenezForteza:2020cve,Bamber:2021knr,Sberna:2021eui}.

\begin{figure*}[th]
  \includegraphics[width=\textwidth]{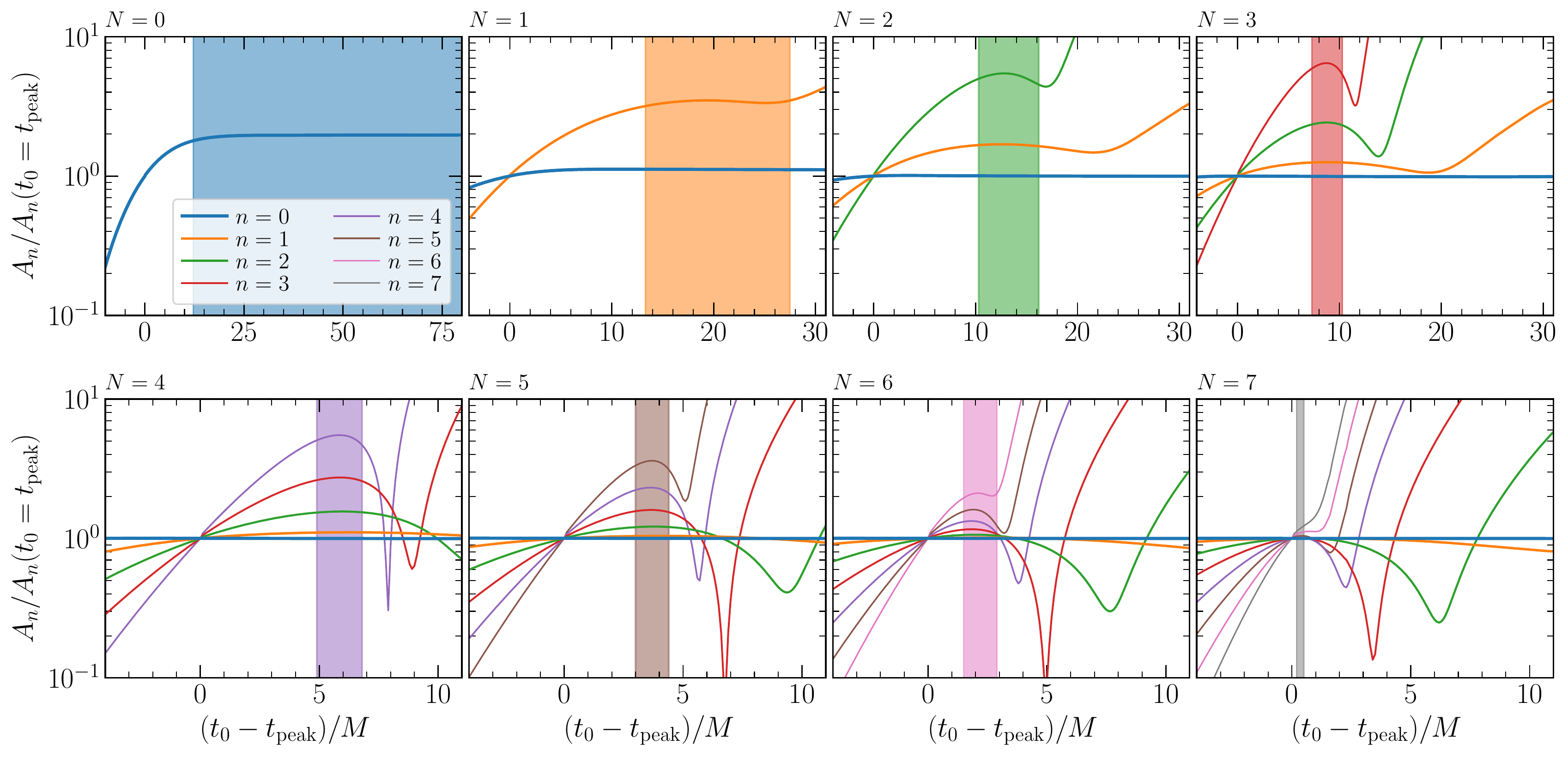}
  \caption{Amplitude $A_n^N(t_0)/A_n^N(t_0=t_{\rm peak})$ of QNMs as a function of the starting time $t_0$ for the SXS:BBH:0305 simulation. The shaded regions show the largest time range such that the amplitude of the highest overtone ($n=N$) is constant within $10\%$.
  }
  \label{A_nN}
\end{figure*}

As noted in Ref.~\cite{Buonanno:2006ui}, the large amount of radiation in a BBH merger ``raises the question as to whether or not the radiated energy and angular momentum are affecting the QNM fits. This issue will, of course, become more significant as the fits are pushed to earlier times.''

In this section we ask two questions: (1) is it really legitimate to describe the whole post-peak waveform as a linear perturbation of the final, stationary Kerr BH? (2) how many overtones can be reliably used to obtain unbiased estimates of the remnant's mass and spin?

If we could indeed ignore the time-evolving background and describe the whole post-peak waveform as a superposition of QNMs from a fixed Kerr background then the overtone amplitudes should be constant, by definition. 
We have seen that this expectation is questionable even in linear theory. In Sec.~\ref{sec:constA} we confirm, perhaps at this point unsurprisingly, that a constant-amplitude overtone superposition does not work in the BH merger case either: the amplitudes of the overtones change significantly when we change the fitting window.

The nonconstancy of the amplitudes is not the only issue with a linear perturbation theory interpretation of post-peak ringdown. Reference~\cite{Giesler:2019uxc} claims that
(i) the inclusion of the fundamental mode and $7$ overtones provides a very accurate description of the ringdown up to the peak strain amplitude, and (ii) including seven overtones significantly reduces the uncertainty in the extracted remnant properties.

In Sec.~\ref{sec:mismatch}  we show that the higher overtones lead to very small mismatches by merely overfitting the waveforms. Furthermore, we argue that these higher overtones try to fit other physics (such as prompt response effects, time variations of the background, and nonlinearities) close to the merger. The addition of several overtones allows for better extraction of the fundamental mode and the first overtone, which are mainly responsible for better estimating remnant properties.
This improved extraction arises because high overtones contributions act as effective terms, ``fitting away'' poorly understood physics.

The fact that higher overtones lead to overfitting is confirmed in Sec.~\ref{sec:epsilon}, where we show that these modes do not play a significant role in extracting the remnant's properties, either. To show this, we swap a particular mode with a random damped exponential. If the ``fake'' mode still fits the waveform with similar or better accuracy, or if it still extracts the remnant properties accurately, we can conclude that the originally ``swapped'' overtone was not really necessary. We use this argument to conclude that overtones with $n\geq 2$ do not significantly contribute to the extraction of the remnant's parameters, and therefore there is no motivation to include them in the modeling.

\subsection{Are overtone amplitudes consistent with the linear model?}
\label{sec:constA}

We start by discussing the inconsistency of the amplitudes obtained by fitting a simple overtone model with fixed frequencies, the \Mx{7} model of Eq.~\eqref{eq:Q_n_model}, to numerical BH merger simulations. 
This inconsistency confirms, from a different perspective, that the asymptotic values of the frequencies should only be used at late times.

In linear BH perturbation theory, the ringdown part of the strain $h$ is given (by definition) by the \Mx{N} model of Eq.~\eqref{eq:QNMmodel}. In this section we focus on the dominant, $\ell=m=2$ component of the radiation. To avoid cluttering, we simplify the notation by setting $A_{n}=A_{22n}$, $\phi_{n}=\phi_{22n}$, and $\omega_{n}=\omega_{22n}$.

Each QNM depends on $4$ real parameters: a real amplitude $A_n$, a real phase $\phi_n$, and a complex QNM frequency $\omega_n= \omega_r^n+i\omega_i^n$. 
The complex frequency is determined by the remnant Kerr BH's mass and spin, while the nature of the perturbations (i.e., the parameters of the binary progenitor) dictates the amplitudes and phases. 
In linear perturbation theory, all of these quantities are \textit{time-independent constants}~\cite{Leaver:1986gd,Andersson:1995zk,Andersson:1996cm,Berti:2006wq,Zhang:2013ksa,Oshita:2021iyn}.

In the \Mx{7} model, the entire post-peak waveform is fitted by this simple model at the cost of including a large number of overtones ($N=7$).
Note that in this model, the complex frequencies are {\em forced} to their asymptotic values: only the amplitudes and phases are free parameters, extracted by fitting the waveform between $t_0$ and $t_f=t_{\rm peak} + 90 M$. 

We want to explore the behavior of amplitudes and phases in the \Mx{7} model we vary $t_0$.
Time variations of the amplitudes or phases with $t_0$ imply departures from the bare-bones ringdown model in linear BH perturbation theory, which can happen for various reasons -- e.g. because the mode frequency has not relaxed to its true value, because the QNM included in the fit is not present in the waveform, or because damped exponentials are actually overfitting some other feature (such as power-law tails, the prompt response, nonlinearities, or numerical noise). 
We will mostly focus on the constancy of the amplitudes, but similar arguments can be made for the phases.

In Fig.~\ref{A_nN}, we plot the amplitudes of different modes as a function of the starting time of the fit. 
Each panel uses a \Mx{N} fitting model with a different number of overtones, from $N=0$ (top left) to $N=7$ (bottom right). As usual, we extrapolate the amplitudes to their values at $t_0=t_{\rm peak}$. 
Let us define (somewhat arbitrarily) the amplitude of a mode to be ``constant'' if it varies by less than $10\%$. 
Vertical colored bands in each panel show the time range in which the amplitude of the highest overtone ($n=N$) is constant according to this criterion.

Consider first the simplest case in which we fit the waveform using only the fundamental mode ($n=N=0$, top left). 
The amplitude of the mode is not constant at early times, and it only saturates to a constant value when the starting time of the fit $t_0 \gtrsim t_{\rm peak} + 12 M$. 
Adding the first overtone ($N=1$) has the effect of further stabilizing the amplitude $A_0$ of the fundamental mode, which now is constant as soon as $t_0\gtrsim t_{\rm peak} + 1 M$. 
This doesn't mean that the two-mode model is a good description close to the peak, because {\em the amplitude of the first overtone} is now rapidly changing in time close to the peak: if we require both mode amplitudes to be constant, we are again forced to consider the late-time portion of the waveform, where $t_0\gtrsim t_{\rm peak} + 13 M$ for $N=1$. 
This confirms previous findings in Ref.~\cite{Bhagwat:2019dtm}~(see the top panel in their Fig.~3). 
Note also that the two-QNM model seems to fail at very late times ($t_0\gtrsim t_{\rm peak} + 28 M$), where $A_1$ blows up very rapidly. 
We interpret this behavior in light of the findings of Sec.~\ref{sec:linearovertones}. In this particular case, the blow-up is due to the free overtone amplitude latching onto a different QNM: as we will show in Sec.~\ref{subsec:agnostic}, the late-time fit is dominated by the $(3,2,0)$ QNM due to spherical-spheroidal mode mixing, and not by the $(2,2,1)$ QNM. 
The exponential blow-up seen here is quite generic. Since we show the mode amplitude rescaled to its peak value (thus incorporating an exponential factor) any term with a time-dependence slower than the expected exponential (e.g., a different QNM, a power-law tail, or numerical artifacts) would give rise to this behavior.

We should also try to systematically quantify fitting errors on the overtone amplitudes.
Without error estimates, one may argue that amplitude variations could be ascribed to the accuracy of the numerical simulations, rather than being due to physical effects (such as the prompt response, tails, or a time-evolving background).
In Appendices~\ref{sec:NRerror} and~\ref{sec:systematics} we exclude this possibility. We show that the finite resolution of the simulation, the extrapolation procedure used to extract the radiation, and the presence of a spurious constant in the signal due to a suboptimal frame choice are subdominant effects that do not significantly affect the inferred amplitudes.
In addition, we will see below that the qualitative behavior of the mode amplitudes is unchanged when we include all known ringdown contributions (such as those coming from spherical-spheroidal mode mixing).
Therefore we conclude that the blow-up is triggered by unmodelled post-merger components, including possible residual nonlinearities from the merger phase.

\begin{figure*}[t]
    \includegraphics[width=\textwidth]{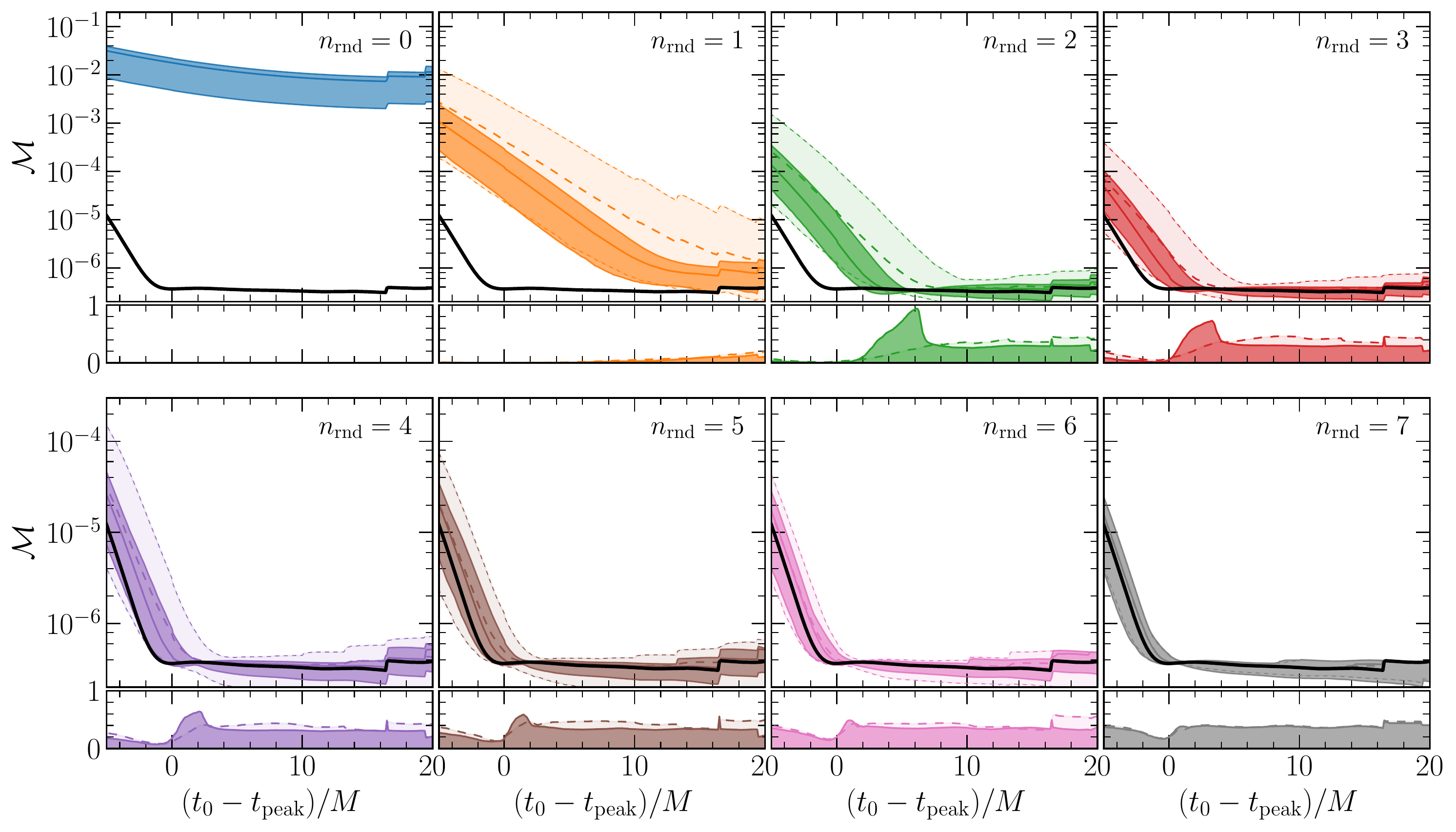}
    \caption{Median and $90\%$ confidence interval of the mismatch ${\mathcal M}$ between ringdown models and the SXS:BBH:0305 waveform as a  function of the starting time of the fit $t_0$. 
    Solid lines and darker shaded areas are found by randomly varying the frequency of a single overtone with $n= n_{\rm rnd}$ from a uniform distribution, as described in the main text.
    Dashed lines and lighter shaded areas correspond to random variations of all overtones with $n\ge n_{\rm rnd}$. 
    The solid black line shows the mismatch for the \Mx{7} model. 
    In the insets below each panel we show the fraction of random samples that outperform the $N=7$ model.}
    \label{fig:mismatch}
\end{figure*}

As we keep adding more and more overtones, we observe some important trends:

\begin{itemize}
\item[(i)] The fundamental mode amplitude $A_0$ stabilizes to constant values at earlier and earlier times: for example, in the \Mx{7} model ($N=7$)  $A_0$ is already constant at $t_0\approx t_{\rm peak} -15 M$. 
This was also noted in \GIST. 
The fundamental mode having constant amplitude before the peak does not necessarily mean that it is excited at times before the peak. 
In fact, there is a more economic explanation: since the fundamental mode is the longest-lived, the amplitude inferred by fitting is largely fixed by the late-time behavior, where the mode is actually present and the amplitude constant.

\item[(ii)] For $N \geq 2$, the highest-overtones amplitudes are ``constant'' only close to a local maximum, and never saturate to a constant value at late times.
This suggests that these modes are not actually contributing to the signal close to the waveform peak, but they are rather overfitting other features of the waveform beyond the simple linear ringdown model (possibly, nonlinear contributions).

\item[(iii)] For $N=1$, $A_1$ is constant (according to our definition) at times $13\lesssim (t_0-t_{\rm peak})/M \lesssim 28$. For $N=2$, $A_1$ is constant in the region $7\lesssim (t_0-t_{\rm peak})/M \lesssim 25$. 
As $N$ increases, two things happen. First of all, the constant-$A_1$ time band moves to earlier times.
Secondly, the region where the highest overtone has approximately constant amplitude keeps shrinking, so that higher-overtone amplitudes are roughly constant in smaller and smaller regions. 
This makes sense, because higher overtones are very short-lived. Consider for example the \Mx{7} model ($N=7$): the width of the time range over which $A_2$, $A_3$, $A_4$, $A_5$, $A_6$, $A_7$ are constant is $5.1$, $2.8$, $1.8$, $1.6$, $0.9$, $0.3$, respectively.
Once the amplitude and phase of the fundamental mode are fixed by the late-time behavior of the waveform, the amplitudes and phases of the higher overtones can be adjusted to fit the rest of the waveform close to the merger. 
\end{itemize}

In our opinion the mode-amplitude evolution shown in Figure~\ref{A_nN} is among the main results of this paper. It has two important implications:
\begin{itemize}
\item[(i)] {\em Overtone models with $N \geq 2$ are unphysical}, because they try to overfit other features of the waveform.
\item[(ii)] Models with at most one overtone ($N\leq 1$) are physical, but {\em they can only be used for meaningful spectroscopy tests at late times.}
\end{itemize}

Below, we show that the constancy of the amplitude is not the only issue: higher-overtone models do not necessarily yield the smallest mismatch with numerical relativity waveforms (Sec.~\ref{sec:mismatch}), nor the best estimate of the remnant's mass and spin (Sec.~\ref{sec:epsilon}).

Finally, we remark that most investigations so far have been devoted to the ``detection problem'': if overtones are present, can we extract them from the waveform?
In Appendix~\ref{sec:frankenstein} we look at the problem of ``false positives,'' i.e.: can a linear-overtone model find spurious evidence for the overtones even when they are {\em not} present?
Somewhat concerningly, we find that the answer is affirmative.

The consistency checks performed here and below are necessary, because ill-posed models can lead us to extracting wrong physics from the numerical simulations.

\subsection{Which overtones are necessary to correctly fit the waveform?}
\label{sec:mismatch}

In the absence of an analytical description of the post-merger phase it is important to avoid overfitting issues when we try to assess the physical contribution of different overtones. For example, we can check that the contribution of overtones cannot be replaced, or even be improved upon, by possibly unphysical low-frequency contributions which are effective at fitting the data.
Indeed, here we show that the mismatch achieved by a superposition of overtones as predicted in GR can be matched, or even surpassed, by adding ``unphysical'' low-frequency damped exponentials with frequency and damping time that do not appear in the spectrum predicted in GR for the final BH~\cite{Bhagwat:2019dtm,Mourier:2020mwa}.
It should be noted that similar ``pseudo-QNMs'' were introduced in the context of the effective-one-body framework precisely to model the rapid transition from the inspiral GW frequency to the post-merger QNM frequency ``plateau''~\cite{Pan:2011gk,Damour:2014yha,Brito:2018rfr}.

\begin{figure*}[t]
    \includegraphics[width=\textwidth]{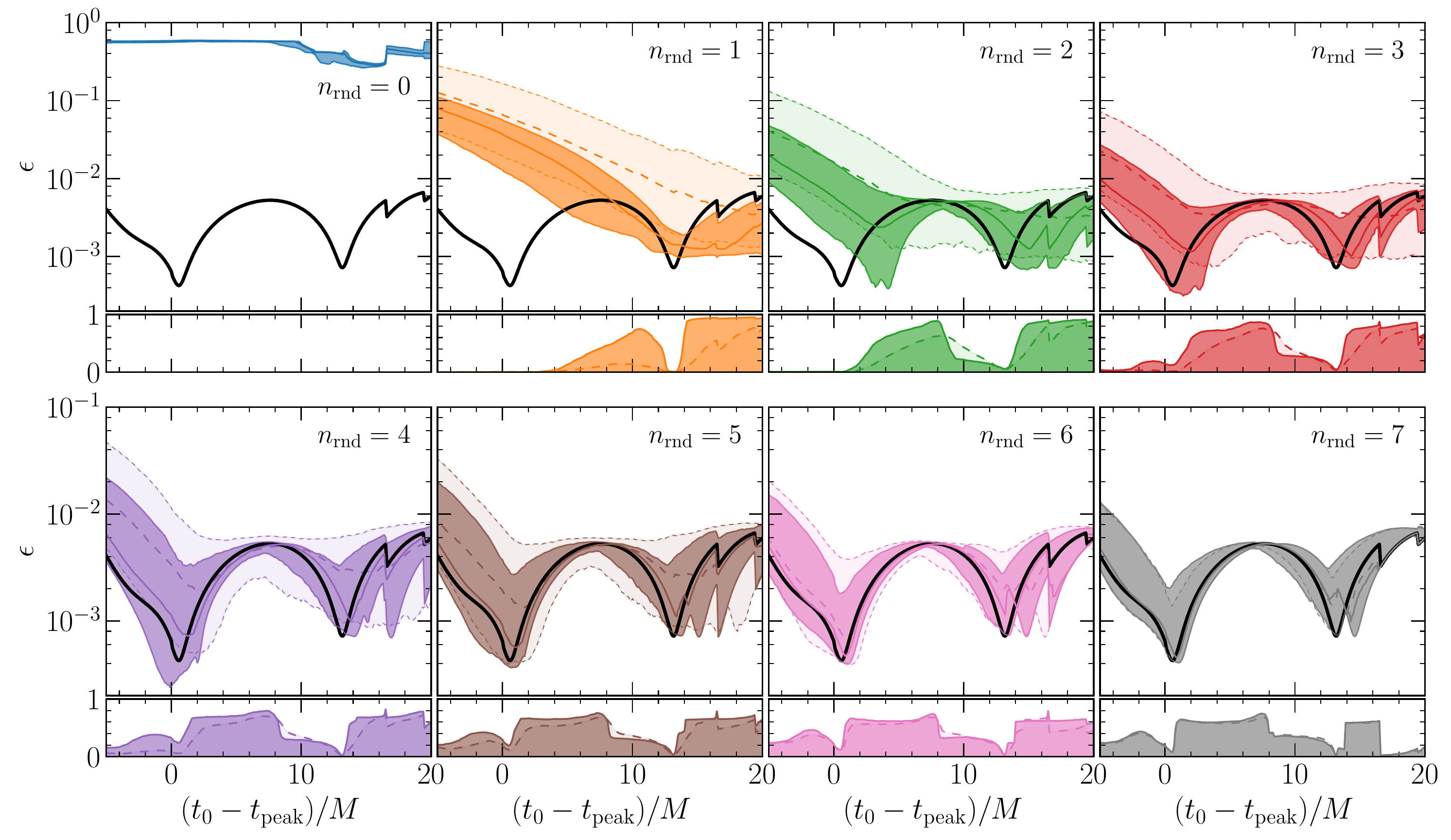}
    \caption{Median and $90\%$ confidence interval of the combined error on the remnant mass and spin $\epsilon$ -- see Eq.~\eqref{eq:epsilondef} -- as a  function of the starting time of the fit $t_0$.
      Solid lines and darker shaded areas are found by randomly varying the frequency of a single overtone with $n= n_{\rm rnd}$ from a uniform distribution, and ignoring the corresponding overtone in the estimation of $\chi_f$ and $M_f$.
      Dashed lines and lighter shaded areas correspond to doing the same for all overtones with $n\ge n_{\rm rnd}$. The solid black line shows the value of $\epsilon$ found when all $N=7$ overtones are used for the estimation of the remnant's properties, as in the \Mx{7} model. In the insets below each panel we show the fraction of random samples that outperform the \Mx{7} model.}
    \label{fig:epsilon}
\end{figure*}

To understand the contribution of each overtone, we fit the waveform using a modified version of the \Mx{7} model:
\begin{align}\label{eq:hrnd}
 Q_7^{n_{\rm rnd}}&= A_{n_{\rm rnd}} e^{-i [\omega_{n_{\rm rnd}} (t - t_{\rm peak}) +  \phi_{n_{\rm rnd}}] }\nonumber\\
 & + \sum_{\mathclap{\substack{n=0\\ n\neq n_{\rm rnd}}}}^{7} A_{n} e^{-i [\omega_{n} (t - t_{\rm peak}) +  \phi_{n} ] } 
\end{align}

This is similar to the \Mx{7} model, except that the complex frequency of the $n=n_{\rm rnd}$ overtone is set to random (and, in general, incorrect) values uniformly sampled in the range $\left[ 0, 2\,{\rm Re}(\omega_{22n)} \right] + i \left[ 0, 2\,{\rm Im}(\omega_{22n}) \right]$. 
The aim of this exercise is to estimate which overtones are overfitting the waveform, if any. 
If a random damped exponentials can fit the waveform better than the \Mx{7} model, we can conclude that the overtone being randomized is not ``special,'' and it is merely overfitting the waveform.

In each panel of Fig.~\ref{fig:mismatch} we randomly draw $10^3$ complex frequencies for the randomized overtone, and we show the $90\%$ confidence level of the corresponding mismatches using dark-shaded bands. 
The thick black line shows the mismatches in the (unrandomized) \Mx{7} model. We also show the $90\%$ confidence level of mismatches computed by randomizing {\em all overtones with $n>n_{\rm rnd}$} -- as opposed to Eq.~(\ref{eq:hrnd}), where only $n=n_{\rm rnd}$ is randomized -- using lighter-shaded bands.
The top-left panel shows that the fundamental mode plays a key role in fitting the ringdown: the fit becomes very poor if we substitute it with a damped exponential with random frequency. 
The first overtone is also crucial in fitting the early part of the waveform, confirming the findings of Ref.~\cite{Baibhav:2017jhs}.
However, as we randomize the frequencies of the higher overtones, we observe that the gap between the colored bands and the black line (the \Mx{7} model) decreases, and the two start overlapping. 
The inset below each panel shows the fraction of random samples that outperform the \Mx{7} model.
For example, at $t_0=t_{\rm peak} + 6 M$, more than $90\%$ of randomly drawn damped sinusoids that replace $n=2$ yield smaller mismatches than the \Mx{7} model.
If all overtones with $n\geq n_{\rm rnd}=2$ were replaced by random frequencies, $2-20\%$ of the samples would still yield better fits than the \Mx{7} model for $t_0-t_{\rm peak}=0$--$5M$.
The performance of higher overtones is even worse. If $n=3$ is replaced, as many as $75\%$ of random samples fit better than the \Mx{7} model at $t_0 \approx t_{\rm peak}+ 3 M$. Similarly, for $n_{\rm rnd}=4-7$, $15\%$ to $66\%$ of the random samples perform better than the \Mx{7} model for  $t_0-t_{\rm peak} = 0$--$5 M$. 

In conclusion, random frequencies often yield better results than higher overtones for $N \geq 2$. This is further evidence that higher overtones play the role of unphysical low-frequency components, which are nonetheless effective at fitting the early-time signal. Recall that the real part of the frequency of the $n$th overtone, $\omega_r^n$, decreases with $n$ and that, at the end of the inspiral, the $\ell=m=2$ component of the signal has frequency $\omega_{\rm peak} \simeq 0.65 \, \omega_{220}$, increasing monotonically towards $\omega_{220}$ post-merger. This is why overtones naturally latch onto lower-frequency components of the signal (see Sec.~\ref{subsec:lowfreqmerger} for a discussion).

\begin{figure*}[t]
    \includegraphics[width=\textwidth]{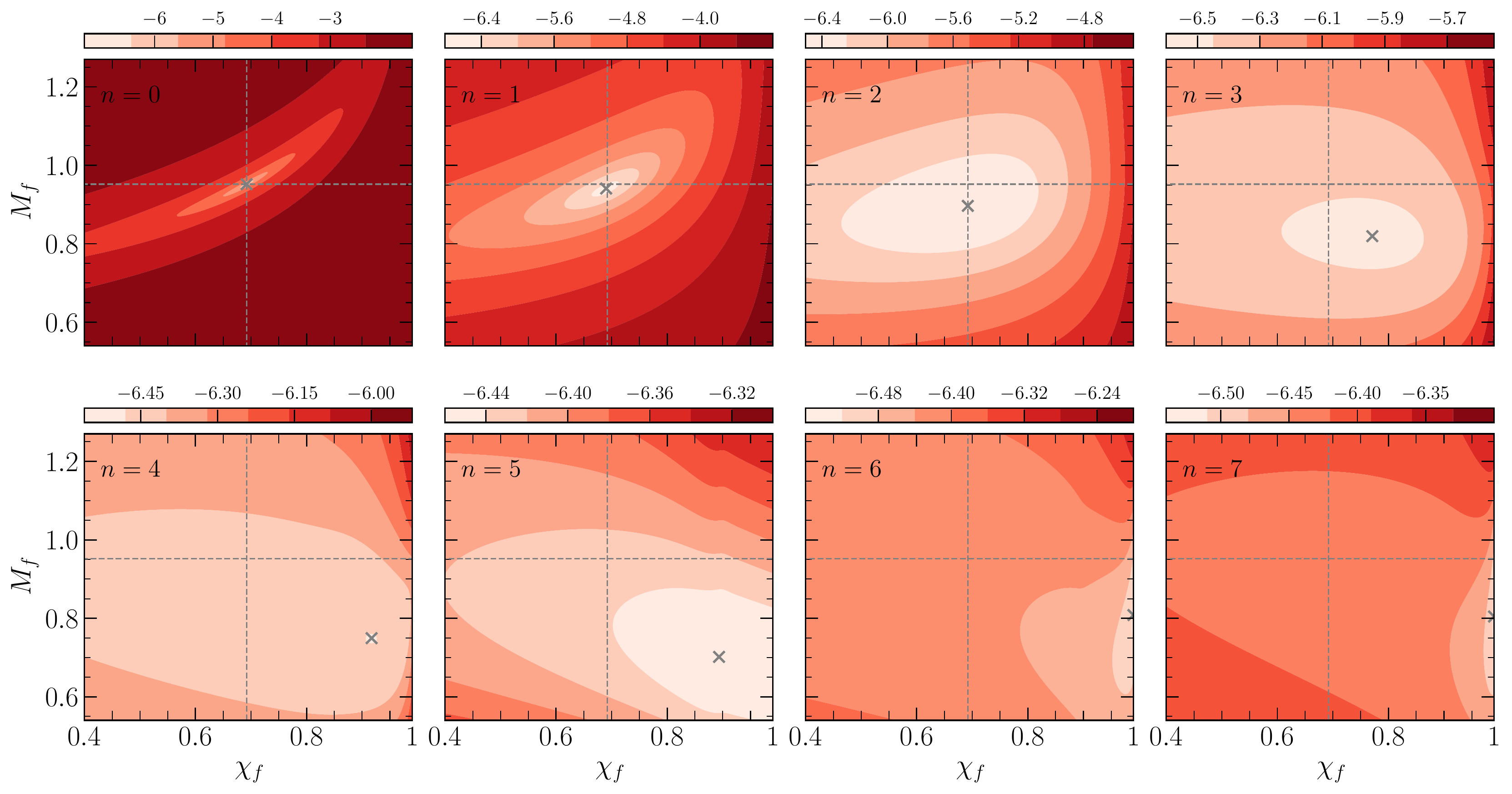}
    \caption{Logarithm of the mismatch, ${\log \mathcal M}$, evaluated at $t_0 = t_{\rm peak}$ in the $(M_f\,,\chi_f)$ plane. Starting from the \Mx{7} model, we vary the mass and spin of overtone $n$, while all other overtones are set to their ``true'' asymptotic values. The crosshair identifies the true value of the remnant's mass and spin. Note that the color maps in each panel correspond to very different ranges. The cross marks the $(M_f\,,\chi_f)$ values minimizing the mismatch.}
    \label{fig:mismatchaM}
\end{figure*}

\subsection{Which overtones are necessary to correctly extract the remnant's properties?}
\label{sec:epsilon}

It has been claimed that the inclusion of higher overtones not only fits the waveform better, but also yields a more accurate estimation of the remnant (dimensionless) spin $\chi_f$ and mass $M_f$. 
What is the reason for this improvement in mass and spin estimation?
Do higher overtones really contribute to the mass and spin measurement, as often claimed in the literature?
One less explored alternative is the possibility that higher overtones are instead just ``fitting out'' nonlinear effects, thus allowing for more accurate extraction of the fundamental mode and of the first overtone.
This would also lead to a better estimation of the remnant's mass and spin. 

To address this question, we perform a gedankenexperiment similar to the one in Sec.~\ref{sec:mismatch}.
We fit the waveform assuming that the complex frequencies of each overtone depend on the remnant's mass and spin as predicted by linear perturbation theory within GR.
Once again, we select one specific overtone $n_{\rm rnd}$ and replace it by (incorrect) complex frequencies drawn randomly in the range $\left[ 0, 2\,{\rm Re}(\omega_{22n)} \right] + i \left[ 0, 2\,{\rm Im}(\omega_{22n}) \right]$:
\begin{align}\label{eq:hrnd2}
 Q_7^{n_{\rm rnd}}(\chi_f, M_f)&=A_{n_{\rm rnd}} e^{-i [\omega_{n_{\rm rnd}} (t - t_{\rm peak}) +  \phi_{n_{\rm rnd}}]}\nn\\
 & + \sum\limits_{\mathclap{\substack{n=0 \\ \hspace{0.4cm} n\neq n_{\rm rnd}}}}^{7}  A_{n} e^{-i [\omega_{n}(\chi_f, M_f) (t - t_{\rm peak}) +  \phi_{n}] } \,.
\end{align}

In this model the mode with index $n=n_{\rm rnd}$ is not used in the estimation of $\chi_f$ and $M_f$.
To quantify the accuracy in extracting the remnant's spin and mass, following \GIST, we compute the quantity
\be\label{eq:epsilondef}
\epsilon =\sqrt{\left(\delta \chi_f\right)^2+ \left(\frac{\delta M_f}{M}\right)^2}\,,
\ee
where $\delta \chi_f$ and $\delta M_f$ are the differences between the inferred value of these quantities and their asymptotic, ``exact'' Kerr values.

In Fig.~\ref{fig:epsilon} we show the median and $90\%$ confidence interval of $\epsilon$ as a  function of the starting time of the fit $t_0$.
The solid black line uses all $N=7$ overtones to estimate the remnant's properties.
In each panel, solid lines and darker colors refer to randomly varying the $n= n_{\rm rnd}$ overtone frequency from a uniform distribution and ignoring the corresponding overtone in the estimation of $\chi_f$ and $M_f$.
Dashed lines and lighter colors were computed by randomizing all overtones with $n\ge n_{\rm  rnd}$.

The conclusions of this exercise are completely consistent with the ones drawn from Fig.~\ref{fig:mismatch}.
In fact, the top left panel shows that the fundamental mode carries most of the information. The second top panel from the left confirms that the first overtone ($n=1$) is crucial to get a good estimate of $\chi_f$ and $M_f$ close to the peak, again confirming the findings of Ref.~\cite{Baibhav:2017jhs}. At later times the first overtone does not matter as much, and we observe more overlap between the orange bands and the black line. The role of higher overtones in parameter estimation is much more marginal. For example, at $t_0=t_{\rm peak}+ 2 M$, $15\%$ of the random complex frequencies that replace $n=2$ yield better results than the second overtone. As we randomize higher overtones ($n_{\rm  rnd}=3$-$7$), the number of samples that do better than the \Mx{7} model ranges between $55\%$ and $70\%$. Note also that the value of $\epsilon$ from \Mx{7} model has a maximum when $t_0-t_{\rm peak} \approx 7$-$8 M$: at those times, the \Mx{7} model is worse than most random samples. If all overtones $n\geq n_{\rm rnd}=2$ are randomized, nearly $35\%$ of the random samples do better than the \Mx{7} model at $t_0=t_{\rm peak}+5 M$; this fraction rises to $66\%$ at $t_0=t_{\rm peak}+8.4 M$. The fact that random frequencies often do better than high-overtone models is further evidence that most of those overtones are unlikely to be physical.

The results above can be understood considering that each overtone introduces two additional fitting parameters, and that $\omega^n_r$ decreases with $n$. 
These properties allow to easily ``fit away'' the earlier nonlinear and low-frequency part of the post-merger signal, making the fundamental mode and the first overtone more easily resolvable. 
This, in turn, improves the estimation of the remnant mass and spin, which are still constrained to the correct Kerr value by the late-time behavior of the signal. 
It is not a surprise that models with several parameters are prone to overfitting\footnote{As Enrico Fermi told Freeman Dyson in a famous encounter~\cite{2004Natur.427..297D}: ``I remember my friend Johnny von Neumann used to say, with four parameters I can fit an elephant, and with five I can make him wiggle his trunk.''} (in fact, this concern was also raised by \GIST). 
However, Reference~\cite{Giesler:2019uxc} concluded that {\em all} overtones are physical by varying all of them while keeping the fundamental mode fixed. 
This approach does not estimate the significance of each individual overtone.
Our analysis shows that as we increase the overtone number, the relevance of each overtone in estimating the remnant's mass and spin decreases. 
Another difference between our results and those of \GIST \, is that they varied the complex frequencies by less than $20\%$ in one dimension: they set $\omega_{\rm rnd}=\omega(1+\delta)$ with $\delta\in \mathbb{R}$. 
Our experiment is more conservative and general, because the real and imaginary parts of the complex QNM frequencies are sampled independently and allowed to vary by $100\%$.

To further our understanding of which overtones are physical, we perform one more test. We first compute the mismatch at $t_0=t_{\rm peak}$ for the \Mx{7} model. Then we vary the frequency of a single QNM by assuming that it depends on $(M_f,\,\chi_f)$ as predicted in GR and changing $(M_f,\,\chi_f)$, while we keep all other modes fixed to their ``correct'' asymptotic values.

In Fig.~\ref{fig:mismatchaM} we show the mismatch at the peak as a function of $M_f$ and $\chi_f$. Each panel was computed by varying the corresponding overtone.
For $n = 0$ the mismatch has a sharp minimum at the true value of the remnant's mass and spin (top left panel), as expected. As $n$ increases, the minimum of the mismatch extends over a much larger region in the $(M_f,\,\chi_f)$ plane. More remarkably, the mismatch for higher overtones has a minimum at values of $(M_f,\,\chi_f)$ that {\em do not} coincide with the true remnant quantities.
For $n=0,\,1$ the true $(M_f,\,\chi_f)$ coincide with the minimum in the figure with an accuracy of $\epsilon\lesssim 0.01$.
For $n=2$, the remnant mass and spin at the minimum mismatch deviate from the true values by $\epsilon\approx0.05$, and the error is dominated by the inaccuracy in $M_f$.
The minimum mismatches for $n>2$ occur for values of $(M_f,\,\chi_f)$ which differ by $\epsilon>0.1$ from the true values, and these minima tend to cluster at near-extremal values of the spin ($\chi_f \to1$). Furthermore, as expected, the variations of the mismatch with $(M_f,\,\chi_f)$ are much milder as we move to higher overtones.
This is consistent with Fig.~\ref{fig:mismatch}, where the $90\%$ confidence levels in the mismatch are very narrow for higher overtones, even though their complex frequencies are allowed to vary by as much as $100\%$.

Reference~\cite{Giesler:2019uxc} used a similar test to claim support for the \Mx{7} model, because the minimum mismatch for $N=7$ occurs at the true remnant spin and mass. However, Figure~\ref{fig:mismatchaM} implies that the minimization of the mismatch is almost entirely due to the modes with $n=0$ and $n=1$. When considered individually, the higher overtones play a negligible role. Even worse, they minimize the mismatch at the wrong values of the remnant's parameters.

In this section we have used the \Mx{7} fitting model to facilitate comparison with \GIST. This model does not take into account spherical-spheroidal mode mixing. As we will see below, this is a substantial effect. By repeating the experiments discussed in this section with mode mixing taken into account, we have verified that the extraction of the first overtone can significantly improve at late times: for example the amplitude $A_{221}$ would not diverge at late times (as it does in Fig.~\ref{A_nN} of Sec.~\ref{sec:constA}), but remain constant. Similarly, the mismatch ${\cal M}$ (Sec.~\ref{sec:mismatch}) and the mass/spin measurement error $\epsilon$ (Sec.~\ref{sec:epsilon}) associated with $n=1$ would improve at late times. However, including the spherical-spheroidal mixing does not yield visible improvements for overtones with $n>1$.

\begin{figure*}[t]
    \includegraphics[width=0.99\textwidth]{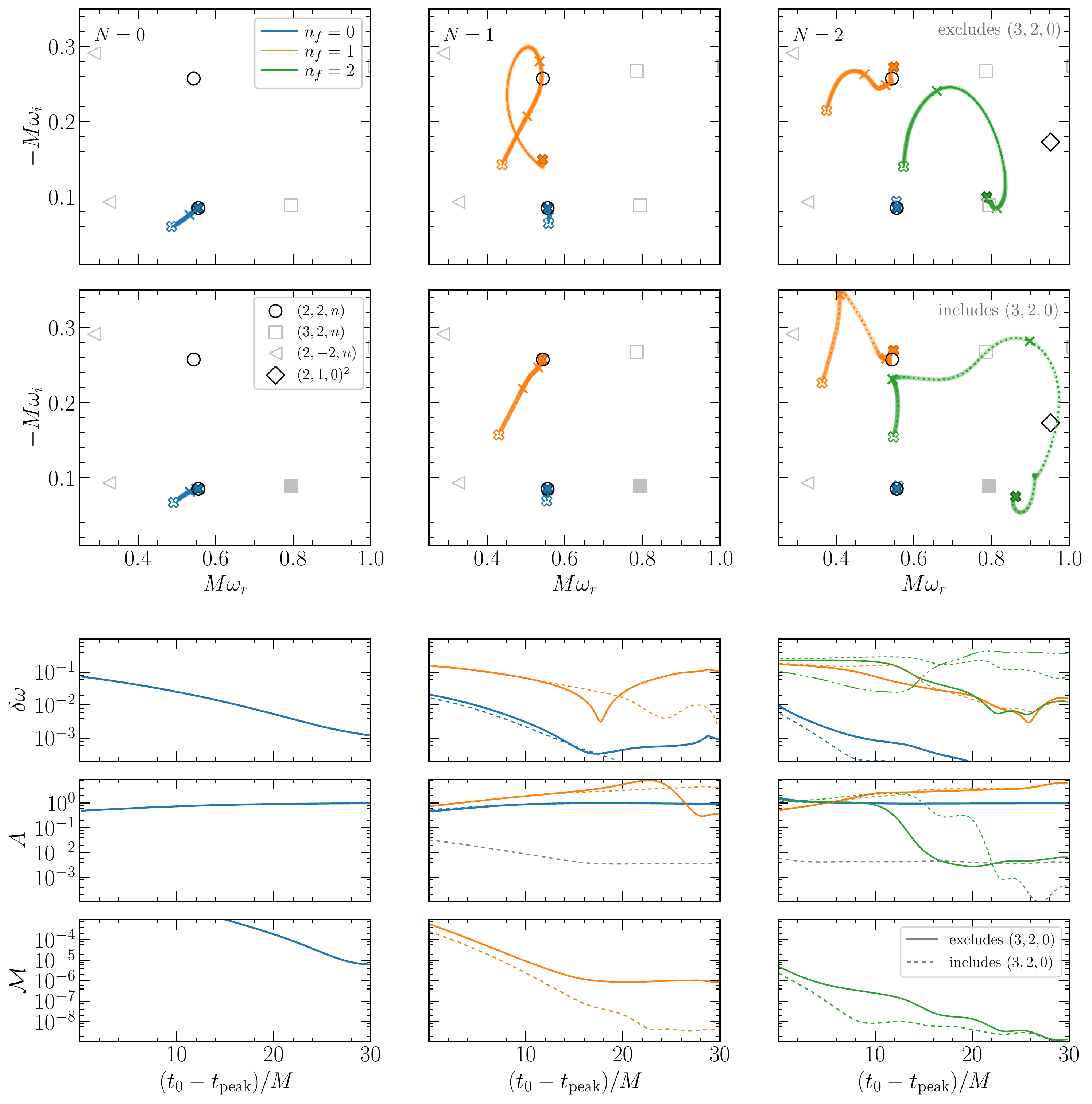}
    \caption{
    Fits of the SXS:BBH:0305 waveform with the \Mr{N} and \Mrm{N} models.
    First panel: the extracted frequencies 
    $( \omega_r, \, - \omega_i )$ for $N=0$ (left), $N=1$ (middle) and $N=2$ (right), with $t_0$ ranging from $t_{\rm peak}$ (empty cross) to $t_{\rm peak}+30 M$ (filled cross). 
    We also mark $t_{\rm peak}+10 M$ and $t_{\rm peak}+20 M$ by crosses.
    Second panel: same as first panel, but when adding a mode with complex frequency fixed to $\omega_{320}$, indicated by a filled gray square.
    Third, fourth and fifth panels show, respectively, $\delta\omega$, $A$ and ${\mathcal M}$ as a function of $t_0$. 
    For $n_f=0,1$, $\delta \omega$ is determined compared to $\omega_{\rm ref}=\omega_{22n_f}$, while for $n_f=2$, $\omega_{\rm ref}=\omega_{320}$. Solid lines represent the  \Mr{N} model while dotted lines represent the \Mrm{N} model. 
    With the green dot-dot-dashed line, we also plot $\delta\omega$ with $\omega_{\rm ref}=\omega_{221}$ when fitted by the \Mrm{N} model. 
    We also plot the $A_{320}$ with a dashed gray line.}
    \label{fig:wri}
\end{figure*}

{Although we have demonstrated the presence of several pathologies in the linear model with $N=7$ overtones, and therefore shown that it is unphysical, this does not imply that overtones are absent altogether. Lack of evidence is not evidence of absence: the overtones may still be present in the signal, but they may be hard to extract because additional physics (including prompt response, tails, and nonlinear effects) is present in the waveform. By accounting for these effects, and in particular for post-merger nonlinearities, it may still be possible to extract the overtones. In Appendix~\ref{sec:NRtoy} we show that, indeed, the overtones can be correctly extracted from a pure linear model with  7 overtones, at least when numerical errors are negligible. Numerical errors increase the minimum mismatch to $O(10^{-9})$, and compromise our ability to extract the overtones. Thus, it may still be possible to successfully extract the overtones with improved NR waveforms and a better understanding of nonlinearities.}

\section{Extracting complex frequencies from the waveform}
\label{sec:howmany}

One approach to look for the presence of a QNM is to fix the mode's complex frequency (or assume that it depends on mass and spin as predicted in GR), and then check if adding that mode improves the mismatch with the numerical waveform (Sec.~\ref{sec:mismatch}) or leads to better accuracy in the remnant parameters (Sec.~\ref{sec:epsilon}). As we discussed, this approach can lead to overfitting.

In this section we take a more agnostic route in the spirit of the original BH spectroscopy proposal. To prevent overfitting, we do not impose \textit{a priori} that modes are present in the numerical data. 
Instead of fixing the complex frequencies of the damped exponentials, we keep them free (see the analogous treatment of the linearized case in Sec.~\ref{sec:linearovertones}).
This is a much stronger test (it is effectively the original BH spectroscopy proposal at infinite SNR), and therefore it should lead to more robust conclusions about which modes are truly present in the data.  

We will show that many of the fitted damped exponentials robustly converge towards known QNM frequencies, naturally selecting the physical modes contributing to the ringdown signal; that is essential to include spherical-spheroidal mode-mixing to identify the correct modes; and that the agnostic fits occasionally lead to some surprises. 
To improve readability, in this section we focus on the dominant $(2,\,2)$ multipole. We defer a discussion of higher multipoles to Appendix~\ref{sec:harmonics}.

\subsection{An agnostic fit}
\label{subsec:agnostic}

We start off by asking the question: {\em how many QNM frequencies can we extract without assuming any (no-hair theorem enforced) relation between them?}

{We fit the complex frequencies and amplitudes using a standard Levenberg–Marquardt algorithm.  The large number of fitting parameters implies the possibility of multiple local minima for the parameter values resulting from the fit, and therefore different initial values may result in different solutions. To address this issue, we employ an iterative approach.
\begin{enumerate}
    \item In the initial iteration, we perform the fit using $1000$ different random initial parameter values for every value of $t_0$. The initial values of the amplitudes are drawn from a log-uniform distribution in the range $[ 10^{-5}, 10^3]$, while phases are drawn uniformly in $[0, 2\pi]$, $M\omega_r$ is drawn uniformly in $[0, 5 \Re(M\omega_{\ell, m, 0})]$, and $M\omega_i$ is drawn uniformly in $[0, 5 \Im(M\omega_{\ell, m, 0})]$. This yields $1000$ solutions, out of which we select those that exhibit the smallest mismatch with the NR waveform, for each $t_0$. 
    \item In the subsequent iteration, we perform the fitting process again for each value of $t_0$, just like in the initial iteration. However, this time, instead of starting with random initial parameter values, we use the initial conditions that correspond to all solutions obtained in the previous iteration. To clarify this step, suppose the previous iteration resulted in unique solutions for $300$ different values of $t_0$. Now, for the current iteration, we fit the parameters again for each specific $t_0$, using initial guesses that equal the solutions obtained in the previous step. In other words, those $300$ distinct solutions are employed as initial guesses for every single $t_0$. Consequently, we will often encounter different solutions for each of these values of $t_0$. Once more, we select the solutions that have the smallest mismatches with the NR waveform.
    \item We repeat this procedure in the same range of $t_0$ until the mismatch for all points is within $1\%$ of their values from the previous iteration. The number of iterations necessary to achieve convergence is typically less than three for small values of $N_f$; when $N_f$ is large (e.g., \(N_f > 5\)), as many as eight iterations might be required.
\end{enumerate}
}

In the top row of Fig.~\ref{fig:wri} we fit the waveform with model \Mr{N} (see Table~\ref{tab:models}), i.e., we extract the complex frequencies of all modes from the fit. We plot the complex frequencies $\omega_{n_f}$ (where the subscript ``f'' stands for ``free'') extracted from the numerical waveforms for $N=0$ (left panel), $N=1$ (middle panel) and $N=2$ (right panel).
In the second row we use model $\Mrm{N}$ instead: in addition to the $N$ free-frequency modes, we also include one mode with frequency fixed at the value of the fundamental mode with $(\ell=3,\,m=2)$, that could (and indeed does) contaminate the $\ell=m=2$ mode because of spherical-spheroidal mode mixing.

In each of the six panels of the top two rows, we show for reference: (i) the known $(2,\,2,\,n)$ overtone frequencies as black, hollow circles; (ii) the ``mirror modes'' corresponding to the known $(2,\,-2,\,n)$ overtone frequencies as gray, hollow {triangles}; and (iii) the known $(3,\,2,\,n)$ overtone frequencies as gray, hollow squares. When the fundamental mode with $(\ell=3,\,m=2)$ is fixed (i.e., in model $\Mrm{N}$) the $(3,\,2,\,0)$ is shown as a gray, filled square, as a reminder that it is already included in the waveform and we are not fitting for its complex frequency.

In the panels below we plot various diagnostic quantities, namely: the ``frequency error'' $\delta \omega$ relative to the expected QNM frequency $\omega_{\rm ref}$, as defined in Eq.~\eqref{eq:deltaomega} (third row); the mode amplitudes $A_{n_f}$ (fourth row); and the mismatches ${\mathcal M}$ of the fits (fifth and bottom row).

Let us analyze these results starting from the left column. If we fit the waveform with only one damped exponential ($N=0$), we see that the extracted mode saturates to the fundamental mode $n=0$. 
At the peak ($t_0 -t_{\rm peak}=0$) the free mode $n_f=0$ deviates by $\delta\omega<0.1$ from $\omega_{\rm ref}=\omega_{220}$. The disagreement improves as the BH relaxes to its stationary end state, crossing $\delta\omega=10^{-2}$ at $t_0 -t_{\rm peak}\simeq 15 M$, and reaching $\delta\omega<10^{-3}$ at $t_0 -t_{\rm peak}=30 M$.

\begin{figure*}[t]
    \includegraphics[width=\textwidth]{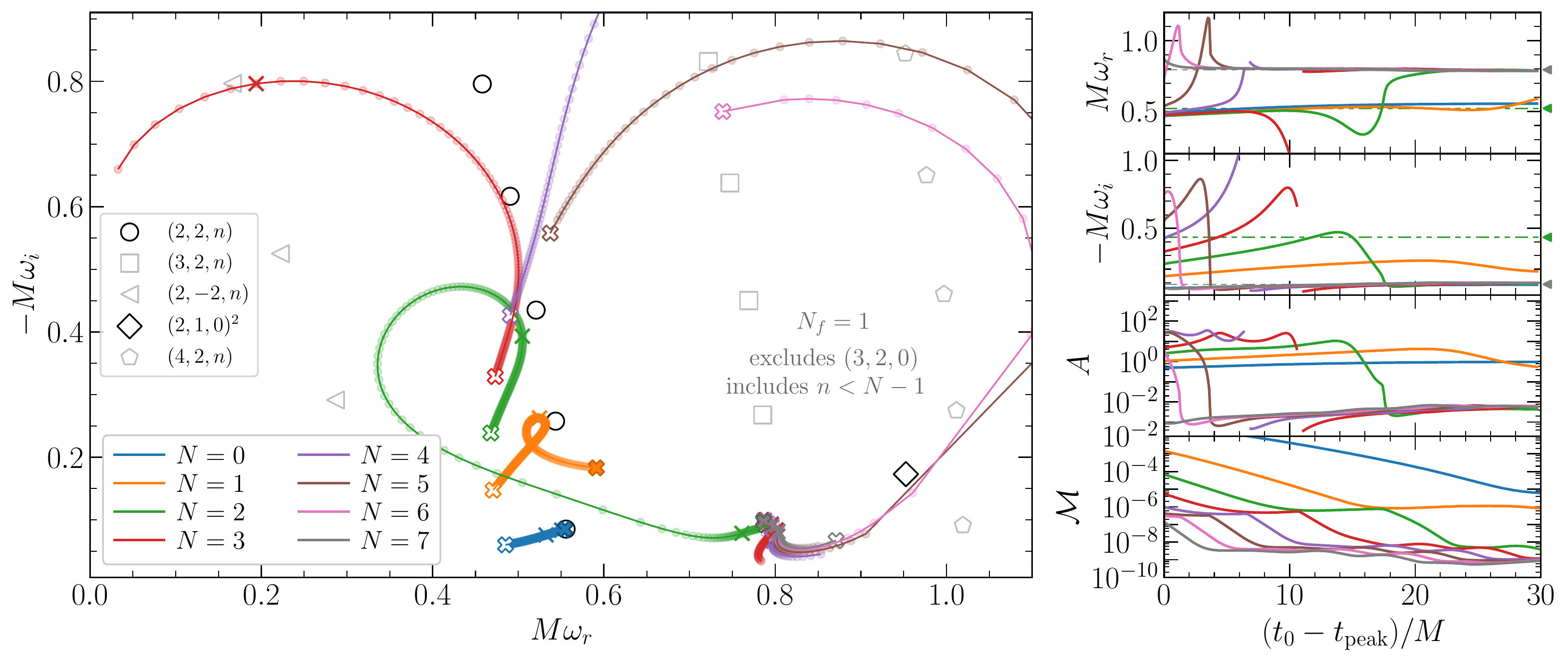}
    \caption{ 
    Fits of the SXS:BBH:0305 waveform with the \Mxr{N}{1} models, with $n=0, \dots, N-1$ fixed-frequency modes, plus one free-frequency mode.
    Left panel: extracted frequencies 
    $(M \omega_r, \, -M \omega_i )$ for $N=0-7$, with $t_0$ ranging from $t_{\rm peak}$ (empty cross) to $t_{\rm peak}+30 M$ (filled cross). 
    We also mark $t_{\rm peak}+10 M$ and $t_{\rm peak}+20 M$ by crosses.
    We mark the QNMs $(2,2,n)$ (black circles), $(2,-2,n)$ (gray triangles),  $(3,2,n)$ (gray squares),  $(4,2,n)$ (gray pentagon) and $(2,1,0)(2,1,0)$ (black cross). 
    Right panels plot $M \omega_{r}$, $-M \omega_{i}$, $A$ and ${\mathcal M}$ as a function of $t_0$. 
    We also plot $M \omega_{320}$ as a gray dot-dot-dashed line and gray triangle marker.
    }
    \label{fig:wri1}
\end{figure*}

If we add one more free mode (model \Mr{1}, central panels) we observe that the fundamental mode is extracted much more accurately:
the solid blue lines in the third row show that at $t_0 -t_{\rm peak}=5M$ it deviates from the expected value by $\delta\omega<0.01$, with $\delta\omega \simeq 3\times 10^{-4}$ at $t_0 -t_{\rm peak}=17 M$. 
The first overtone (solid orange lines) is extracted much more poorly, with $\delta\omega=0.2$  at $t_0 =t_{\rm peak}$ and $\delta\omega<0.1$ after $\approx 6 M $. 
Interestingly, there is a minimum ($\delta\omega \approx 0.003$) at about $t_0 -t_{\rm peak}=18 M$, after which the free mode starts deviating again from the ``true'' $n=1$ overtone. After $27 M$ the deviation is $\delta\omega>0.1$. 
This seems to indicate that some other component is spoiling the recovery of the $n=1$ overtone.

To confirm this conjecture, we add one more free mode (model \Mr{2}, right panels). 
Now the fundamental mode is always extracted extremely well, with $\delta\omega<0.01$, and even $\delta\omega<10^{-5}$ at late times.
Both of the remaining free frequencies ($n_f=1,2$) start off in the vicinity of $n=1$ at $t_0 =t_{\rm peak}$.
At later times, $n_f=1$ (solid orange line) moves towards $n=1$, with a minimum of $\delta\omega = 6\times10^{-4}$ at $26M$. 
However, $n_f=2$ (solid green line) behaves very differently, and at late times it converges towards the $(3,\,2,\,0)$ mode, with a relative deviation of $\delta\omega=0.01$ at $27 M$.
This contamination, expected because of spherical-spheroidal mode mixing~\cite{Berti:2005gp,Buonanno:2006ui,Kelly:2012nd,Berti:2014fga}, explains why the \Mr{1} model failed to converge to the $n=1$ overtone.
Since the $n=1$ is dominant at early times, the free mode $n_f=1$ latches onto it. The $(3,\,2,\,0)$ mode, despite having ${\mathcal O}(10^3)$ smaller amplitude than the $(2,\,2,\,1)$ mode, decays more slowly and becomes dominant at later times.

In the second row of Fig.~\ref{fig:wri} we ``subtract'' mode-mixing by fixing one damped exponential to the frequency of the $(3,\,2,\,0)$ mode.
Dashed lines in the bottom three rows show the results of fitting the waveform with this ``fixed mode-mixing'', \Mrm{N} model.

The left column shows that the addition of the $(3,2,0)$ mode has a negligible effect on \Mrm{0}. 
For \Mrm{1} (central column), including mode mixing further improves the extraction of the fundamental mode at late times, with $\delta\omega=10^{-5}$ at $t_0 -t_{\rm peak}=25M$. More importantly, the free mode with $n_f=1$ now robustly converges towards $\omega_{221}$, crossing the predicted value around $24 M$, with $\delta \omega \in [10^{-3}, 10^{-2}]$ for $t_0 -t_{\rm peak}\in [24,30]M$. 
With the \Mrm{2} model (right column), we still observe that $n_f=1$ saturates toward $\omega_{221}$ at late times, with $\delta\omega<0.01$ between $22-27M$. 
At early times, somewhat surprisingly, the $\omega_{221}$ overtone is ``picked up'' by the second free mode ($n_f=2$), which at later times drifts off towards the right of the complex plane, passing by the $\omega_{321}$ mode and even  showing non-conclusive hints of a possible presence of the nonlinear mode sourced by the square of the $\omega_{2,1,0}$ mode.
The presence of these two modes in the simulations is plausible, but none of them has amplitude large enough to be confidently identified.

The fixed-frequency component of \Mrm{1} has a consistent amplitude $A_{320} \sim 4\times10^{-3}$ for $t_0 -t_{\rm peak} \geq 16M$.
A similar value is recovered with \Mrm{2} at even earlier times. This is indication that the additional free component is fitting away some contamination that pollutes the recovery of $A_{320}$ at earlier times.

\begin{figure*}[t]
    \includegraphics[width=\textwidth]{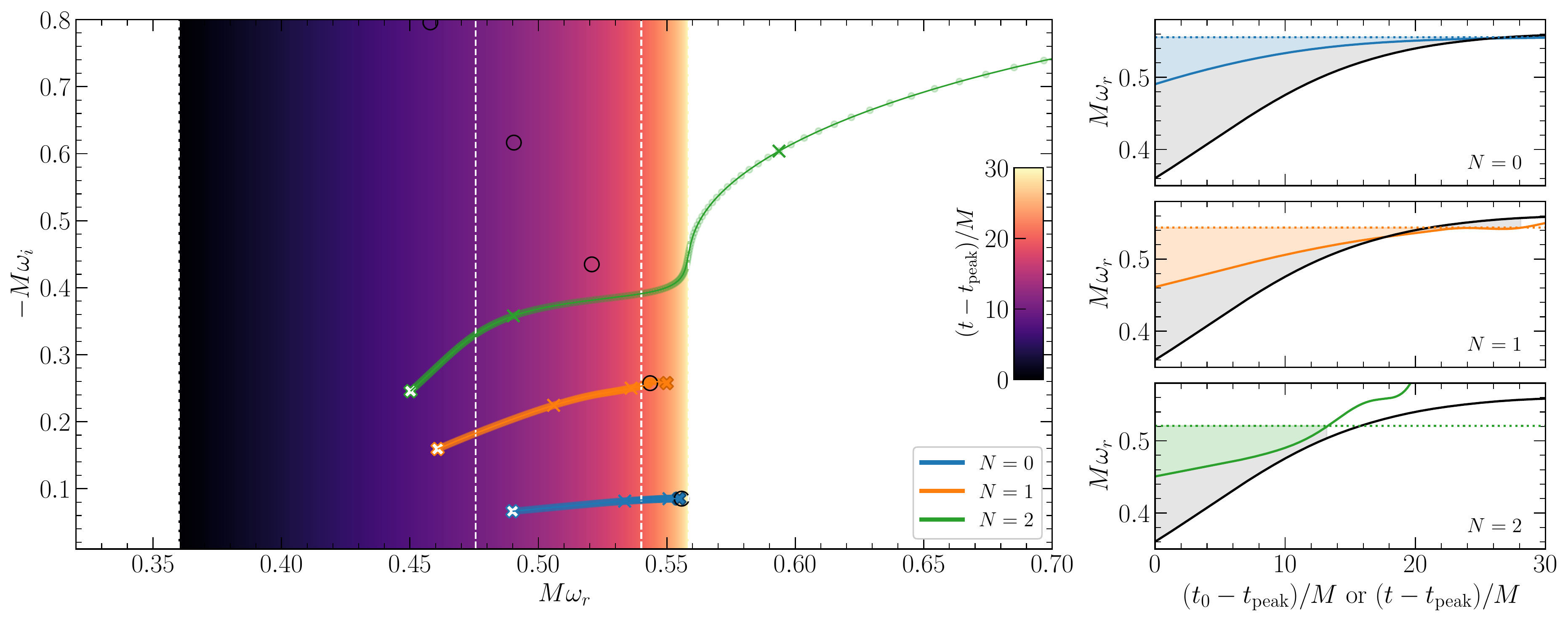}
    \caption{
    Fits performed using the \Mxrm{N}{1} model with $N=0,\,1,\,2$. 
    Left panel: same as Fig.~\ref{fig:wri1}. We also show how $M\omega_{\rm GW}$ evolves (x-axis) as a function of time (see color bar). 
    White dashed lines indicate $t=t_{\rm peak}+10 M$ and $t=t_{\rm peak}+20 M$ in the instantaneous frequency evolution.
    The panels on the right show $M\omega_{r}$ extracted with $N=0,\,1,\,2$ (top, middle and bottom, respectively) as a function of the starting time $t_0$ in solid colored lines). 
    For comparison, we also plot $M\omega_{\rm GW}(t)$ (black line) and the real part of the expected QNM frequencies (colored horizontal dotted lines).}
    \label{fig:freq_evol}
\end{figure*}

Another noteworthy feature of Fig.~\ref{fig:wri} is that {\em mismatches improve significantly when we keep the frequencies free.} 
For example, at $t_0=t_{\rm peak}$, \Mr{1} (\Mr{2}) shows an improvement by one (two) orders of magnitude in the mismatch compared to \Mx{1} (\Mx{2}). 
In fact, the \Mr{1} (\Mr{2}) models even have better mismatches than the fixed-frequency models \Mx{2} (\Mx{4}).
As we discuss in Appendix~\ref{sec:systematics}, \Mr{3} (a model with three free frequencies) yields smaller mismatches than the \Mx{7}.
The inclusion of mode mixing has a dramatic effect on mismatches at late times. With the \Mr{1} model (two free modes), the mismatch saturates at $O(10^{-7})$; when we also include mode mixing in model \Mrm{1}, the mismatches are as low as $O(10^{-9})$. A similar improvement occurs going from \Mr{1} to \Mr{2}, when the second free mode ($n_f=2$) saturates to the $(3,\,2,\,0)$ mode frequency.

So far we have used two classes of fitting models (\Mr{N} or \Mrm{N}) in which all complex frequencies are free, except for the $(3,2,0)$ mode-mixing component in \Mrm{N}. 
The results in Fig.~\ref{fig:wri} show that the free frequencies capture the modes $(2,2,0)$, $(2,2,1)$ and $(3,2,0)$, but {\em there is no evidence for higher overtones ($n>1$)}.

Could the absence of ``detectable'' higher overtones be an artifact of the use of free-frequency models? To exclude this hypothesis, we now fit the $(2,\,2)$ mode by the \Mxr{N}{1} model. As a reminder, this means that one mode frequency is free ($N_f=1$), but $N$ mode frequencies ($n=0,\,\dots,\,N-1$) are fixed to the predicted overtone values. For $N=0$, model \Mxr{0}{1} consists (trivially) of a single, free-frequency damped exponential.

In Fig.~\ref{fig:wri1} (which can be compared to the toy-model investigations of Fig.~\ref{fig:fitfreqtoys}) we plot the frequencies, amplitudes, and mismatches extracted in this way.
For model \Mxr{0}{1}, the free mode ($n_f=0$) converges to $n=0$ at late times (blue lines). 
This late-time convergence is lost when $N>0$.
For example, model \Mxr{1}{1} is reminiscent of Fig.~\ref{fig:wri}: the free mode frequency ``flies by'' the $(2,\,2,\,1)$ overtone at times $t_0 -t_{\rm peak} \in [10-20] M$, but then it drifts to higher frequencies (orange lines). 

Interestingly, when $N>1$ the free frequency always tends to the $(3,\,2,\,0)$ mode-mixing QNM, with $\delta\omega$ as low as $O(10^{-3})$ at some point after $20M$. 
The free mode gets close to the expected overtone frequency at intermediate times only for $N=2$ (green line) and $N=3$ (red line).
As $N$ increases, the free mode latches onto $\omega_{320}$ at earlier and earlier times: the free frequency gets within $\delta\omega<0.1$ of the $(3,\,2,\,0)$ mode after $17M$, $11M$ and $7M$ for $N=2$, $N=3$ and $N=4$, respectively.

Finally, in Fig.~\ref{fig:freq_evol} we repeat the same experiment, but this time we also fix the $(3,\,2,\,0)$ mode, i.e. we fit the $(2,2)$ mode by the \Mxrm{N}{1} model.
The results are qualitatively similar to Fig.~\ref{fig:wri1}: only the fundamental mode and the first overtone saturate to the expected QNM frequencies. By including mode mixing, we manage to get the $N=2$ fitting model relatively closer to the expected $n=2$ overtone at $t_0 -t_{\rm peak} \sim 13 M$. However this mode does not asymptote to the $n=2$ overtone at late time, and instead it keeps drifting to large values of $M\omega_r$. 
In the next section, we give a physical interpretation to the observed low-frequency behavior of the free modes.

We have studied three different ``agnostic'' fitting models (Figs.~\ref{fig:wri}, \ref{fig:wri1} and \ref{fig:freq_evol}). All three models support an important conclusion: {\em the only identifiable physical modes in the $\ell=m=2$ mode of the radiation are $(2,\,2,\,0)$, $(2,\,2,\,1)$ and $(3,\,2,\,0)$.} Higher overtones ($N>1$) cannot be robustly identified by free-frequency fitting.

\subsection{Are we fitting the low-frequency merger signal?}
\label{subsec:lowfreqmerger}

In Fig.~\ref{fig:wri},~\ref{fig:wri1} and~\ref{fig:freq_evol},  we observe a clear pattern (that becomes progressively clearer as we remove ``undesired'' physics, such as spherical-spheroidal mode mixing): at least one free mode starts from low $M\omega_r$ and evolves to higher $M\omega_r$, often (but not always) saturating to one of the expected QNM frequencies.  For example, in Fig.~\ref{fig:wri1} the free mode in \Mxr{N}{1} for $N=1,\,2,\,3$ always starts at $M\omega_r\simeq 0.47$ and evolves in time (at least initially) to higher $M\omega_r$.  This behavior is reminiscent of the well-known ``chirping'' that characterizes the binary's inspiral. To make the comparison more quantitative, we can compute the instantaneous GW frequency~\cite{Buonanno:2006ui,Berti:2007fi}:
\be
\omega_{\rm GW} =\f{\dot{h}_{22}(t)}{h_{22}(t)}\,.
\ee

The black solid line in the right panels of Fig.~\ref{fig:freq_evol} shows how $M\omega_{\rm GW}$ evolves in time. We also plot, for comparison, the values of $M\omega_r$ extracted using the \Mxrm{N}{1} models. Note that the time axes are different: $M\omega_{\rm GW}$ is the instantaneous frequency at time $t$, while  $M\omega_{r}$ is plotted as a function of the starting time $t_0$ of the fitting window (the ending point is $t_f=t_{\rm peak}+90 M$). This is why $M\omega_{\rm GW}<M\omega_r$: $M\omega_r$ contains information about physics in the range $[t_0,\,t_f]$, and as such it includes late-time contributions to $M\omega_{\rm GW}$.

At late times, the real frequencies $M\omega_r$ extracted using the \Mxrm{0}{1} model (top right panel) and the \Mxrm{1}{1} model (middle right panel) approach the real frequency of the $n=0$ and $n=1$ overtones, respectively. As we remarked earlier, this is not true for model \Mxrm{2}{1}: $M\omega_r$ keeps growing after $t_0-t_{\rm peak} \gtrsim 13 M$, instead of saturating to the $n=2$ overtone. In Appendix~\ref{sec:systematics}, we argue that this large-$M\omega_r$ evolution of the \Mxrm{2}{1} model is not physical.

These findings are consistent with other results in the literature. By adding even lower frequency contributions (specifically, modes corresponding to counter-rotating perturbations, which should not be excited unless the binary has large spins antialigned with the orbital angular momentum~\cite{Bernuzzi:2010ty,Barausse:2011kb,Li:2021wgz}),  Ref.~\cite{Dhani:2020nik} found that a linear QNM model with a large number of overtones yields low mismatches even {\em before} the waveform peak, where a linearized description should not be valid.

These results suggest that once the ``physical'' ringdown modes (typically the fundamental mode, the first overtone, and the mode-mixing contribution) have been fitted for, {\em additional free modes have a tendency to simply track the GW frequency chirp at early times.} Higher overtones have low frequencies, and this observation is consistent with the empirical fact that overtones do such a good job at effectively reducing the mismatch close to the peak.

{It is important to emphasize again that lack of evidence is not evidence of absence: while we are unable to extract higher overtones even when including a large number of free modes, this does not imply the absence of overtones altogether. The extraction of overtones is challenging due to the presence of additional physics and nonlinear effects. It may still be possible to achieve it by incorporating a more comprehensive understanding of all post-merger nonlinearities, and allowing for the dynamical character of the time-varying background.}

\subsection{Extracting BH properties in the presence of nonlinearities}
\label{subsec:massspinest}

\begin{figure}[t]
    \includegraphics[width=\columnwidth]{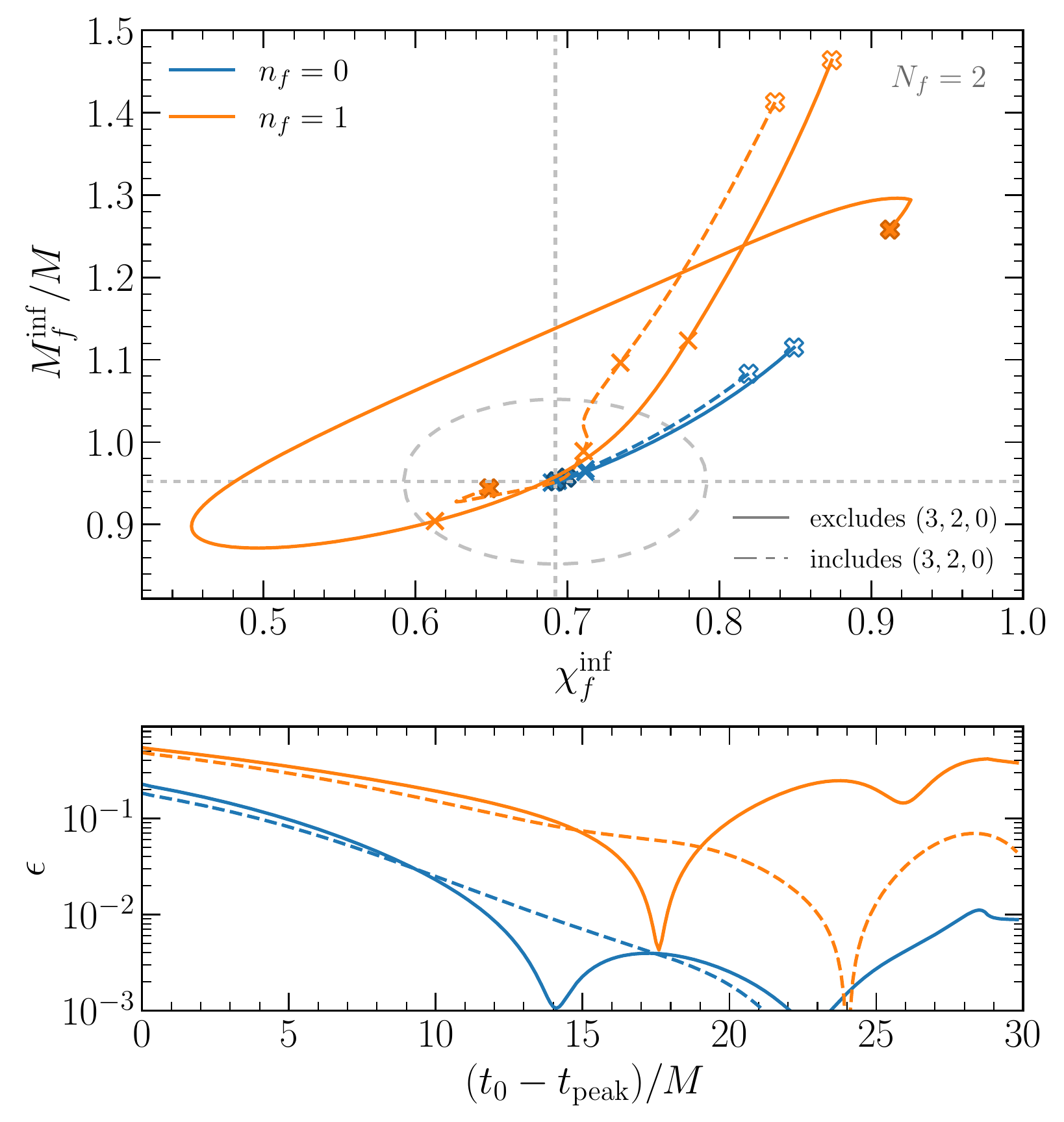}
    \caption{Remnant mass and spin $(M_f^{\rm inf},\,,\chi_f^{\rm inf})$ inferred using \Mr{1} [solid lines] and \Mrm{1} [dashed lines]) for $t_0$ ranging from $t_{\rm peak}$ (hollow crosses) to $t_{\rm peak}+30 M$ (filled crosses). We also mark $t_{\rm peak}+10 M$ and $t_{\rm peak}+20 M$ by crosses. Mass and spin are calculated directly from $\omega_{n_f}$, without assuming any no-hair theorem enforced relation between the modes. The gray vertical (horizontal) lines mark the true $\chi_f$ ($M_f$). The dashed ellipse corresponds to $\epsilon=0.1$.}
    \label{fig:aM_inf}
\end{figure}

The comprehensive fitting experiments performed so far show that higher overtones are hard to extract from numerical waveforms, and they are often unlikely to be physical. In the infinite SNR limit, {\em how do frequencies inferred by fitting free damped exponentials translate into mass and spin values?} This is the important question to ask in the spirit of the original, agnostic BH spectroscopy proposal. The answer depends on the starting time of the fit, and it is affected by the inclusion of mode-mixing contributions from the $(3,2,0)$ mode.

Consider first model \Mr{1}, with two free complex frequencies. 
We will identify the $n_f=0$ component with the $(2,\,2,\,0)$ mode and the $n_f=1$ component with the $(2,\,2,\,1)$ mode. A posteriori, we now assume that these damped exponentials correspond to the GR QNMs. Then we can invert the (complex) functional relationships $\omega_{\ell m n} (M_f, \chi_f)$ provided in Ref.~\cite{Berti:2005ys} separately for the $(2,2,0)$ and $(2,2,1)$ modes, as follows:
\begin{align}
\chi_f^{\rm inf}&={\mathcal Q}_{\ell m n}^{-1}\left(-\f{\omega_r}{2\omega_i}\right),\nn\\
M_f^{\rm inf}&= \f{M \omega_r}{{\rm Re}  [\omega_{\ell m n} (\chi_f^{\rm inf}, M)] }\,.
\end{align}
Here ${\mathcal Q}_{\ell m n}^{-1}$ is the inverse of the monotonic function relating the quality factor and the spin, ${\mathcal Q}_{\ell m n}(\chi_f)$~\cite{Berti:2005ys}.
If the modes can be robustly inferred from the data, the mass and spin inferred from each of these two inversions should coincide. 
In fact, performing this exercise with three or more ``measured'' parameters is a null test of GR.

We already know from Fig.~\ref{fig:wri} that in a fit with two free modes, neither of the extracted frequencies at $t_0=t_{\rm peak}$ coincides with the first overtone:
the difference between the ``recovered first overtone'' and the true $(2,2,1)$ mode is $\delta \omega>0.1$. 
These recovered frequencies will inevitably give wrong estimates for the mass and spin of the remnant. 

Let us assume that the frequencies extracted at the peak with model \Mr{1} are physical QNMs.\footnote{This assumption is implicit in the discussion of Ref.~\cite{Isi:2019aib}. However, the low SNR of GW150914 implies that the measurement is dominated by statistical errors, and therefore the inferred mass and spin can be compatible with GR. Our discussion here focuses on systematic errors, that will dominate at large SNR.}
In Fig.~\ref{fig:aM_inf} we compute the remnant mass and spin inferred from each overtone under this assumption.
The $(M_f,\chi_f)$ values inferred by assuming that the free mode $n_f=0$ corresponds to the $(2,2,0)$ mode converge monotonically towards the remnant BH value as $t_0$ increases, with $\epsilon<0.01$ after $t_0 \gtrsim 11M$ (solid blue line in the bottom panel). The addition of mode mixing improves the recovered values only after $t_0 \gtrsim 11M$ (dashed blue line).

On the other hand, the values obtained through the $n_f=1$ mode assuming that it corresponds to the $(2,2,1)$ mode are significantly biased ($\epsilon>0.1$) until $t_0 \sim 15M$. Around $t_0 \sim 18M$, and for a short interval $\Delta t \sim 1M$ around it, $\epsilon < 0.01$, but the accuracy in the inferred mass and spin decreases at later times. 
The late-time inference of mass and spin with this second mode can be marginally improved by including mode mixing  (model \Mrm{1}): this keeps $\epsilon\lesssim 0.1$ even at late times.
In any case, this experiment clarifies that {\em even in the infinite SNR limit, it is not easy to get percent-level estimates of the remnant parameters with the first overtone in an agnostic search.}

The hollow crosses in the top panel of Fig.~\ref{fig:aM_inf} show that when the fit starts at the peak, both the mass and the spin are significantly overestimated. 
Note also that in this exercise we find mass and spin by inverting each of the two modes independently. This is different from Ref.~\cite{Isi:2019aib}, where {\em both} frequencies were assumed to correspond to their GR values for a given $(M_f,\chi_f)$.  

This exercise has an interesting implication for data analysis. At the relatively low SNR of GW150914, the errors on the recovered parameters are so large that the \Mr{1} model gives complex frequency estimates compatible with the true quasinormal frequencies~\cite{Isi:2019aib,Finch:2022ynt}.
However our slightly generalized version of the test,
where we relax the assumption that all modes are related to mass and spin as in GR, reveals that the \Mr{1} model is effective at reducing the mismatch at the peak, but it does not reproduce the physical QNM content of signal.

Reference~\cite{Isi:2019aib} claimed evidence for the first overtone, in part, because the \Mr{1} model yields better estimates of the remnant mass and spin than the \Mr{0} model at the peak. As we have shown in Fig.~\ref{fig:wri}, the complex frequency extracted with the \Mr{0} model at the peak is indeed biased because of the low-frequency merger signal.  It is true, as claimed in Ref.~\cite{Isi:2019aib}, that the \Mr{1} model yields better estimates of the remnant's parameters. However this is not because the additional exponential gives a good estimate of the first overtone frequency, but simply because it fits away the low-frequency merger. This allows the {\em fundamental mode} to be extracted more accurately, improving the remnant mass and spin estimate.

\begin{figure*}[t]
    \includegraphics[width=\textwidth]{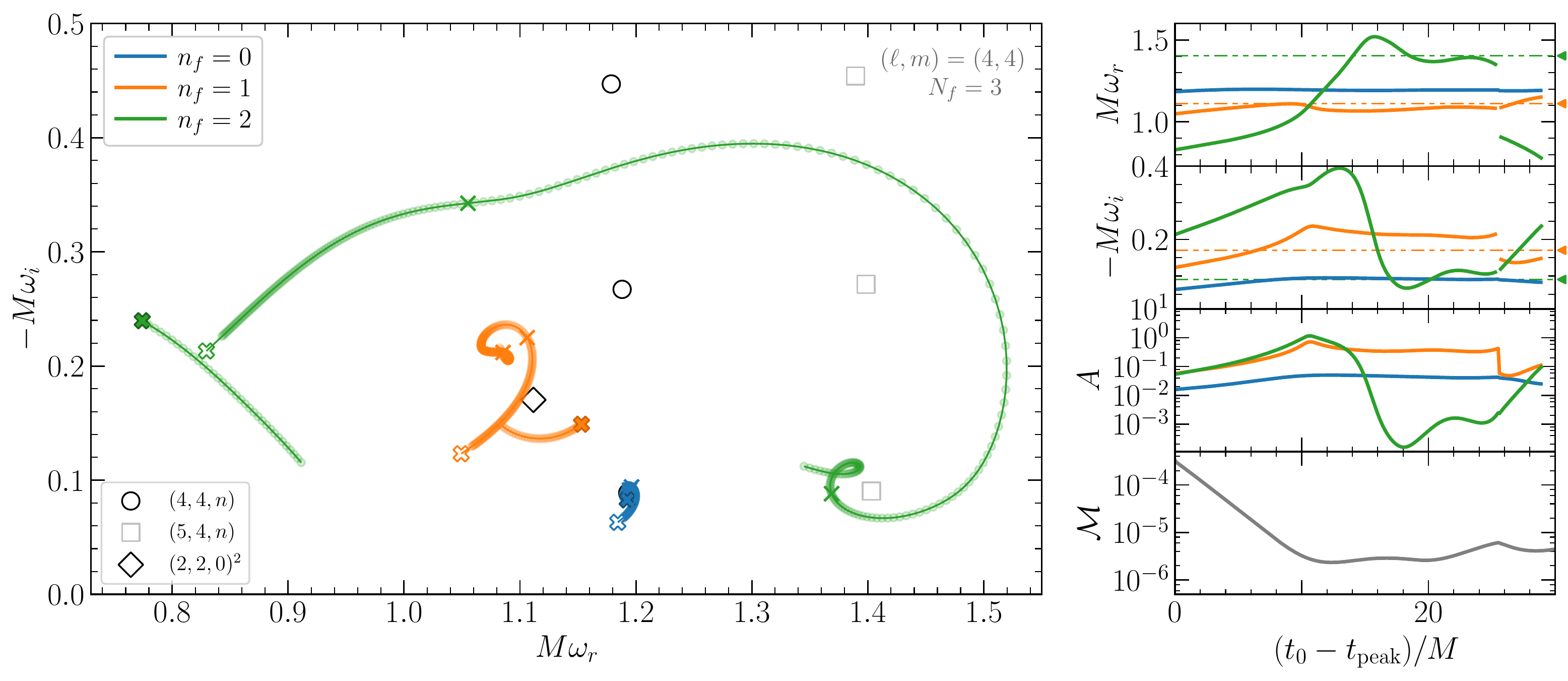}
    \caption{Fits of the $(4,\,4)$ multipole using the \Mr{2} model. 
    Left panel: the extracted frequencies 
    $(M\omega_r, \, - M\omega_i )$ for $n_f=0,\,1,\,2$, with $t_0$ ranging from $t_{\rm peak}$ (empty cross) to $t_{\rm peak}+30 M$ (filled cross); other crosses correspond to times $t_{\rm peak}+10 M$ and $t_{\rm peak}+20 M$.
    We mark the QNMs $(4,4,n)$ (black circles), $(5,4,n)$ (gray squares) and the quadratic mode $(2,2,0)(2,2,0)$ (black diamond). 
    Right panels: $M\omega_{r}$, $-M\omega_{i}$, $A$ and ${\mathcal M}$ as functions of $t_0$. The horizontal orange (green) dot-dot-dashed lines and triangle markers correspond to the expected frequencies for $M\omega_{220\times220}$ ($M\omega_{540}$).}
    \label{fig:wri3_44}
\end{figure*}

While the \Mr{1} model does not allow us to extract the first overtone at $t_0=t_{\rm peak}$, leading to large deviations in the inferred values of $(M_f,\,\chi_f)$, one might argue that the addition of more free modes could improve the situation.
This is not the case: the agnostic models we explored do not lead to an unbiased extraction of the first overtone at the peak. This is illustrated, for example, by Fig.~\ref{fig:wri4} in Appendix~\ref{sec:systematics}, where we fit the $(2,\,2)$ multipole with as many as $5$ damped sinusoids while also including mode mixing. Even with such a large number of modes, and despite the mismatch ${\mathcal M}\sim{\mathcal O}(10^{-9})$ being so small that one runs into the risk of fitting numerical noise, we cannot extract the first overtone at $t_0=t_{\rm peak}$.

It is by now common practice to fit GW data with the fundamental mode plus one or two overtones, treating the remnant mass and spin (along with the mode amplitude and phase) as free parameters. Many studies claim evidence for an overtone when (i) fits with either one or two overtones lead to values of $(M_f,\,\chi_f)$ which are in better agreement with the values extracted from the full waveform than fits using only the fundamental mode; and (ii) the amplitude of the additional overtone (or overtones) is nonzero~\cite{Isi:2019aib,Isi:2021iql,Finch:2021qph,Isi:2022mhy,Finch:2022ynt,Ma:2023vvr,Ma:2023cwe}. As we argued in Sec.~\ref{sec:postpeaklinearity},  the fact that adding QNMs leads to a better extraction of mass and spin does not imply that these additional QNMs are physical. This is confirmed, e.g., by the plots in the central column ($N=1$) of Fig.~\ref{fig:wri}. By adding an extra mode (orange line) we can better fit the low-frequency merger signal at early times, which leads to a more accurate extraction of the fundamental mode and improves the estimate of $(M_f, \, \chi_f)$ -- but not because of any physical contribution from the first overtone. It should also be clear that a nonzero amplitude of the overtone at the peak is not sufficient to claim detection of the overtone, as the additional QNM is simply (over)fitting the low-frequency merger signal. To make any credible detection claims, overtones should be extracted in an agnostic fashion.

There is another important caveat. In this whole analysis we focused on the dominant $(2, 2)$ mode, but real data will also contain multipoles such as (4, 4), which are usually dominant over the $(3,2,0)$ mode (with amplitude $A_{320}\sim4\times10^{-3}$). This will inevitably complicate the problem in ways that should be addressed in future work. 

The main goal of this section is to point out two important limitations of BH spectroscopy using only overtones: (1) at early times, we do not observe the excitation of the first overtone; instead, the second mode in a free-frequency fit tries to pick up the low-frequency merger signal.
(2) while the fundamental mode gives reliable estimates of the remnant mass and spin at late times, the first overtone never allows for percent-level estimates of the mass and spin; it may allow for $\sim 10\%$ estimates of the mass and spin at late time for loud enough signals, but only if we carefully take into account mode mixing and higher multipoles.

\subsection{Higher multipoles: overtones or nonlinearities? The example of $(\ell, m)=(4,4)$}
\label{sec:lm44}

It is instructive to apply the techniques we deployed for the $(2,2)$ multipole to higher multipoles of the radiation. To improve readability, we discuss the full waveform including higher multipoles in Appendix~\ref{sec:sumharmonics}. There we observe that:
(1) BH spectroscopy with overtones {\em at early times} (assuming they are excited) requires good modeling of the low-frequency merger signal; 
(2) BH spectroscopy with overtones {\em at late times} should be done after accounting for other multipoles, which can often be more robustly extracted from the waveform.

Here we single out a particularly interesting example: the $(4,4)$ multipole, which is known to contain a quadratic mode $(2,2,0)(2,2,0)$ sourced by the square of the $(2,2,0)$ mode~\cite{Ma:2022wpv,Mitman:2022qdl, Cheung:2022rbm}. 
We fit this multipole with the \Mr{2} model ($N_f=3$ free modes). In Fig.~\ref{fig:wri3_44} we plot the three extracted frequencies, their amplitudes, and the corresponding mismatches. 

The fundamental mode $(4,4,0)$ can be extracted with $\delta\omega<0.01$ after $t_0-t_{\rm peak}>5 M$, and an average $\delta\omega\sim \mathcal{O}(10^{-3})$ in the time range shown. 
The quadratic mode $(2,2,0)(2,2,0)$ can be identified with $\delta\omega\sim 10^{-2}-10^{-1}$~\cite{Cheung:2022rbm,Mitman:2022qdl}. 
Its frequency is fairly well recovered, while the damping time tends to be overestimated compared to the predicted value.  
The third free mode starts off at lower $M\omega_r$, evolving towards the $(5,4,0)$ mode in the time range $t_0-t_{\rm peak}=15-25M$, and moving to lower frequencies again at later times. 
We do not find any evidence of excitation of the first overtone. 
This conclusion holds even if we fix the lowest modes, $(4,4,0)$ or $(5,4,0)$, to their values as predicted in GR.

The main message of this analysis is clear: {\em the nonlinear mode is easier to recover in a free-frequency fit than the first linear overtone.}

\begin{figure*}[t]
    \includegraphics[width=\textwidth]{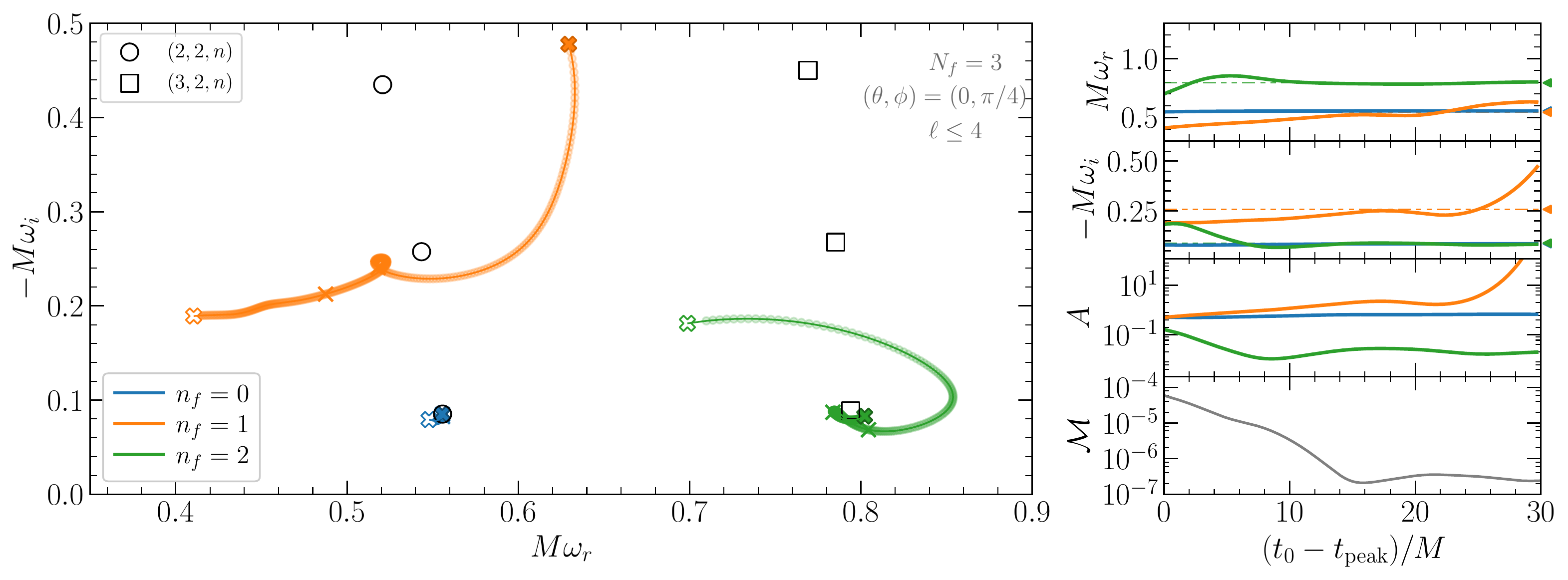}
    \caption{Fitting $h(t, \theta, \phi)$ at  $(\theta, \phi)= (0, \pi/4)$ using the \Mr{2} model ($N_f=3$).  Left panel: the extracted frequencies 
    $( M \omega_r, \, - M \omega_i )$ for $n_f=0,\,1,\,2$, with $t_0$ ranging from $t_{\rm peak}$ (empty cross) to $t_{\rm peak}+30 M$ (filled cross). We also mark $t_{\rm peak}+10 M$ and $t_{\rm peak}+20 M$ by crosses.
     Right panels: $M \omega_{r}$,  $-M \omega_{i}$, $A$ and ${\mathcal M}$ as a function of $t_0$. We show $M \omega_{220}$, $M \omega_{221}$, $M \omega_{320}$ as blue, orange and green dot-dot-dashed horizontal lines and triangle markers.}
    \label{fig:wri3_sum}
\end{figure*}

\begin{figure*}[t]
    \includegraphics[width=\textwidth]{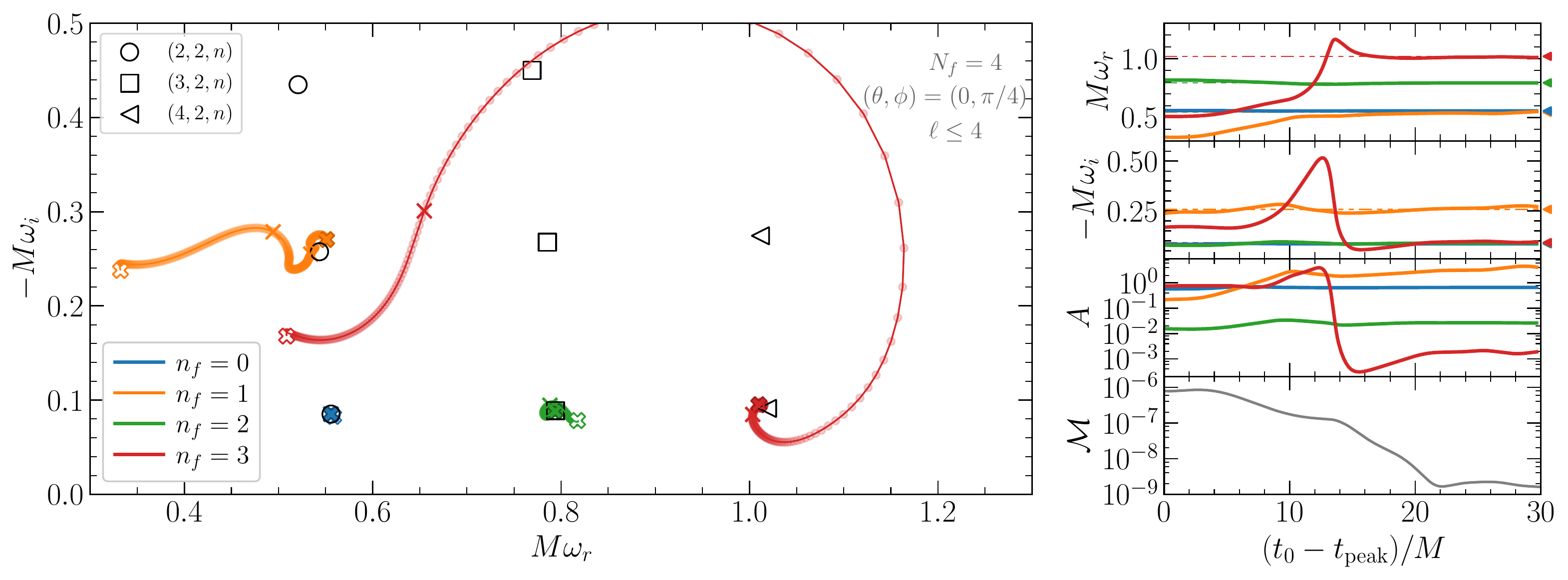}
    \caption{Fitting $h(t, \theta, \phi)$ at  $(\theta, \phi)= (0, \pi/4)$ using the \Mr{3} model ($N_f=4$).  Left panel: the extracted frequencies 
    $(M \omega_r, \, - M \omega_i )$ for $n_f=0,\,1,\,2,\,3$, with $t_0$ ranging from $t_{\rm peak}$ (empty cross) to $t_{\rm peak}+30 M$ (filled cross). We also mark $t_{\rm peak}+10 M$ and $t_{\rm peak}+20 M$ by crosses.
     Right panels: $M \omega_{r}$, $-M \omega_{i}$, $A$ and ${\mathcal M}$ as a function of $t_0$. We show $M \omega_{220}$, $M \omega_{221}$, $M \omega_{320}$, $M \omega_{420}$ as blue, orange, green and red dot-dot-dashed horizontal lines and triangle markers.}
    \label{fig:wri4_sum}
\end{figure*}

\section{Can the first overtone be extracted in the presence of subdominant multipoles?}
\label{sec:sumharmonics}

In Sec.~\ref{subsec:agnostic} we discussed how spherical-spheroidal mixing of the $(3,\,2)$ multipole affects the extraction of the first overtone from the dominant $(2,\,2)$ multipole. 
We concluded that at least $N_f=3$ free modes are required to extract the first overtone and that it is easier to extract (long-lived) fundamental modes than fast decaying overtones, even if the latter have a much larger amplitude. 
In real-world data analysis problems, the waveform will in general be a superposition of several multipoles.
We now ask: how many free modes would be necessary to extract the first overtone $(2,\,2,\,1)$ when the actual waveform is a superposition of several multipoles?

We fit the waveform
\be
h(t, \theta, \phi)=\sum_{\ell m} h_{\ell m}(t) {}_{-2}Y_{lm}(\theta ,\phi )
\ee
using the \Mr{N} model. 
For simplicity we focus on $\ell\leq4$. We quantify the relative contribution of subdominant multipoles through the ratio
\be
R_{\ell m} =\f{\max_{t} (h_{\ell m}(t) {}_{-2}Y_{lm}(\theta ,\phi ))}{\max_{t} (h_{22}(t) {}_{-2}Y_{22}(\theta ,\phi ))}\,.
\ee

We consider two scenarios: $(\theta, \phi)= (0, \pi/4)$ (``GW150914-like scenario'') and $(\theta, \phi)= (\pi/9, \pi/4)$ (``higher-multipole scenario'').

\subsection{Face-on, GW150914-like scenario}

If for simplicity we assume $(\theta, \phi)= (0, \pi/4)$, a face-on configuration compatible with the orientation measured for GW150914~\cite{LIGOScientific:2016vlm}, we find that the only observable spherical multipoles are $(2,\,2)$, $(3,\,2)$ and $(4,\,2)$, with $R_{32}=0.08$ and $R_{42}=0.006$.

In Fig.~\ref{fig:wri3_sum} we show the extracted frequencies, along with their amplitudes and the mismatch, obtained by fitting $h(t, 0, \pi/4)$ with the \Mr{2} model (i.e. $N_f=3$ free modes). 
While we recover the $(3,\,2,\,0)$ mode, {\em the first overtone $(2,\,2,\,1)$ cannot be reliably extracted.}
This is somewhat surprising, because the \Mr{2} model was sufficient to extract the first overtone from the $\ell=m=2$ multipole (compare the central row of Fig.~\ref{fig:wri}).

In Fig.~\ref{fig:wri4_sum} we make the obvious next step, and we try again to fit $h(t, 0, \pi/4)$, but this time with the \Mr{3} model. 
We now find that the first overtone (orange line) can be identified with sufficient accuracy at $\sim t_{\rm peak} + 12 M$. However, the $(3,2,0)$ mode (green line) can be extracted more accurately and at earlier times, and therefore this may be a more natural ``second mode'' to choose when performing tests of GR.
Besides (as expected) the recovery of the $(3,2,0)$ mode improves as we include more free modes in the fit, moving from \Mr{2} to \Mr{3}.
As usual, this is because the additional free mode fits other early-time contributions which, if not accounted for, can harm the recovery of the $(3,2,0)$ mode.
In fact, the fourth free mode in the \Mr{3} model recovers the $(4,2,0)$ spherical-spheroidal mixing mode frequency (red line) at about the same $t_0$ at which the $(2,2,1)$ mode is found, and with comparable accuracy. The $(4,2)$ multipole -- despite having a small relative amplitude, $R_{42}={\mathcal O}(10^{-3})$ -- can affect the extraction of the $(2,2,1)$ overtone.

{\em Even in the relatively optimistic case of a face-on binary, when only the $(3,2)$ and $(4,2)$ multipoles significantly contribute to the strain, the extraction of the first overtone $(2,\,2,\,1)$ relies on the successful extraction of the (long-lived) fundamental modes of subdominant multipoles.}

\begin{figure*}[t]
    \includegraphics[width=\textwidth]{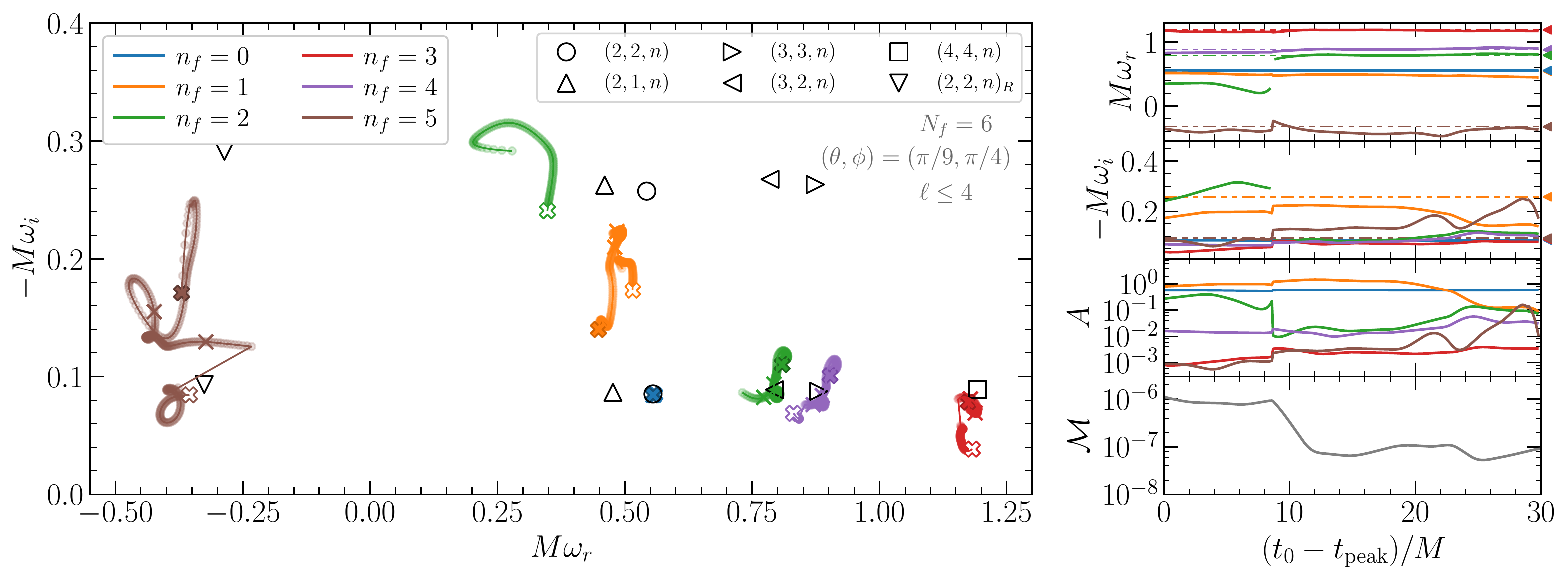}
    \caption{ Fitting $h(t, \theta, \phi)$ at  $(\theta, \phi)= (\pi/9, \pi/4)$ using the \Mr{5} model ($N_f=6$).  Left panel: the extracted frequencies 
    $( M \omega_r, \, - M \omega_i )$ for $n_f=[0,\,\dots,\,5]$, with $t_0$ ranging from $t_{\rm peak}$ (empty cross) to $t_{\rm peak}+30 M$ (filled cross). We also mark $t_{\rm peak}+10 M$ and $t_{\rm peak}+20 M$ by crosses.
   Right panels : $M \omega_{r}$, $-M \omega_{i}$, $A$ and ${\mathcal M}$ as a function of $t_0$. We also plot $M \omega_{220}$, $M \omega_{221}$, $M \omega_{320}$, $M \omega_{330}$, $M \omega_{440}$, and the retrograde mode $M \omega_{220,\mathrm{R}}=-M \omega^*_{2-20}$ as blue, orange, green, purple, red and brown dot-dot-dashed lines and triangle markers.}
    \label{fig:wri5_sum}
\end{figure*}

\subsection{A non-face-on binary}

The face-on scenario considered above is idealized and somewhat optimistic, since the assumption that $\theta=0$ artificially removes many multipolar components. For more generic binary orientations we would expect more multipoles to contribute to the strain, and therefore the recovery of the first overtone should be more difficult. 

We now consider a more realistic case where the inclination is not exactly face-on: $(\theta, \phi)= (\pi/9, \pi/4)$. 
Now the dominant multipoles have $R_{32}=0.06$, $R_{33}=0.03$, $R_{44}=0.01$, $R_{21}=0.01$, and in addition there are $5$ multipoles with $10^{-3}<R_{\ell m}<10^{-2}$.

In Fig.~\ref{fig:wri5_sum} we show the extracted frequencies, amplitudes and mismatches when we fit $h(t, \pi/9, \pi/4)$ with the \Mr{5} model. 
Now the first overtone cannot be extracted even when we include $N_f=6$ free modes: the only modes that can be cleanly extracted are $(2,2,0)$, $(3,3,0)$, $(3,2,0)$ and $(4,4,0)$. 
The free mode labeled by $n_f=1$ (orange line) approaches the $(2,1,0)$ mode, but the model fails to extract this mode at times $t_0-t_{\rm peak}<30 M$.  The $n_f=5$ free mode (brown line) tries to fit a ``retrograde'' mode, that we label by $(2,2,0)_R$. Here, following Ref.~\cite{MaganaZertuche:2021syq}, we call ``retrograde'' those QNMs that have frequency $M\omega_{\ell mn,\mathrm{R}}=-M \omega^*_{\ell -m n}$. In Appendices~\ref{sec:systematics} and~\ref{subsec:lm21} we show that retrograde modes are present also in the $(2,2)$ and $(2, 1)$ multipoles, respectively.

The case we studied in this section is still an optimistic scenario for overtone detection for two main reasons.

First of all, GW150914-like signals have nearly equal masses, so that the subdominant multipoles have relatively small amplitudes. Even in this ``optimal'' case, the extraction of the $(2,2,1)$ overtone is severely affected by higher multipoles. The extraction of the $(2,2,1)$ overtone will be more difficult (while the extraction of subdominant multipoles will become easier) for binary configurations with asymmetric masses and large spins, where higher multipoles are more excited. For this reason, asymmetric systems are better targets for BH spectroscopy.

Secondly, if we assume random orientations, more than $94\%$ of the sources have $\theta>\pi/9$. As we increase $\theta$, the contribution of subdominant multipoles increases, and isolating the $(2,2,1)$ overtone gets harder. In fact, we have experimented and found that the $(2,2,1)$ overtone cannot be extracted with good accuracy at even smaller inclinations, such as $\theta=5\degree (=\pi/36)$.

{\em We conclude that in  practical scenarios, it would be hard to extract the first overtone from GW150914-lke signals.}
This is because at the times $t_0\gtrsim t_{\rm peak} + 12 M$ when the $(2,\,2,\,1)$ mode is recovered by the fits, the fundamental modes of subdominant multipoles are usually dominant.
{\em Even when the overtone can be confidently identified, the fundamental modes of subdominant multipoles are probably more reliable for performing BH spectroscopy tests.}

These considerations ignore detector noise. Depending on the mass and spin of the remnant, noise could further reduce the ``effective'' amplitude of subdominant mutipoles, impeding their extraction from the data. Our inability to extract these subdominant multipoles can significantly affect the extraction of the $(2,2,1)$ overtone.

\section{Conclusions and discussion}
\label{sec:discussion}

The examples discussed so far show that the inclusion of a large number of overtones in BH spectroscopy has no physical justification. Claims that one or more overtones can be observed at the peak of the waveform are theoretically inconsistent with the analysis of NR waveforms. In fact, the analysis of the previous section leads to an unavoidable conclusion: {\em BH spectroscopy using overtones is either unfeasible or suboptimal compared to spectroscopy using additional multipoles.}

\subsection{Observational implications}

Independently of the current debate on the statistical significance of high-overtone fits to the GW150914 data~\cite{Isi:2019aib,Isi:2021iql,Cotesta:2022pci,Isi:2022mhy,Finch:2022ynt,Ma:2023vvr,Ma:2023cwe}, our results demonstrate that fits with multiple modes starting at $t_{\rm peak}$ do not correspond to the physical excitation of the QNMs of the remnant, but are simply effective at overfitting the low-frequency components of the waveform near merger.
As shown in Appendix~\ref{sec:frankenstein}, these models would give rise to false evidence in favor of overtones even when the post-peak signal is unphysical and overtones should not be present.
Any interpretation of overtone measurements close to the peak (i.e., outside the regime of validity of linear perturbation theory) is devoid of physical meaning and leads to unreliable BH spectroscopy tests.

Tests based on early-time overtone models are conceptually equivalent to parametrized tests where low-frequency effective components are used to model the near-merger signal.
Such tests have already been performed by the LIGO-Virgo collaboration at the time of the discovery of GW150914~\cite{LIGOScientific:2016lio}, and they have been extended to the GW catalogs in subsequent analyses~\cite{LIGOScientific:2019fpa,LIGOScientific:2020tif,LIGOScientific:2021sio}.
More recently, modifications to merger quantities were considered within an effective-one-body model~\cite{Maggio:2022hre}.
These are certainly valuable investigations, but they cannot be regarded as BH spectroscopy tests: they do not probe directly the excitation of the QNMs of the remnant BH, but rather the nonlinear portion of the signal, whose structure is not analytically understood.

These arguments also apply to tests of the Hawking area law based on overtone measurements~\cite{Isi:2020tac}. For GW150914-like binaries, these tests are a time-domain reformulation of the inspiral-merger-ringdown consistency tests and they are equivalent to frequency-domain tests, since frequency leakage between pre- and post-merger is negligible (as shown in Ref.~\cite{Ghosh:2017gfp}).

\subsection{Pseudospectral instabilities and nonlinearities}

Our analysis raises some interesting questions. Why are we unable to extract higher overtones even in linear perturbation theory? Why are the overtone frequencies that we infer by fitting time-domain signals different from the expected values?

Some answers to these questions could come from recent work on the pseudospectrum.
Nollert demonstrated a long time ago that small changes in the potential can destabilize the whole QNM spectrum~\cite{Nollert:1996rf} (see also~\cite{Barausse:2014tra}). Recent calculations of the so-called ``pseudospectrum'' in the context of BH physics put these ideas on a more solid mathematical foundation, showing that the overtones are more easily destabilized than the fundamental mode (while claiming that the fundamental mode is, in fact, stable under certain classes of perturbations)~\cite{Jaramillo:2020tuu}. Later work showed that even the fundamental mode is not immune from spectral instabilities~\cite{Cheung:2021bol}. In practice, this is unlikely to affect our ability to do BH spectroscopy {\em with the fundamental mode} in observationally relevant scenarios~\cite{Berti:2022xfj}.

What is still unclear is whether spectral instabilities affect overtones, and to what extent. The intrinsic nonlinearity of GR implies that the linear wave equations are {\em always} affected by second-order terms that can be treated as perturbations~\cite{Gleiser:1995gx,Campanelli:1998jv,Brizuela:2009qd,Ioka:2007ak,Nakano:2007cj,Pazos:2010xf,Loutrel:2020wbw,Ripley:2020xby}.
There has been some encouraging progress in our understanding of nonlinearities in BH physics~\cite{Bosch:2016vcp,Bosch:2019anc,Sberna:2021eui,Lagos:2022otp}.
However several open questions remain. Are the perturbations induced by nonlinear effects sufficient to destabilize most overtones in the context of comparable-mass BH mergers? Can the spectral instability of the overtones explain our failure to find them in time-domain waveforms? Do these spectral instabilities explain why overtones are so hard to identify?

\subsection{Amplitude parametrizations and Occam penalties}

Besides being physically inconsistent for reasons we have already discussed, multiple overtone models with free amplitudes are problematic from the point of view of data analysis.
Depending on the SNR of the signal, it may be necessary to include several overtones to obtain unbiased inference. 
However, each QNM comes with two additional free parameters: an amplitude and a phase. As pointed out in Ref.~\cite{Bustillo:2020buq}, a large number of parameters generally decreases the model evidence due to the associated Occam penalty. Reference~\cite{Bustillo:2020buq} concluded that even for extremely loud signals, the Occam penalty would not allow us to use overtones to test the ``no-hair'' hypothesis. It also reported a significant increase in the Bayesian evidence if one uses phenomenological templates rather than an agnostic superposition of damped sinusoids.

In principle, this problem can be alleviated by parametrizing the overtone amplitudes as function of the parameters of the progenitors.
For example, reference~\cite{London:2014cma} presented a fit of the first overtone amplitude as a function of binary parameters for nonspinning progenitors valid at times $t_0 > 10 \, M_f$ (see also Ref.~\cite{London:2018gaq}).
However, later work found that when we try to push such ``global'' fits to earlier times, the result is not robust under variations of $t_0$: see e.g. Figs.~3 and 4 of Ref.~\cite{JimenezForteza:2020cve}.
Even if we can model the overtone amplitudes as functions of the properties of the remnant progenitors, measuring several overtone frequencies in real data may be impractical. Fisher matrix estimates suggest that it is easier to obtain evidence for multiple modes using higher angular multipoles rather than overtones~\cite{JimenezForteza:2020cve}. 
These conclusions are in disagreement with Ref.~\cite{Isi:2021iql}, which however assumed the overtone model to be valid at the peak and employed a different detection criterion. 

In future work we will systematically rank the most important QNMs in the spectrum as functions of the progenitor parameters, investigate strategies for a robust modeling and measurement of multiple modes, and revisit previous BH spectroscopy horizon estimates~\cite{Berti:2016lat,Ota:2021ypb}.

\subsection{Weak and strong ``no-hair'' tests}

Recent observational implementations of BH spectroscopy tests are quite different in nature~\cite{Carullo:2019flw,Isi:2019aib,LIGOScientific:2020tif,Isi:2021iql,Capano:2021etf,LIGOScientific:2021sio,Cotesta:2022pci,Isi:2022mhy,Finch:2022ynt}.
Since the work of Ref.~\cite{Gossan:2011ha}, it has been customary to perform ``no-hair'' tests not by directly measuring multiple QNM frequencies -- as envisioned in the original BH spectroscopy proposals~\cite{Detweiler:1980gk, Dreyer:2003bv, Berti:2005ys} -- but rather by measuring the mass and the spin of the BH (the impact of charge is currently negligible~\cite{Carullo:2021oxn}) together with a single parametric deviation, e.g., $\delta\omega_{221}$.

This parametrization is sensible. The use of physical degrees of freedom reduces degeneracies, decreases the number of parameters in the test to three  (as opposed to the four real parameters that correspond to measuring two complex frequencies), and consequently decreases the SNR required for the test.
However, the parametrization only allows for a ``weak'' BH spectroscopy test, because we must assume \textit{a priori} which modes are present in the data. As such, it must be applied with great care.
This ``weak'' version of the test could be consistent with GR just because it enforces the presence of (spurious) low-frequency modes, which are effective at fitting nonlinearities or other physical artifacts near the peak.

Conversely, the original BH spectroscopy proposal relies on a ``strong,'' more conservative version of the test: we should perform an agnostic search for multiple damped exponentials, and only \textit{a posteriori} identify them with a given set of modes, provided that the results of the agnostic frequency search are robust enough.
This ``strong'' version was used in most of the fitting experiments reported in this manuscript, it does not impose the presence of a given mode in the data, and (while harder to implement) it leads to more robust conclusions.

The key take-away message for the purpose of this discussion is that if the mode is actually present in the data, in the large-SNR limit it must be possible to replace the ``weak'' form of the test by its ``strong'' version. 
As we have shown in Sec.~\ref{subsec:massspinest} above, if we carry out BH spectroscopy tests with one overtone close to $t_{\rm peak}$, \textit{this does not happen even in the infinite SNR limit}, i.e., when we fit state-of-the-art NR simulations. The opposite is true for tests that use the fundamental modes corresponding to different multipoles at later times.
In either form of the test, using the fundamental modes has an additional observational advantage: at times $t_0$ significantly larger than $t_{\rm peak}$, the statistical uncertainty on the starting time has a much smaller impact.
For sufficiently large SNR, fundamental modes with different values of $(\ell,\,m)$ can be robustly identified over multiple cycles, and errors $\Delta t_0 \sim {\cal O}(1)\, M_f$ have negligible impact on their recovery.
On the contrary, the starting time uncertainty has a large impact on overtones-based tests near the peak~\cite{Cotesta:2022pci, Finch:2022ynt}.

\subsection{Beyond-Kerr parametrizations: effective detection templates versus physical parametrizations}

Any BH spectroscopy tests relying on templates with a single parametric deviation, such as those in Ref.~\cite{LIGOScientific:2020tif}, are unphysical.
This is because virtually all modified-gravity theories are expected to change all of the QNM frequencies, not just one~\cite{Tattersall:2017erk,Tattersall:2018map,Cardoso:2019mqo,McManus:2019ulj, Volkel:2022aca, Li:2022pcy, Hussain:2022ins} (this discussion applies also to parametrized tests in the inspiral: see e.g. Sec.~II of Ref.~\cite{Carullo:2021dui}).
However, it is important to make a distinction between two different goals: the {\em detection} of beyond-GR effects, and the identification of their origin.

We can safely use unphysical models as effective ``detection templates'' with the minimal number of parameters required to detect hypothetical deviations from GR. 
While using multiple parameters may become beneficial in the context of next-generation detectors~\cite{Gupta:2020lxa,Datta:2022izc}, adding too many parameters can introduce correlations, increase the SNR required for a detection of beyond-GR effects, and decrease the sensitivity of the search. 
In addition, single-parameter templates have been shown to be effective at recovering deviations which affect multiple parameters~\cite{Johnson-McDaniel:2021yge}.
If future observations using these simple parametric tests show any hints of deviations from GR, they should be followed up with better motivated templates to identify their origin.

One proposal to bypass this problem in ringdown tests is to use parametrized frameworks such as {\sc ParSpec}~\cite{Maselli:2019mjd,Carullo:2021dui}. The idea is to perform a perturbative expansion around GR, in which corrections to the individual QNM frequencies are proportional to the (small) couplings that appear in the action of beyond-GR theories.
This is a sensible parametrization that introduces additional information, and increases the sensitivity of searches for new physics at fixed SNR~\cite{Carullo:2021dui}.
The sensitivity can be further increased for specific beyond-GR theories~\cite{Silva:2022srr}, which can be easily embedded within a {\sc ParSpec}-like framework in the small-coupling limit.
A valid objection to this approach is that {\sc ParSpec}-like frameworks, being intrinsically perturbative, could miss nonperturbative deviations from the GR spectrum.
This is true but perhaps unlikely, given mounting evidence that the observed GW signals must be quite close to the predictions of GR.
It may be advisable to run two parallel searches: a more sensitive, {\sc ParSpec}-like search for perturbative deviations, and a less sensitive (but more general) search for nonperturbative signatures.

\subsection*{Note added}
{Soon after this manuscript appeared on the arXiv, a preprint by Nee et al.~\cite{Nee:2023osy} studied the excitation of overtones by comparing Regge-Wheeler and P\"oschl-Teller potentials. Their findings are similar to those presented in Sec.~\ref{sec:linearovertones}: ``\ldots large overtone numbers may [...] remove non-quasi-normal mode contributions that are relevant at early times of a ringdown, but do not necessarily correspond to the physical excitation of modes of the system''~\cite{Nee:2023osy}.}

\acknowledgments

We are grateful to Swetha Bhagwat, Yanbei Chen, Kyriakos Destounis, Will Farr, Pedro Ferreira, Xisco Jim\'enez Forteza, Matt Giesler, Lam Hui, Max Isi, Danny Laghi, Macarena Lagos, Paul Lasky, Lionel London, Sizheng Ma, Keefe Mitman, Paolo Pani, Rodrigo Panosso-Macedo, Lorenzo Pierini, Leo Stein, Saul Teukolsky, Eric Thrane and Sam Wong for helpful discussions. 
M.H.Y.C. is a Croucher Scholar supported by the Croucher Foundation.
E.B., R.C. and M.H.Y.C. are supported by NSF Grants No. AST-2006538, PHY-2207502, PHY-090003 and PHY20043, and NASA Grants No. 19-ATP19-0051, 20-LPS20-0011 and 21-ATP21-0010.
This research project was conducted using computational resources at the Maryland Advanced Research Computing Center (MARCC). 
V.C.\ is a Villum Investigator and a DNRF Chair, supported by VILLUM Foundation (grant no. VIL37766) and the DNRF Chair program (grant no. DNRF162) by the Danish National Research Foundation. V.C.\ acknowledges financial support provided under the European
Union's H2020 ERC Advanced Grant ``Black holes: gravitational engines of discovery'' grant agreement
no.\ Gravitas–101052587. Views and opinions expressed are, however, those of the author only and do not necessarily reflect those of the European Union or the European Research Council. Neither the European Union nor the granting authority can be held responsible for them.
G.C. acknowledges support by the Della Riccia Foundation under an Early Career Scientist Fellowship.
G.C. acknowledges funding from the European Union’s Horizon 2020 research and innovation program under the Marie Sklodowska-Curie grant agreement No. 847523 ‘INTERACTIONS’, from the Villum Investigator program supported by the VILLUM Foundation (grant no. VIL37766) and the DNRF Chair program (grant no. DNRF162) by the Danish National Research Foundation.
F.D. acknowledges financial support provided by FCT/Portugal through grant No. SFRH/BD/143657/2019.
This project has received funding from the European Union's Horizon 2020 research and innovation programme under the Marie Sklodowska-Curie grant agreement No 101007855.
We thank FCT for financial support through Project~No.~UIDB/00099/2020.
We acknowledge financial support provided by FCT/Portugal through grants PTDC/MAT-APL/30043/2017 and PTDC/FIS-AST/7002/2020.
Part of E.B.’s work was performed at the Aspen Center for Physics, which is supported by National Science Foundation grant PHY-1607611.
The authors would like to acknowledge networking support by the GWverse COST Action CA16104, ``Black holes, gravitational waves and fundamental physics''.

\appendix

\section{Ambiguities in the definition of ringdown starting time and
  ringdown excitation}
\label{app:Green}

A ringdown waveform is formally {\it defined} as the superposition of damped exponential given in Eq.~\eqref{eq:model}. One of the main issues we address in this paper is the following: is this definition {\em ever} a good approximation of a given waveform, and to what level of approximation?

In this appendix we use a simple toy model and some examples from linear BH perturbation theory to clear up possible misunderstandings about (i) the very possibility to define a ringdown ``starting time,'' and (ii) the meaning of ``ringdown excitation.''

\subsection{A Green's function toy model}

The first observation is that conservative systems described by linear wave equations can always be described in terms of their eigenmodes. However, it is well known that a simple description in terms of constant-amplitude eigenmodes may not be adequate even in such systems. Consider for illustration the following wave equation in one spatial dimension, explored already in Refs.~\cite{ICTP_Cardoso,Lagos:2022otp}:
\be
\frac{\partial^2\Psi}{\partial x^2}-\frac{\partial^2\Psi}{\partial t^2}-2V_0\delta(x)\Psi=0\,.
\ee
Through a Laplace transform, $\psi(s,x)=\int_0^\infty e^{-s t} \Psi(t,x)dt$, we can transform the partial differential equation into the ordinary non-homogeneous equation
\beq
\frac{d^2\psi}{d x^2}+\left[\omega^2-2V_0\delta(x)\right]\psi&=&-\frac{\partial \Psi}{\partial t}\Big|_{t=0} +i\omega \Psi \Big|_{t=0}\nonumber\\
&\equiv& S\,,\label{non_simple}
\eeq
where we have defined the frequency $\omega=is$ and a source term $S$. Define the ``left'' homogeneous solution as that satisfying outgoing boundary conditions at large negative values of $x$,
\be
\psi_L=
\begin{cases}
  e^{-i\omega x}\,, \quad \quad \quad \quad \quad \quad \quad\,\,x\to-\infty\\
  A_{\rm in}e^{-i\omega x}+A_{\rm out}e^{i\omega x}\,,\quad x\to +\infty\\
\end{cases}
\ee
and the ``right'' homogeneous solution as
\be 
\psi_R=e^{i\omega x}\,, \quad x\to+\infty\,.
\ee
The (constant) Wronskian $W$ of these two solutions is $W\equiv \psi_L\psi'_R-\psi'_L\psi_R =2i\omega A_{\rm in}$. By demanding continuity of $\psi_L$ and a discontinuity in its derivative consistent with a $\delta$ function, we find
\beq
&&A_{\rm in}+A_{\rm out}=1\,,\\
&&A_{\rm out}-A_{\rm in}+1=\frac{2V_0}{i\omega}\,,
\eeq
which is equivalent to 
\be
A_{\rm out}=-i\frac{V_0}{\omega}\,,\quad A_{\rm in}=1+i\frac{V_0}{\omega}\,.
\ee
The Wronskian vanishes at the single zero $\omega=-iV_0$, which is therefore a QNM frequency, and the only one supported by this potential.

The solution of Eq.~\eqref{non_simple} satisfying outgoing boundary conditions as  $x\to \infty$ is
\be
\psi=e^{i\omega x}\int dx \frac{\psi_L S}{W}\,.
\ee

Consider for simplicity an initial pulse $S=\delta(x-x_0)$. The integral then yields
\be
\psi=\frac{e^{i\omega (x-x_0)}}{2}+e^{i\omega (x+x_0)}\frac{A_{\rm out}}{2A_{\rm in}}\,.
\ee
This is the full solution in the Fourier plane. To get the time-domain solution, we calculate the inverse,
\beq
2\pi\Psi&=&\int d\omega e^{-i\omega t}\psi \\
&=&\int d\omega\frac{e^{-i\omega (x-x_0-t)}}{2}+\int d\omega e^{-i\omega (x+x_0-t)}\frac{A_{\rm out}}{2A_{\rm in}}\,.\nonumber
\eeq

\begin{figure}[t]
    \includegraphics[width=\columnwidth]{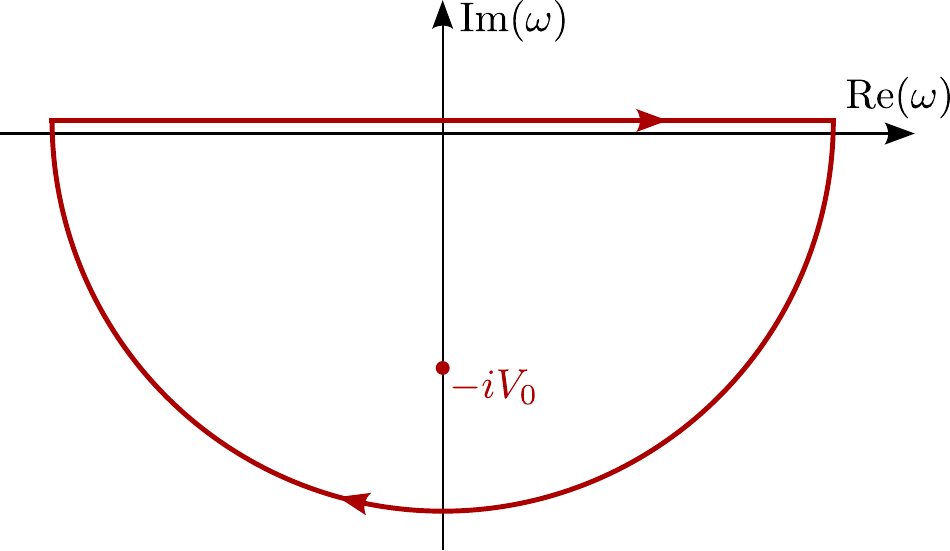}
    \caption{Integration contour to get the time response in the presence of a $\delta$-function potential and localized initial data.}
    \label{fig:contour_deltaV}
  \end{figure}
  
But $2\pi \delta(x-a)=\int d\omega e^{i\omega(x-a)}$, thus we have 
\beq
\Psi=\frac{1}{2}\delta(x-x_0-t)+\frac{1}{4\pi}\int d\omega e^{-i\omega (x+x_0-t)}\frac{-iV_0}{\omega+iV_0}\,.\nonumber
\eeq
The remaining integral is performed in the complex plane, by closing the contour of integration as shown in Fig.~\ref{fig:contour_deltaV}. There are no branch cuts.
For $t<x_0+x$, we can close the contour in the upper half of the complex plane. Since there are no poles in the upper half plane, the integral vanishes and we get
\be
\Psi=\frac{1}{2}\delta(x-x_0-t)\,,\quad t<x_0+x\,.
\ee
This is the {\em prompt response}: in flat space, signals travel at the speed of light (here $c=1$) on the light cone. Thus, the contribution above corresponds to a signal emitted from the source at $x_0$
and reaching the observer at $t=x-x_0$.

\begin{figure}[t]
    \includegraphics[width=\columnwidth]{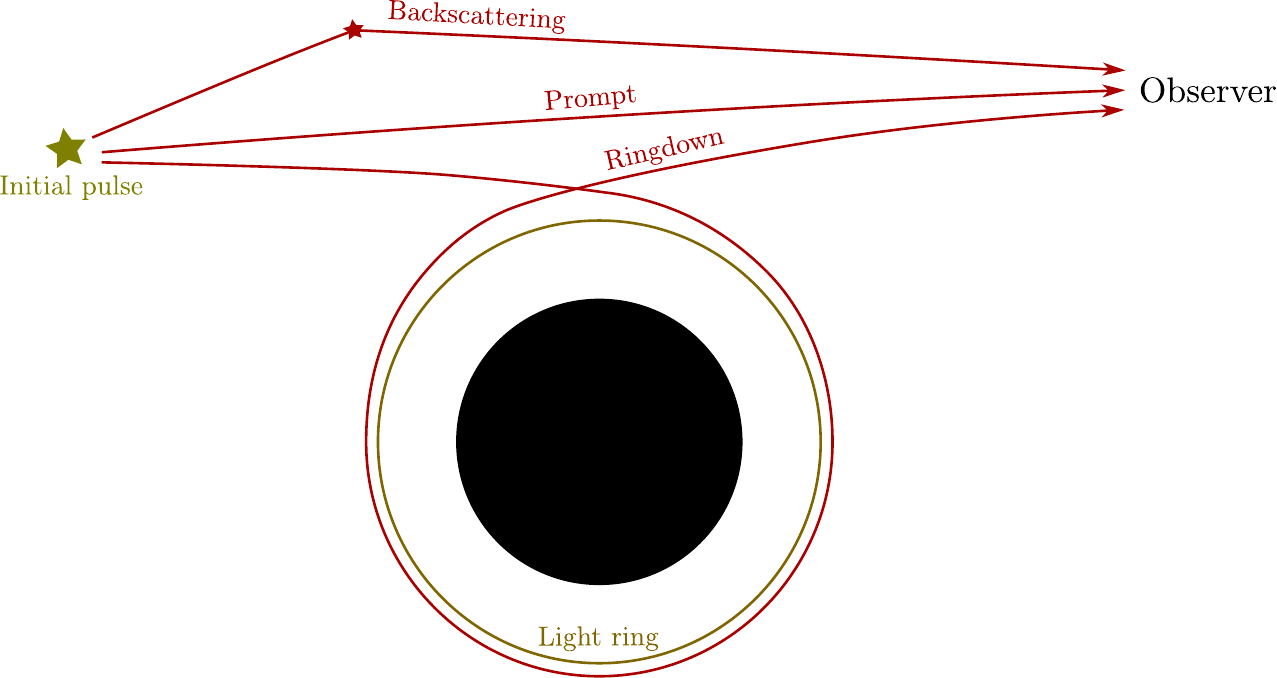}
    \caption{Schematic representation of the radiation emitted by sources moving around BHs and their path towards far-away observers. The yellow circle represents the light ring.}
    \label{fig:signal_phases}
\end{figure}

\begin{figure*}[ht!]
    \includegraphics[width=\columnwidth]{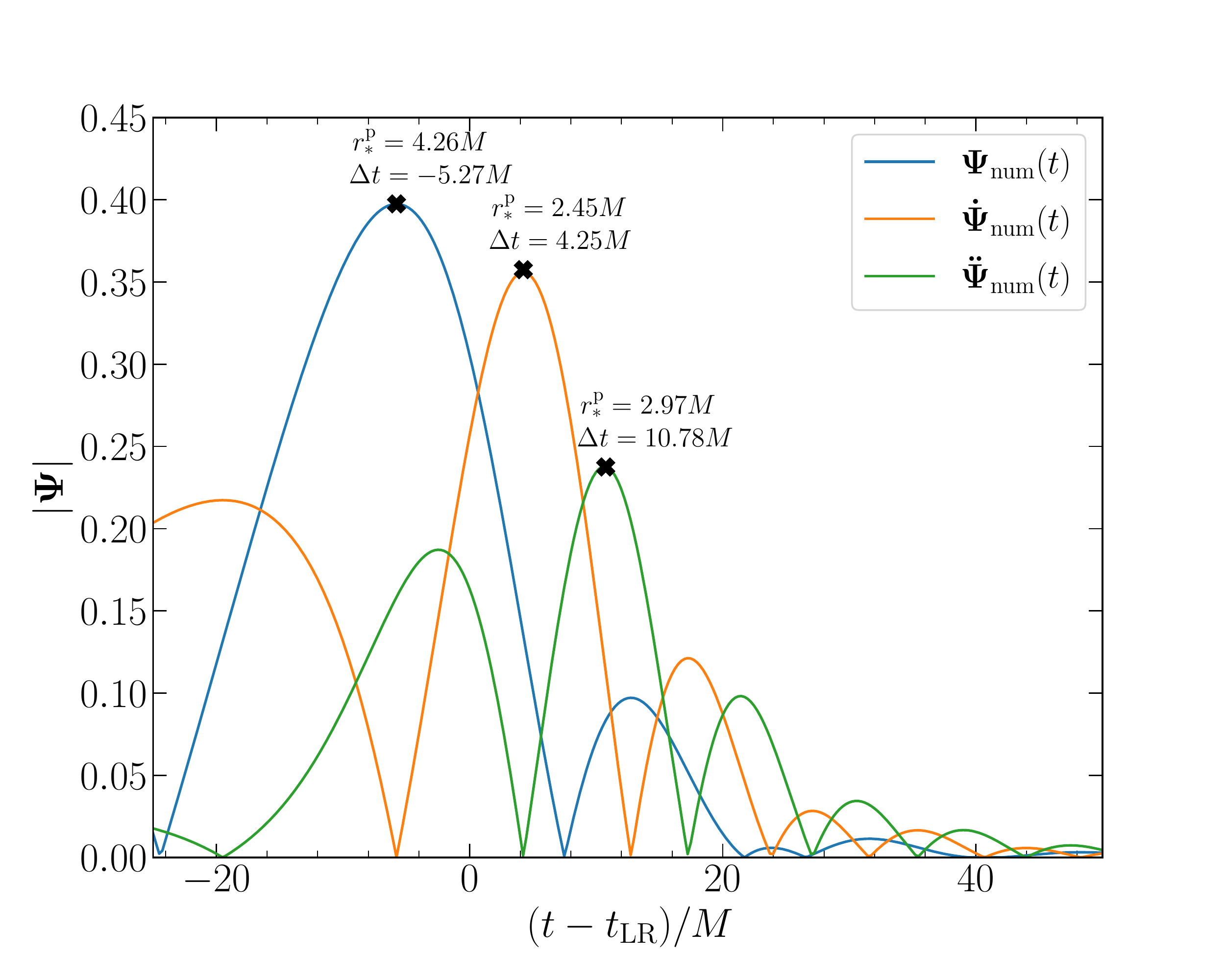}
    \includegraphics[width=\columnwidth]{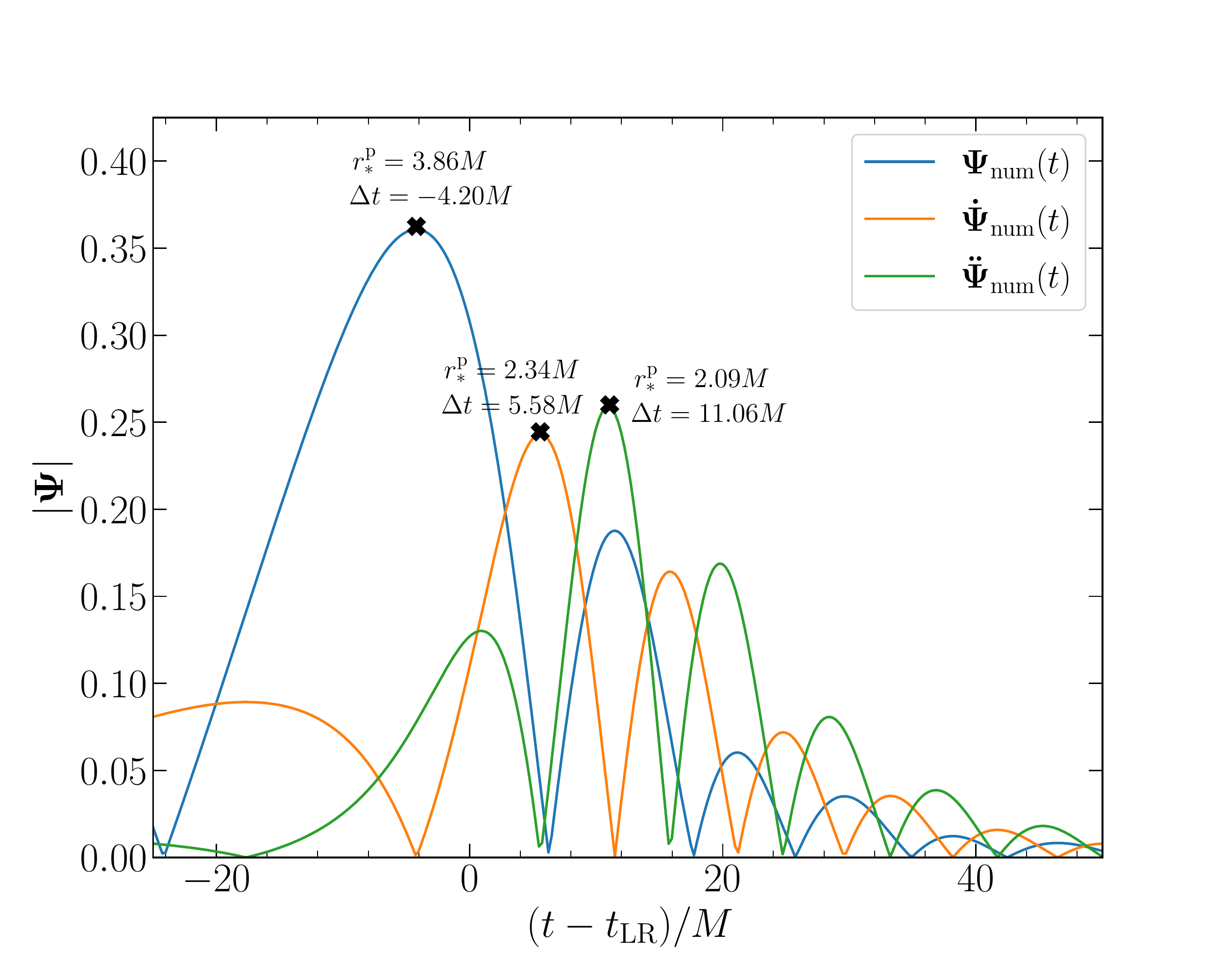} 
    \caption{Left panel: time evolution of the $\ell=2$ mode for the Zerilli function $\Psi$ (and its first and second time derivatives) extracted at null infinity, for initial data given by Eqs~\eqref{eq:ID_Zerilli}-\eqref{eq:ID_Zerilli1} with $r_*^{(0)} = 50M$ and $\sigma = 5.0M$. Here, $t_\text{LR}$ corresponds to the instant when the gaussian crosses the light ring, located at $r_\text{LR}=3M$. The first burst occurs slightly before $t_\text{LR}$, when part of the front of the wavepacket reaches the peak of the gravitational potential (which is close to the light ring), leading to the direct emission of radiation. The following peaks of $\Psi$ occur at time intervals close to half of the light-ring period  $T_\text{LR}/2 \sim 16M$ and correspond to the ringdown phase, where trapped waves at the light ring leak out both to infinity and into the BH.
    Note that the peaks in the time derivatives of the Zerilli function occur later.
    Right panel: same, but for $r_*^{(0)} = 50M$ and $\sigma = 0.5M$. Now the first burst of radiation appears later, since the gaussian is narrower. The ringdown is more excited, i.e., the relative magnitude of the second peak of $\Psi$ with respect to the first is larger. Thus, wide gaussians suppress the ringdown.}
    \label{fig:MaxZerilli}
\end{figure*}

For $t>x_0+x$, we must close the contour in the lower half of the complex plane. Taking the residue at the pole, we find
\be
\Psi=\frac{1}{2}\delta(x-x_0-t)+\frac{V_0}{2}e^{-V_0[t-(x+x_0)]}\,,\quad t>x_0+x\,.
\ee
Note that the second term peaks at $t=x_0+x$. It corresponds to the initial burst traveling to the potential barrier (which takes a time $x_0$) and then back to the observer at $x$, emerging after interaction with the potential barrier. The signal is exponentially suppressed afterwards at a rate determined by the single, purely damped QNM frequency.

In the simple scenario above we assumed sharp, localized initial data. The effect of a broadened pulse can be understood by adding an extra Dirac delta, so that the source $S=\delta(x-x_0)+\delta(x-x_1)$ with $x_1>x_0$. It is now easy to see that
\beq
\Psi&=&\frac{1}{2}\delta(x-x_0-t)+\frac{1}{2}\delta(x-x_1-t)\,,\quad t<x_0+x\\
\Psi&=&\frac{1}{2}\delta(x-x_0-t)+\frac{1}{2}\delta(x-x_1-t)+\frac{V_0}{2}e^{-V_0(t-(x+x_0))}\,,\nonumber\\
&&\qquad    \quad \qquad \quad \qquad \quad x_1+x>t>x_0+x \\
\Psi&=&\frac{1}{2}\delta(x-x_0-t)+\frac{1}{2}\delta(x-x_1-t)\nonumber\\
&+&\frac{V_0}{2}e^{-V_0(t-x)}\left(e^{V_0x_0}+e^{V_0x_1}\right)\,,\qquad \quad  t>x_1+x
\eeq
We see then two ``ringdown'' stages of different amplitudes, driven of course by the ``extended'' initial data. The implication is clear: {\em a constant-in-time QNM amplitude description is not valid for generic initial data, even when we consider conservative systems.}

In curved spacetimes, this picture is further complicated by the absence of propagation on the light cone: backscattering off the spacetime curvature produces tails, beyond the effects already described. A depiction of what happens when a source is emitting close to a BH is shown in Fig.~\ref{fig:signal_phases}. There is direct, prompt emission from the source to the observer, the analog of propagation on the light cone in Minkowski. Concurrent with this process, waves are continuously backscattered off the spacetime curvature, a low-frequency effect that is also present for massive fields or for spacetimes of odd dimensionality (see e.g.~\cite{Alsing:2011er,Cardoso:2002pa}).
In addition, light is trapped close to the potential barrier (i.e., close to the BH light ring) and continuously re-emitted. The sum of all these effects produces the signal seen by an observer. Once the source crosses the BH event horizon, direct emission ceases to exist, but re-emission by the light ring and backscattering can still occur.

There are two important lessons to be learned from this simple toy model: (1) {\em the signal is never expected to be described by a pure ringdown waveform} because other components (including prompt emission and backscattering) will always be present; (2) even in a Minkowski background and in the linear regime, {\em the ``ringdown starting time'' is not a well-defined quantity.} 

We further clarify this point below by considering a curved background.
The relative importance of the different emission components, and their effect in searching for damped sinusoids, is explored in more detail in Section~\ref{sec:linearovertones} of the main text with a toy model.

\subsection{When is ringdown excited?}

In a BH spacetime the situation becomes even murkier. To stress how nontrivial it is to define the ringdown starting time (or the ``ringdown stage,'' more in general), we repeat the classic Gedankenexperiment first performed by Vishveshwara~\cite{Vishveshwara:1970zz}.

We let a gaussian wavepacket scatter off a nonrotating BH, and we compare the location of the peaks of the signal with the respective position of the center of the gaussian at the emission of the direct signal, i.e., at time $t_\text{peak} = t + r^\text{ext}_* -  r^\text{p}_*$. Here $r^\text{ext}_*$ is the extraction location on the numerical grid (in terms of the Schwarzschild tortoise coordinate), $r ^{(0)}_*$ is the initial (tortoise) radius of the gaussian center, and $r^\text{p}_*=r^{(0)}_* - t$ is the radius of the center of the gaussian at time $t$.
This can be achieved by evolving the Zerilli equation~\cite{Zerilli:1970se} with initial data
\beq
\Psi(t=0,r) &=& 0 \, , \label{eq:ID_Zerilli}\\
\partial_t \Psi(t=0,r) &=& \exp\left[-(r-r ^{(0)}_*)^2/(2\sigma^2)\right]  \label{eq:ID_Zerilli1}\,.
\eeq

The resulting time evolution of the Zerilli function
and of its first two time derivatives are shown in Fig.~\ref{fig:MaxZerilli} for a quadrupolar ($\ell=2$) perturbation. 
The two panels refer to gaussian wavepackets with different widths ($\sigma=5.0M$ on the left, and $\sigma=0.5M$ on the right). 
It is apparent that different physical quantities have different peak times. 
The first peak in $\Psi$ (as extracted at null infinity) occurs {\em slightly before} the gaussian crosses the light ring, and it corresponds to direct radiation excited from the peak of the gravitational potential. 
The broader gaussian excites the burst first, because the ``leading edge'' of the wavepacket reaches the peak first when compared to the narrower gaussian.
However, the zero of $\ddot \Psi$, which roughly corresponds to the peak of energy emission because $\dot{E}\propto \left|\dot{\Psi}\right|^2$, occurs {\em shortly after} the gaussian has crossed the light-ring.
This can be interpreted in terms of the leakage of waves that are quasi-trapped at the light ring on unstable circular orbits, i.e., as ``the beginning of the ringdown.'' 
This interpretation is supported by the time difference between the peaks of $\Psi$, which is comparable to half the light-ring period $T_\text{LR}/2 \sim 16M $. 
Note also that for the larger gaussian, the ringdown phase is suppressed: the second peak of $\Psi$ has smaller relative magnitude with respect to the first peak of direct radiation, due to the interaction of the infalling wavepacket with radiation escaping to infinity.

\begin{figure}[t]
    \includegraphics[width=\columnwidth]{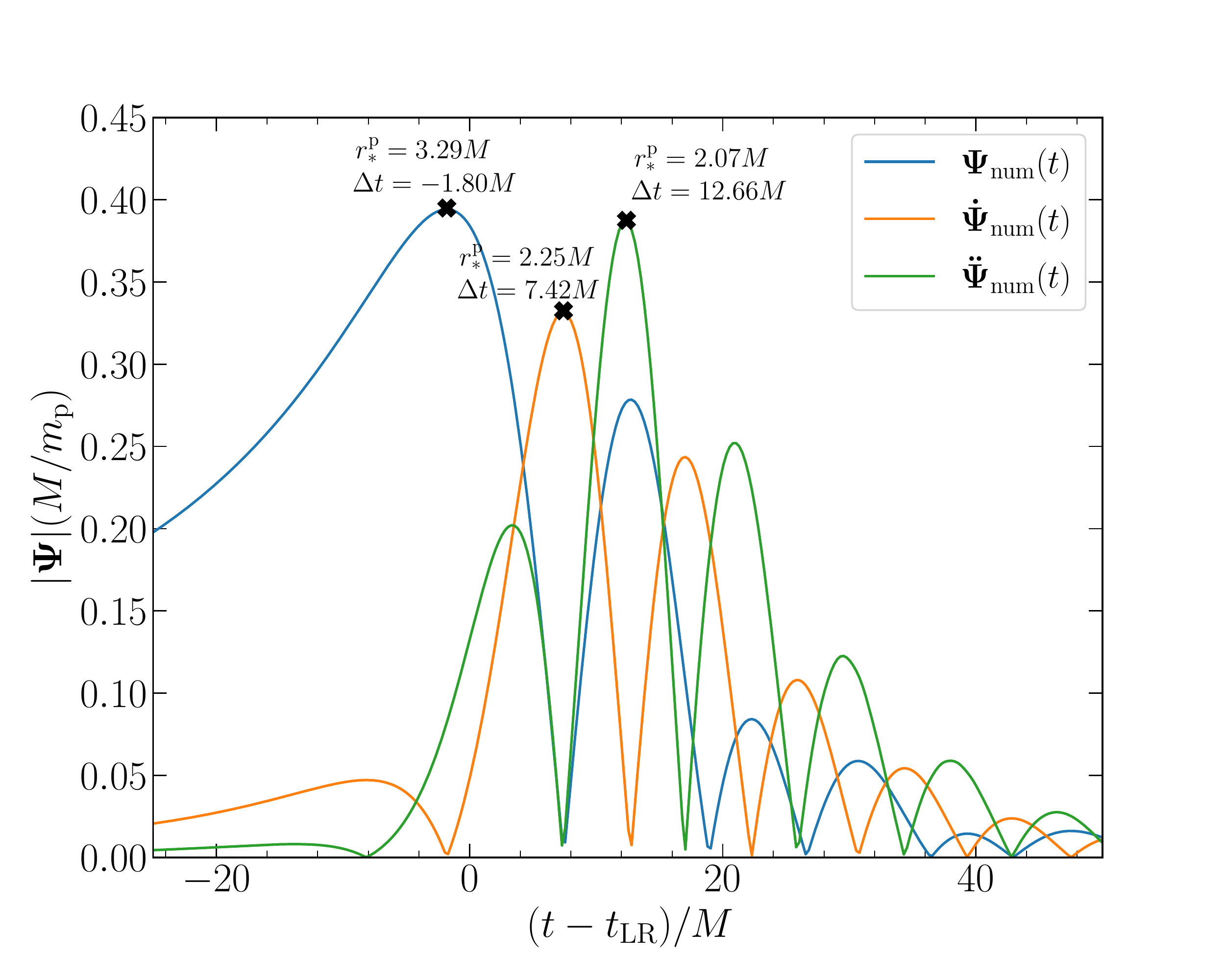}
    \caption{Same as Fig.~\ref{fig:MaxZerilli}, but instead of a gaussian wavepacket we let a point particle fall radially onto the BH from rest at infinity. The waveforms are scaled by the ratio between the particle mass $m_p$ and the BH mass $M$. Results are qualitatively similar to the gaussian case.}
    \label{fig:MaxZerilliPP}
\end{figure}

In Fig.~\ref{fig:MaxZerilliPP} we show that the results are similar when we consider the radial infall of a point particle from rest at infinity, rather than the scattering of a gaussian wavepacket.

However, it would be wrong to read too much into these data. 
The light ring is constantly being hit with forward-beamed radiation, thus GWs from the light ring are in principle present as soon as the radiation has time to interact with it. 
The intensity of these waves becomes larger when the particle gets closer, but there is some inherent delay time, since the waves are trapped in quasicircular motion. 
Therefore there is a large uncertainty in trying to associate the peak of the waveform, or of the energy flux, to a particular particle position. 
For example, in the point particle case, it is far-fetched to claim that the flux peak occurs when the particle is close to the horizon ($r^\text{p}_*=2.07 \rm M$) {\it because} of the particle being close to the horizon: the peak is at that location because the particle has traveled some distance before the GWs it emitted close to the light ring start leaking out. For much more detailed studies of ringdown excitation in the point-particle limit, see e.g. Refs.~\cite{Hadar:2009ip,Hadar:2011vj,Hadar:2014dpa,Hadar:2015xpa,Folacci:2018cic,Apte:2019txp,Lim:2019xrb,Hughes:2019zmt,Lim:2022veo,Oshita:2022pkc,Oshita:2022yry,Rom:2022uvv}.

The punchline of this discussion is that {\em even for linear perturbations of a Schwarzschild spacetime, the ringdown starting time is not a well-defined quantity.} It depends on which quantity we monitor (differing e.g. for the wave amplitude and energy flux), and there is no mathematical or physical basis to claim a well-defined instant as ``the'' ringdown starting time.

\begin{figure}[t]
    \includegraphics[width=\columnwidth]{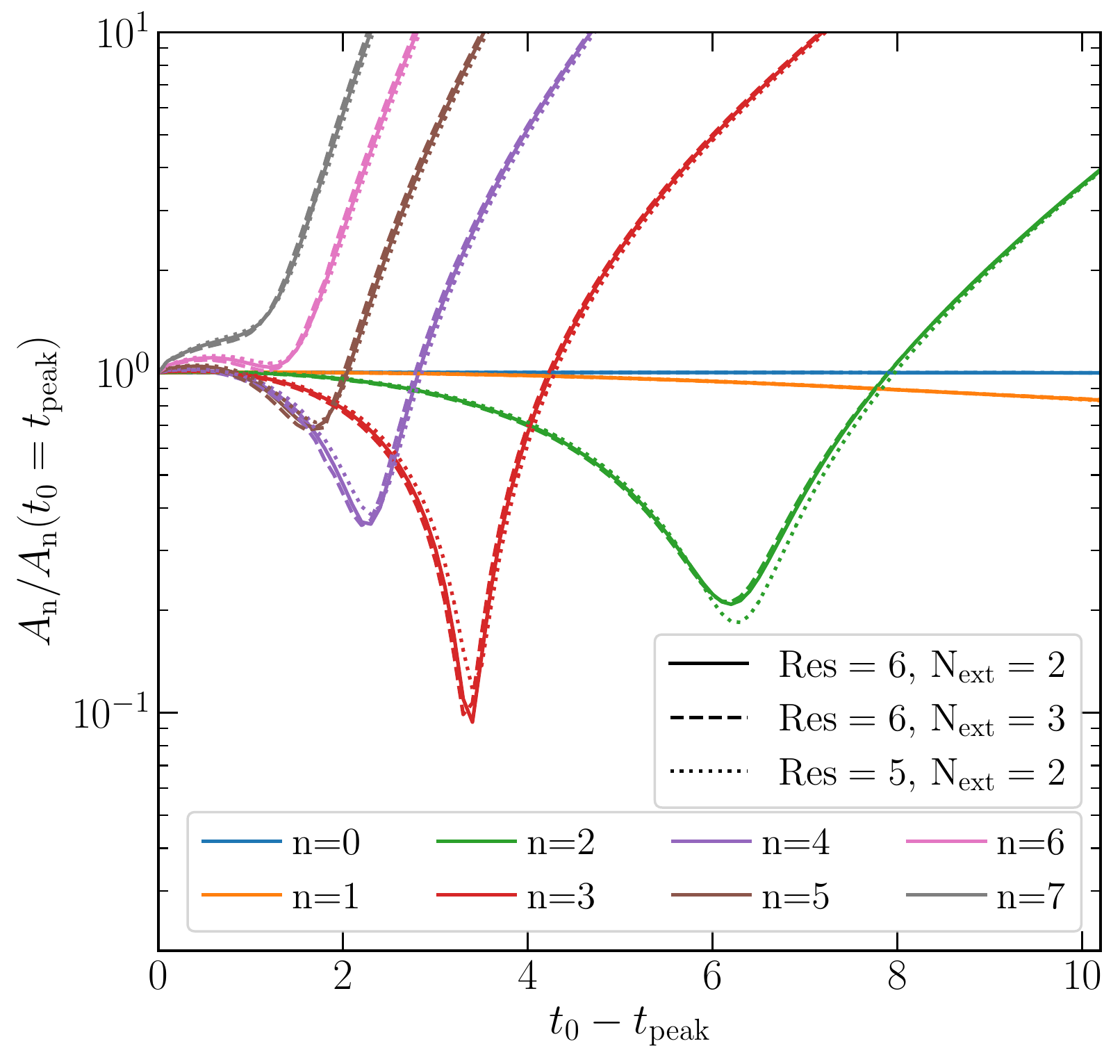}
    \caption{Same as the bottom-right panel with $N = 7$ in Fig.~\ref{A_nN}, but using the NR waveform with (i) the second-highest resolution level $\mathrm{Lev} = 5$ and $N_\mathrm{ext} = 2$ (dotted lines), and (ii) the highest resolution level $\mathrm{Lev} = 6$ but with $N_\mathrm{ext} = 3$ (dash-dotted lines). Solid lines correspond to the fits in Fig.~\ref{A_nN}, which use the reference NR simulation with resolution level $\mathrm{Lev} = 6$ and $N_\mathrm{ext} = 2$.}
    \label{fig:NRerr_amp}
\end{figure}

\section{Impact of numerical relativity errors on ringdown amplitude fits}
\label{sec:NRerror}

Two important sources of error in the SXS NR waveforms analyzed here are the finite numerical resolution of the simulations (henceforth resolution error) and the GW extraction procedure, which involves extrapolation to infinity of the waveforms computed at finite radii (henceforth extrapolation error). Our goal in this appendix is to understand to what extent these errors affect our ringdown amplitude fits (see Ref.~\cite{Boyle:2019kee} for a more detailed discussion of numerical errors).

The resolution in the {\sc SpEC} code depends on the number of points and subdomains in the discrete spacetime grid. 
These are controlled by a tolerance parameter in the adaptive mesh refinement (AMR) algorithm, which modifies the number of points and subdomains when the local measurement of the error is larger than the tolerance. 
In converged NR simulations the discretization error decreases when lowering the AMR tolerance parameter, which corresponds to increasing the resolution level. 
Hence, a common way to gauge the impact of the resolution error in NR simulations is to compare the two highest resolution levels of the same simulation. 

The extrapolation error is caused by the fact that the NR waveforms are not computed directly at future null infinity $\mathscr{I}^+$.\footnote{NR waveforms can be extracted at  $\mathscr{I}^+$ using Cauchy Characteristic Extraction~\cite{Bishop:1996gt, Babiuc:2005pg, Winicour:2008vpn}. Unfortunately, at the moment the availability of these NR simulations is limited.}
Hence they have to be extrapolated to $\mathscr{I}^+$ using a polynomial fit of the strain, produced using its value computed at different extrapolation radii. 
Since polynomials of different order $N_\mathrm{ext}$ will give slightly different extrapolations of the waveform at $\mathscr{I}^+$, it is common to estimate the impact of the extrapolation error by comparing the waveform extrapolated using polynomials with different values of $N_\mathrm{ext}$. 
The reference extrapolation order for the merger and ringdown part of the waveform is $N_\mathrm{ext} = 2$ (see Sec. 2.4.1 of Ref.~\cite{Boyle:2019kee}). 
This waveform is typically compared with the same waveform extrapolated using an $N_\mathrm{ext} = 3$ polynomial to estimate the extrapolation error.

We want to assess the effect of these two sources of error on the fits of the post-merger NR waveform performed in Secs.~\ref{sec:postpeaklinearity} and \ref{sec:howmany}. 
In particular, we are interested in understanding whether the variation of the amplitude of the modes when fitting the waveform starting at different initial times $t_0$ could be due to this error. 

In Fig.~\ref{A_nN} of the main text we use the reference NR waveform at the highest resolution level ($\mathrm{Lev} = 6$) with $N_\mathrm{ext} = 2$. 
In Fig.~\ref{fig:NRerr_amp} we perform again the same fit of the post-merger NR waveform using a model with $N = 7$ overtones, but this time we compare with (i) the NR waveform with the second highest resolution level ($\mathrm{Lev} = 5$) and $N_\mathrm{ext} = 2$, as well as (ii) the NR waveform with $\mathrm{Lev} = 6$, but $N_\mathrm{ext} = 3$. 
We find that using NR waveforms with different resolution levels and different extrapolation orders makes almost no visible difference in the fits. These sources of error in the NR simulations cannot explain the variation of the fitted amplitudes of the modes when varying $t_0$.

\section{Effect of a spurious constant}
\label{sec:systematics}

\begin{figure}[t]
    \includegraphics[width=\columnwidth]{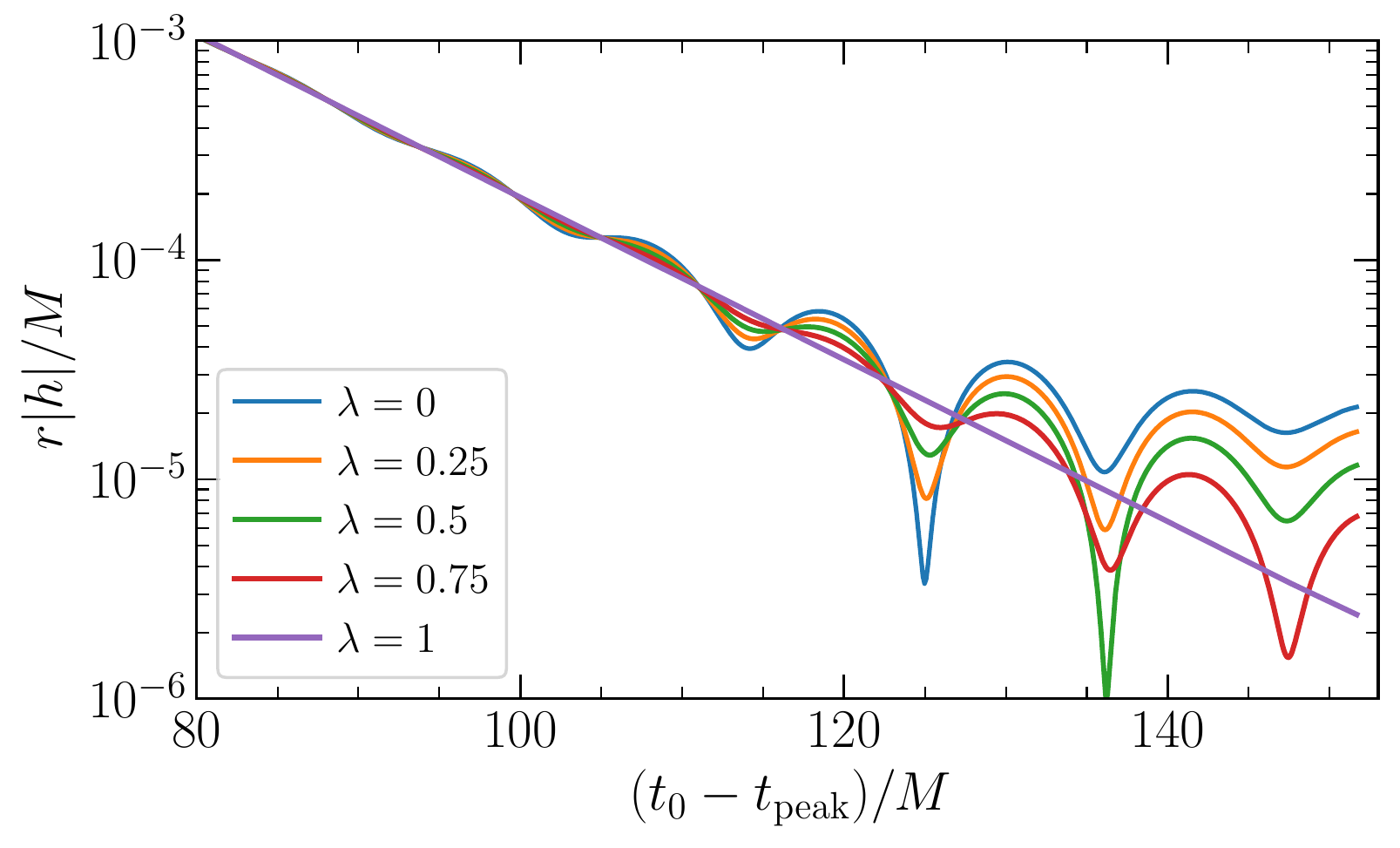}
    \caption{$r |h_\lambda|/M$: Amplitudes of \texttt{SXS:BBH:0305} when the spurious constant is not subtracted perfectly: see Eq.~(\ref{eq:hlambda}).}
    \label{fig:hlamda}
\end{figure}

\begin{figure*}[t]
    \includegraphics[width=\textwidth]{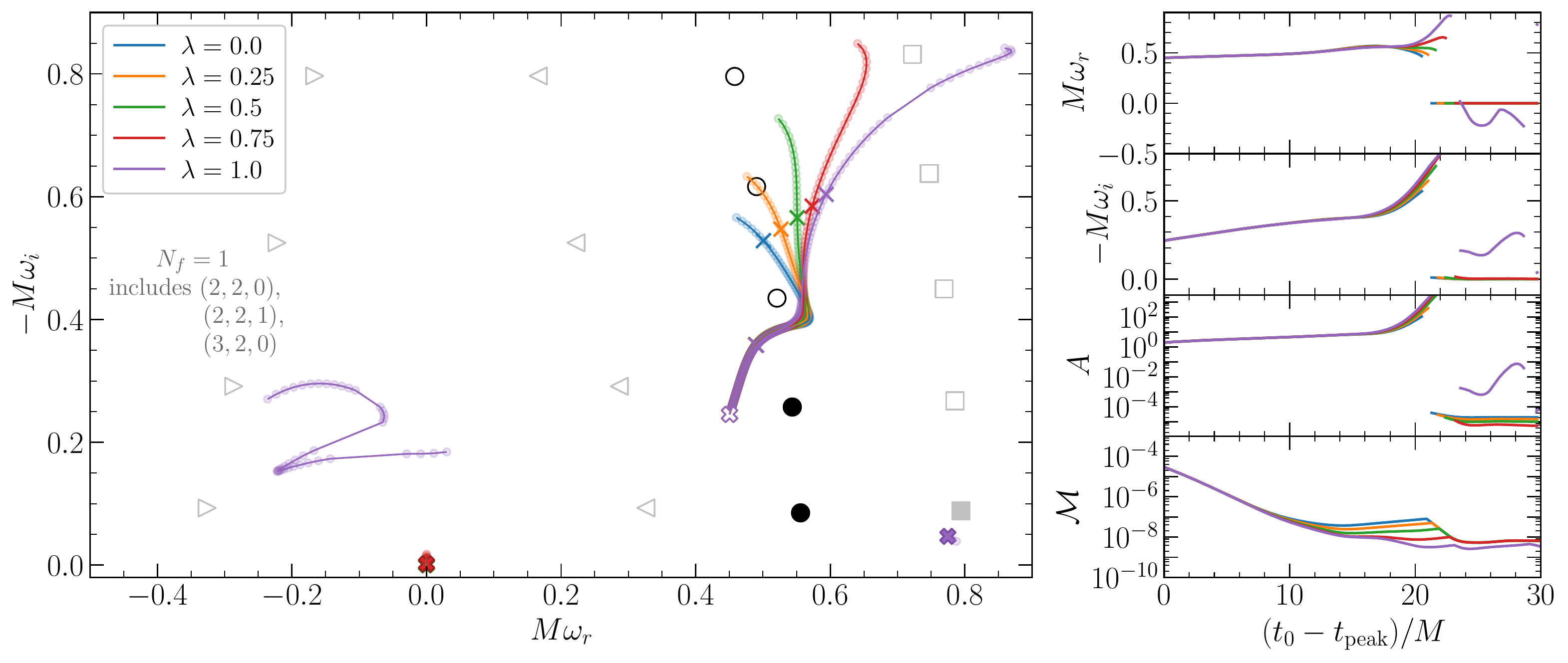}
    \caption{Fitting $h_\lambda$ with the \Mxrm{2}{1} model: one free mode in addition to $(2,2,0)$, $(2,2,1)$, $(3,2,0)$ (which are all marked by filled symbols in the left panel). Left panel: extracted frequencies 
    $(M\omega_r, \, - M\omega_i )$ for $n_f=0$, with $t_0$ ranging from $t_{\rm peak}$ (empty cross) to $t_{\rm peak}+30 M$ (filled cross). We also mark $t_{\rm peak}+10 M$ and $t_{\rm peak}+20 M$ by crosses.
    We denote known QNMs as follows: $(2,2,n)$ (black circles), $(2,-2,n)$ (gray left-pointing triangles),  $(2,2,n)_R$ (gray right-pointing triangles), $(3,2,n)$ (gray squares). Right panels: evolution of $M\omega_{r}$, $-M\omega_{i}$, $A$ and ${\mathcal M}$ as functions $t_0$.}
    \label{fig:wrilambda}
\end{figure*}

\begin{figure*}[t]
    \includegraphics[width=\textwidth]{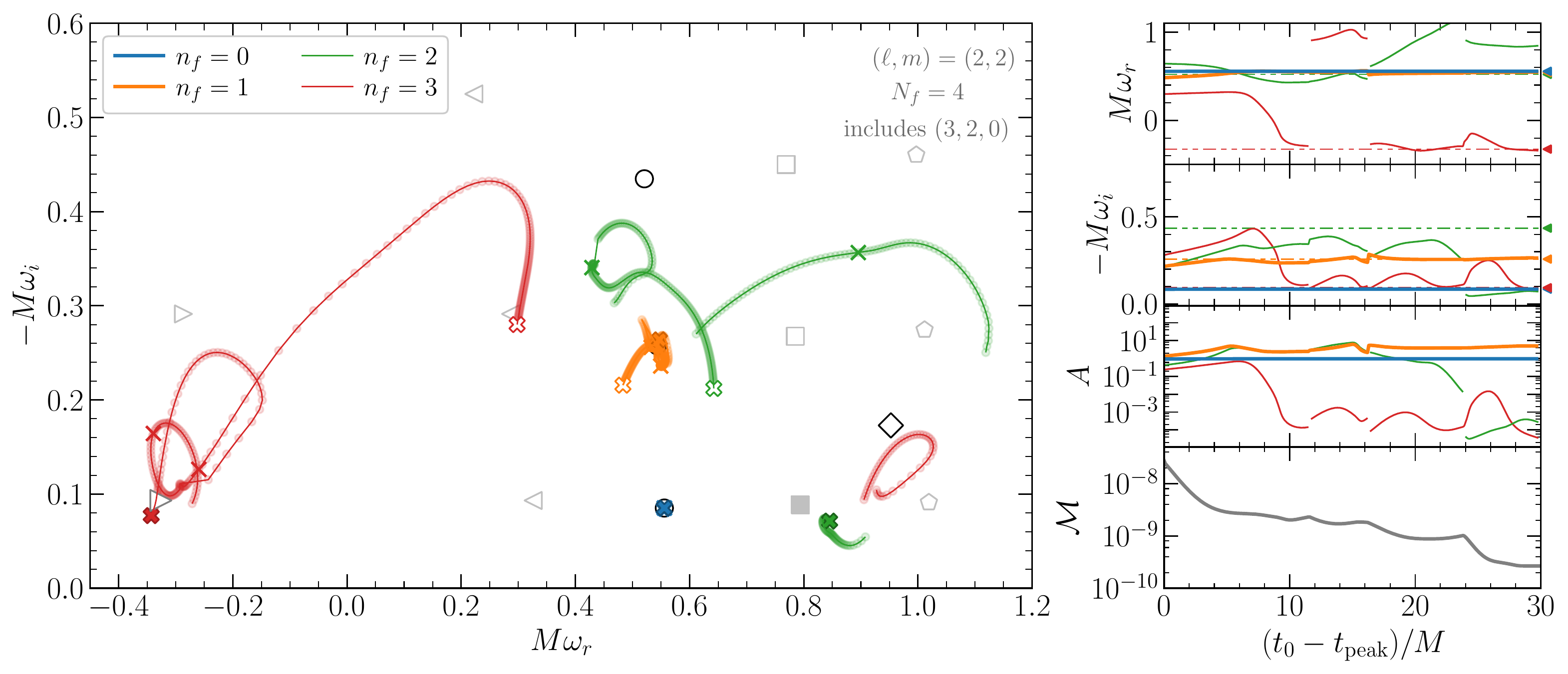}
    \caption{Fits with the \Mrm{3} model, i.e., four free modes ($N_f=4$)  in addition to  $(3,2,0)$ (marked by a filled symbol in the left panel). Left panel: extracted frequencies 
    $(M\omega_r, \, - M\omega_i )$ for $n_f=0,\,1,\,2,\,3$, with $t_0$ ranging from $t_{\rm peak}$ (empty cross) to $t_{\rm peak}+30 M$ (filled cross). We also mark $t_{\rm peak}+10 M$ and $t_{\rm peak}+20 M$ by crosses.
    We denote known QNMs as follows: $(2,2,n)$ (black circles), $(2,-2,n)$ (gray left-pointing triangles), $(2,2,n)_R$ (gray right-pointing triangles), $(3,2,n)$ (gray squares),  $(4,2,n)$ (gray pentagons). Right panels: evolution of $M\omega_{r}$, $-M\omega_{i}$, $A$ and ${\mathcal M}$ as functions of $t_0$. We also plot $M\omega_{220}$, $M\omega_{221}$, $M\omega_{222}$, $M\omega_{22n,\,R}=-M\omega_{2-22}^*$ using blue, orange, green and red dot-dashed horizontal lines, and triangle markers in the same colors.}
    \label{fig:wri4}
\end{figure*}

In Sec.~\ref{sec:howmany} we pointed out that in the \Mxrm{3}{1} model, the free mode is merely fitting the late-merger GW frequency at early times $t_0-t_{\rm peak}<15 M$, while at late times ($t_0-t_{\rm peak}>15 M$) it is unphysical, and it is likely fitting noise. Here we justify the second statement.

The waveform \texttt{SXS:BBH:0305} used in this study contains a spurious constant (see also Ref.~\cite{Pereira:2022kqn}). 
To determine this constant we subtract the fundamental mode $A_{220} e^{-i (\omega_{220} t+ \phi_{220})}$, which can be very accurately determined, from the late-time ringdown, with the result 
\be
c_n=-(0.8+1.8 \ {i\mkern1mu}) \times10^{-5} 
\ee

In this section we gauge the impact of this term on our fits, using waveforms where the constant has not been perfectly subtracted:
\be\label{eq:hlambda}
h_\lambda = h_{22}^{\rm SXS} -\lambda\,c_n\,.
\ee
Here, $h_{22}^{\rm SXS}$ is the original extrapolated waveform provided by the SXS collaboration, and we have introduced a continuous parameter $\lambda$ such that $\lambda=0$ gives the waveform from which the spurious constant $c_n$ has not been subtracted, while $\lambda=1$ corresponds to perfect subtraction. 

In Fig.~\ref{fig:hlamda}, we plot the waveform amplitude for different values of $\lambda$. 
The original waveform ($\lambda=0$, blue line) does not show an exponential decay at late times, when $c_n$ dominates over the fundamental mode. 
Once we extract $c_n$ perfectly ($\lambda=1$, purple line), we recover the ``pure'' exponential decay due to the $(2,\,2,\,0)$ mode. 

Next, we fit $h_\lambda$ with the \Mxrm{2}{1} model. Recall this means one free mode in addition to the $(2,2,0)$, $(2,2,1)$, $(3,2,0)$ fixed modes. 
In Fig.~\ref{fig:wrilambda} we plot the extracted complex frequency, the amplitude of the free mode and the mismatch, as usual. 
The early-time behavior is the same for all waveforms, being controlled by the late-merger GW frequency. 
However the late-time behavior is very sensitive to $\lambda$.  
For $t_0-t_{\rm peak}=15 M$--$20 M$, the imaginary part of the free mode increases for all $\lambda$, but the trajectory of the real part of the frequency in the complex plane depends on $\lambda$.
Whenever $\lambda \neq 1$ we find that $M\omega_r$ drops to zero for $t_0-t_{\rm peak}\gtrsim 20 M$:
the free mode is fitting a spurious constant with value $(1-\lambda) c_n$. 
This is a consistency check of our fitting procedure, and shows that we can extract even very small spurious noise components ($\sim {\mathcal O}(10^{-6})$).

The situation is different for $\lambda=1$, when the spurious noise has been subtracted: the free mode now finds evidence for a retrograde mode with $M\omega_r<0$. 

The impact of the spurious constant and the presence of retrograde modes can be better understood by adding multiple free modes. 
In Fig.~\ref{fig:wri4} we fit the $\lambda=1$ waveform with model \Mrm{3}: $N_f=4$ free modes in addition to the fixed $(3,2,0)$ mode.
We observe that the fundamental mode can be extracted nearly perfectly $\delta\omega\sim {\mathcal O}(10^{-5})$ and the $n_f=1$ mode converges to the first overtone, with $\delta\omega\sim {\mathcal O}(10^{-2})$.
However the behavior of the $n_f=2,\,3$ modes is very sensitive to numerical noise. 
The mismatch when fitting with \Mrm{3} is $\mathcal{M}\sim \mathcal{O} (10^{-9})$, which is much smaller than differences between waveforms computed at different resolutions, further indicating the possibility than one or two free modes could be simply fitting numerical noise. 
However, the real frequency $M\omega_r$ of the $n_f=3$ mode is negative and indicates the presence of a retrograde mode (shown by a down-pointing triangle) around $t_0-t_{\rm peak}=10 M$, with an amplitude of ${\mathcal O}(10^{-4})$. 
We also see that {\em none of the free modes coincide with higher overtones ($n>1$). Once again, it is not possible to extract the first overtone at $t_0=t_{\rm peak}$.}
We conclude that the presence of the spurious constant has only a small impact at very late times, and it does not affect the extraction of overtones near the peak.
  
\section{A toy model for SXS:BBH:0305}
\label{sec:NRtoy}

\begin{figure*}[t]
  \includegraphics[width=\textwidth]{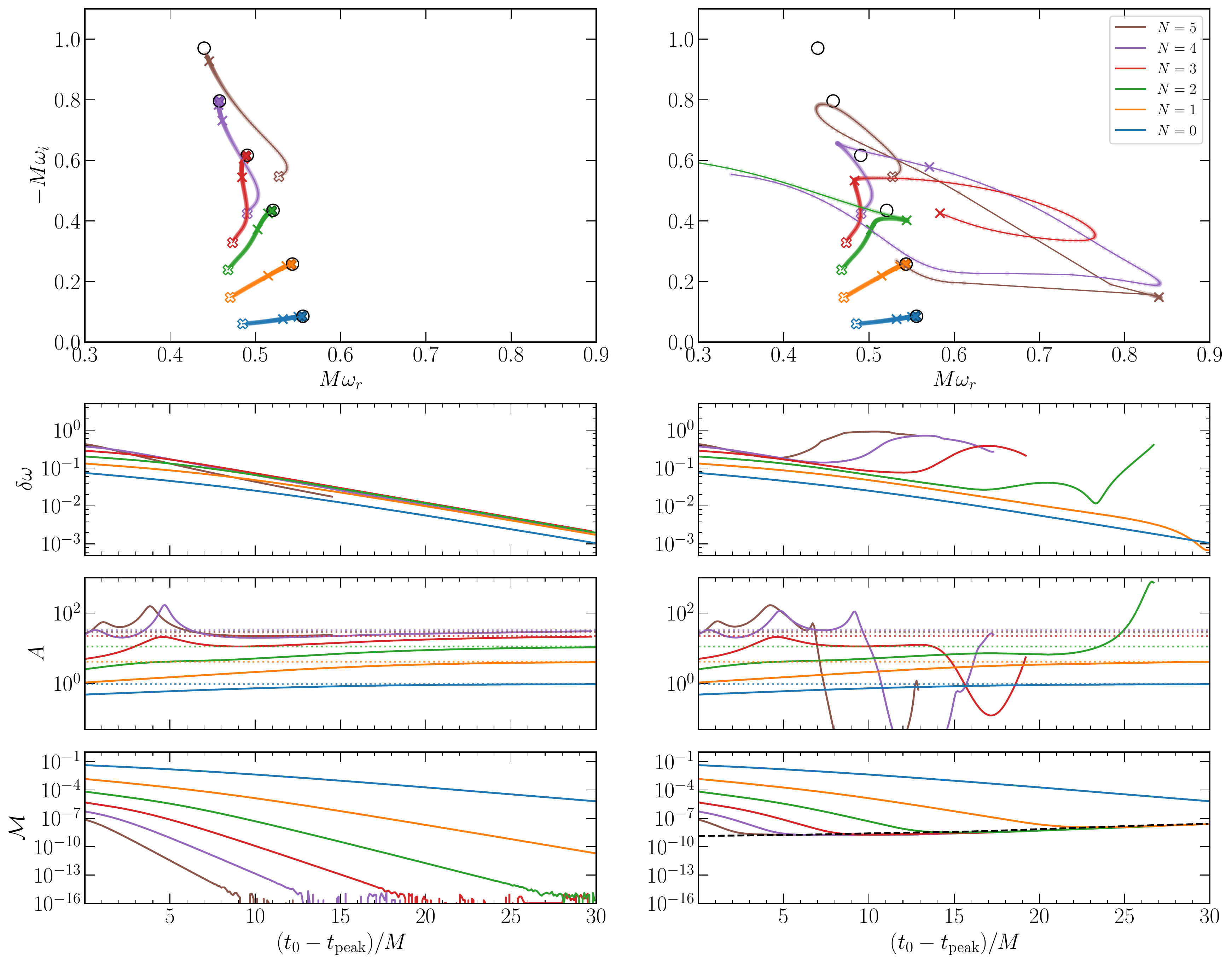}
  \caption{Fits of the $\mathcal{Q}_7(t)$ (left column) and $\mathcal{Q}_7^e(t)$ (right column) toy waveforms with model \Mxr{N}{1}.
  First row: evolution of the best-fit frequency of the highest overtone with $t_0$. 
  Second row: difference between the fitted free frequency and the predicted value for the corresponding overtone, as defined in Eq.~\eqref{eq:deltaomega}.
  Third row: amplitude of the free overtone in each fit. The injected amplitudes of the toy waveforms are indicated by horizontal dotted lines.
  Bottom row: mismatch between the best-fit waveform and the actual waveform for each fit.
  In the bottom-right panel we also show with a black dashed line the mismatch between $\mathcal{Q}_7(t)$ and $\mathcal{Q}_7^e(t)$, i.e., the approximate mismatch induced by the injected noise floor.
  As in Fig.~\ref{fig:fitfreqtoys} we remove spurious data points. This is why some of the curves end prematurely.}
  \label{fig:fitfreqtoysSXS}
\end{figure*}

\begin{figure*}[t]
  \includegraphics[width=\textwidth]{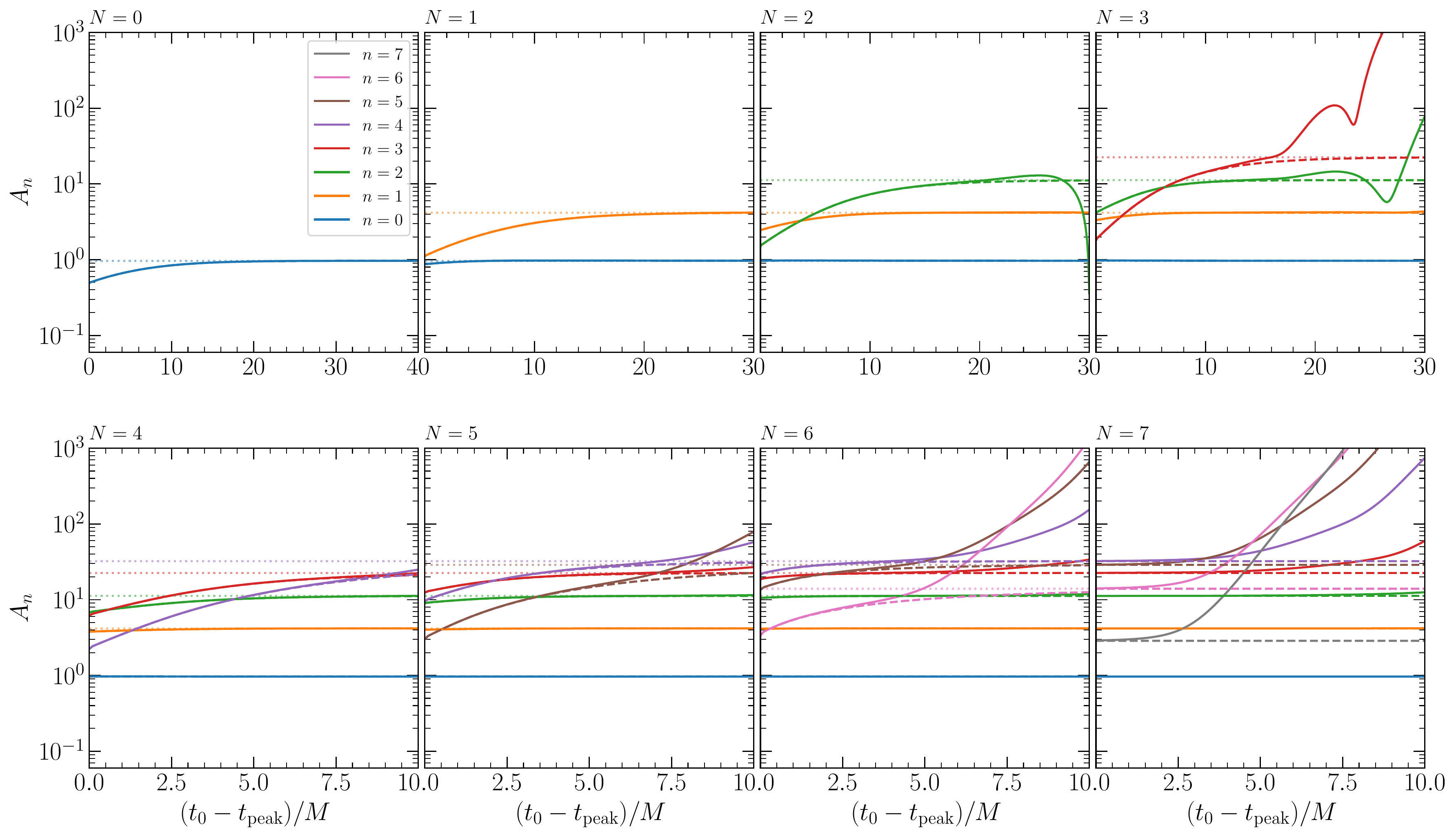}
  \caption{The amplitude of the overtones for the $\mathcal{Q}_7(t)$ (dashed) and $\mathcal{Q}_7^e(t)$ (solid) toy waveforms, similar to Fig.~\ref{fig:fitamplRW}.
  Each panel correspond to a fit with a different number of overtones in the fitting model. All mode frequencies are fixed to their predicted values.
  Faint, dotted horizontal lines are the injected values of the amplitude for each mode.
  Note that the time range of the plots is different across different panels.}
  \label{fig:toyampsSXS}
\end{figure*}

In Sec.~\ref{sec:linearovertones} we examined toy model waveforms to argue that the overtone model may not be a good description of the ringdown as early as the peak {\em even in linear theory}.
While we expect the same conclusions to hold for NR waveforms, the results in Sec.~\ref{sec:linearovertones} should not be carelessly used as a benchmark, because
there are some important (and not obvious) differences between the linear and NR waveforms.
Firstly, the linear model we explored refers to the Zerilli and Regge-Wheeler functions. A linear combination of these functions yields the Newman-Penrose scalar $\Psi$, while the NR waveforms refer to the strain $h$. The two quantities are connected by a second derivative in time, $\Psi \sim \ddot{h}$, meaning that the peak of $\Psi$ and $h$ may differ by a few $M$ (see e.g. the discussion in Appendix~\ref{app:Green}).
Secondly, the linear models we considered are strictly real, because the real and imaginary parts of the Regge-Wheeler equation Eq.~\eqref{equation_regge_wheeler} decouple and we specified real initial conditions, while
the complex NR waveforms must be found as a linear superposition of \emph{complex} damped exponentials as in Eq.~\eqref{eq:QNMmodel}.
This technical difference could affect the quality of the fits.
In fact, in this section we will see that fits for the NR waveforms often give smoother results than the linear waveforms across different starting times. This is because the full phase evolution of the waveform is encoded in the complex waveform, and this typically improves the quality of fits in terms of complex exponentials -- essentially spinning phasors (with an exponentially decaying amplitude) in the complex plane.
Thirdly, while the Price power-law tail is apparent in the linear waveform, to our knowledge there is no conclusive evidence of its presence in the NR waveforms.
While we could demonstrate that the exponential blow up of the QNM amplitudes at late times for the linear waveforms is not driven by numerical noise, it is uncertain whether the same conclusion applies to NR waveforms.

\begin{figure}[t]
    \includegraphics[width=\columnwidth]{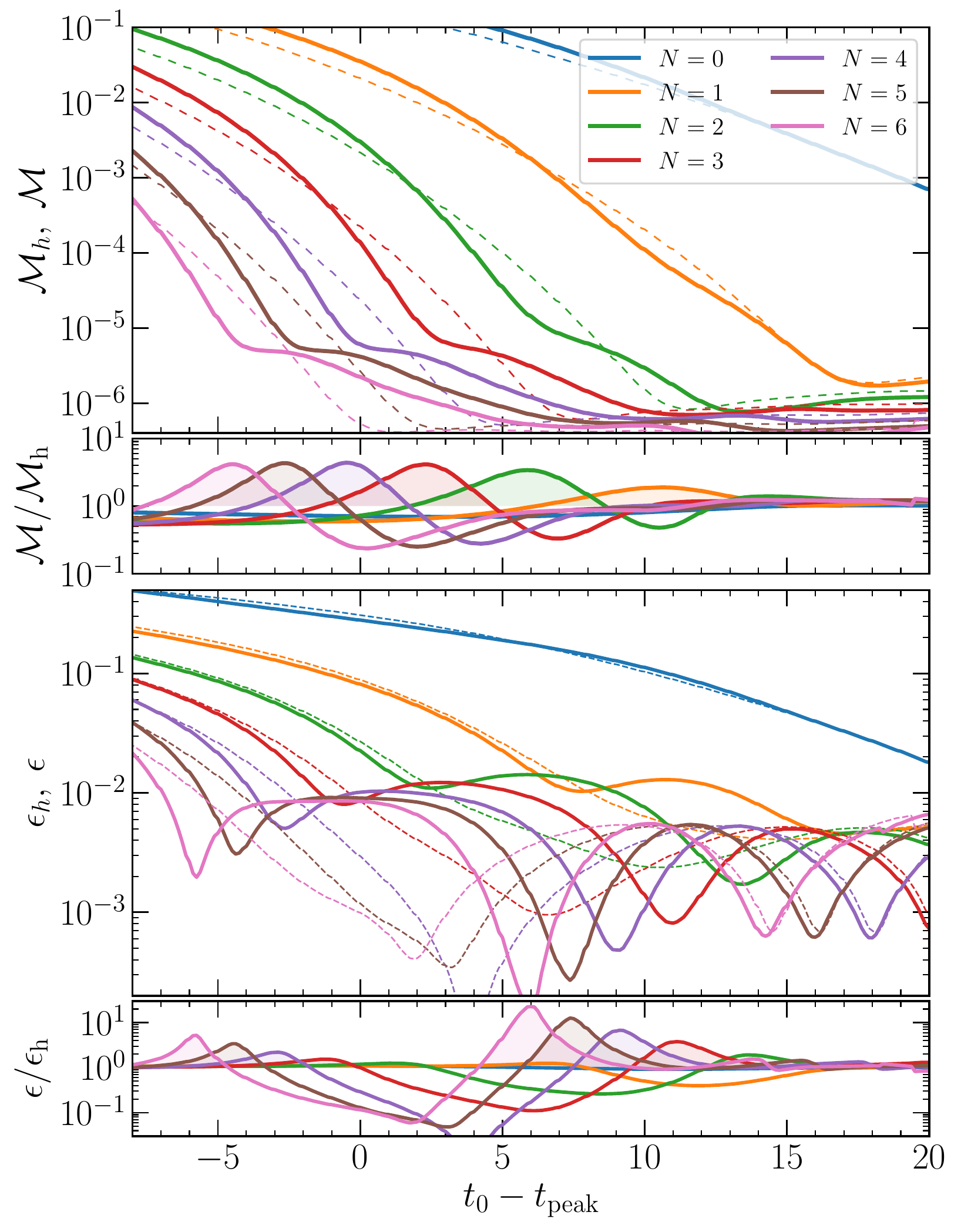}
    \caption{Mismatch and mass/spin errors (first and third panel from the top, respectively) for the hybrid waveform (solid line) and SXS:BBH:0305 (dashed line). We also plot the ratio of the corresponding quantities for the hybrid waveform (solid line) and SXS:BBH:0305 (dashed line), shading the region where the overtones fit the hybrid waveform better.}
    \label{fig:frankenstein}
\end{figure}

To explore the effect of these differences,
in the spirit of Sec.~\ref{sec:linearovertones} we construct two toy waveforms:
\begin{align}
    \mathcal{Q}_7(t) &\equiv \sum\limits_{n=0}^{7}  A_{n} e^{-i [\omega_{n} (t- t_{\rm peak}  +\phi_n]}, \\
    \mathcal{Q}_7^e(t) &\equiv  \mathcal{Q}_7(t) + e(t),
\end{align}
where $\mathcal{Q}_7(t)$ is a clean combination of $7$ overtones, and $\mathcal{Q}_7^e(t)$ contains in addition the estimated numerical error $e(t)$.
These toy waveforms are constructed to mimic the $\ell=m=2$ multipole ringdown waveform of the SXS:BBH:0305 simulation (resolution ${\rm Lev} = 6$, extrapolation order $N = 2$: see Appendix~\ref{sec:NRerror}), so we use the first seven overtone frequencies of its remnant BH, and the injected values of $A_n$ and $\phi_n$ are chosen to be equal to those obtained by fitting the SXS:BBH:0305 waveform starting at the peak with $7$ overtones.
The curly symbol $\mathcal{Q}_7(t)$ is a reminder that, contrary to the real waveforms in Sec.~\ref{sec:linearovertones}, these waveforms are complex.
The error is approximated as the difference between the ${\rm Lev} = 6$ and ${\rm Lev} = 5$ waveforms minimized over a relative time and phase difference.

We fit these toy waveforms by a linear combination of complex QNMs.
We first repeat the free-frequency search of Sec.~\ref{sec:linearovertones} by adding more overtones to our fitting model while leaving only the highest overtone frequency free.
As shown in Fig.~\ref{fig:fitfreqtoysSXS}, the free mode converges to the expected overtone frequencies of the clean toy waveform $ \mathcal{Q}_7(t)$ at least for $N \leq 5$. Once we add noise in $\mathcal{Q}_7^e(t)$, the fit fails to converge unless $N \leq 1$.
This proves that {\em even if there were no nonlinearities near the peak and the amplitudes of the overtones were constant starting at the peak, a subdominant contamination to the waveform could affect our ability to extract the correct overtone frequencies.}
However we should be able to recover the first overtone at late enough starting time, contrary to the linear waveform case, where we cannot extract any overtone unless we use a very specific starting time. In fact, this is just what we observe in Fig.~\ref{fig:wri1} for the SXS:BBH:0305 waveform.

Next, we repeat the fits with fixed frequencies, also as in Sec.~\ref{sec:linearovertones}.
Fig.~\ref{fig:toyampsSXS} shows the amplitude of the overtones as a function of $t_0$ for both $\mathcal{Q}_7(t)$ (dashed lines) and $\mathcal{Q}_7^e(t)$ (solid lines).
Adding more overtones stabilizes the lower overtone amplitudes. For $N = 7$ there is a brief period of time where all overtone amplitudes are constant even near the peak, even for the $\mathcal{Q}_7^e(t)$ toy waveform.
While the agnostic (free-frequency) mode search would fail for $N \geq 3$, if the waveform could be modeled as QNMs modulo small contamination, we should expect the amplitudes of the overtones to be flat for a brief moment.
For example, with an $N = 7$ fit model (bottom right panel of Fig.~\ref{fig:toyampsSXS}),
the amplitudes of the $n = 2$ and $3$ overtones are approximately constant over a time range $\sim 10 M$ and $\sim 5 M$, respectively. 
The corresponding range of flatness (if any) in Fig.~\ref{A_nN} is significantly shorter. Therefore the numerical error alone cannot explain why we fail to consistently model SXS:BBH:0305 as a combination of overtones starting at the peak.

\section{An unphysical hybrid waveform}
\label{sec:frankenstein}

In Sec.~\ref{sec:postpeaklinearity} we show that the overtones found when fitting the early post-peak BBH waveforms are unphysical.  In this appendix, we confirm this finding by producing an unphysical post-peak BBH waveform, and showing that the overtones can still fit it with similar accuracy. 

We produce the unphysical post-peak BBH waveform by hybridizing the $\ell=m=2$ mode of our prototypical NR waveform SXS:BBH:0305 (henceforth $h_2$) with another waveform in the catalog, SXS:BBH:0220 (henceforth $h_1$).  The latter represents an equal-mass BH binary with dimensionless spins $\chi_1=-0.4$ and  $\chi_2=-0.8$ antialigned with the orbital angular momentum.  We first align the two waveforms such that $t = 0 M$ correspond to the peak of their $\ell=m=2$ mode. Then we construct the hybrid waveform by smoothly blending their amplitudes $A_1$ and $A_2$ and their frequencies $\omega_1$ and $\omega_2$ as follows:
\begin{align}
A_\mathrm{hyb}(t) &= A_1(t) W(t) + A_2(t) \left(1-W(t)\right),\nonumber\\
\omega_\mathrm{hyb}(t) &= \omega_1(t)  W(t) +\omega_2(t) \left(1-W(t)\right),
\end{align}
where $W(t)$ is a window function defined as
\begin{equation}
W(t)=
    \begin{dcases}
        1 & t \leq t_1\\
	\mathrm{S}\left(\frac{t_2 - t_1}{t-t_1} + \frac{t_2 - t_1}{t-t_2}\right) &  t_1 < t < t_2\\
        0 & t \geq t_2,\\
    \end{dcases}
\end{equation}
where $S(t)$ is the logistic function, $t_1 = t_{\rm peak}$ and $t_2 = t_{\rm peak}+20 M$.  With this definition, the hybrid waveform exactly matches $h_1$ ($h_2$) for $t\leq t_{\rm peak}$ ($t\geq t_{\rm peak}+ 20M$), and it is an unphysical combination of the two in the region $t_1 < t < t_2$.

We fit the hybrid waveform by the \Mx{N} model,  with QNM frequencies set to those of the BH remnant of the SXS:BBH:0305 waveform.  In the upper panel of Fig.~\ref{fig:frankenstein}, we show the mismatch $\mathcal{M}_\mathrm{h}$ between the fits and the hybrid waveform as a function of $t_0 - t_\mathrm{peak}$,  where $t_\mathrm{peak}$ is the peak of the hybrid waveform.  For comparison,  we also show the mismatch $\mathcal{M}$ between SXS:BBH:0305 and the \Mx{N} model fits.  Using more and more overtones improves the fit of the hybrid waveform in the unphysical region $(t_{\rm peak} <t_0< t_{\rm peak} + 20M)$ as much as it does it for the SXS:BBH:0305 waveform.  In fact, in some ranges of $t_0$ the fit of the hybrid waveform in the unphysical region has an even {\em smaller} mismatch than the fit of the SXS:BBH:0305 waveform. This is clear from the second panel of the figure,  where we show the ratio $\mathcal{M}/\mathcal{M}_\mathrm{h}$. 

We find that the overtones typically add in counter-phase in the fit of the hybrid waveform, just as they do when fitting SXS:BBH:0305. The extension of the overtone model up to the peak requires very large, presumably unphysical amplitudes, which however can be fine-tuned to cancel out. Modes in counterphase are very effective at (pathologically) canceling each other. This is precisely what is required to overfit the early post-merger signal. 

Interestingly, including overtones in the fit of the post-peak hybrid waveform also improves the measurement of the remnant spin $\chi_f$ and mass $M_f$.  We prove this by repeating the fit of this waveform with models \Mx{N} with free $\chi_f$ and $M_f$ (in addition to the free amplitudes and phases of the modes). We determine the deviation of the fitted $\chi_f$ and $M_f$ from their asymptotic values using the quantity $\epsilon_\mathrm{h}$, defined in analogy with Eq.~\eqref{eq:epsilondef}.  Smaller values of $\epsilon_\mathrm{h}$ indicate a better measurement of $\chi_f$ and $M_f$.

In the third panel of Fig.~\ref{fig:frankenstein},  we show $\epsilon_\mathrm{h}$ as a function of $t_0 - t_\mathrm{peak}$.  For comparison we also plot the value of $\epsilon$ obtained when fitting the \Mx{N} models against the SXS:BBH:0305 waveform.
The results confirm what we found by computing the mismatch: $\epsilon_\mathrm{h}$ gets smaller when adding a larger number of overtones,  even when we start the fit at values of $t_0$ for which the hybrid waveform is unphysical.  The bottom panel of Fig.~\ref{fig:frankenstein} shows that  $\epsilon_\mathrm{h}$ is even smaller than $\epsilon$ in some ranges of $t_0$.  The fact that using the overtones allows us to improve the measurement of $\chi_f$ and $M_f$ even for the unphysical hybrid waveform is clear evidence that their role is to match whichever early post-peak waveform,  allowing the dominant mode to correctly fit the late post-peak waveform, which really carries information about the remnant BH properties.

\begin{figure*}[t]
    \includegraphics[width=\textwidth]{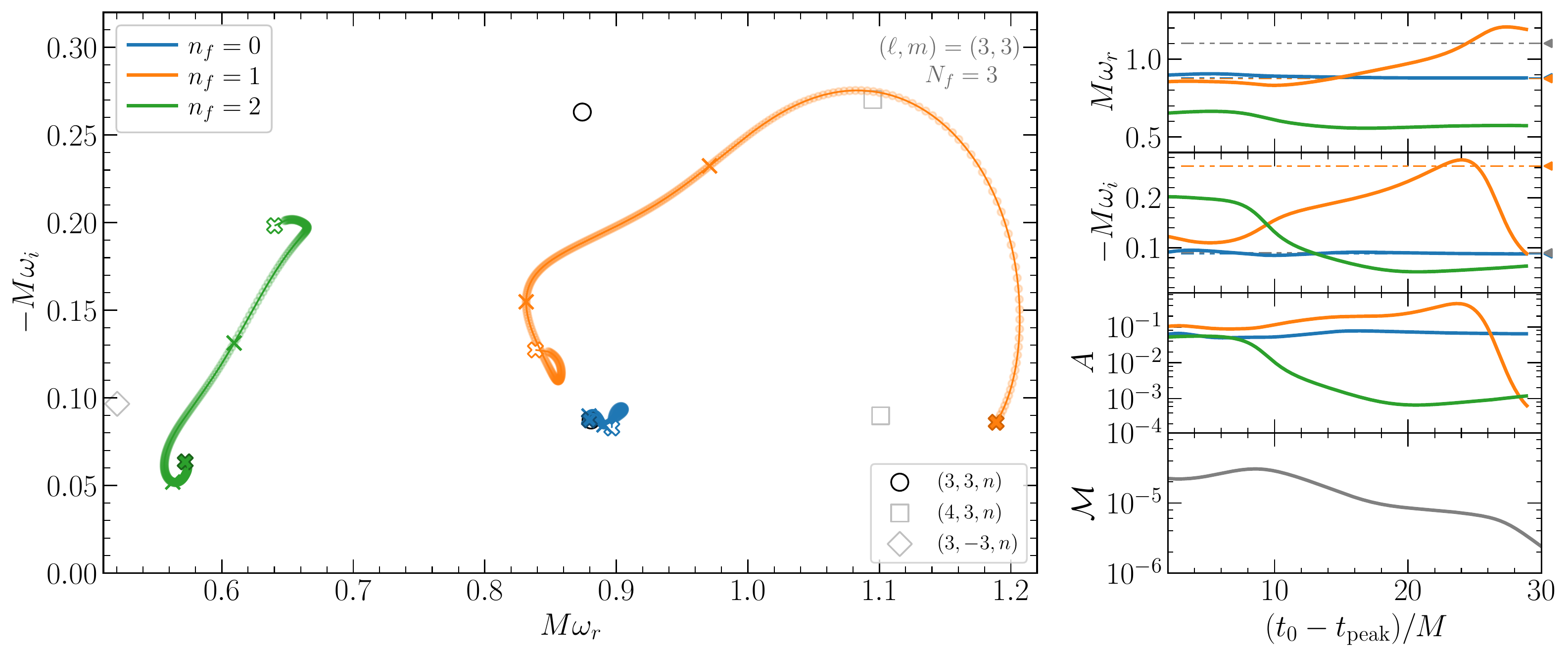}
    \caption{Fits of the $(3,\,3)$ multipole with the \Mr{2} model. Left panel: the extracted frequencies 
    $(M\omega_r, \, - M\omega_i )$ for $n_f=0,\,1,\,2$, with $t_0$ ranging from $t_{\rm peak}$ (empty cross) to $t_{\rm peak}+30 M$ (filled cross); other crosses correspond to times $t_{\rm peak}+10 M$ and $t_{\rm peak}+20 M$.
    We also mark the QNMs $(3,3,n)$ (black circles), $(3,-3,n)$ (gray diamonds) and $(4,3,n)$ (gray squares), for reference. 
    Right panels: $M\omega_{r}$, $M\omega_{i}$, $A$ and ${\mathcal M}$ as functions of $t_0$. Horizontal gray (orange) dot-dot-dashed lines and triangle markers correspond to the expected frequencies for $M\omega_{430}$ ($M\omega_{331}$).}
    \label{fig:wri3_33}
\end{figure*}

\begin{figure*}[t]
    \includegraphics[width=\textwidth]{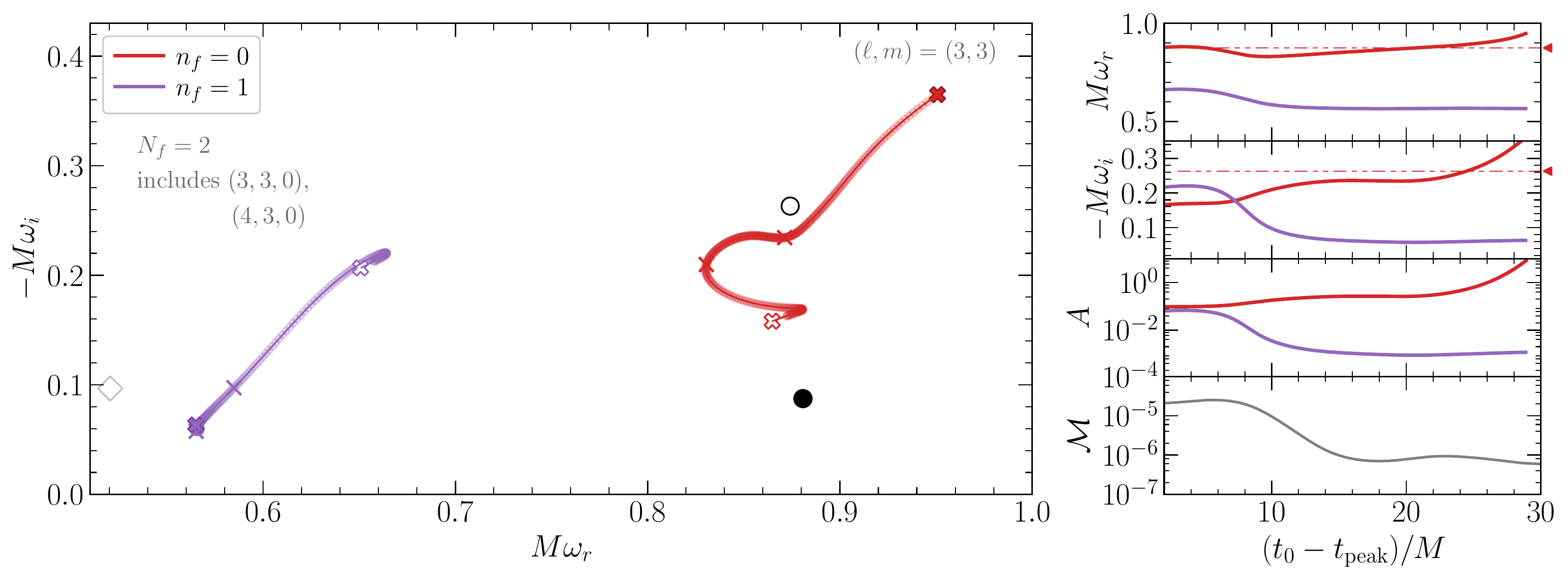}
    \caption{Fits of the $(3,\,3)$ multipole using the \Mxrm{2}{2} model, i.e., two free modes ($N_f=2$) in addition to two additional modes with frequencies fixed to the $(3,3,0)$ and $(4,3,0)$ values. Left panel: the extracted frequencies 
    $(M\omega_r, \, -M\omega_i )$ for $n_f=0,\,1$ and $N_f=2$, for $t_0$ ranging from $t_{\rm peak}$ (empty cross) to $t_{\rm peak}+30 M$ (filled cross); other crosses correspond to times $t_{\rm peak}+10 M$ and $t_{\rm peak}+20 M$.
    We mark the QNMs $(3,3,n)$ (black circles) and $(3,-3,n)$ (gray diamonds). 
    Right panels: $M\omega_{r}$, $M\omega_{i}$, $A$ and ${\mathcal M}$ as functions of $t_0$. The horizontal red dot-dot-dashed line and triangle marker corresponds to the expected frequency for $M\omega_{331}$.}
    \label{fig:wri2_33}
\end{figure*}

\begin{figure*}[t]
    \includegraphics[width=\textwidth]{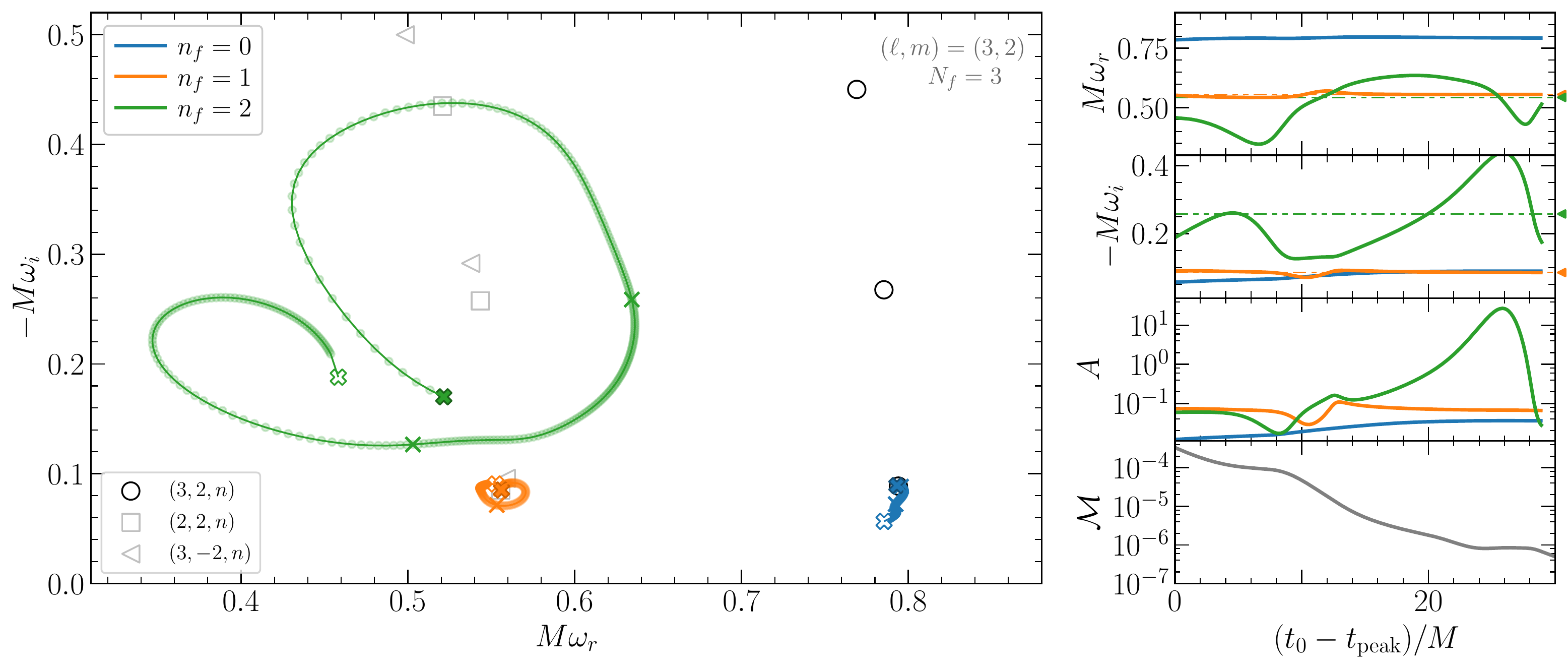}
    \caption{Fits of the $(3,\,2)$ multipole with the \Mr{2} model. 
    Left panel: the extracted frequencies 
    $(M\omega_r, \, -M\omega_i )$ for $n_f=0,\,1,\,2$  for $t_0$ ranging from $t_{\rm peak}$ (empty cross) to $t_{\rm peak}+30 M$ (filled cross); other crosses correspond to times $t_{\rm peak}+10 M$ and $t_{\rm peak}+20 M$.
    We mark the QNMs $(3,2,n)$ (black circles), $(3,-2,n)$ (gray triangle), and $(2,2,n)$ (gray squares). Note that the marker for the $(2,2,0)$ mode almost exactly overlaps with that of the $(3,-2,0)$ mode.
    Right panels: $M\omega_{r}$, $-M\omega_{i}$, $A$ and ${\mathcal M}$ as functions of $t_0$. The horizontal orange (green) dot-dot-dashed lines and triangle markers correspond to the expected frequencies for $M\omega_{220}$ ($M\omega_{221}$).}
    \label{fig:wri3_32}
\end{figure*}

\begin{figure*}[t]
    \includegraphics[width=\textwidth]{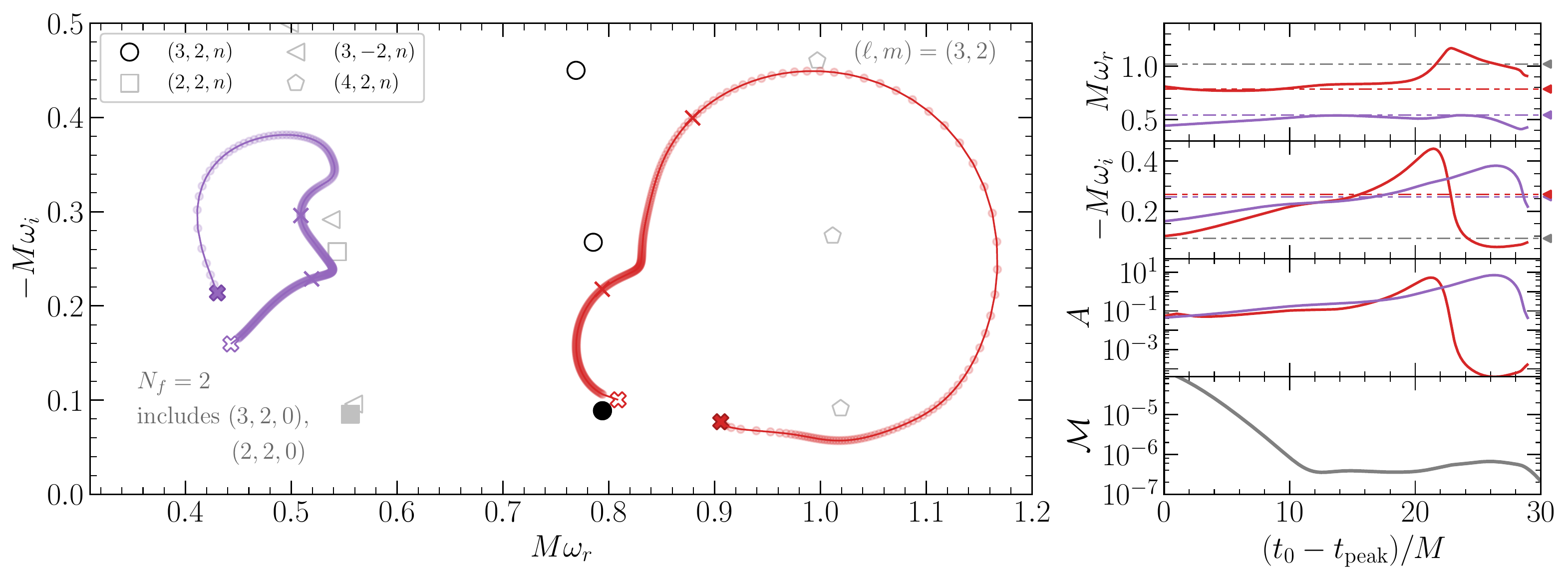}
    \caption{Fits of the $(3,\,2)$ multipole using the \Mxrm{2}{2} model: two free modes ($N_f=2$) in addition to $(2, 2, 0)$ and $(3,2,0)$.  Left panel: the extracted frequencies 
    $(M\omega_r, \, - M\omega_i )$ for $n_f=0,\,1$ and $N_f=2$, for $t_0$ ranging from $t_{\rm peak}$ (empty cross) to $t_{\rm peak}+30 M$ (filled cross); other crosses correspond to times $t_{\rm peak}+10 M$ and $t_{\rm peak}+20 M$.
    We mark the QNMs $(3,2,n)$ (black circles), $(3,-2,n)$ (gray triangles), $(2,2,n)$ (gray squares) and $(4,2,n)$ (gray pentagons).
    Right panels: $M\omega_{r}$, $-M\omega_{i}$, $A$ and ${\mathcal M}$ as functions of $t_0$. The horizontal gray, purple, and red dot-dot-dashed lines and gray triangle markers correspond to the expected frequencies for $M\omega_{420}$, $M\omega_{221}$, and $M\omega_{321}$, respectively.}
    \label{fig:wri2_32}
\end{figure*}

\begin{figure*}[t]
    \includegraphics[width=\textwidth]{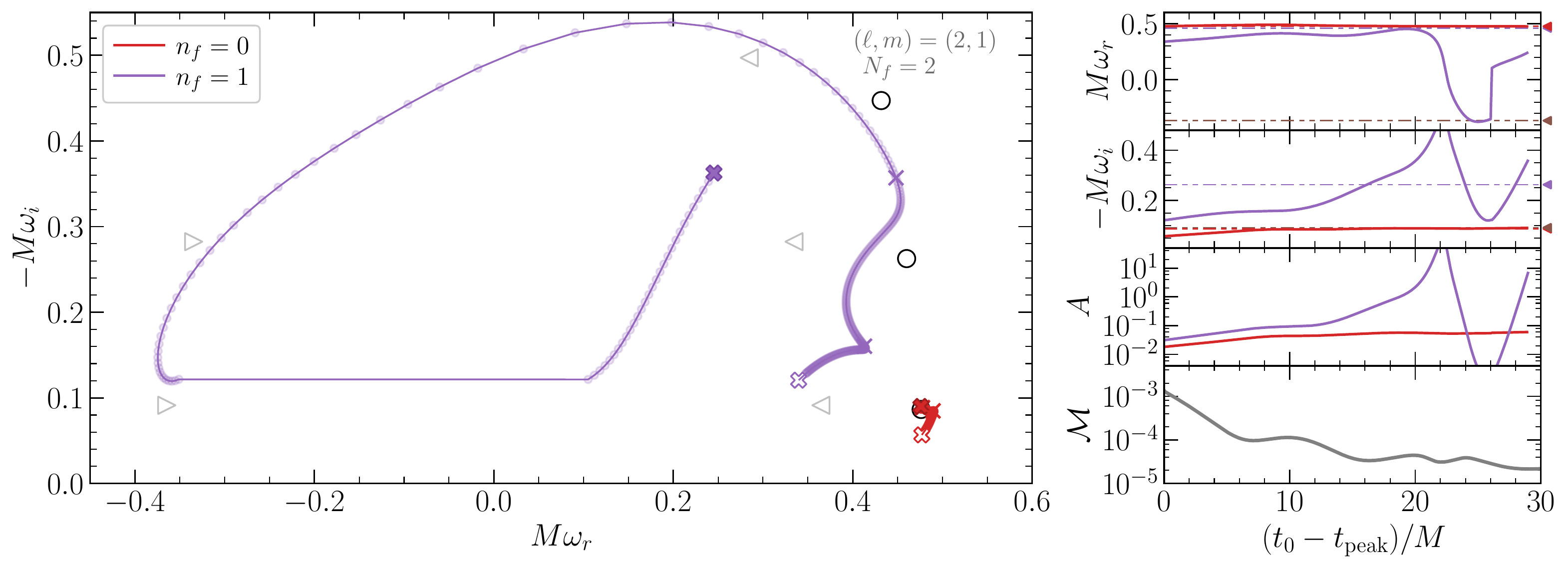}
    \caption{Fits of the $(2,\,1)$ multipole using the \Mr{1} model.  
    Left panel: the extracted frequencies 
    $(M\omega_r, \, -M\omega_i )$ for $n_f=0,\,1$ with $t_0$ ranging from $t_{\rm peak}$ (empty cross) to $t_{\rm peak}+30 M$ (filled cross); other crosses are at $t_{\rm peak}+10 M$ and $t_{\rm peak}+20 M$.
    We mark the known QNM frequencies as follows: $(2,1,n)$ (black circles), $(2,-1,n)$ (gray left-pointing triangles),  $(2,1,n)_R$ (gray right-pointing triangles).
    Right panels: $M\omega_{r}$, $M\omega_{i}$, $A$ and ${\mathcal M}$ as functions of $t_0$. 
    Horizontal red, purple and brown dot-dot-dashed lines and triangle markers are the expected frequencies for the modes $M\omega_{210}$, $M\omega_{211}$ and  $-M\omega_{2-11}^*$, respectively.}
    \label{fig:wri2_21}
\end{figure*}

\section{Higher multipoles}
\label{sec:harmonics}

In Sec.~\ref{sec:lm44} we focused on the $(4,4)$ subdominant multipole, noting that (quite remarkably) the nonlinear mode is easier to extract than the first linear overtone.

In this section we further analyze the $(3,3)$, $(3,2)$, and $(2, 1)$ subdominant multipoles to understand which modes can be extracted from the numerical waveforms, and whether observations that apply to the $(2,2)$ mode generalize to other multipoles of the radiation.

\subsection{$(\ell, m)=(3,3)$}

We fit the $(3,3)$ multipole with the {\Mr{2}} model. In Fig.~\ref{fig:wri3_33} we plot the extracted frequencies, amplitudes, and mismatches. 
The fundamental mode $(3,3,0)$ (blue lines) can be extracted with $\delta\omega<0.01$ as long as $t_0-t_{\rm peak}\gtrsim 10 M$. 
The second free mode ($n_f=1$, orange line) shows a tendency to approach the $(4,3,0)$ mode-mixing QNM, and it still seems to be changing in frequency and amplitude after $t_0-t_{\rm peak}>30 M$. 
The third free mode ($n_f=2$, green line) moves between two QNM frequencies: one (at early times) with large $|M\omega_i|$ and large amplitude, and one at late times with small $|M\omega_i|$ and small amplitude. 

As shown in Fig.~\ref{fig:wri2_33}, this mode is present even when we fit with the \Mrm{2} model, i.e., we add $N_f=2$ free modes along with two fixed frequencies corresponding to the modes $(3,3,0)$ and $(4,3,0)$. 
This model also finds a mode moving between two QNM frequencies: one with $A\simeq 0.053$ and $M\omega\simeq 0.65-0.20i$ at early times ($t_{0}-t_{\rm peak}<5 M$), and one with $A\simeq 0.0012$ and $M\omega\simeq 0.56-0.063i$ at late times ($t_{0}-t_{\rm peak}>15 M$). The free mode with $n_f=0$ in Fig.~\ref{fig:wri2_33} finds a frequency close to the $(3,3,1)$ mode, with ``closest approach'' happening at $t_0-t_{\rm peak} \approx 25 M$.

\subsection{ $(\ell, m)=(3,2)$}

We start by fitting the $(3,2)$ multipole with the free-frequency \Mr{2} model.
We plot the three extracted frequencies, amplitudes and mismatches in Fig.~\ref{fig:wri3_32}. 
The fundamental mode $(3,2,0)$ can be extracted with $\delta\omega<0.01$ when $t_0-t_{\rm peak}\gtrsim 12 M$ (blue line). 
The spherical-spheroidal mixing contamination from the $(2,2,0)$ mode can also be accurately identified, with $\delta\omega\lesssim 10^{-2}$  in the time range we consider (note that in Fig.~\ref{fig:wri3_32} the $(2,2,0)$ mode almost exactly overlaps with the $(3,-2,0)$ mode, although no evidence is found for this additional mode).
The third free mode circles around the $(2,2,1)$ overtone, but never converges to any known QNM frequency value. 

In Fig.~\ref{fig:wri2_32} we try better extract some of the modes by repeating the fit with the {\Mxrm{2}{2}} template, i.e., two fixed modes [$(3,2,0)$ and $(2,2,0)$] and two free modes. 
The $n_f=0$ mode (red line) moves towards the $(3,2,1)$ overtone, but then drifts towards the $(4,2,0)$ QNM due to mode-mixing contamination after $t_0-t_{\rm peak}>25 M$.
The $n_f=1$ mode hovers around to the $(2,2,1)$ overtone, but never gets really close to it.

In conclusion, mode mixing contamination seems to prevent an accurate recovery of the first overtone with any of the free-frequency fits that we attempted.

\subsection{ $(\ell, m)=(2,1)$}
\label{subsec:lm21}

In Fig.~\ref{fig:wri2_21} we plot the two extracted frequencies, amplitudes and mismatches found by fitting the $(2,1)$ multipole with the \Mr{1} model. 
As usual, the fundamental mode $(2,1,0)$ can be extracted with $\delta\omega<0.01$ after $t_0-t_{\rm peak}>5 M$. We also find some evidence of the retrograde mode $(2,1,0)_R$.
If we repeat the fit with the \Mr{2} model the fits are generally very noisy, and there is no robust evidence for the first overtone or for spherical-spheroidal mixing modes.

\bibliography{ref}

\begin{thebibliography}{207}%
\makeatletter
\providecommand \@ifxundefined [1]{%
 \@ifx{#1\undefined}
}%
\providecommand \@ifnum [1]{%
 \ifnum #1\expandafter \@firstoftwo
 \else \expandafter \@secondoftwo
 \fi
}%
\providecommand \@ifx [1]{%
 \ifx #1\expandafter \@firstoftwo
 \else \expandafter \@secondoftwo
 \fi
}%
\providecommand \natexlab [1]{#1}%
\providecommand \enquote  [1]{``#1''}%
\providecommand \bibnamefont  [1]{#1}%
\providecommand \bibfnamefont [1]{#1}%
\providecommand \citenamefont [1]{#1}%
\providecommand \href@noop [0]{\@secondoftwo}%
\providecommand \href [0]{\begingroup \@sanitize@url \@href}%
\providecommand \@href[1]{\@@startlink{#1}\@@href}%
\providecommand \@@href[1]{\endgroup#1\@@endlink}%
\providecommand \@sanitize@url [0]{\catcode `\\12\catcode `\$12\catcode
  `\&12\catcode `\#12\catcode `\^12\catcode `\_12\catcode `\%12\relax}%
\providecommand \@@startlink[1]{}%
\providecommand \@@endlink[0]{}%
\providecommand \url  [0]{\begingroup\@sanitize@url \@url }%
\providecommand \@url [1]{\endgroup\@href {#1}{\urlprefix }}%
\providecommand \urlprefix  [0]{URL }%
\providecommand \Eprint [0]{\href }%
\providecommand \doibase [0]{http://dx.doi.org/}%
\providecommand \selectlanguage [0]{\@gobble}%
\providecommand \bibinfo  [0]{\@secondoftwo}%
\providecommand \bibfield  [0]{\@secondoftwo}%
\providecommand \translation [1]{[#1]}%
\providecommand \BibitemOpen [0]{}%
\providecommand \bibitemStop [0]{}%
\providecommand \bibitemNoStop [0]{.\EOS\space}%
\providecommand \EOS [0]{\spacefactor3000\relax}%
\providecommand \BibitemShut  [1]{\csname bibitem#1\endcsname}%
\let\auto@bib@innerbib\@empty
\bibitem [{\citenamefont {Regge}\ and\ \citenamefont
  {Wheeler}(1957)}]{Regge:1957td}%
  \BibitemOpen
  \bibfield  {author} {\bibinfo {author} {\bibfnamefont {T.}~\bibnamefont
  {Regge}}\ and\ \bibinfo {author} {\bibfnamefont {J.~A.}\ \bibnamefont
  {Wheeler}},\ }\href {\doibase 10.1103/PhysRev.108.1063} {\bibfield  {journal}
  {\bibinfo  {journal} {Phys. Rev.}\ }\textbf {\bibinfo {volume} {108}},\
  \bibinfo {pages} {1063} (\bibinfo {year} {1957})}\BibitemShut {NoStop}%
\bibitem [{\citenamefont {Zerilli}(1970{\natexlab{a}})}]{Zerilli:1970se}%
  \BibitemOpen
  \bibfield  {author} {\bibinfo {author} {\bibfnamefont {F.~J.}\ \bibnamefont
  {Zerilli}},\ }\href {\doibase 10.1103/PhysRevLett.24.737} {\bibfield
  {journal} {\bibinfo  {journal} {Phys. Rev. Lett.}\ }\textbf {\bibinfo
  {volume} {24}},\ \bibinfo {pages} {737} (\bibinfo {year}
  {1970}{\natexlab{a}})}\BibitemShut {NoStop}%
\bibitem [{\citenamefont {Zerilli}(1970{\natexlab{b}})}]{Zerilli:1970wzz}%
  \BibitemOpen
  \bibfield  {author} {\bibinfo {author} {\bibfnamefont {F.~J.}\ \bibnamefont
  {Zerilli}},\ }\href {\doibase 10.1103/PhysRevD.2.2141} {\bibfield  {journal}
  {\bibinfo  {journal} {Phys. Rev. D}\ }\textbf {\bibinfo {volume} {2}},\
  \bibinfo {pages} {2141} (\bibinfo {year} {1970}{\natexlab{b}})}\BibitemShut
  {NoStop}%
\bibitem [{\citenamefont {Vishveshwara}(1970)}]{Vishveshwara:1970zz}%
  \BibitemOpen
  \bibfield  {author} {\bibinfo {author} {\bibfnamefont {C.~V.}\ \bibnamefont
  {Vishveshwara}},\ }\href {\doibase 10.1038/227936a0} {\bibfield  {journal}
  {\bibinfo  {journal} {Nature}\ }\textbf {\bibinfo {volume} {227}},\ \bibinfo
  {pages} {936} (\bibinfo {year} {1970})}\BibitemShut {NoStop}%
\bibitem [{\citenamefont {Nollert}(1999)}]{Nollert:1999ji}%
  \BibitemOpen
  \bibfield  {author} {\bibinfo {author} {\bibfnamefont {H.-P.}\ \bibnamefont
  {Nollert}},\ }\href {\doibase 10.1088/0264-9381/16/12/201} {\bibfield
  {journal} {\bibinfo  {journal} {Class. Quant. Grav.}\ }\textbf {\bibinfo
  {volume} {16}},\ \bibinfo {pages} {R159} (\bibinfo {year}
  {1999})}\BibitemShut {NoStop}%
\bibitem [{\citenamefont {Kokkotas}\ and\ \citenamefont
  {Schmidt}(1999)}]{Kokkotas:1999bd}%
  \BibitemOpen
  \bibfield  {author} {\bibinfo {author} {\bibfnamefont {K.~D.}\ \bibnamefont
  {Kokkotas}}\ and\ \bibinfo {author} {\bibfnamefont {B.~G.}\ \bibnamefont
  {Schmidt}},\ }\href {\doibase 10.12942/lrr-1999-2} {\bibfield  {journal}
  {\bibinfo  {journal} {Living Rev. Rel.}\ }\textbf {\bibinfo {volume} {2}},\
  \bibinfo {pages} {2} (\bibinfo {year} {1999})},\ \Eprint
  {http://arxiv.org/abs/gr-qc/9909058} {arXiv:gr-qc/9909058} \BibitemShut
  {NoStop}%
\bibitem [{\citenamefont {Berti}\ \emph {et~al.}(2009)\citenamefont {Berti},
  \citenamefont {Cardoso},\ and\ \citenamefont {Starinets}}]{Berti:2009kk}%
  \BibitemOpen
  \bibfield  {author} {\bibinfo {author} {\bibfnamefont {E.}~\bibnamefont
  {Berti}}, \bibinfo {author} {\bibfnamefont {V.}~\bibnamefont {Cardoso}}, \
  and\ \bibinfo {author} {\bibfnamefont {A.~O.}\ \bibnamefont {Starinets}},\
  }\href {\doibase 10.1088/0264-9381/26/16/163001} {\bibfield  {journal}
  {\bibinfo  {journal} {Class. Quant. Grav.}\ }\textbf {\bibinfo {volume}
  {26}},\ \bibinfo {pages} {163001} (\bibinfo {year} {2009})},\ \Eprint
  {http://arxiv.org/abs/0905.2975} {arXiv:0905.2975 [gr-qc]} \BibitemShut
  {NoStop}%
\bibitem [{\citenamefont {Konoplya}\ and\ \citenamefont
  {Zhidenko}(2011)}]{Konoplya:2011qq}%
  \BibitemOpen
  \bibfield  {author} {\bibinfo {author} {\bibfnamefont {R.~A.}\ \bibnamefont
  {Konoplya}}\ and\ \bibinfo {author} {\bibfnamefont {A.}~\bibnamefont
  {Zhidenko}},\ }\href {\doibase 10.1103/RevModPhys.83.793} {\bibfield
  {journal} {\bibinfo  {journal} {Rev. Mod. Phys.}\ }\textbf {\bibinfo {volume}
  {83}},\ \bibinfo {pages} {793} (\bibinfo {year} {2011})},\ \Eprint
  {http://arxiv.org/abs/1102.4014} {arXiv:1102.4014 [gr-qc]} \BibitemShut
  {NoStop}%
\bibitem [{\citenamefont {Press}(1971)}]{Press:1971wr}%
  \BibitemOpen
  \bibfield  {author} {\bibinfo {author} {\bibfnamefont {W.~H.}\ \bibnamefont
  {Press}},\ }\href {\doibase 10.1086/180849} {\bibfield  {journal} {\bibinfo
  {journal} {Astrophys. J. Lett.}\ }\textbf {\bibinfo {volume} {170}},\
  \bibinfo {pages} {L105} (\bibinfo {year} {1971})}\BibitemShut {NoStop}%
\bibitem [{\citenamefont {Davis}\ \emph {et~al.}(1971)\citenamefont {Davis},
  \citenamefont {Ruffini}, \citenamefont {Press},\ and\ \citenamefont
  {Price}}]{Davis:1971gg}%
  \BibitemOpen
  \bibfield  {author} {\bibinfo {author} {\bibfnamefont {M.}~\bibnamefont
  {Davis}}, \bibinfo {author} {\bibfnamefont {R.}~\bibnamefont {Ruffini}},
  \bibinfo {author} {\bibfnamefont {W.~H.}\ \bibnamefont {Press}}, \ and\
  \bibinfo {author} {\bibfnamefont {R.~H.}\ \bibnamefont {Price}},\ }\href
  {\doibase 10.1103/PhysRevLett.27.1466} {\bibfield  {journal} {\bibinfo
  {journal} {Phys. Rev. Lett.}\ }\textbf {\bibinfo {volume} {27}},\ \bibinfo
  {pages} {1466} (\bibinfo {year} {1971})}\BibitemShut {NoStop}%
\bibitem [{\citenamefont {Chandrasekhar}\ and\ \citenamefont
  {Detweiler}(1975)}]{Chandrasekhar:1975zza}%
  \BibitemOpen
  \bibfield  {author} {\bibinfo {author} {\bibfnamefont {S.}~\bibnamefont
  {Chandrasekhar}}\ and\ \bibinfo {author} {\bibfnamefont {S.~L.}\ \bibnamefont
  {Detweiler}},\ }\href {\doibase 10.1098/rspa.1975.0112} {\bibfield  {journal}
  {\bibinfo  {journal} {Proc. Roy. Soc. Lond. A}\ }\textbf {\bibinfo {volume}
  {344}},\ \bibinfo {pages} {441} (\bibinfo {year} {1975})}\BibitemShut
  {NoStop}%
\bibitem [{\citenamefont {Ferrari}\ and\ \citenamefont
  {Mashhoon}(1984)}]{Ferrari:1984zz}%
  \BibitemOpen
  \bibfield  {author} {\bibinfo {author} {\bibfnamefont {V.}~\bibnamefont
  {Ferrari}}\ and\ \bibinfo {author} {\bibfnamefont {B.}~\bibnamefont
  {Mashhoon}},\ }\href {\doibase 10.1103/PhysRevD.30.295} {\bibfield  {journal}
  {\bibinfo  {journal} {Phys. Rev. D}\ }\textbf {\bibinfo {volume} {30}},\
  \bibinfo {pages} {295} (\bibinfo {year} {1984})}\BibitemShut {NoStop}%
\bibitem [{\citenamefont {Mashhoon}(1985)}]{Mashhoon:1985cya}%
  \BibitemOpen
  \bibfield  {author} {\bibinfo {author} {\bibfnamefont {B.}~\bibnamefont
  {Mashhoon}},\ }\href {\doibase 10.1103/PhysRevD.31.290} {\bibfield  {journal}
  {\bibinfo  {journal} {Phys. Rev. D}\ }\textbf {\bibinfo {volume} {31}},\
  \bibinfo {pages} {290} (\bibinfo {year} {1985})}\BibitemShut {NoStop}%
\bibitem [{\citenamefont {Blome}\ and\ \citenamefont
  {Mashhoon}(1984)}]{BLOME1984231}%
  \BibitemOpen
  \bibfield  {author} {\bibinfo {author} {\bibfnamefont {H.-J.}\ \bibnamefont
  {Blome}}\ and\ \bibinfo {author} {\bibfnamefont {B.}~\bibnamefont
  {Mashhoon}},\ }\href {\doibase https://doi.org/10.1016/0375-9601(84)90769-2}
  {\bibfield  {journal} {\bibinfo  {journal} {Physics Letters A}\ }\textbf
  {\bibinfo {volume} {100}},\ \bibinfo {pages} {231} (\bibinfo {year}
  {1984})}\BibitemShut {NoStop}%
\bibitem [{\citenamefont {Schutz}\ and\ \citenamefont
  {Will}(1985)}]{Schutz:1985km}%
  \BibitemOpen
  \bibfield  {author} {\bibinfo {author} {\bibfnamefont {B.~F.}\ \bibnamefont
  {Schutz}}\ and\ \bibinfo {author} {\bibfnamefont {C.~M.}\ \bibnamefont
  {Will}},\ }\href {\doibase 10.1086/184453} {\bibfield  {journal} {\bibinfo
  {journal} {Astrophys. J. Lett.}\ }\textbf {\bibinfo {volume} {291}},\
  \bibinfo {pages} {L33} (\bibinfo {year} {1985})}\BibitemShut {NoStop}%
\bibitem [{\citenamefont {Teukolsky}(1972)}]{Teukolsky:1972my}%
  \BibitemOpen
  \bibfield  {author} {\bibinfo {author} {\bibfnamefont {S.~A.}\ \bibnamefont
  {Teukolsky}},\ }\href {\doibase 10.1103/PhysRevLett.29.1114} {\bibfield
  {journal} {\bibinfo  {journal} {Phys. Rev. Lett.}\ }\textbf {\bibinfo
  {volume} {29}},\ \bibinfo {pages} {1114} (\bibinfo {year}
  {1972})}\BibitemShut {NoStop}%
\bibitem [{\citenamefont {Teukolsky}(1973)}]{Teukolsky:1973ha}%
  \BibitemOpen
  \bibfield  {author} {\bibinfo {author} {\bibfnamefont {S.~A.}\ \bibnamefont
  {Teukolsky}},\ }\href {\doibase 10.1086/152444} {\bibfield  {journal}
  {\bibinfo  {journal} {Astrophys. J.}\ }\textbf {\bibinfo {volume} {185}},\
  \bibinfo {pages} {635} (\bibinfo {year} {1973})}\BibitemShut {NoStop}%
\bibitem [{\citenamefont {Press}\ and\ \citenamefont
  {Teukolsky}(1973)}]{Press:1973zz}%
  \BibitemOpen
  \bibfield  {author} {\bibinfo {author} {\bibfnamefont {W.~H.}\ \bibnamefont
  {Press}}\ and\ \bibinfo {author} {\bibfnamefont {S.~A.}\ \bibnamefont
  {Teukolsky}},\ }\href {\doibase 10.1086/152445} {\bibfield  {journal}
  {\bibinfo  {journal} {Astrophys. J.}\ }\textbf {\bibinfo {volume} {185}},\
  \bibinfo {pages} {649} (\bibinfo {year} {1973})}\BibitemShut {NoStop}%
\bibitem [{\citenamefont {Teukolsky}\ and\ \citenamefont
  {Press}(1974)}]{Teukolsky:1974yv}%
  \BibitemOpen
  \bibfield  {author} {\bibinfo {author} {\bibfnamefont {S.~A.}\ \bibnamefont
  {Teukolsky}}\ and\ \bibinfo {author} {\bibfnamefont {W.~H.}\ \bibnamefont
  {Press}},\ }\href {\doibase 10.1086/153180} {\bibfield  {journal} {\bibinfo
  {journal} {Astrophys. J.}\ }\textbf {\bibinfo {volume} {193}},\ \bibinfo
  {pages} {443} (\bibinfo {year} {1974})}\BibitemShut {NoStop}%
\bibitem [{\citenamefont {Detweiler}(1977)}]{Detweiler:1977gy}%
  \BibitemOpen
  \bibfield  {author} {\bibinfo {author} {\bibfnamefont {S.~L.}\ \bibnamefont
  {Detweiler}},\ }\href {\doibase 10.1098/rspa.1977.0005} {\bibfield  {journal}
  {\bibinfo  {journal} {Proc. Roy. Soc. Lond. A}\ }\textbf {\bibinfo {volume}
  {352}},\ \bibinfo {pages} {381} (\bibinfo {year} {1977})}\BibitemShut
  {NoStop}%
\bibitem [{\citenamefont {Detweiler}(1980)}]{Detweiler:1980gk}%
  \BibitemOpen
  \bibfield  {author} {\bibinfo {author} {\bibfnamefont {S.~L.}\ \bibnamefont
  {Detweiler}},\ }\href {\doibase 10.1086/158109} {\bibfield  {journal}
  {\bibinfo  {journal} {Astrophys. J.}\ }\textbf {\bibinfo {volume} {239}},\
  \bibinfo {pages} {292} (\bibinfo {year} {1980})}\BibitemShut {NoStop}%
\bibitem [{\citenamefont {Jaff{\'e}}(1934)}]{Jaffe1934}%
  \BibitemOpen
  \bibfield  {author} {\bibinfo {author} {\bibfnamefont {G.}~\bibnamefont
  {Jaff{\'e}}},\ }\href {\doibase 10.1007/BF01333263} {\bibfield  {journal}
  {\bibinfo  {journal} {Zeitschrift f{\"u}r Physik}\ }\textbf {\bibinfo
  {volume} {87}},\ \bibinfo {pages} {535} (\bibinfo {year} {1934})}\BibitemShut
  {NoStop}%
\bibitem [{\citenamefont {{Baber}}\ and\ \citenamefont
  {{Hass{\'e}}}(1935)}]{1935PCPS...31..564B}%
  \BibitemOpen
  \bibfield  {author} {\bibinfo {author} {\bibfnamefont {W.~G.}\ \bibnamefont
  {{Baber}}}\ and\ \bibinfo {author} {\bibfnamefont {H.~R.}\ \bibnamefont
  {{Hass{\'e}}}},\ }\href {\doibase 10.1017/S0305004100013566} {\bibfield
  {journal} {\bibinfo  {journal} {Proceedings of the Cambridge Philosophical
  Society}\ }\textbf {\bibinfo {volume} {31}},\ \bibinfo {pages} {564}
  (\bibinfo {year} {1935})}\BibitemShut {NoStop}%
\bibitem [{\citenamefont {Leaver}(1985)}]{Leaver:1985ax}%
  \BibitemOpen
  \bibfield  {author} {\bibinfo {author} {\bibfnamefont {E.~W.}\ \bibnamefont
  {Leaver}},\ }\href {\doibase 10.1098/rspa.1985.0119} {\bibfield  {journal}
  {\bibinfo  {journal} {Proc. Roy. Soc. Lond. A}\ }\textbf {\bibinfo {volume}
  {402}},\ \bibinfo {pages} {285} (\bibinfo {year} {1985})}\BibitemShut
  {NoStop}%
\bibitem [{\citenamefont {Echeverria}(1989)}]{Echeverria:1989hg}%
  \BibitemOpen
  \bibfield  {author} {\bibinfo {author} {\bibfnamefont {F.}~\bibnamefont
  {Echeverria}},\ }\href {\doibase 10.1103/PhysRevD.40.3194} {\bibfield
  {journal} {\bibinfo  {journal} {Phys. Rev. D}\ }\textbf {\bibinfo {volume}
  {40}},\ \bibinfo {pages} {3194} (\bibinfo {year} {1989})}\BibitemShut
  {NoStop}%
\bibitem [{\citenamefont {Finn}(1992)}]{Finn:1992wt}%
  \BibitemOpen
  \bibfield  {author} {\bibinfo {author} {\bibfnamefont {L.~S.}\ \bibnamefont
  {Finn}},\ }\href {\doibase 10.1103/PhysRevD.46.5236} {\bibfield  {journal}
  {\bibinfo  {journal} {Phys. Rev. D}\ }\textbf {\bibinfo {volume} {46}},\
  \bibinfo {pages} {5236} (\bibinfo {year} {1992})},\ \Eprint
  {http://arxiv.org/abs/gr-qc/9209010} {arXiv:gr-qc/9209010} \BibitemShut
  {NoStop}%
\bibitem [{\citenamefont {{Thorne}}(1987)}]{1987thyg.book..330T}%
  \BibitemOpen
  \bibfield  {author} {\bibinfo {author} {\bibfnamefont {K.~S.}\ \bibnamefont
  {{Thorne}}},\ }in\ \href@noop {} {\emph {\bibinfo {booktitle} {Three Hundred
  Years of Gravitation}}}\ (\bibinfo {year} {1987})\ pp.\ \bibinfo {pages}
  {330--458}\BibitemShut {NoStop}%
\bibitem [{\citenamefont {Hod}(1998)}]{Hod:1998vk}%
  \BibitemOpen
  \bibfield  {author} {\bibinfo {author} {\bibfnamefont {S.}~\bibnamefont
  {Hod}},\ }\href {\doibase 10.1103/PhysRevLett.81.4293} {\bibfield  {journal}
  {\bibinfo  {journal} {Phys. Rev. Lett.}\ }\textbf {\bibinfo {volume} {81}},\
  \bibinfo {pages} {4293} (\bibinfo {year} {1998})},\ \Eprint
  {http://arxiv.org/abs/gr-qc/9812002} {arXiv:gr-qc/9812002} \BibitemShut
  {NoStop}%
\bibitem [{\citenamefont {Dreyer}(2003)}]{Dreyer:2002vy}%
  \BibitemOpen
  \bibfield  {author} {\bibinfo {author} {\bibfnamefont {O.}~\bibnamefont
  {Dreyer}},\ }\href {\doibase 10.1103/PhysRevLett.90.081301} {\bibfield
  {journal} {\bibinfo  {journal} {Phys. Rev. Lett.}\ }\textbf {\bibinfo
  {volume} {90}},\ \bibinfo {pages} {081301} (\bibinfo {year} {2003})},\
  \Eprint {http://arxiv.org/abs/gr-qc/0211076} {arXiv:gr-qc/0211076}
  \BibitemShut {NoStop}%
\bibitem [{\citenamefont {Dreyer}\ \emph {et~al.}(2004)\citenamefont {Dreyer},
  \citenamefont {Kelly}, \citenamefont {Krishnan}, \citenamefont {Finn},
  \citenamefont {Garrison},\ and\ \citenamefont
  {Lopez-Aleman}}]{Dreyer:2003bv}%
  \BibitemOpen
  \bibfield  {author} {\bibinfo {author} {\bibfnamefont {O.}~\bibnamefont
  {Dreyer}}, \bibinfo {author} {\bibfnamefont {B.~J.}\ \bibnamefont {Kelly}},
  \bibinfo {author} {\bibfnamefont {B.}~\bibnamefont {Krishnan}}, \bibinfo
  {author} {\bibfnamefont {L.~S.}\ \bibnamefont {Finn}}, \bibinfo {author}
  {\bibfnamefont {D.}~\bibnamefont {Garrison}}, \ and\ \bibinfo {author}
  {\bibfnamefont {R.}~\bibnamefont {Lopez-Aleman}},\ }\href {\doibase
  10.1088/0264-9381/21/4/003} {\bibfield  {journal} {\bibinfo  {journal}
  {Class. Quant. Grav.}\ }\textbf {\bibinfo {volume} {21}},\ \bibinfo {pages}
  {787} (\bibinfo {year} {2004})},\ \Eprint
  {http://arxiv.org/abs/gr-qc/0309007} {arXiv:gr-qc/0309007} \BibitemShut
  {NoStop}%
\bibitem [{\citenamefont {Kokkotas}\ \emph {et~al.}(2001)\citenamefont
  {Kokkotas}, \citenamefont {Apostolatos},\ and\ \citenamefont
  {Andersson}}]{Kokkotas:1999mn}%
  \BibitemOpen
  \bibfield  {author} {\bibinfo {author} {\bibfnamefont {K.~D.}\ \bibnamefont
  {Kokkotas}}, \bibinfo {author} {\bibfnamefont {T.~A.}\ \bibnamefont
  {Apostolatos}}, \ and\ \bibinfo {author} {\bibfnamefont {N.}~\bibnamefont
  {Andersson}},\ }\href {\doibase 10.1046/j.1365-8711.2001.03945.x} {\bibfield
  {journal} {\bibinfo  {journal} {Mon. Not. Roy. Astron. Soc.}\ }\textbf
  {\bibinfo {volume} {320}},\ \bibinfo {pages} {307} (\bibinfo {year}
  {2001})},\ \Eprint {http://arxiv.org/abs/gr-qc/9901072} {arXiv:gr-qc/9901072}
  \BibitemShut {NoStop}%
\bibitem [{\citenamefont {Cunningham}\ \emph {et~al.}(1978)\citenamefont
  {Cunningham}, \citenamefont {Price},\ and\ \citenamefont
  {Moncrief}}]{Cunningham:1978zfa}%
  \BibitemOpen
  \bibfield  {author} {\bibinfo {author} {\bibfnamefont {C.~T.}\ \bibnamefont
  {Cunningham}}, \bibinfo {author} {\bibfnamefont {R.~H.}\ \bibnamefont
  {Price}}, \ and\ \bibinfo {author} {\bibfnamefont {V.}~\bibnamefont
  {Moncrief}},\ }\href {\doibase 10.1086/156413} {\bibfield  {journal}
  {\bibinfo  {journal} {Astrophys. J.}\ }\textbf {\bibinfo {volume} {224}},\
  \bibinfo {pages} {643} (\bibinfo {year} {1978})}\BibitemShut {NoStop}%
\bibitem [{\citenamefont {Ferrari}\ and\ \citenamefont
  {Ruffini}(1981)}]{Ferrari:1981dh}%
  \BibitemOpen
  \bibfield  {author} {\bibinfo {author} {\bibfnamefont {V.}~\bibnamefont
  {Ferrari}}\ and\ \bibinfo {author} {\bibfnamefont {R.}~\bibnamefont
  {Ruffini}},\ }\href {\doibase 10.1016/0370-2693(81)90930-8} {\bibfield
  {journal} {\bibinfo  {journal} {Phys. Lett. B}\ }\textbf {\bibinfo {volume}
  {98}},\ \bibinfo {pages} {381} (\bibinfo {year} {1981})}\BibitemShut
  {NoStop}%
\bibitem [{\citenamefont {Stark}\ and\ \citenamefont
  {Piran}(1985)}]{Stark:1985da}%
  \BibitemOpen
  \bibfield  {author} {\bibinfo {author} {\bibfnamefont {R.~F.}\ \bibnamefont
  {Stark}}\ and\ \bibinfo {author} {\bibfnamefont {T.}~\bibnamefont {Piran}},\
  }\href {\doibase 10.1103/PhysRevLett.55.891} {\bibfield  {journal} {\bibinfo
  {journal} {Phys. Rev. Lett.}\ }\textbf {\bibinfo {volume} {55}},\ \bibinfo
  {pages} {891} (\bibinfo {year} {1985})},\ \bibinfo {note} {[Erratum:
  Phys.Rev.Lett. 56, 97 (1986)]}\BibitemShut {NoStop}%
\bibitem [{\citenamefont {Berti}\ \emph {et~al.}(2018)\citenamefont {Berti},
  \citenamefont {Yagi}, \citenamefont {Yang},\ and\ \citenamefont
  {Yunes}}]{Berti:2018vdi}%
  \BibitemOpen
  \bibfield  {author} {\bibinfo {author} {\bibfnamefont {E.}~\bibnamefont
  {Berti}}, \bibinfo {author} {\bibfnamefont {K.}~\bibnamefont {Yagi}},
  \bibinfo {author} {\bibfnamefont {H.}~\bibnamefont {Yang}}, \ and\ \bibinfo
  {author} {\bibfnamefont {N.}~\bibnamefont {Yunes}},\ }\href {\doibase
  10.1007/s10714-018-2372-6} {\bibfield  {journal} {\bibinfo  {journal} {Gen.
  Rel. Grav.}\ }\textbf {\bibinfo {volume} {50}},\ \bibinfo {pages} {49}
  (\bibinfo {year} {2018})},\ \Eprint {http://arxiv.org/abs/1801.03587}
  {arXiv:1801.03587 [gr-qc]} \BibitemShut {NoStop}%
\bibitem [{\citenamefont {Cardoso}\ and\ \citenamefont
  {Pani}(2019)}]{Cardoso:2019rvt}%
  \BibitemOpen
  \bibfield  {author} {\bibinfo {author} {\bibfnamefont {V.}~\bibnamefont
  {Cardoso}}\ and\ \bibinfo {author} {\bibfnamefont {P.}~\bibnamefont {Pani}},\
  }\href {\doibase 10.1007/s41114-019-0020-4} {\bibfield  {journal} {\bibinfo
  {journal} {Living Rev. Rel.}\ }\textbf {\bibinfo {volume} {22}},\ \bibinfo
  {pages} {4} (\bibinfo {year} {2019})},\ \Eprint
  {http://arxiv.org/abs/1904.05363} {arXiv:1904.05363 [gr-qc]} \BibitemShut
  {NoStop}%
\bibitem [{\citenamefont {Berti}\ \emph
  {et~al.}(2006{\natexlab{a}})\citenamefont {Berti}, \citenamefont {Cardoso},\
  and\ \citenamefont {Casals}}]{Berti:2005gp}%
  \BibitemOpen
  \bibfield  {author} {\bibinfo {author} {\bibfnamefont {E.}~\bibnamefont
  {Berti}}, \bibinfo {author} {\bibfnamefont {V.}~\bibnamefont {Cardoso}}, \
  and\ \bibinfo {author} {\bibfnamefont {M.}~\bibnamefont {Casals}},\ }\href
  {\doibase 10.1103/PhysRevD.73.109902} {\bibfield  {journal} {\bibinfo
  {journal} {Phys. Rev. D}\ }\textbf {\bibinfo {volume} {73}},\ \bibinfo
  {pages} {024013} (\bibinfo {year} {2006}{\natexlab{a}})},\ \bibinfo {note}
  {[Erratum: Phys.Rev.D 73, 109902 (2006)]},\ \Eprint
  {http://arxiv.org/abs/gr-qc/0511111} {arXiv:gr-qc/0511111} \BibitemShut
  {NoStop}%
\bibitem [{\citenamefont {Berti}\ and\ \citenamefont
  {Klein}(2014)}]{Berti:2014fga}%
  \BibitemOpen
  \bibfield  {author} {\bibinfo {author} {\bibfnamefont {E.}~\bibnamefont
  {Berti}}\ and\ \bibinfo {author} {\bibfnamefont {A.}~\bibnamefont {Klein}},\
  }\href {\doibase 10.1103/PhysRevD.90.064012} {\bibfield  {journal} {\bibinfo
  {journal} {Phys. Rev. D}\ }\textbf {\bibinfo {volume} {90}},\ \bibinfo
  {pages} {064012} (\bibinfo {year} {2014})},\ \Eprint
  {http://arxiv.org/abs/1408.1860} {arXiv:1408.1860 [gr-qc]} \BibitemShut
  {NoStop}%
\bibitem [{\citenamefont {Leaver}(1986)}]{Leaver:1986gd}%
  \BibitemOpen
  \bibfield  {author} {\bibinfo {author} {\bibfnamefont {E.~W.}\ \bibnamefont
  {Leaver}},\ }\href {\doibase 10.1103/PhysRevD.34.384} {\bibfield  {journal}
  {\bibinfo  {journal} {Phys. Rev. D}\ }\textbf {\bibinfo {volume} {34}},\
  \bibinfo {pages} {384} (\bibinfo {year} {1986})}\BibitemShut {NoStop}%
\bibitem [{\citenamefont {Berti}\ \emph
  {et~al.}(2006{\natexlab{b}})\citenamefont {Berti}, \citenamefont {Cardoso},\
  and\ \citenamefont {Will}}]{Berti:2005ys}%
  \BibitemOpen
  \bibfield  {author} {\bibinfo {author} {\bibfnamefont {E.}~\bibnamefont
  {Berti}}, \bibinfo {author} {\bibfnamefont {V.}~\bibnamefont {Cardoso}}, \
  and\ \bibinfo {author} {\bibfnamefont {C.~M.}\ \bibnamefont {Will}},\ }\href
  {\doibase 10.1103/PhysRevD.73.064030} {\bibfield  {journal} {\bibinfo
  {journal} {Phys. Rev. D}\ }\textbf {\bibinfo {volume} {73}},\ \bibinfo
  {pages} {064030} (\bibinfo {year} {2006}{\natexlab{b}})},\ \Eprint
  {http://arxiv.org/abs/gr-qc/0512160} {arXiv:gr-qc/0512160} \BibitemShut
  {NoStop}%
\bibitem [{RDw()}]{RDwebsites}%
  \BibitemOpen
  \href@noop {} {}\bibinfo {note} {{Webpages with Mathematica notebooks and
  numerical quasinormal mode tables: \\
  \url{https://pages.jh.edu/eberti2/ringdown/} \\
  \url{https://centra.tecnico.ulisboa.pt/network/grit/files/} \\
  \url{https://paolopani.weebly.com/notebooks.html} }}\BibitemShut {NoStop}%
\bibitem [{\citenamefont {Andersson}(1995)}]{Andersson:1995zk}%
  \BibitemOpen
  \bibfield  {author} {\bibinfo {author} {\bibfnamefont {N.}~\bibnamefont
  {Andersson}},\ }\href {\doibase 10.1103/PhysRevD.51.353} {\bibfield
  {journal} {\bibinfo  {journal} {Phys. Rev. D}\ }\textbf {\bibinfo {volume}
  {51}},\ \bibinfo {pages} {353} (\bibinfo {year} {1995})}\BibitemShut
  {NoStop}%
\bibitem [{\citenamefont {Andersson}(1997)}]{Andersson:1996cm}%
  \BibitemOpen
  \bibfield  {author} {\bibinfo {author} {\bibfnamefont {N.}~\bibnamefont
  {Andersson}},\ }\href {\doibase 10.1103/PhysRevD.55.468} {\bibfield
  {journal} {\bibinfo  {journal} {Phys. Rev. D}\ }\textbf {\bibinfo {volume}
  {55}},\ \bibinfo {pages} {468} (\bibinfo {year} {1997})},\ \Eprint
  {http://arxiv.org/abs/gr-qc/9607064} {arXiv:gr-qc/9607064} \BibitemShut
  {NoStop}%
\bibitem [{\citenamefont {Berti}\ and\ \citenamefont
  {Cardoso}(2006)}]{Berti:2006wq}%
  \BibitemOpen
  \bibfield  {author} {\bibinfo {author} {\bibfnamefont {E.}~\bibnamefont
  {Berti}}\ and\ \bibinfo {author} {\bibfnamefont {V.}~\bibnamefont
  {Cardoso}},\ }\href {\doibase 10.1103/PhysRevD.74.104020} {\bibfield
  {journal} {\bibinfo  {journal} {Phys. Rev. D}\ }\textbf {\bibinfo {volume}
  {74}},\ \bibinfo {pages} {104020} (\bibinfo {year} {2006})},\ \Eprint
  {http://arxiv.org/abs/gr-qc/0605118} {arXiv:gr-qc/0605118} \BibitemShut
  {NoStop}%
\bibitem [{\citenamefont {Zhang}\ \emph {et~al.}(2013)\citenamefont {Zhang},
  \citenamefont {Berti},\ and\ \citenamefont {Cardoso}}]{Zhang:2013ksa}%
  \BibitemOpen
  \bibfield  {author} {\bibinfo {author} {\bibfnamefont {Z.}~\bibnamefont
  {Zhang}}, \bibinfo {author} {\bibfnamefont {E.}~\bibnamefont {Berti}}, \ and\
  \bibinfo {author} {\bibfnamefont {V.}~\bibnamefont {Cardoso}},\ }\href
  {\doibase 10.1103/PhysRevD.88.044018} {\bibfield  {journal} {\bibinfo
  {journal} {Phys. Rev. D}\ }\textbf {\bibinfo {volume} {88}},\ \bibinfo
  {pages} {044018} (\bibinfo {year} {2013})},\ \Eprint
  {http://arxiv.org/abs/1305.4306} {arXiv:1305.4306 [gr-qc]} \BibitemShut
  {NoStop}%
\bibitem [{\citenamefont {Oshita}(2021)}]{Oshita:2021iyn}%
  \BibitemOpen
  \bibfield  {author} {\bibinfo {author} {\bibfnamefont {N.}~\bibnamefont
  {Oshita}},\ }\href {\doibase 10.1103/PhysRevD.104.124032} {\bibfield
  {journal} {\bibinfo  {journal} {Phys. Rev. D}\ }\textbf {\bibinfo {volume}
  {104}},\ \bibinfo {pages} {124032} (\bibinfo {year} {2021})},\ \Eprint
  {http://arxiv.org/abs/2109.09757} {arXiv:2109.09757 [gr-qc]} \BibitemShut
  {NoStop}%
\bibitem [{\citenamefont {Lagos}\ and\ \citenamefont
  {Hui}(2022)}]{Lagos:2022otp}%
  \BibitemOpen
  \bibfield  {author} {\bibinfo {author} {\bibfnamefont {M.}~\bibnamefont
  {Lagos}}\ and\ \bibinfo {author} {\bibfnamefont {L.}~\bibnamefont {Hui}},\
  }\href@noop {} {\  (\bibinfo {year} {2022})},\ \Eprint
  {http://arxiv.org/abs/2208.07379} {arXiv:2208.07379 [gr-qc]} \BibitemShut
  {NoStop}%
\bibitem [{\citenamefont {Flanagan}\ and\ \citenamefont
  {Hughes}(1998)}]{Flanagan:1997sx}%
  \BibitemOpen
  \bibfield  {author} {\bibinfo {author} {\bibfnamefont {E.~E.}\ \bibnamefont
  {Flanagan}}\ and\ \bibinfo {author} {\bibfnamefont {S.~A.}\ \bibnamefont
  {Hughes}},\ }\href {\doibase 10.1103/PhysRevD.57.4535} {\bibfield  {journal}
  {\bibinfo  {journal} {Phys. Rev. D}\ }\textbf {\bibinfo {volume} {57}},\
  \bibinfo {pages} {4535} (\bibinfo {year} {1998})},\ \Eprint
  {http://arxiv.org/abs/gr-qc/9701039} {arXiv:gr-qc/9701039} \BibitemShut
  {NoStop}%
\bibitem [{\citenamefont {Berti}\ \emph
  {et~al.}(2006{\natexlab{c}})\citenamefont {Berti}, \citenamefont {Cardoso},\
  and\ \citenamefont {Will}}]{Berti:2006hb}%
  \BibitemOpen
  \bibfield  {author} {\bibinfo {author} {\bibfnamefont {E.}~\bibnamefont
  {Berti}}, \bibinfo {author} {\bibfnamefont {V.}~\bibnamefont {Cardoso}}, \
  and\ \bibinfo {author} {\bibfnamefont {C.~M.}\ \bibnamefont {Will}},\ }\href
  {\doibase 10.1063/1.2348047} {\bibfield  {journal} {\bibinfo  {journal} {AIP
  Conf. Proc.}\ }\textbf {\bibinfo {volume} {848}},\ \bibinfo {pages} {687}
  (\bibinfo {year} {2006}{\natexlab{c}})},\ \Eprint
  {http://arxiv.org/abs/gr-qc/0601077} {arXiv:gr-qc/0601077} \BibitemShut
  {NoStop}%
\bibitem [{\citenamefont {Pretorius}(2005)}]{Pretorius:2005gq}%
  \BibitemOpen
  \bibfield  {author} {\bibinfo {author} {\bibfnamefont {F.}~\bibnamefont
  {Pretorius}},\ }\href {\doibase 10.1103/PhysRevLett.95.121101} {\bibfield
  {journal} {\bibinfo  {journal} {Phys. Rev. Lett.}\ }\textbf {\bibinfo
  {volume} {95}},\ \bibinfo {pages} {121101} (\bibinfo {year} {2005})},\
  \Eprint {http://arxiv.org/abs/gr-qc/0507014} {arXiv:gr-qc/0507014}
  \BibitemShut {NoStop}%
\bibitem [{\citenamefont {Campanelli}\ \emph {et~al.}(2006)\citenamefont
  {Campanelli}, \citenamefont {Lousto}, \citenamefont {Marronetti},\ and\
  \citenamefont {Zlochower}}]{Campanelli:2005dd}%
  \BibitemOpen
  \bibfield  {author} {\bibinfo {author} {\bibfnamefont {M.}~\bibnamefont
  {Campanelli}}, \bibinfo {author} {\bibfnamefont {C.~O.}\ \bibnamefont
  {Lousto}}, \bibinfo {author} {\bibfnamefont {P.}~\bibnamefont {Marronetti}},
  \ and\ \bibinfo {author} {\bibfnamefont {Y.}~\bibnamefont {Zlochower}},\
  }\href {\doibase 10.1103/PhysRevLett.96.111101} {\bibfield  {journal}
  {\bibinfo  {journal} {Phys. Rev. Lett.}\ }\textbf {\bibinfo {volume} {96}},\
  \bibinfo {pages} {111101} (\bibinfo {year} {2006})},\ \Eprint
  {http://arxiv.org/abs/gr-qc/0511048} {arXiv:gr-qc/0511048} \BibitemShut
  {NoStop}%
\bibitem [{\citenamefont {Baker}\ \emph {et~al.}(2006)\citenamefont {Baker},
  \citenamefont {Centrella}, \citenamefont {Choi}, \citenamefont {Koppitz},\
  and\ \citenamefont {van Meter}}]{Baker:2005vv}%
  \BibitemOpen
  \bibfield  {author} {\bibinfo {author} {\bibfnamefont {J.~G.}\ \bibnamefont
  {Baker}}, \bibinfo {author} {\bibfnamefont {J.}~\bibnamefont {Centrella}},
  \bibinfo {author} {\bibfnamefont {D.-I.}\ \bibnamefont {Choi}}, \bibinfo
  {author} {\bibfnamefont {M.}~\bibnamefont {Koppitz}}, \ and\ \bibinfo
  {author} {\bibfnamefont {J.}~\bibnamefont {van Meter}},\ }\href {\doibase
  10.1103/PhysRevLett.96.111102} {\bibfield  {journal} {\bibinfo  {journal}
  {Phys. Rev. Lett.}\ }\textbf {\bibinfo {volume} {96}},\ \bibinfo {pages}
  {111102} (\bibinfo {year} {2006})},\ \Eprint
  {http://arxiv.org/abs/gr-qc/0511103} {arXiv:gr-qc/0511103} \BibitemShut
  {NoStop}%
\bibitem [{\citenamefont {Buonanno}\ \emph {et~al.}(2007)\citenamefont
  {Buonanno}, \citenamefont {Cook},\ and\ \citenamefont
  {Pretorius}}]{Buonanno:2006ui}%
  \BibitemOpen
  \bibfield  {author} {\bibinfo {author} {\bibfnamefont {A.}~\bibnamefont
  {Buonanno}}, \bibinfo {author} {\bibfnamefont {G.~B.}\ \bibnamefont {Cook}},
  \ and\ \bibinfo {author} {\bibfnamefont {F.}~\bibnamefont {Pretorius}},\
  }\href {\doibase 10.1103/PhysRevD.75.124018} {\bibfield  {journal} {\bibinfo
  {journal} {Phys. Rev. D}\ }\textbf {\bibinfo {volume} {75}},\ \bibinfo
  {pages} {124018} (\bibinfo {year} {2007})},\ \Eprint
  {http://arxiv.org/abs/gr-qc/0610122} {arXiv:gr-qc/0610122} \BibitemShut
  {NoStop}%
\bibitem [{\citenamefont {Berti}\ \emph
  {et~al.}(2007{\natexlab{a}})\citenamefont {Berti}, \citenamefont {Cardoso},
  \citenamefont {Gonzalez}, \citenamefont {Sperhake}, \citenamefont {Hannam},
  \citenamefont {Husa},\ and\ \citenamefont {Bruegmann}}]{Berti:2007fi}%
  \BibitemOpen
  \bibfield  {author} {\bibinfo {author} {\bibfnamefont {E.}~\bibnamefont
  {Berti}}, \bibinfo {author} {\bibfnamefont {V.}~\bibnamefont {Cardoso}},
  \bibinfo {author} {\bibfnamefont {J.~A.}\ \bibnamefont {Gonzalez}}, \bibinfo
  {author} {\bibfnamefont {U.}~\bibnamefont {Sperhake}}, \bibinfo {author}
  {\bibfnamefont {M.}~\bibnamefont {Hannam}}, \bibinfo {author} {\bibfnamefont
  {S.}~\bibnamefont {Husa}}, \ and\ \bibinfo {author} {\bibfnamefont
  {B.}~\bibnamefont {Bruegmann}},\ }\href {\doibase 10.1103/PhysRevD.76.064034}
  {\bibfield  {journal} {\bibinfo  {journal} {Phys. Rev. D}\ }\textbf {\bibinfo
  {volume} {76}},\ \bibinfo {pages} {064034} (\bibinfo {year}
  {2007}{\natexlab{a}})},\ \Eprint {http://arxiv.org/abs/gr-qc/0703053}
  {arXiv:gr-qc/0703053} \BibitemShut {NoStop}%
\bibitem [{\citenamefont {Abbott}\ \emph
  {et~al.}(2016{\natexlab{a}})\citenamefont {Abbott} \emph
  {et~al.}}]{LIGOScientific:2016aoc}%
  \BibitemOpen
  \bibfield  {author} {\bibinfo {author} {\bibfnamefont {B.~P.}\ \bibnamefont
  {Abbott}} \emph {et~al.} (\bibinfo {collaboration} {LIGO Scientific,
  Virgo}),\ }\href {\doibase 10.1103/PhysRevLett.116.061102} {\bibfield
  {journal} {\bibinfo  {journal} {Phys. Rev. Lett.}\ }\textbf {\bibinfo
  {volume} {116}},\ \bibinfo {pages} {061102} (\bibinfo {year}
  {2016}{\natexlab{a}})},\ \Eprint {http://arxiv.org/abs/1602.03837}
  {arXiv:1602.03837 [gr-qc]} \BibitemShut {NoStop}%
\bibitem [{\citenamefont {Aasi}\ \emph {et~al.}(2015)\citenamefont {Aasi} \emph
  {et~al.}}]{LIGOScientific:2014pky}%
  \BibitemOpen
  \bibfield  {author} {\bibinfo {author} {\bibfnamefont {J.}~\bibnamefont
  {Aasi}} \emph {et~al.} (\bibinfo {collaboration} {LIGO Scientific}),\ }\href
  {\doibase 10.1088/0264-9381/32/7/074001} {\bibfield  {journal} {\bibinfo
  {journal} {Class. Quant. Grav.}\ }\textbf {\bibinfo {volume} {32}},\ \bibinfo
  {pages} {074001} (\bibinfo {year} {2015})},\ \Eprint
  {http://arxiv.org/abs/1411.4547} {arXiv:1411.4547 [gr-qc]} \BibitemShut
  {NoStop}%
\bibitem [{\citenamefont {Acernese}\ \emph {et~al.}(2015)\citenamefont
  {Acernese} \emph {et~al.}}]{VIRGO:2014yos}%
  \BibitemOpen
  \bibfield  {author} {\bibinfo {author} {\bibfnamefont {F.}~\bibnamefont
  {Acernese}} \emph {et~al.} (\bibinfo {collaboration} {VIRGO}),\ }\href
  {\doibase 10.1088/0264-9381/32/2/024001} {\bibfield  {journal} {\bibinfo
  {journal} {Class. Quant. Grav.}\ }\textbf {\bibinfo {volume} {32}},\ \bibinfo
  {pages} {024001} (\bibinfo {year} {2015})},\ \Eprint
  {http://arxiv.org/abs/1408.3978} {arXiv:1408.3978 [gr-qc]} \BibitemShut
  {NoStop}%
\bibitem [{\citenamefont {Akutsu}\ \emph {et~al.}(2021)\citenamefont {Akutsu}
  \emph {et~al.}}]{KAGRA:2020tym}%
  \BibitemOpen
  \bibfield  {author} {\bibinfo {author} {\bibfnamefont {T.}~\bibnamefont
  {Akutsu}} \emph {et~al.} (\bibinfo {collaboration} {KAGRA}),\ }\href
  {\doibase 10.1093/ptep/ptaa125} {\bibfield  {journal} {\bibinfo  {journal}
  {PTEP}\ }\textbf {\bibinfo {volume} {2021}},\ \bibinfo {pages} {05A101}
  (\bibinfo {year} {2021})},\ \Eprint {http://arxiv.org/abs/2005.05574}
  {arXiv:2005.05574 [physics.ins-det]} \BibitemShut {NoStop}%
\bibitem [{\citenamefont {Abbott}\ \emph
  {et~al.}(2019{\natexlab{a}})\citenamefont {Abbott} \emph
  {et~al.}}]{LIGOScientific:2018mvr}%
  \BibitemOpen
  \bibfield  {author} {\bibinfo {author} {\bibfnamefont {B.~P.}\ \bibnamefont
  {Abbott}} \emph {et~al.} (\bibinfo {collaboration} {LIGO Scientific,
  Virgo}),\ }\href {\doibase 10.1103/PhysRevX.9.031040} {\bibfield  {journal}
  {\bibinfo  {journal} {Phys. Rev. X}\ }\textbf {\bibinfo {volume} {9}},\
  \bibinfo {pages} {031040} (\bibinfo {year} {2019}{\natexlab{a}})},\ \Eprint
  {http://arxiv.org/abs/1811.12907} {arXiv:1811.12907 [astro-ph.HE]}
  \BibitemShut {NoStop}%
\bibitem [{\citenamefont {Abbott}\ \emph
  {et~al.}(2021{\natexlab{a}})\citenamefont {Abbott} \emph
  {et~al.}}]{LIGOScientific:2020ibl}%
  \BibitemOpen
  \bibfield  {author} {\bibinfo {author} {\bibfnamefont {R.}~\bibnamefont
  {Abbott}} \emph {et~al.} (\bibinfo {collaboration} {LIGO Scientific,
  Virgo}),\ }\href {\doibase 10.1103/PhysRevX.11.021053} {\bibfield  {journal}
  {\bibinfo  {journal} {Phys. Rev. X}\ }\textbf {\bibinfo {volume} {11}},\
  \bibinfo {pages} {021053} (\bibinfo {year} {2021}{\natexlab{a}})},\ \Eprint
  {http://arxiv.org/abs/2010.14527} {arXiv:2010.14527 [gr-qc]} \BibitemShut
  {NoStop}%
\bibitem [{\citenamefont {Abbott}\ \emph
  {et~al.}(2021{\natexlab{b}})\citenamefont {Abbott} \emph
  {et~al.}}]{LIGOScientific:2021usb}%
  \BibitemOpen
  \bibfield  {author} {\bibinfo {author} {\bibfnamefont {R.}~\bibnamefont
  {Abbott}} \emph {et~al.} (\bibinfo {collaboration} {LIGO Scientific,
  VIRGO}),\ }\href@noop {} {\  (\bibinfo {year} {2021}{\natexlab{b}})},\
  \Eprint {http://arxiv.org/abs/2108.01045} {arXiv:2108.01045 [gr-qc]}
  \BibitemShut {NoStop}%
\bibitem [{\citenamefont {Abbott}\ \emph
  {et~al.}(2021{\natexlab{c}})\citenamefont {Abbott} \emph
  {et~al.}}]{LIGOScientific:2021djp}%
  \BibitemOpen
  \bibfield  {author} {\bibinfo {author} {\bibfnamefont {R.}~\bibnamefont
  {Abbott}} \emph {et~al.} (\bibinfo {collaboration} {LIGO Scientific, VIRGO,
  KAGRA}),\ }\href@noop {} {\  (\bibinfo {year} {2021}{\natexlab{c}})},\
  \Eprint {http://arxiv.org/abs/2111.03606} {arXiv:2111.03606 [gr-qc]}
  \BibitemShut {NoStop}%
\bibitem [{\citenamefont {Nitz}\ \emph {et~al.}(2019)\citenamefont {Nitz},
  \citenamefont {Capano}, \citenamefont {Nielsen}, \citenamefont {Reyes},
  \citenamefont {White}, \citenamefont {Brown},\ and\ \citenamefont
  {Krishnan}}]{Nitz:2018imz}%
  \BibitemOpen
  \bibfield  {author} {\bibinfo {author} {\bibfnamefont {A.~H.}\ \bibnamefont
  {Nitz}}, \bibinfo {author} {\bibfnamefont {C.}~\bibnamefont {Capano}},
  \bibinfo {author} {\bibfnamefont {A.~B.}\ \bibnamefont {Nielsen}}, \bibinfo
  {author} {\bibfnamefont {S.}~\bibnamefont {Reyes}}, \bibinfo {author}
  {\bibfnamefont {R.}~\bibnamefont {White}}, \bibinfo {author} {\bibfnamefont
  {D.~A.}\ \bibnamefont {Brown}}, \ and\ \bibinfo {author} {\bibfnamefont
  {B.}~\bibnamefont {Krishnan}},\ }\href {\doibase 10.3847/1538-4357/ab0108}
  {\bibfield  {journal} {\bibinfo  {journal} {Astrophys. J.}\ }\textbf
  {\bibinfo {volume} {872}},\ \bibinfo {pages} {195} (\bibinfo {year}
  {2019})},\ \Eprint {http://arxiv.org/abs/1811.01921} {arXiv:1811.01921
  [gr-qc]} \BibitemShut {NoStop}%
\bibitem [{\citenamefont {Nitz}\ \emph {et~al.}(2020)\citenamefont {Nitz},
  \citenamefont {Dent}, \citenamefont {Davies}, \citenamefont {Kumar},
  \citenamefont {Capano}, \citenamefont {Harry}, \citenamefont {Mozzon},
  \citenamefont {Nuttall}, \citenamefont {Lundgren},\ and\ \citenamefont
  {T\'apai}}]{Nitz:2020oeq}%
  \BibitemOpen
  \bibfield  {author} {\bibinfo {author} {\bibfnamefont {A.~H.}\ \bibnamefont
  {Nitz}}, \bibinfo {author} {\bibfnamefont {T.}~\bibnamefont {Dent}}, \bibinfo
  {author} {\bibfnamefont {G.~S.}\ \bibnamefont {Davies}}, \bibinfo {author}
  {\bibfnamefont {S.}~\bibnamefont {Kumar}}, \bibinfo {author} {\bibfnamefont
  {C.~D.}\ \bibnamefont {Capano}}, \bibinfo {author} {\bibfnamefont
  {I.}~\bibnamefont {Harry}}, \bibinfo {author} {\bibfnamefont
  {S.}~\bibnamefont {Mozzon}}, \bibinfo {author} {\bibfnamefont
  {L.}~\bibnamefont {Nuttall}}, \bibinfo {author} {\bibfnamefont
  {A.}~\bibnamefont {Lundgren}}, \ and\ \bibinfo {author} {\bibfnamefont
  {M.}~\bibnamefont {T\'apai}},\ }\href {\doibase 10.3847/1538-4357/ab733f}
  {\bibfield  {journal} {\bibinfo  {journal} {Astrophys. J.}\ }\textbf
  {\bibinfo {volume} {891}},\ \bibinfo {pages} {123} (\bibinfo {year}
  {2020})},\ \Eprint {http://arxiv.org/abs/1910.05331} {arXiv:1910.05331
  [astro-ph.HE]} \BibitemShut {NoStop}%
\bibitem [{\citenamefont {Nitz}\ \emph {et~al.}(2021)\citenamefont {Nitz},
  \citenamefont {Capano}, \citenamefont {Kumar}, \citenamefont {Wang},
  \citenamefont {Kastha}, \citenamefont {Sch\"afer}, \citenamefont
  {Dhurkunde},\ and\ \citenamefont {Cabero}}]{Nitz:2021uxj}%
  \BibitemOpen
  \bibfield  {author} {\bibinfo {author} {\bibfnamefont {A.~H.}\ \bibnamefont
  {Nitz}}, \bibinfo {author} {\bibfnamefont {C.~D.}\ \bibnamefont {Capano}},
  \bibinfo {author} {\bibfnamefont {S.}~\bibnamefont {Kumar}}, \bibinfo
  {author} {\bibfnamefont {Y.-F.}\ \bibnamefont {Wang}}, \bibinfo {author}
  {\bibfnamefont {S.}~\bibnamefont {Kastha}}, \bibinfo {author} {\bibfnamefont
  {M.}~\bibnamefont {Sch\"afer}}, \bibinfo {author} {\bibfnamefont
  {R.}~\bibnamefont {Dhurkunde}}, \ and\ \bibinfo {author} {\bibfnamefont
  {M.}~\bibnamefont {Cabero}},\ }\href {\doibase 10.3847/1538-4357/ac1c03}
  {\bibfield  {journal} {\bibinfo  {journal} {Astrophys. J.}\ }\textbf
  {\bibinfo {volume} {922}},\ \bibinfo {pages} {76} (\bibinfo {year} {2021})},\
  \Eprint {http://arxiv.org/abs/2105.09151} {arXiv:2105.09151 [astro-ph.HE]}
  \BibitemShut {NoStop}%
\bibitem [{\citenamefont {Venumadhav}\ \emph {et~al.}(2020)\citenamefont
  {Venumadhav}, \citenamefont {Zackay}, \citenamefont {Roulet}, \citenamefont
  {Dai},\ and\ \citenamefont {Zaldarriaga}}]{Venumadhav:2019lyq}%
  \BibitemOpen
  \bibfield  {author} {\bibinfo {author} {\bibfnamefont {T.}~\bibnamefont
  {Venumadhav}}, \bibinfo {author} {\bibfnamefont {B.}~\bibnamefont {Zackay}},
  \bibinfo {author} {\bibfnamefont {J.}~\bibnamefont {Roulet}}, \bibinfo
  {author} {\bibfnamefont {L.}~\bibnamefont {Dai}}, \ and\ \bibinfo {author}
  {\bibfnamefont {M.}~\bibnamefont {Zaldarriaga}},\ }\href {\doibase
  10.1103/PhysRevD.101.083030} {\bibfield  {journal} {\bibinfo  {journal}
  {Phys. Rev. D}\ }\textbf {\bibinfo {volume} {101}},\ \bibinfo {pages}
  {083030} (\bibinfo {year} {2020})},\ \Eprint
  {http://arxiv.org/abs/1904.07214} {arXiv:1904.07214 [astro-ph.HE]}
  \BibitemShut {NoStop}%
\bibitem [{\citenamefont {Zackay}\ \emph {et~al.}(2021)\citenamefont {Zackay},
  \citenamefont {Dai}, \citenamefont {Venumadhav}, \citenamefont {Roulet},\
  and\ \citenamefont {Zaldarriaga}}]{Zackay:2019btq}%
  \BibitemOpen
  \bibfield  {author} {\bibinfo {author} {\bibfnamefont {B.}~\bibnamefont
  {Zackay}}, \bibinfo {author} {\bibfnamefont {L.}~\bibnamefont {Dai}},
  \bibinfo {author} {\bibfnamefont {T.}~\bibnamefont {Venumadhav}}, \bibinfo
  {author} {\bibfnamefont {J.}~\bibnamefont {Roulet}}, \ and\ \bibinfo {author}
  {\bibfnamefont {M.}~\bibnamefont {Zaldarriaga}},\ }\href {\doibase
  10.1103/PhysRevD.104.063030} {\bibfield  {journal} {\bibinfo  {journal}
  {Phys. Rev. D}\ }\textbf {\bibinfo {volume} {104}},\ \bibinfo {pages}
  {063030} (\bibinfo {year} {2021})},\ \Eprint
  {http://arxiv.org/abs/1910.09528} {arXiv:1910.09528 [astro-ph.HE]}
  \BibitemShut {NoStop}%
\bibitem [{\citenamefont {Abbott}\ \emph
  {et~al.}(2019{\natexlab{b}})\citenamefont {Abbott} \emph
  {et~al.}}]{LIGOScientific:2019fpa}%
  \BibitemOpen
  \bibfield  {author} {\bibinfo {author} {\bibfnamefont {B.~P.}\ \bibnamefont
  {Abbott}} \emph {et~al.} (\bibinfo {collaboration} {LIGO Scientific,
  Virgo}),\ }\href {\doibase 10.1103/PhysRevD.100.104036} {\bibfield  {journal}
  {\bibinfo  {journal} {Phys. Rev. D}\ }\textbf {\bibinfo {volume} {100}},\
  \bibinfo {pages} {104036} (\bibinfo {year} {2019}{\natexlab{b}})},\ \Eprint
  {http://arxiv.org/abs/1903.04467} {arXiv:1903.04467 [gr-qc]} \BibitemShut
  {NoStop}%
\bibitem [{\citenamefont {Abbott}\ \emph
  {et~al.}(2021{\natexlab{d}})\citenamefont {Abbott} \emph
  {et~al.}}]{LIGOScientific:2020tif}%
  \BibitemOpen
  \bibfield  {author} {\bibinfo {author} {\bibfnamefont {R.}~\bibnamefont
  {Abbott}} \emph {et~al.} (\bibinfo {collaboration} {LIGO Scientific,
  Virgo}),\ }\href {\doibase 10.1103/PhysRevD.103.122002} {\bibfield  {journal}
  {\bibinfo  {journal} {Phys. Rev. D}\ }\textbf {\bibinfo {volume} {103}},\
  \bibinfo {pages} {122002} (\bibinfo {year} {2021}{\natexlab{d}})},\ \Eprint
  {http://arxiv.org/abs/2010.14529} {arXiv:2010.14529 [gr-qc]} \BibitemShut
  {NoStop}%
\bibitem [{\citenamefont {Abbott}\ \emph
  {et~al.}(2021{\natexlab{e}})\citenamefont {Abbott} \emph
  {et~al.}}]{LIGOScientific:2021sio}%
  \BibitemOpen
  \bibfield  {author} {\bibinfo {author} {\bibfnamefont {R.}~\bibnamefont
  {Abbott}} \emph {et~al.} (\bibinfo {collaboration} {LIGO Scientific, VIRGO,
  KAGRA}),\ }\href@noop {} {\  (\bibinfo {year} {2021}{\natexlab{e}})},\
  \Eprint {http://arxiv.org/abs/2112.06861} {arXiv:2112.06861 [gr-qc]}
  \BibitemShut {NoStop}%
\bibitem [{\citenamefont {Berti}\ \emph
  {et~al.}(2007{\natexlab{b}})\citenamefont {Berti}, \citenamefont {Cardoso},
  \citenamefont {Cardoso},\ and\ \citenamefont {Cavaglia}}]{Berti:2007zu}%
  \BibitemOpen
  \bibfield  {author} {\bibinfo {author} {\bibfnamefont {E.}~\bibnamefont
  {Berti}}, \bibinfo {author} {\bibfnamefont {J.}~\bibnamefont {Cardoso}},
  \bibinfo {author} {\bibfnamefont {V.}~\bibnamefont {Cardoso}}, \ and\
  \bibinfo {author} {\bibfnamefont {M.}~\bibnamefont {Cavaglia}},\ }\href
  {\doibase 10.1103/PhysRevD.76.104044} {\bibfield  {journal} {\bibinfo
  {journal} {Phys. Rev. D}\ }\textbf {\bibinfo {volume} {76}},\ \bibinfo
  {pages} {104044} (\bibinfo {year} {2007}{\natexlab{b}})},\ \Eprint
  {http://arxiv.org/abs/0707.1202} {arXiv:0707.1202 [gr-qc]} \BibitemShut
  {NoStop}%
\bibitem [{\citenamefont {Baibhav}\ \emph {et~al.}(2018)\citenamefont
  {Baibhav}, \citenamefont {Berti}, \citenamefont {Cardoso},\ and\
  \citenamefont {Khanna}}]{Baibhav:2017jhs}%
  \BibitemOpen
  \bibfield  {author} {\bibinfo {author} {\bibfnamefont {V.}~\bibnamefont
  {Baibhav}}, \bibinfo {author} {\bibfnamefont {E.}~\bibnamefont {Berti}},
  \bibinfo {author} {\bibfnamefont {V.}~\bibnamefont {Cardoso}}, \ and\
  \bibinfo {author} {\bibfnamefont {G.}~\bibnamefont {Khanna}},\ }\href
  {\doibase 10.1103/PhysRevD.97.044048} {\bibfield  {journal} {\bibinfo
  {journal} {Phys. Rev. D}\ }\textbf {\bibinfo {volume} {97}},\ \bibinfo
  {pages} {044048} (\bibinfo {year} {2018})},\ \Eprint
  {http://arxiv.org/abs/1710.02156} {arXiv:1710.02156 [gr-qc]} \BibitemShut
  {NoStop}%
\bibitem [{\citenamefont {Kamaretsos}\ \emph
  {et~al.}(2012{\natexlab{a}})\citenamefont {Kamaretsos}, \citenamefont
  {Hannam}, \citenamefont {Husa},\ and\ \citenamefont
  {Sathyaprakash}}]{Kamaretsos:2011um}%
  \BibitemOpen
  \bibfield  {author} {\bibinfo {author} {\bibfnamefont {I.}~\bibnamefont
  {Kamaretsos}}, \bibinfo {author} {\bibfnamefont {M.}~\bibnamefont {Hannam}},
  \bibinfo {author} {\bibfnamefont {S.}~\bibnamefont {Husa}}, \ and\ \bibinfo
  {author} {\bibfnamefont {B.~S.}\ \bibnamefont {Sathyaprakash}},\ }\href
  {\doibase 10.1103/PhysRevD.85.024018} {\bibfield  {journal} {\bibinfo
  {journal} {Phys. Rev. D}\ }\textbf {\bibinfo {volume} {85}},\ \bibinfo
  {pages} {024018} (\bibinfo {year} {2012}{\natexlab{a}})},\ \Eprint
  {http://arxiv.org/abs/1107.0854} {arXiv:1107.0854 [gr-qc]} \BibitemShut
  {NoStop}%
\bibitem [{\citenamefont {Kamaretsos}\ \emph
  {et~al.}(2012{\natexlab{b}})\citenamefont {Kamaretsos}, \citenamefont
  {Hannam},\ and\ \citenamefont {Sathyaprakash}}]{Kamaretsos:2012bs}%
  \BibitemOpen
  \bibfield  {author} {\bibinfo {author} {\bibfnamefont {I.}~\bibnamefont
  {Kamaretsos}}, \bibinfo {author} {\bibfnamefont {M.}~\bibnamefont {Hannam}},
  \ and\ \bibinfo {author} {\bibfnamefont {B.}~\bibnamefont {Sathyaprakash}},\
  }\href {\doibase 10.1103/PhysRevLett.109.141102} {\bibfield  {journal}
  {\bibinfo  {journal} {Phys. Rev. Lett.}\ }\textbf {\bibinfo {volume} {109}},\
  \bibinfo {pages} {141102} (\bibinfo {year} {2012}{\natexlab{b}})},\ \Eprint
  {http://arxiv.org/abs/1207.0399} {arXiv:1207.0399 [gr-qc]} \BibitemShut
  {NoStop}%
\bibitem [{\citenamefont {London}\ \emph {et~al.}(2014)\citenamefont {London},
  \citenamefont {Shoemaker},\ and\ \citenamefont {Healy}}]{London:2014cma}%
  \BibitemOpen
  \bibfield  {author} {\bibinfo {author} {\bibfnamefont {L.}~\bibnamefont
  {London}}, \bibinfo {author} {\bibfnamefont {D.}~\bibnamefont {Shoemaker}}, \
  and\ \bibinfo {author} {\bibfnamefont {J.}~\bibnamefont {Healy}},\ }\href
  {\doibase 10.1103/PhysRevD.90.124032} {\bibfield  {journal} {\bibinfo
  {journal} {Phys. Rev. D}\ }\textbf {\bibinfo {volume} {90}},\ \bibinfo
  {pages} {124032} (\bibinfo {year} {2014})},\ \bibinfo {note} {[Erratum:
  Phys.Rev.D 94, 069902 (2016)]},\ \Eprint {http://arxiv.org/abs/1404.3197}
  {arXiv:1404.3197 [gr-qc]} \BibitemShut {NoStop}%
\bibitem [{\citenamefont {Bhagwat}\ \emph {et~al.}(2016)\citenamefont
  {Bhagwat}, \citenamefont {Brown},\ and\ \citenamefont
  {Ballmer}}]{Bhagwat:2016ntk}%
  \BibitemOpen
  \bibfield  {author} {\bibinfo {author} {\bibfnamefont {S.}~\bibnamefont
  {Bhagwat}}, \bibinfo {author} {\bibfnamefont {D.~A.}\ \bibnamefont {Brown}},
  \ and\ \bibinfo {author} {\bibfnamefont {S.~W.}\ \bibnamefont {Ballmer}},\
  }\href {\doibase 10.1103/PhysRevD.94.084024} {\bibfield  {journal} {\bibinfo
  {journal} {Phys. Rev. D}\ }\textbf {\bibinfo {volume} {94}},\ \bibinfo
  {pages} {084024} (\bibinfo {year} {2016})},\ \bibinfo {note} {[Erratum:
  Phys.Rev.D 95, 069906 (2017)]},\ \Eprint {http://arxiv.org/abs/1607.07845}
  {arXiv:1607.07845 [gr-qc]} \BibitemShut {NoStop}%
\bibitem [{\citenamefont {Thrane}\ \emph {et~al.}(2017)\citenamefont {Thrane},
  \citenamefont {Lasky},\ and\ \citenamefont {Levin}}]{Thrane:2017lqn}%
  \BibitemOpen
  \bibfield  {author} {\bibinfo {author} {\bibfnamefont {E.}~\bibnamefont
  {Thrane}}, \bibinfo {author} {\bibfnamefont {P.~D.}\ \bibnamefont {Lasky}}, \
  and\ \bibinfo {author} {\bibfnamefont {Y.}~\bibnamefont {Levin}},\ }\href
  {\doibase 10.1103/PhysRevD.96.102004} {\bibfield  {journal} {\bibinfo
  {journal} {Phys. Rev. D}\ }\textbf {\bibinfo {volume} {96}},\ \bibinfo
  {pages} {102004} (\bibinfo {year} {2017})},\ \Eprint
  {http://arxiv.org/abs/1706.05152} {arXiv:1706.05152 [gr-qc]} \BibitemShut
  {NoStop}%
\bibitem [{\citenamefont {London}(2020)}]{London:2018gaq}%
  \BibitemOpen
  \bibfield  {author} {\bibinfo {author} {\bibfnamefont {L.~T.}\ \bibnamefont
  {London}},\ }\href {\doibase 10.1103/PhysRevD.102.084052} {\bibfield
  {journal} {\bibinfo  {journal} {Phys. Rev. D}\ }\textbf {\bibinfo {volume}
  {102}},\ \bibinfo {pages} {084052} (\bibinfo {year} {2020})},\ \Eprint
  {http://arxiv.org/abs/1801.08208} {arXiv:1801.08208 [gr-qc]} \BibitemShut
  {NoStop}%
\bibitem [{\citenamefont {Baibhav}\ and\ \citenamefont
  {Berti}(2019)}]{Baibhav:2018rfk}%
  \BibitemOpen
  \bibfield  {author} {\bibinfo {author} {\bibfnamefont {V.}~\bibnamefont
  {Baibhav}}\ and\ \bibinfo {author} {\bibfnamefont {E.}~\bibnamefont
  {Berti}},\ }\href {\doibase 10.1103/PhysRevD.99.024005} {\bibfield  {journal}
  {\bibinfo  {journal} {Phys. Rev. D}\ }\textbf {\bibinfo {volume} {99}},\
  \bibinfo {pages} {024005} (\bibinfo {year} {2019})},\ \Eprint
  {http://arxiv.org/abs/1809.03500} {arXiv:1809.03500 [gr-qc]} \BibitemShut
  {NoStop}%
\bibitem [{\citenamefont {Baibhav}\ \emph {et~al.}(2020)\citenamefont
  {Baibhav}, \citenamefont {Berti},\ and\ \citenamefont
  {Cardoso}}]{Baibhav:2020tma}%
  \BibitemOpen
  \bibfield  {author} {\bibinfo {author} {\bibfnamefont {V.}~\bibnamefont
  {Baibhav}}, \bibinfo {author} {\bibfnamefont {E.}~\bibnamefont {Berti}}, \
  and\ \bibinfo {author} {\bibfnamefont {V.}~\bibnamefont {Cardoso}},\ }\href
  {\doibase 10.1103/PhysRevD.101.084053} {\bibfield  {journal} {\bibinfo
  {journal} {Phys. Rev. D}\ }\textbf {\bibinfo {volume} {101}},\ \bibinfo
  {pages} {084053} (\bibinfo {year} {2020})},\ \Eprint
  {http://arxiv.org/abs/2001.10011} {arXiv:2001.10011 [gr-qc]} \BibitemShut
  {NoStop}%
\bibitem [{\citenamefont {Cook}(2020)}]{Cook:2020otn}%
  \BibitemOpen
  \bibfield  {author} {\bibinfo {author} {\bibfnamefont {G.~B.}\ \bibnamefont
  {Cook}},\ }\href {\doibase 10.1103/PhysRevD.102.024027} {\bibfield  {journal}
  {\bibinfo  {journal} {Phys. Rev. D}\ }\textbf {\bibinfo {volume} {102}},\
  \bibinfo {pages} {024027} (\bibinfo {year} {2020})},\ \Eprint
  {http://arxiv.org/abs/2004.08347} {arXiv:2004.08347 [gr-qc]} \BibitemShut
  {NoStop}%
\bibitem [{\citenamefont {Jim\'enez~Forteza}\ \emph {et~al.}(2020)\citenamefont
  {Jim\'enez~Forteza}, \citenamefont {Bhagwat}, \citenamefont {Pani},\ and\
  \citenamefont {Ferrari}}]{JimenezForteza:2020cve}%
  \BibitemOpen
  \bibfield  {author} {\bibinfo {author} {\bibfnamefont {X.}~\bibnamefont
  {Jim\'enez~Forteza}}, \bibinfo {author} {\bibfnamefont {S.}~\bibnamefont
  {Bhagwat}}, \bibinfo {author} {\bibfnamefont {P.}~\bibnamefont {Pani}}, \
  and\ \bibinfo {author} {\bibfnamefont {V.}~\bibnamefont {Ferrari}},\ }\href
  {\doibase 10.1103/PhysRevD.102.044053} {\bibfield  {journal} {\bibinfo
  {journal} {Phys. Rev. D}\ }\textbf {\bibinfo {volume} {102}},\ \bibinfo
  {pages} {044053} (\bibinfo {year} {2020})},\ \Eprint
  {http://arxiv.org/abs/2005.03260} {arXiv:2005.03260 [gr-qc]} \BibitemShut
  {NoStop}%
\bibitem [{\citenamefont {Ota}\ and\ \citenamefont
  {Chirenti}(2022)}]{Ota:2021ypb}%
  \BibitemOpen
  \bibfield  {author} {\bibinfo {author} {\bibfnamefont {I.}~\bibnamefont
  {Ota}}\ and\ \bibinfo {author} {\bibfnamefont {C.}~\bibnamefont {Chirenti}},\
  }\href {\doibase 10.1103/PhysRevD.105.044015} {\bibfield  {journal} {\bibinfo
   {journal} {Phys. Rev. D}\ }\textbf {\bibinfo {volume} {105}},\ \bibinfo
  {pages} {044015} (\bibinfo {year} {2022})},\ \Eprint
  {http://arxiv.org/abs/2108.01774} {arXiv:2108.01774 [gr-qc]} \BibitemShut
  {NoStop}%
\bibitem [{\citenamefont {Li}\ \emph {et~al.}(2022)\citenamefont {Li},
  \citenamefont {Sun}, \citenamefont {Lo}, \citenamefont {Payne},\ and\
  \citenamefont {Chen}}]{Li:2021wgz}%
  \BibitemOpen
  \bibfield  {author} {\bibinfo {author} {\bibfnamefont {X.}~\bibnamefont
  {Li}}, \bibinfo {author} {\bibfnamefont {L.}~\bibnamefont {Sun}}, \bibinfo
  {author} {\bibfnamefont {R.~K.~L.}\ \bibnamefont {Lo}}, \bibinfo {author}
  {\bibfnamefont {E.}~\bibnamefont {Payne}}, \ and\ \bibinfo {author}
  {\bibfnamefont {Y.}~\bibnamefont {Chen}},\ }\href {\doibase
  10.1103/PhysRevD.105.024016} {\bibfield  {journal} {\bibinfo  {journal}
  {Phys. Rev. D}\ }\textbf {\bibinfo {volume} {105}},\ \bibinfo {pages}
  {024016} (\bibinfo {year} {2022})},\ \Eprint
  {http://arxiv.org/abs/2110.03116} {arXiv:2110.03116 [gr-qc]} \BibitemShut
  {NoStop}%
\bibitem [{\citenamefont {Maga\~na Zertuche}\ \emph {et~al.}(2022)\citenamefont
  {Maga\~na Zertuche} \emph {et~al.}}]{MaganaZertuche:2021syq}%
  \BibitemOpen
  \bibfield  {author} {\bibinfo {author} {\bibfnamefont {L.}~\bibnamefont
  {Maga\~na Zertuche}} \emph {et~al.},\ }\href {\doibase
  10.1103/PhysRevD.105.104015} {\bibfield  {journal} {\bibinfo  {journal}
  {Phys. Rev. D}\ }\textbf {\bibinfo {volume} {105}},\ \bibinfo {pages}
  {104015} (\bibinfo {year} {2022})},\ \Eprint
  {http://arxiv.org/abs/2110.15922} {arXiv:2110.15922 [gr-qc]} \BibitemShut
  {NoStop}%
\bibitem [{\citenamefont {Gossan}\ \emph {et~al.}(2012)\citenamefont {Gossan},
  \citenamefont {Veitch},\ and\ \citenamefont {Sathyaprakash}}]{Gossan:2011ha}%
  \BibitemOpen
  \bibfield  {author} {\bibinfo {author} {\bibfnamefont {S.}~\bibnamefont
  {Gossan}}, \bibinfo {author} {\bibfnamefont {J.}~\bibnamefont {Veitch}}, \
  and\ \bibinfo {author} {\bibfnamefont {B.~S.}\ \bibnamefont
  {Sathyaprakash}},\ }\href {\doibase 10.1103/PhysRevD.85.124056} {\bibfield
  {journal} {\bibinfo  {journal} {Phys. Rev. D}\ }\textbf {\bibinfo {volume}
  {85}},\ \bibinfo {pages} {124056} (\bibinfo {year} {2012})},\ \Eprint
  {http://arxiv.org/abs/1111.5819} {arXiv:1111.5819 [gr-qc]} \BibitemShut
  {NoStop}%
\bibitem [{\citenamefont {Meidam}\ \emph {et~al.}(2014)\citenamefont {Meidam},
  \citenamefont {Agathos}, \citenamefont {Van Den~Broeck}, \citenamefont
  {Veitch},\ and\ \citenamefont {Sathyaprakash}}]{Meidam:2014jpa}%
  \BibitemOpen
  \bibfield  {author} {\bibinfo {author} {\bibfnamefont {J.}~\bibnamefont
  {Meidam}}, \bibinfo {author} {\bibfnamefont {M.}~\bibnamefont {Agathos}},
  \bibinfo {author} {\bibfnamefont {C.}~\bibnamefont {Van Den~Broeck}},
  \bibinfo {author} {\bibfnamefont {J.}~\bibnamefont {Veitch}}, \ and\ \bibinfo
  {author} {\bibfnamefont {B.~S.}\ \bibnamefont {Sathyaprakash}},\ }\href
  {\doibase 10.1103/PhysRevD.90.064009} {\bibfield  {journal} {\bibinfo
  {journal} {Phys. Rev. D}\ }\textbf {\bibinfo {volume} {90}},\ \bibinfo
  {pages} {064009} (\bibinfo {year} {2014})},\ \Eprint
  {http://arxiv.org/abs/1406.3201} {arXiv:1406.3201 [gr-qc]} \BibitemShut
  {NoStop}%
\bibitem [{\citenamefont {Punturo}\ \emph {et~al.}(2010)\citenamefont {Punturo}
  \emph {et~al.}}]{Punturo:2010zz}%
  \BibitemOpen
  \bibfield  {author} {\bibinfo {author} {\bibfnamefont {M.}~\bibnamefont
  {Punturo}} \emph {et~al.},\ }\href {\doibase 10.1088/0264-9381/27/19/194002}
  {\bibfield  {journal} {\bibinfo  {journal} {Class. Quant. Grav.}\ }\textbf
  {\bibinfo {volume} {27}},\ \bibinfo {pages} {194002} (\bibinfo {year}
  {2010})}\BibitemShut {NoStop}%
\bibitem [{\citenamefont {Bhagwat}\ \emph
  {et~al.}(2020{\natexlab{a}})\citenamefont {Bhagwat}, \citenamefont {Cabero},
  \citenamefont {Capano}, \citenamefont {Krishnan},\ and\ \citenamefont
  {Brown}}]{Bhagwat:2019bwv}%
  \BibitemOpen
  \bibfield  {author} {\bibinfo {author} {\bibfnamefont {S.}~\bibnamefont
  {Bhagwat}}, \bibinfo {author} {\bibfnamefont {M.}~\bibnamefont {Cabero}},
  \bibinfo {author} {\bibfnamefont {C.~D.}\ \bibnamefont {Capano}}, \bibinfo
  {author} {\bibfnamefont {B.}~\bibnamefont {Krishnan}}, \ and\ \bibinfo
  {author} {\bibfnamefont {D.~A.}\ \bibnamefont {Brown}},\ }\href {\doibase
  10.1103/PhysRevD.102.024023} {\bibfield  {journal} {\bibinfo  {journal}
  {Phys. Rev. D}\ }\textbf {\bibinfo {volume} {102}},\ \bibinfo {pages}
  {024023} (\bibinfo {year} {2020}{\natexlab{a}})},\ \Eprint
  {http://arxiv.org/abs/1910.13203} {arXiv:1910.13203 [gr-qc]} \BibitemShut
  {NoStop}%
\bibitem [{\citenamefont {Borhanian}\ \emph {et~al.}(2020)\citenamefont
  {Borhanian}, \citenamefont {Arun}, \citenamefont {Pfeiffer},\ and\
  \citenamefont {Sathyaprakash}}]{Borhanian:2019kxt}%
  \BibitemOpen
  \bibfield  {author} {\bibinfo {author} {\bibfnamefont {S.}~\bibnamefont
  {Borhanian}}, \bibinfo {author} {\bibfnamefont {K.~G.}\ \bibnamefont {Arun}},
  \bibinfo {author} {\bibfnamefont {H.~P.}\ \bibnamefont {Pfeiffer}}, \ and\
  \bibinfo {author} {\bibfnamefont {B.~S.}\ \bibnamefont {Sathyaprakash}},\
  }\href {\doibase 10.1088/1361-6382/ab6a21} {\bibfield  {journal} {\bibinfo
  {journal} {Class. Quant. Grav.}\ }\textbf {\bibinfo {volume} {37}},\ \bibinfo
  {pages} {065006} (\bibinfo {year} {2020})},\ \Eprint
  {http://arxiv.org/abs/1901.08516} {arXiv:1901.08516 [gr-qc]} \BibitemShut
  {NoStop}%
\bibitem [{\citenamefont {Cabero}\ \emph {et~al.}(2020)\citenamefont {Cabero},
  \citenamefont {Westerweck}, \citenamefont {Capano}, \citenamefont {Kumar},
  \citenamefont {Nielsen},\ and\ \citenamefont {Krishnan}}]{Cabero:2019zyt}%
  \BibitemOpen
  \bibfield  {author} {\bibinfo {author} {\bibfnamefont {M.}~\bibnamefont
  {Cabero}}, \bibinfo {author} {\bibfnamefont {J.}~\bibnamefont {Westerweck}},
  \bibinfo {author} {\bibfnamefont {C.~D.}\ \bibnamefont {Capano}}, \bibinfo
  {author} {\bibfnamefont {S.}~\bibnamefont {Kumar}}, \bibinfo {author}
  {\bibfnamefont {A.~B.}\ \bibnamefont {Nielsen}}, \ and\ \bibinfo {author}
  {\bibfnamefont {B.}~\bibnamefont {Krishnan}},\ }\href {\doibase
  10.1103/PhysRevD.101.064044} {\bibfield  {journal} {\bibinfo  {journal}
  {Phys. Rev. D}\ }\textbf {\bibinfo {volume} {101}},\ \bibinfo {pages}
  {064044} (\bibinfo {year} {2020})},\ \Eprint
  {http://arxiv.org/abs/1911.01361} {arXiv:1911.01361 [gr-qc]} \BibitemShut
  {NoStop}%
\bibitem [{\citenamefont {Berti}\ \emph {et~al.}(2016)\citenamefont {Berti},
  \citenamefont {Sesana}, \citenamefont {Barausse}, \citenamefont {Cardoso},\
  and\ \citenamefont {Belczynski}}]{Berti:2016lat}%
  \BibitemOpen
  \bibfield  {author} {\bibinfo {author} {\bibfnamefont {E.}~\bibnamefont
  {Berti}}, \bibinfo {author} {\bibfnamefont {A.}~\bibnamefont {Sesana}},
  \bibinfo {author} {\bibfnamefont {E.}~\bibnamefont {Barausse}}, \bibinfo
  {author} {\bibfnamefont {V.}~\bibnamefont {Cardoso}}, \ and\ \bibinfo
  {author} {\bibfnamefont {K.}~\bibnamefont {Belczynski}},\ }\href {\doibase
  10.1103/PhysRevLett.117.101102} {\bibfield  {journal} {\bibinfo  {journal}
  {Phys. Rev. Lett.}\ }\textbf {\bibinfo {volume} {117}},\ \bibinfo {pages}
  {101102} (\bibinfo {year} {2016})},\ \Eprint
  {http://arxiv.org/abs/1605.09286} {arXiv:1605.09286 [gr-qc]} \BibitemShut
  {NoStop}%
\bibitem [{\citenamefont {Carullo}\ \emph {et~al.}(2018)\citenamefont {Carullo}
  \emph {et~al.}}]{Carullo:2018sfu}%
  \BibitemOpen
  \bibfield  {author} {\bibinfo {author} {\bibfnamefont {G.}~\bibnamefont
  {Carullo}} \emph {et~al.},\ }\href {\doibase 10.1103/PhysRevD.98.104020}
  {\bibfield  {journal} {\bibinfo  {journal} {Phys. Rev. D}\ }\textbf {\bibinfo
  {volume} {98}},\ \bibinfo {pages} {104020} (\bibinfo {year} {2018})},\
  \Eprint {http://arxiv.org/abs/1805.04760} {arXiv:1805.04760 [gr-qc]}
  \BibitemShut {NoStop}%
\bibitem [{\citenamefont {Del~Pozzo}\ and\ \citenamefont
  {Nagar}(2017)}]{DelPozzo:2016kmd}%
  \BibitemOpen
  \bibfield  {author} {\bibinfo {author} {\bibfnamefont {W.}~\bibnamefont
  {Del~Pozzo}}\ and\ \bibinfo {author} {\bibfnamefont {A.}~\bibnamefont
  {Nagar}},\ }\href {\doibase 10.1103/PhysRevD.95.124034} {\bibfield  {journal}
  {\bibinfo  {journal} {Phys. Rev. D}\ }\textbf {\bibinfo {volume} {95}},\
  \bibinfo {pages} {124034} (\bibinfo {year} {2017})},\ \Eprint
  {http://arxiv.org/abs/1606.03952} {arXiv:1606.03952 [gr-qc]} \BibitemShut
  {NoStop}%
\bibitem [{\citenamefont {Carullo}\ \emph {et~al.}(2019)\citenamefont
  {Carullo}, \citenamefont {Del~Pozzo},\ and\ \citenamefont
  {Veitch}}]{Carullo:2019flw}%
  \BibitemOpen
  \bibfield  {author} {\bibinfo {author} {\bibfnamefont {G.}~\bibnamefont
  {Carullo}}, \bibinfo {author} {\bibfnamefont {W.}~\bibnamefont {Del~Pozzo}},
  \ and\ \bibinfo {author} {\bibfnamefont {J.}~\bibnamefont {Veitch}},\ }\href
  {\doibase 10.1103/PhysRevD.99.123029, 10.1103/PhysRevD.100.089903} {\bibfield
   {journal} {\bibinfo  {journal} {Phys. Rev.}\ }\textbf {\bibinfo {volume}
  {D99}},\ \bibinfo {pages} {123029} (\bibinfo {year} {2019})},\ \bibinfo
  {note} {[Erratum: Phys. Rev.D100,no.8,089903(2019)]},\ \Eprint
  {http://arxiv.org/abs/1902.07527} {arXiv:1902.07527 [gr-qc]} \BibitemShut
  {NoStop}%
\bibitem [{\citenamefont {Isi}\ \emph {et~al.}(2019)\citenamefont {Isi},
  \citenamefont {Giesler}, \citenamefont {Farr}, \citenamefont {Scheel},\ and\
  \citenamefont {Teukolsky}}]{Isi:2019aib}%
  \BibitemOpen
  \bibfield  {author} {\bibinfo {author} {\bibfnamefont {M.}~\bibnamefont
  {Isi}}, \bibinfo {author} {\bibfnamefont {M.}~\bibnamefont {Giesler}},
  \bibinfo {author} {\bibfnamefont {W.~M.}\ \bibnamefont {Farr}}, \bibinfo
  {author} {\bibfnamefont {M.~A.}\ \bibnamefont {Scheel}}, \ and\ \bibinfo
  {author} {\bibfnamefont {S.~A.}\ \bibnamefont {Teukolsky}},\ }\href {\doibase
  10.1103/PhysRevLett.123.111102} {\bibfield  {journal} {\bibinfo  {journal}
  {Phys. Rev. Lett.}\ }\textbf {\bibinfo {volume} {123}},\ \bibinfo {pages}
  {111102} (\bibinfo {year} {2019})},\ \Eprint
  {http://arxiv.org/abs/1905.00869} {arXiv:1905.00869 [gr-qc]} \BibitemShut
  {NoStop}%
\bibitem [{\citenamefont {Carullo}\ \emph {et~al.}()\citenamefont {Carullo},
  \citenamefont {Pozzo}, \citenamefont {Laghi}, \citenamefont {Isi},\ and\
  \citenamefont {Veitch}}]{pyRing}%
  \BibitemOpen
  \bibfield  {author} {\bibinfo {author} {\bibfnamefont {G.}~\bibnamefont
  {Carullo}}, \bibinfo {author} {\bibfnamefont {W.~D.}\ \bibnamefont {Pozzo}},
  \bibinfo {author} {\bibfnamefont {D.}~\bibnamefont {Laghi}}, \bibinfo
  {author} {\bibfnamefont {M.}~\bibnamefont {Isi}}, \ and\ \bibinfo {author}
  {\bibfnamefont {J.}~\bibnamefont {Veitch}},\ }\href@noop {} {\enquote
  {\bibinfo {title} {\texttt{pyRing}: a time-domain ringdown analysis python
  package},}\ }\bibinfo {howpublished}
  {\url{https://git.ligo.org/lscsoft/pyring}}\BibitemShut {NoStop}%
\bibitem [{\citenamefont {Abbott}\ \emph
  {et~al.}(2020{\natexlab{a}})\citenamefont {Abbott} \emph
  {et~al.}}]{LIGOScientific:2020iuh}%
  \BibitemOpen
  \bibfield  {author} {\bibinfo {author} {\bibfnamefont {R.}~\bibnamefont
  {Abbott}} \emph {et~al.} (\bibinfo {collaboration} {LIGO Scientific,
  Virgo}),\ }\href {\doibase 10.1103/PhysRevLett.125.101102} {\bibfield
  {journal} {\bibinfo  {journal} {Phys. Rev. Lett.}\ }\textbf {\bibinfo
  {volume} {125}},\ \bibinfo {pages} {101102} (\bibinfo {year}
  {2020}{\natexlab{a}})},\ \Eprint {http://arxiv.org/abs/2009.01075}
  {arXiv:2009.01075 [gr-qc]} \BibitemShut {NoStop}%
\bibitem [{\citenamefont {Abbott}\ \emph
  {et~al.}(2020{\natexlab{b}})\citenamefont {Abbott} \emph
  {et~al.}}]{LIGOScientific:2020ufj}%
  \BibitemOpen
  \bibfield  {author} {\bibinfo {author} {\bibfnamefont {R.}~\bibnamefont
  {Abbott}} \emph {et~al.} (\bibinfo {collaboration} {LIGO Scientific,
  Virgo}),\ }\href {\doibase 10.3847/2041-8213/aba493} {\bibfield  {journal}
  {\bibinfo  {journal} {Astrophys. J. Lett.}\ }\textbf {\bibinfo {volume}
  {900}},\ \bibinfo {pages} {L13} (\bibinfo {year} {2020}{\natexlab{b}})},\
  \Eprint {http://arxiv.org/abs/2009.01190} {arXiv:2009.01190 [astro-ph.HE]}
  \BibitemShut {NoStop}%
\bibitem [{\citenamefont {Carullo}\ \emph {et~al.}(2022)\citenamefont
  {Carullo}, \citenamefont {Laghi}, \citenamefont {Johnson-McDaniel},
  \citenamefont {Del~Pozzo}, \citenamefont {Dias}, \citenamefont {Godazgar},\
  and\ \citenamefont {Santos}}]{Carullo:2021oxn}%
  \BibitemOpen
  \bibfield  {author} {\bibinfo {author} {\bibfnamefont {G.}~\bibnamefont
  {Carullo}}, \bibinfo {author} {\bibfnamefont {D.}~\bibnamefont {Laghi}},
  \bibinfo {author} {\bibfnamefont {N.~K.}\ \bibnamefont {Johnson-McDaniel}},
  \bibinfo {author} {\bibfnamefont {W.}~\bibnamefont {Del~Pozzo}}, \bibinfo
  {author} {\bibfnamefont {O.~J.~C.}\ \bibnamefont {Dias}}, \bibinfo {author}
  {\bibfnamefont {M.}~\bibnamefont {Godazgar}}, \ and\ \bibinfo {author}
  {\bibfnamefont {J.~E.}\ \bibnamefont {Santos}},\ }\href {\doibase
  10.1103/PhysRevD.105.062009} {\bibfield  {journal} {\bibinfo  {journal}
  {Phys. Rev. D}\ }\textbf {\bibinfo {volume} {105}},\ \bibinfo {pages}
  {062009} (\bibinfo {year} {2022})},\ \Eprint
  {http://arxiv.org/abs/2109.13961} {arXiv:2109.13961 [gr-qc]} \BibitemShut
  {NoStop}%
\bibitem [{\citenamefont {Carullo}\ \emph {et~al.}(2021)\citenamefont
  {Carullo}, \citenamefont {Laghi}, \citenamefont {Veitch},\ and\ \citenamefont
  {Del~Pozzo}}]{Carullo:2021yxh}%
  \BibitemOpen
  \bibfield  {author} {\bibinfo {author} {\bibfnamefont {G.}~\bibnamefont
  {Carullo}}, \bibinfo {author} {\bibfnamefont {D.}~\bibnamefont {Laghi}},
  \bibinfo {author} {\bibfnamefont {J.}~\bibnamefont {Veitch}}, \ and\ \bibinfo
  {author} {\bibfnamefont {W.}~\bibnamefont {Del~Pozzo}},\ }\href {\doibase
  10.1103/PhysRevLett.126.161102} {\bibfield  {journal} {\bibinfo  {journal}
  {Phys. Rev. Lett.}\ }\textbf {\bibinfo {volume} {126}},\ \bibinfo {pages}
  {161102} (\bibinfo {year} {2021})},\ \Eprint
  {http://arxiv.org/abs/2103.06167} {arXiv:2103.06167 [gr-qc]} \BibitemShut
  {NoStop}%
\bibitem [{\citenamefont {Laghi}\ \emph {et~al.}(2021)\citenamefont {Laghi},
  \citenamefont {Carullo}, \citenamefont {Veitch},\ and\ \citenamefont
  {Del~Pozzo}}]{Laghi:2020rgl}%
  \BibitemOpen
  \bibfield  {author} {\bibinfo {author} {\bibfnamefont {D.}~\bibnamefont
  {Laghi}}, \bibinfo {author} {\bibfnamefont {G.}~\bibnamefont {Carullo}},
  \bibinfo {author} {\bibfnamefont {J.}~\bibnamefont {Veitch}}, \ and\ \bibinfo
  {author} {\bibfnamefont {W.}~\bibnamefont {Del~Pozzo}},\ }\href {\doibase
  10.1088/1361-6382/abde19} {\bibfield  {journal} {\bibinfo  {journal} {Class.
  Quant. Grav.}\ }\textbf {\bibinfo {volume} {38}},\ \bibinfo {pages} {095005}
  (\bibinfo {year} {2021})},\ \Eprint {http://arxiv.org/abs/2011.03816}
  {arXiv:2011.03816 [gr-qc]} \BibitemShut {NoStop}%
\bibitem [{\citenamefont {Maselli}\ \emph {et~al.}(2020)\citenamefont
  {Maselli}, \citenamefont {Pani}, \citenamefont {Gualtieri},\ and\
  \citenamefont {Berti}}]{Maselli:2019mjd}%
  \BibitemOpen
  \bibfield  {author} {\bibinfo {author} {\bibfnamefont {A.}~\bibnamefont
  {Maselli}}, \bibinfo {author} {\bibfnamefont {P.}~\bibnamefont {Pani}},
  \bibinfo {author} {\bibfnamefont {L.}~\bibnamefont {Gualtieri}}, \ and\
  \bibinfo {author} {\bibfnamefont {E.}~\bibnamefont {Berti}},\ }\href
  {\doibase 10.1103/PhysRevD.101.024043} {\bibfield  {journal} {\bibinfo
  {journal} {Phys. Rev. D}\ }\textbf {\bibinfo {volume} {101}},\ \bibinfo
  {pages} {024043} (\bibinfo {year} {2020})},\ \Eprint
  {http://arxiv.org/abs/1910.12893} {arXiv:1910.12893 [gr-qc]} \BibitemShut
  {NoStop}%
\bibitem [{\citenamefont {Carullo}(2021)}]{Carullo:2021dui}%
  \BibitemOpen
  \bibfield  {author} {\bibinfo {author} {\bibfnamefont {G.}~\bibnamefont
  {Carullo}},\ }\href {\doibase 10.1103/PhysRevD.103.124043} {\bibfield
  {journal} {\bibinfo  {journal} {Phys. Rev. D}\ }\textbf {\bibinfo {volume}
  {103}},\ \bibinfo {pages} {124043} (\bibinfo {year} {2021})},\ \Eprint
  {http://arxiv.org/abs/2102.05939} {arXiv:2102.05939 [gr-qc]} \BibitemShut
  {NoStop}%
\bibitem [{\citenamefont {Isi}\ and\ \citenamefont {Farr}(2021)}]{Isi:2021iql}%
  \BibitemOpen
  \bibfield  {author} {\bibinfo {author} {\bibfnamefont {M.}~\bibnamefont
  {Isi}}\ and\ \bibinfo {author} {\bibfnamefont {W.~M.}\ \bibnamefont {Farr}},\
  }\href@noop {} {\  (\bibinfo {year} {2021})},\ \Eprint
  {http://arxiv.org/abs/2107.05609} {arXiv:2107.05609 [gr-qc]} \BibitemShut
  {NoStop}%
\bibitem [{\citenamefont {Giesler}\ \emph {et~al.}(2019)\citenamefont
  {Giesler}, \citenamefont {Isi}, \citenamefont {Scheel},\ and\ \citenamefont
  {Teukolsky}}]{Giesler:2019uxc}%
  \BibitemOpen
  \bibfield  {author} {\bibinfo {author} {\bibfnamefont {M.}~\bibnamefont
  {Giesler}}, \bibinfo {author} {\bibfnamefont {M.}~\bibnamefont {Isi}},
  \bibinfo {author} {\bibfnamefont {M.~A.}\ \bibnamefont {Scheel}}, \ and\
  \bibinfo {author} {\bibfnamefont {S.}~\bibnamefont {Teukolsky}},\ }\href
  {\doibase 10.1103/PhysRevX.9.041060} {\bibfield  {journal} {\bibinfo
  {journal} {Phys. Rev. X}\ }\textbf {\bibinfo {volume} {9}},\ \bibinfo {pages}
  {041060} (\bibinfo {year} {2019})},\ \Eprint
  {http://arxiv.org/abs/1903.08284} {arXiv:1903.08284 [gr-qc]} \BibitemShut
  {NoStop}%
\bibitem [{\citenamefont {Ota}\ and\ \citenamefont
  {Chirenti}(2020)}]{Ota:2019bzl}%
  \BibitemOpen
  \bibfield  {author} {\bibinfo {author} {\bibfnamefont {I.}~\bibnamefont
  {Ota}}\ and\ \bibinfo {author} {\bibfnamefont {C.}~\bibnamefont {Chirenti}},\
  }\href {\doibase 10.1103/PhysRevD.101.104005} {\bibfield  {journal} {\bibinfo
   {journal} {Phys. Rev. D}\ }\textbf {\bibinfo {volume} {101}},\ \bibinfo
  {pages} {104005} (\bibinfo {year} {2020})},\ \Eprint
  {http://arxiv.org/abs/1911.00440} {arXiv:1911.00440 [gr-qc]} \BibitemShut
  {NoStop}%
\bibitem [{\citenamefont {Bhagwat}\ \emph
  {et~al.}(2020{\natexlab{b}})\citenamefont {Bhagwat}, \citenamefont {Forteza},
  \citenamefont {Pani},\ and\ \citenamefont {Ferrari}}]{Bhagwat:2019dtm}%
  \BibitemOpen
  \bibfield  {author} {\bibinfo {author} {\bibfnamefont {S.}~\bibnamefont
  {Bhagwat}}, \bibinfo {author} {\bibfnamefont {X.~J.}\ \bibnamefont
  {Forteza}}, \bibinfo {author} {\bibfnamefont {P.}~\bibnamefont {Pani}}, \
  and\ \bibinfo {author} {\bibfnamefont {V.}~\bibnamefont {Ferrari}},\ }\href
  {\doibase 10.1103/PhysRevD.101.044033} {\bibfield  {journal} {\bibinfo
  {journal} {Phys. Rev. D}\ }\textbf {\bibinfo {volume} {101}},\ \bibinfo
  {pages} {044033} (\bibinfo {year} {2020}{\natexlab{b}})},\ \Eprint
  {http://arxiv.org/abs/1910.08708} {arXiv:1910.08708 [gr-qc]} \BibitemShut
  {NoStop}%
\bibitem [{\citenamefont {Bustillo}\ \emph {et~al.}(2021)\citenamefont
  {Bustillo}, \citenamefont {Lasky},\ and\ \citenamefont
  {Thrane}}]{Bustillo:2020buq}%
  \BibitemOpen
  \bibfield  {author} {\bibinfo {author} {\bibfnamefont {J.~C.}\ \bibnamefont
  {Bustillo}}, \bibinfo {author} {\bibfnamefont {P.~D.}\ \bibnamefont {Lasky}},
  \ and\ \bibinfo {author} {\bibfnamefont {E.}~\bibnamefont {Thrane}},\ }\href
  {\doibase 10.1103/PhysRevD.103.024041} {\bibfield  {journal} {\bibinfo
  {journal} {Phys. Rev. D}\ }\textbf {\bibinfo {volume} {103}},\ \bibinfo
  {pages} {024041} (\bibinfo {year} {2021})},\ \Eprint
  {http://arxiv.org/abs/2010.01857} {arXiv:2010.01857 [gr-qc]} \BibitemShut
  {NoStop}%
\bibitem [{\citenamefont {Okounkova}(2020)}]{Okounkova:2020vwu}%
  \BibitemOpen
  \bibfield  {author} {\bibinfo {author} {\bibfnamefont {M.}~\bibnamefont
  {Okounkova}},\ }\href@noop {} {\  (\bibinfo {year} {2020})},\ \Eprint
  {http://arxiv.org/abs/2004.00671} {arXiv:2004.00671 [gr-qc]} \BibitemShut
  {NoStop}%
\bibitem [{\citenamefont {Mourier}\ \emph {et~al.}(2021)\citenamefont
  {Mourier}, \citenamefont {Jim\'enez~Forteza}, \citenamefont {Pook-Kolb},
  \citenamefont {Krishnan},\ and\ \citenamefont {Schnetter}}]{Mourier:2020mwa}%
  \BibitemOpen
  \bibfield  {author} {\bibinfo {author} {\bibfnamefont {P.}~\bibnamefont
  {Mourier}}, \bibinfo {author} {\bibfnamefont {X.}~\bibnamefont
  {Jim\'enez~Forteza}}, \bibinfo {author} {\bibfnamefont {D.}~\bibnamefont
  {Pook-Kolb}}, \bibinfo {author} {\bibfnamefont {B.}~\bibnamefont {Krishnan}},
  \ and\ \bibinfo {author} {\bibfnamefont {E.}~\bibnamefont {Schnetter}},\
  }\href {\doibase 10.1103/PhysRevD.103.044054} {\bibfield  {journal} {\bibinfo
   {journal} {Phys. Rev. D}\ }\textbf {\bibinfo {volume} {103}},\ \bibinfo
  {pages} {044054} (\bibinfo {year} {2021})},\ \Eprint
  {http://arxiv.org/abs/2010.15186} {arXiv:2010.15186 [gr-qc]} \BibitemShut
  {NoStop}%
\bibitem [{\citenamefont {Dhani}(2021)}]{Dhani:2020nik}%
  \BibitemOpen
  \bibfield  {author} {\bibinfo {author} {\bibfnamefont {A.}~\bibnamefont
  {Dhani}},\ }\href {\doibase 10.1103/PhysRevD.103.104048} {\bibfield
  {journal} {\bibinfo  {journal} {Phys. Rev. D}\ }\textbf {\bibinfo {volume}
  {103}},\ \bibinfo {pages} {104048} (\bibinfo {year} {2021})},\ \Eprint
  {http://arxiv.org/abs/2010.08602} {arXiv:2010.08602 [gr-qc]} \BibitemShut
  {NoStop}%
\bibitem [{\citenamefont {Dhani}\ and\ \citenamefont
  {Sathyaprakash}(2021)}]{Dhani:2021vac}%
  \BibitemOpen
  \bibfield  {author} {\bibinfo {author} {\bibfnamefont {A.}~\bibnamefont
  {Dhani}}\ and\ \bibinfo {author} {\bibfnamefont {B.~S.}\ \bibnamefont
  {Sathyaprakash}},\ }\href@noop {} {\  (\bibinfo {year} {2021})},\ \Eprint
  {http://arxiv.org/abs/2107.14195} {arXiv:2107.14195 [gr-qc]} \BibitemShut
  {NoStop}%
\bibitem [{\citenamefont {Finch}\ and\ \citenamefont
  {Moore}(2021{\natexlab{a}})}]{Finch:2021iip}%
  \BibitemOpen
  \bibfield  {author} {\bibinfo {author} {\bibfnamefont {E.}~\bibnamefont
  {Finch}}\ and\ \bibinfo {author} {\bibfnamefont {C.~J.}\ \bibnamefont
  {Moore}},\ }\href {\doibase 10.1103/PhysRevD.103.084048} {\bibfield
  {journal} {\bibinfo  {journal} {Phys. Rev. D}\ }\textbf {\bibinfo {volume}
  {103}},\ \bibinfo {pages} {084048} (\bibinfo {year} {2021}{\natexlab{a}})},\
  \Eprint {http://arxiv.org/abs/2102.07794} {arXiv:2102.07794 [gr-qc]}
  \BibitemShut {NoStop}%
\bibitem [{\citenamefont {Finch}\ and\ \citenamefont
  {Moore}(2021{\natexlab{b}})}]{Finch:2021qph}%
  \BibitemOpen
  \bibfield  {author} {\bibinfo {author} {\bibfnamefont {E.}~\bibnamefont
  {Finch}}\ and\ \bibinfo {author} {\bibfnamefont {C.~J.}\ \bibnamefont
  {Moore}},\ }\href {\doibase 10.1103/PhysRevD.104.123034} {\bibfield
  {journal} {\bibinfo  {journal} {Phys. Rev. D}\ }\textbf {\bibinfo {volume}
  {104}},\ \bibinfo {pages} {123034} (\bibinfo {year} {2021}{\natexlab{b}})},\
  \Eprint {http://arxiv.org/abs/2108.09344} {arXiv:2108.09344 [gr-qc]}
  \BibitemShut {NoStop}%
\bibitem [{\citenamefont {Forteza}\ \emph {et~al.}(2023)\citenamefont
  {Forteza}, \citenamefont {Bhagwat}, \citenamefont {Kumar},\ and\
  \citenamefont {Pani}}]{Forteza:2022tgq}%
  \BibitemOpen
  \bibfield  {author} {\bibinfo {author} {\bibfnamefont {X.~J.}\ \bibnamefont
  {Forteza}}, \bibinfo {author} {\bibfnamefont {S.}~\bibnamefont {Bhagwat}},
  \bibinfo {author} {\bibfnamefont {S.}~\bibnamefont {Kumar}}, \ and\ \bibinfo
  {author} {\bibfnamefont {P.}~\bibnamefont {Pani}},\ }\href {\doibase
  10.1103/PhysRevLett.130.021001} {\bibfield  {journal} {\bibinfo  {journal}
  {Phys. Rev. Lett.}\ }\textbf {\bibinfo {volume} {130}},\ \bibinfo {pages}
  {021001} (\bibinfo {year} {2023})},\ \Eprint
  {http://arxiv.org/abs/2205.14910} {arXiv:2205.14910 [gr-qc]} \BibitemShut
  {NoStop}%
\bibitem [{\citenamefont {Forteza}\ and\ \citenamefont
  {Mourier}(2021)}]{Forteza:2021wfq}%
  \BibitemOpen
  \bibfield  {author} {\bibinfo {author} {\bibfnamefont {X.~J.}\ \bibnamefont
  {Forteza}}\ and\ \bibinfo {author} {\bibfnamefont {P.}~\bibnamefont
  {Mourier}},\ }\href {\doibase 10.1103/PhysRevD.104.124072} {\bibfield
  {journal} {\bibinfo  {journal} {Phys. Rev. D}\ }\textbf {\bibinfo {volume}
  {104}},\ \bibinfo {pages} {124072} (\bibinfo {year} {2021})},\ \Eprint
  {http://arxiv.org/abs/2107.11829} {arXiv:2107.11829 [gr-qc]} \BibitemShut
  {NoStop}%
\bibitem [{\citenamefont {Jaramillo}\ and\ \citenamefont
  {Krishnan}(2022)}]{Jaramillo:2022oqn}%
  \BibitemOpen
  \bibfield  {author} {\bibinfo {author} {\bibfnamefont {J.~L.}\ \bibnamefont
  {Jaramillo}}\ and\ \bibinfo {author} {\bibfnamefont {B.}~\bibnamefont
  {Krishnan}},\ }\href@noop {} {\  (\bibinfo {year} {2022})},\ \Eprint
  {http://arxiv.org/abs/2211.03405} {arXiv:2211.03405 [gr-qc]} \BibitemShut
  {NoStop}%
\bibitem [{\citenamefont {Cotesta}\ \emph {et~al.}(2022)\citenamefont
  {Cotesta}, \citenamefont {Carullo}, \citenamefont {Berti},\ and\
  \citenamefont {Cardoso}}]{Cotesta:2022pci}%
  \BibitemOpen
  \bibfield  {author} {\bibinfo {author} {\bibfnamefont {R.}~\bibnamefont
  {Cotesta}}, \bibinfo {author} {\bibfnamefont {G.}~\bibnamefont {Carullo}},
  \bibinfo {author} {\bibfnamefont {E.}~\bibnamefont {Berti}}, \ and\ \bibinfo
  {author} {\bibfnamefont {V.}~\bibnamefont {Cardoso}},\ }\href {\doibase
  10.1103/PhysRevLett.129.111102} {\bibfield  {journal} {\bibinfo  {journal}
  {Phys. Rev. Lett.}\ }\textbf {\bibinfo {volume} {129}},\ \bibinfo {pages}
  {111102} (\bibinfo {year} {2022})},\ \Eprint
  {http://arxiv.org/abs/2201.00822} {arXiv:2201.00822 [gr-qc]} \BibitemShut
  {NoStop}%
\bibitem [{\citenamefont {Isi}\ and\ \citenamefont {Farr}(2022)}]{Isi:2022mhy}%
  \BibitemOpen
  \bibfield  {author} {\bibinfo {author} {\bibfnamefont {M.}~\bibnamefont
  {Isi}}\ and\ \bibinfo {author} {\bibfnamefont {W.~M.}\ \bibnamefont {Farr}},\
  }\href@noop {} {\  (\bibinfo {year} {2022})},\ \Eprint
  {http://arxiv.org/abs/2202.02941} {arXiv:2202.02941 [gr-qc]} \BibitemShut
  {NoStop}%
\bibitem [{\citenamefont {Finch}\ and\ \citenamefont
  {Moore}(2022)}]{Finch:2022ynt}%
  \BibitemOpen
  \bibfield  {author} {\bibinfo {author} {\bibfnamefont {E.}~\bibnamefont
  {Finch}}\ and\ \bibinfo {author} {\bibfnamefont {C.~J.}\ \bibnamefont
  {Moore}},\ }\href {\doibase 10.1103/PhysRevD.106.043005} {\bibfield
  {journal} {\bibinfo  {journal} {Phys. Rev. D}\ }\textbf {\bibinfo {volume}
  {106}},\ \bibinfo {pages} {043005} (\bibinfo {year} {2022})},\ \Eprint
  {http://arxiv.org/abs/2205.07809} {arXiv:2205.07809 [gr-qc]} \BibitemShut
  {NoStop}%
\bibitem [{\citenamefont {Ma}\ \emph {et~al.}(2023{\natexlab{a}})\citenamefont
  {Ma}, \citenamefont {Sun},\ and\ \citenamefont {Chen}}]{Ma:2023vvr}%
  \BibitemOpen
  \bibfield  {author} {\bibinfo {author} {\bibfnamefont {S.}~\bibnamefont
  {Ma}}, \bibinfo {author} {\bibfnamefont {L.}~\bibnamefont {Sun}}, \ and\
  \bibinfo {author} {\bibfnamefont {Y.}~\bibnamefont {Chen}},\ }\href@noop {}
  {\  (\bibinfo {year} {2023}{\natexlab{a}})},\ \Eprint
  {http://arxiv.org/abs/2301.06639} {arXiv:2301.06639 [gr-qc]} \BibitemShut
  {NoStop}%
\bibitem [{\citenamefont {Ma}\ \emph {et~al.}(2023{\natexlab{b}})\citenamefont
  {Ma}, \citenamefont {Sun},\ and\ \citenamefont {Chen}}]{Ma:2023cwe}%
  \BibitemOpen
  \bibfield  {author} {\bibinfo {author} {\bibfnamefont {S.}~\bibnamefont
  {Ma}}, \bibinfo {author} {\bibfnamefont {L.}~\bibnamefont {Sun}}, \ and\
  \bibinfo {author} {\bibfnamefont {Y.}~\bibnamefont {Chen}},\ }\href@noop {}
  {\  (\bibinfo {year} {2023}{\natexlab{b}})},\ \Eprint
  {http://arxiv.org/abs/2301.06705} {arXiv:2301.06705 [gr-qc]} \BibitemShut
  {NoStop}%
\bibitem [{\citenamefont {Pan}\ \emph {et~al.}(2011)\citenamefont {Pan},
  \citenamefont {Buonanno}, \citenamefont {Boyle}, \citenamefont {Buchman},
  \citenamefont {Kidder}, \citenamefont {Pfeiffer},\ and\ \citenamefont
  {Scheel}}]{Pan:2011gk}%
  \BibitemOpen
  \bibfield  {author} {\bibinfo {author} {\bibfnamefont {Y.}~\bibnamefont
  {Pan}}, \bibinfo {author} {\bibfnamefont {A.}~\bibnamefont {Buonanno}},
  \bibinfo {author} {\bibfnamefont {M.}~\bibnamefont {Boyle}}, \bibinfo
  {author} {\bibfnamefont {L.~T.}\ \bibnamefont {Buchman}}, \bibinfo {author}
  {\bibfnamefont {L.~E.}\ \bibnamefont {Kidder}}, \bibinfo {author}
  {\bibfnamefont {H.~P.}\ \bibnamefont {Pfeiffer}}, \ and\ \bibinfo {author}
  {\bibfnamefont {M.~A.}\ \bibnamefont {Scheel}},\ }\href {\doibase
  10.1103/PhysRevD.84.124052} {\bibfield  {journal} {\bibinfo  {journal} {Phys.
  Rev. D}\ }\textbf {\bibinfo {volume} {84}},\ \bibinfo {pages} {124052}
  (\bibinfo {year} {2011})},\ \Eprint {http://arxiv.org/abs/1106.1021}
  {arXiv:1106.1021 [gr-qc]} \BibitemShut {NoStop}%
\bibitem [{\citenamefont {Damour}\ and\ \citenamefont
  {Nagar}(2014)}]{Damour:2014yha}%
  \BibitemOpen
  \bibfield  {author} {\bibinfo {author} {\bibfnamefont {T.}~\bibnamefont
  {Damour}}\ and\ \bibinfo {author} {\bibfnamefont {A.}~\bibnamefont {Nagar}},\
  }\href {\doibase 10.1103/PhysRevD.90.024054} {\bibfield  {journal} {\bibinfo
  {journal} {Phys. Rev. D}\ }\textbf {\bibinfo {volume} {90}},\ \bibinfo
  {pages} {024054} (\bibinfo {year} {2014})},\ \Eprint
  {http://arxiv.org/abs/1406.0401} {arXiv:1406.0401 [gr-qc]} \BibitemShut
  {NoStop}%
\bibitem [{\citenamefont {Brito}\ \emph {et~al.}(2018)\citenamefont {Brito},
  \citenamefont {Buonanno},\ and\ \citenamefont {Raymond}}]{Brito:2018rfr}%
  \BibitemOpen
  \bibfield  {author} {\bibinfo {author} {\bibfnamefont {R.}~\bibnamefont
  {Brito}}, \bibinfo {author} {\bibfnamefont {A.}~\bibnamefont {Buonanno}}, \
  and\ \bibinfo {author} {\bibfnamefont {V.}~\bibnamefont {Raymond}},\ }\href
  {\doibase 10.1103/PhysRevD.98.084038} {\bibfield  {journal} {\bibinfo
  {journal} {Phys. Rev. D}\ }\textbf {\bibinfo {volume} {98}},\ \bibinfo
  {pages} {084038} (\bibinfo {year} {2018})},\ \Eprint
  {http://arxiv.org/abs/1805.00293} {arXiv:1805.00293 [gr-qc]} \BibitemShut
  {NoStop}%
\bibitem [{\citenamefont {Glampedakis}\ \emph {et~al.}(2017)\citenamefont
  {Glampedakis}, \citenamefont {Johnson},\ and\ \citenamefont
  {Kennefick}}]{Glampedakis:2017rar}%
  \BibitemOpen
  \bibfield  {author} {\bibinfo {author} {\bibfnamefont {K.}~\bibnamefont
  {Glampedakis}}, \bibinfo {author} {\bibfnamefont {A.~D.}\ \bibnamefont
  {Johnson}}, \ and\ \bibinfo {author} {\bibfnamefont {D.}~\bibnamefont
  {Kennefick}},\ }\href {\doibase 10.1103/PhysRevD.96.024036} {\bibfield
  {journal} {\bibinfo  {journal} {Phys. Rev. D}\ }\textbf {\bibinfo {volume}
  {96}},\ \bibinfo {pages} {024036} (\bibinfo {year} {2017})},\ \Eprint
  {http://arxiv.org/abs/1702.06459} {arXiv:1702.06459 [gr-qc]} \BibitemShut
  {NoStop}%
\bibitem [{\citenamefont {Zenginoglu}\ and\ \citenamefont
  {Khanna}(2011)}]{Zenginoglu:2011zz}%
  \BibitemOpen
  \bibfield  {author} {\bibinfo {author} {\bibfnamefont {A.}~\bibnamefont
  {Zenginoglu}}\ and\ \bibinfo {author} {\bibfnamefont {G.}~\bibnamefont
  {Khanna}},\ }\href {\doibase 10.1103/PhysRevX.1.021017} {\bibfield  {journal}
  {\bibinfo  {journal} {Phys. Rev. X}\ }\textbf {\bibinfo {volume} {1}},\
  \bibinfo {pages} {021017} (\bibinfo {year} {2011})},\ \Eprint
  {http://arxiv.org/abs/1108.1816} {arXiv:1108.1816 [gr-qc]} \BibitemShut
  {NoStop}%
\bibitem [{\citenamefont {Cardoso}\ \emph {et~al.}(2021)\citenamefont
  {Cardoso}, \citenamefont {Duque},\ and\ \citenamefont
  {Khanna}}]{Cardoso:2021vjq}%
  \BibitemOpen
  \bibfield  {author} {\bibinfo {author} {\bibfnamefont {V.}~\bibnamefont
  {Cardoso}}, \bibinfo {author} {\bibfnamefont {F.}~\bibnamefont {Duque}}, \
  and\ \bibinfo {author} {\bibfnamefont {G.}~\bibnamefont {Khanna}},\ }\href
  {\doibase 10.1103/PhysRevD.103.L081501} {\bibfield  {journal} {\bibinfo
  {journal} {Phys. Rev. D}\ }\textbf {\bibinfo {volume} {103}},\ \bibinfo
  {pages} {L081501} (\bibinfo {year} {2021})},\ \Eprint
  {http://arxiv.org/abs/2101.01186} {arXiv:2101.01186 [gr-qc]} \BibitemShut
  {NoStop}%
\bibitem [{\citenamefont {Krivan}\ \emph {et~al.}(1997)\citenamefont {Krivan},
  \citenamefont {Laguna}, \citenamefont {Papadopoulos},\ and\ \citenamefont
  {Andersson}}]{Krivan:1997hc}%
  \BibitemOpen
  \bibfield  {author} {\bibinfo {author} {\bibfnamefont {W.}~\bibnamefont
  {Krivan}}, \bibinfo {author} {\bibfnamefont {P.}~\bibnamefont {Laguna}},
  \bibinfo {author} {\bibfnamefont {P.}~\bibnamefont {Papadopoulos}}, \ and\
  \bibinfo {author} {\bibfnamefont {N.}~\bibnamefont {Andersson}},\ }\href
  {\doibase 10.1103/PhysRevD.56.3395} {\bibfield  {journal} {\bibinfo
  {journal} {Phys. Rev. D}\ }\textbf {\bibinfo {volume} {56}},\ \bibinfo
  {pages} {3395} (\bibinfo {year} {1997})},\ \Eprint
  {http://arxiv.org/abs/gr-qc/9702048} {arXiv:gr-qc/9702048} \BibitemShut
  {NoStop}%
\bibitem [{\citenamefont {Pazos-Avalos}\ and\ \citenamefont
  {Lousto}(2005)}]{Pazos-Avalos:2004uyd}%
  \BibitemOpen
  \bibfield  {author} {\bibinfo {author} {\bibfnamefont {E.}~\bibnamefont
  {Pazos-Avalos}}\ and\ \bibinfo {author} {\bibfnamefont {C.~O.}\ \bibnamefont
  {Lousto}},\ }\href {\doibase 10.1103/PhysRevD.72.084022} {\bibfield
  {journal} {\bibinfo  {journal} {Phys. Rev. D}\ }\textbf {\bibinfo {volume}
  {72}},\ \bibinfo {pages} {084022} (\bibinfo {year} {2005})},\ \Eprint
  {http://arxiv.org/abs/gr-qc/0409065} {arXiv:gr-qc/0409065} \BibitemShut
  {NoStop}%
\bibitem [{\citenamefont {Zengino\u{g}lu}\ \emph {et~al.}(2014)\citenamefont
  {Zengino\u{g}lu}, \citenamefont {Khanna},\ and\ \citenamefont
  {Burko}}]{Zenginoglu:2012us}%
  \BibitemOpen
  \bibfield  {author} {\bibinfo {author} {\bibfnamefont {A.}~\bibnamefont
  {Zengino\u{g}lu}}, \bibinfo {author} {\bibfnamefont {G.}~\bibnamefont
  {Khanna}}, \ and\ \bibinfo {author} {\bibfnamefont {L.~M.}\ \bibnamefont
  {Burko}},\ }\href {\doibase 10.1007/s10714-014-1672-8} {\bibfield  {journal}
  {\bibinfo  {journal} {Gen. Rel. Grav.}\ }\textbf {\bibinfo {volume} {46}},\
  \bibinfo {pages} {1672} (\bibinfo {year} {2014})},\ \Eprint
  {http://arxiv.org/abs/1208.5839} {arXiv:1208.5839 [gr-qc]} \BibitemShut
  {NoStop}%
\bibitem [{\citenamefont {Burko}\ and\ \citenamefont
  {Khanna}(2014)}]{Burko:2013bra}%
  \BibitemOpen
  \bibfield  {author} {\bibinfo {author} {\bibfnamefont {L.~M.}\ \bibnamefont
  {Burko}}\ and\ \bibinfo {author} {\bibfnamefont {G.}~\bibnamefont {Khanna}},\
  }\href {\doibase 10.1103/PhysRevD.89.044037} {\bibfield  {journal} {\bibinfo
  {journal} {Phys. Rev. D}\ }\textbf {\bibinfo {volume} {89}},\ \bibinfo
  {pages} {044037} (\bibinfo {year} {2014})},\ \Eprint
  {http://arxiv.org/abs/1312.5247} {arXiv:1312.5247 [gr-qc]} \BibitemShut
  {NoStop}%
\bibitem [{\citenamefont {Sundararajan}\ \emph {et~al.}(2007)\citenamefont
  {Sundararajan}, \citenamefont {Khanna},\ and\ \citenamefont
  {Hughes}}]{Sundararajan:2007jg}%
  \BibitemOpen
  \bibfield  {author} {\bibinfo {author} {\bibfnamefont {P.~A.}\ \bibnamefont
  {Sundararajan}}, \bibinfo {author} {\bibfnamefont {G.}~\bibnamefont
  {Khanna}}, \ and\ \bibinfo {author} {\bibfnamefont {S.~A.}\ \bibnamefont
  {Hughes}},\ }\href {\doibase 10.1103/PhysRevD.76.104005} {\bibfield
  {journal} {\bibinfo  {journal} {Phys. Rev. D}\ }\textbf {\bibinfo {volume}
  {76}},\ \bibinfo {pages} {104005} (\bibinfo {year} {2007})},\ \Eprint
  {http://arxiv.org/abs/gr-qc/0703028} {arXiv:gr-qc/0703028} \BibitemShut
  {NoStop}%
\bibitem [{\citenamefont {Cardoso}\ \emph {et~al.}(2022)\citenamefont
  {Cardoso}, \citenamefont {Destounis}, \citenamefont {Duque}, \citenamefont
  {Panosso~Macedo},\ and\ \citenamefont {Maselli}}]{Cardoso:2022whc}%
  \BibitemOpen
  \bibfield  {author} {\bibinfo {author} {\bibfnamefont {V.}~\bibnamefont
  {Cardoso}}, \bibinfo {author} {\bibfnamefont {K.}~\bibnamefont {Destounis}},
  \bibinfo {author} {\bibfnamefont {F.}~\bibnamefont {Duque}}, \bibinfo
  {author} {\bibfnamefont {R.}~\bibnamefont {Panosso~Macedo}}, \ and\ \bibinfo
  {author} {\bibfnamefont {A.}~\bibnamefont {Maselli}},\ }\href {\doibase
  10.1103/PhysRevLett.129.241103} {\bibfield  {journal} {\bibinfo  {journal}
  {Phys. Rev. Lett.}\ }\textbf {\bibinfo {volume} {129}},\ \bibinfo {pages}
  {241103} (\bibinfo {year} {2022})},\ \Eprint
  {http://arxiv.org/abs/2210.01133} {arXiv:2210.01133 [gr-qc]} \BibitemShut
  {NoStop}%
\bibitem [{\citenamefont {Virtanen}\ \emph {et~al.}(2020)\citenamefont
  {Virtanen} \emph {et~al.}}]{2020SciPy-NMeth}%
  \BibitemOpen
  \bibfield  {author} {\bibinfo {author} {\bibfnamefont {P.}~\bibnamefont
  {Virtanen}} \emph {et~al.},\ }\href {\doibase 10.1038/s41592-019-0686-2}
  {\bibfield  {journal} {\bibinfo  {journal} {Nature Methods}\ }\textbf
  {\bibinfo {volume} {17}},\ \bibinfo {pages} {261} (\bibinfo {year}
  {2020})}\BibitemShut {NoStop}%
\bibitem [{\citenamefont {Price}(1972)}]{Price:1971fb}%
  \BibitemOpen
  \bibfield  {author} {\bibinfo {author} {\bibfnamefont {R.~H.}\ \bibnamefont
  {Price}},\ }\href {\doibase 10.1103/PhysRevD.5.2419} {\bibfield  {journal}
  {\bibinfo  {journal} {Phys. Rev. D}\ }\textbf {\bibinfo {volume} {5}},\
  \bibinfo {pages} {2419} (\bibinfo {year} {1972})}\BibitemShut {NoStop}%
\bibitem [{\citenamefont {Jaramillo}\ \emph {et~al.}(2021)\citenamefont
  {Jaramillo}, \citenamefont {Panosso~Macedo},\ and\ \citenamefont
  {Al~Sheikh}}]{Jaramillo:2020tuu}%
  \BibitemOpen
  \bibfield  {author} {\bibinfo {author} {\bibfnamefont {J.~L.}\ \bibnamefont
  {Jaramillo}}, \bibinfo {author} {\bibfnamefont {R.}~\bibnamefont
  {Panosso~Macedo}}, \ and\ \bibinfo {author} {\bibfnamefont {L.}~\bibnamefont
  {Al~Sheikh}},\ }\href {\doibase 10.1103/PhysRevX.11.031003} {\bibfield
  {journal} {\bibinfo  {journal} {Phys. Rev. X}\ }\textbf {\bibinfo {volume}
  {11}},\ \bibinfo {pages} {031003} (\bibinfo {year} {2021})},\ \Eprint
  {http://arxiv.org/abs/2004.06434} {arXiv:2004.06434 [gr-qc]} \BibitemShut
  {NoStop}%
\bibitem [{\citenamefont {Jaramillo}\ \emph {et~al.}(2022)\citenamefont
  {Jaramillo}, \citenamefont {Panosso~Macedo},\ and\ \citenamefont
  {Sheikh}}]{Jaramillo:2021tmt}%
  \BibitemOpen
  \bibfield  {author} {\bibinfo {author} {\bibfnamefont {J.~L.}\ \bibnamefont
  {Jaramillo}}, \bibinfo {author} {\bibfnamefont {R.}~\bibnamefont
  {Panosso~Macedo}}, \ and\ \bibinfo {author} {\bibfnamefont {L.~A.}\
  \bibnamefont {Sheikh}},\ }\href {\doibase 10.1103/PhysRevLett.128.211102}
  {\bibfield  {journal} {\bibinfo  {journal} {Phys. Rev. Lett.}\ }\textbf
  {\bibinfo {volume} {128}},\ \bibinfo {pages} {211102} (\bibinfo {year}
  {2022})},\ \Eprint {http://arxiv.org/abs/2105.03451} {arXiv:2105.03451
  [gr-qc]} \BibitemShut {NoStop}%
\bibitem [{\citenamefont {Cheung}\ \emph
  {et~al.}(2022{\natexlab{a}})\citenamefont {Cheung}, \citenamefont
  {Destounis}, \citenamefont {Macedo}, \citenamefont {Berti},\ and\
  \citenamefont {Cardoso}}]{Cheung:2021bol}%
  \BibitemOpen
  \bibfield  {author} {\bibinfo {author} {\bibfnamefont {M.~H.-Y.}\
  \bibnamefont {Cheung}}, \bibinfo {author} {\bibfnamefont {K.}~\bibnamefont
  {Destounis}}, \bibinfo {author} {\bibfnamefont {R.~P.}\ \bibnamefont
  {Macedo}}, \bibinfo {author} {\bibfnamefont {E.}~\bibnamefont {Berti}}, \
  and\ \bibinfo {author} {\bibfnamefont {V.}~\bibnamefont {Cardoso}},\ }\href
  {\doibase 10.1103/PhysRevLett.128.111103} {\bibfield  {journal} {\bibinfo
  {journal} {Phys. Rev. Lett.}\ }\textbf {\bibinfo {volume} {128}},\ \bibinfo
  {pages} {111103} (\bibinfo {year} {2022}{\natexlab{a}})},\ \Eprint
  {http://arxiv.org/abs/2111.05415} {arXiv:2111.05415 [gr-qc]} \BibitemShut
  {NoStop}%
\bibitem [{\citenamefont {Berti}\ \emph {et~al.}(2022)\citenamefont {Berti},
  \citenamefont {Cardoso}, \citenamefont {Cheung}, \citenamefont {Di~Filippo},
  \citenamefont {Duque}, \citenamefont {Martens},\ and\ \citenamefont
  {Mukohyama}}]{Berti:2022xfj}%
  \BibitemOpen
  \bibfield  {author} {\bibinfo {author} {\bibfnamefont {E.}~\bibnamefont
  {Berti}}, \bibinfo {author} {\bibfnamefont {V.}~\bibnamefont {Cardoso}},
  \bibinfo {author} {\bibfnamefont {M.~H.-Y.}\ \bibnamefont {Cheung}}, \bibinfo
  {author} {\bibfnamefont {F.}~\bibnamefont {Di~Filippo}}, \bibinfo {author}
  {\bibfnamefont {F.}~\bibnamefont {Duque}}, \bibinfo {author} {\bibfnamefont
  {P.}~\bibnamefont {Martens}}, \ and\ \bibinfo {author} {\bibfnamefont
  {S.}~\bibnamefont {Mukohyama}},\ }\href {\doibase
  10.1103/PhysRevD.106.084011} {\bibfield  {journal} {\bibinfo  {journal}
  {Phys. Rev. D}\ }\textbf {\bibinfo {volume} {106}},\ \bibinfo {pages}
  {084011} (\bibinfo {year} {2022})},\ \Eprint
  {http://arxiv.org/abs/2205.08547} {arXiv:2205.08547 [gr-qc]} \BibitemShut
  {NoStop}%
\bibitem [{\citenamefont {Sberna}\ \emph {et~al.}(2022)\citenamefont {Sberna},
  \citenamefont {Bosch}, \citenamefont {East}, \citenamefont {Green},\ and\
  \citenamefont {Lehner}}]{Sberna:2021eui}%
  \BibitemOpen
  \bibfield  {author} {\bibinfo {author} {\bibfnamefont {L.}~\bibnamefont
  {Sberna}}, \bibinfo {author} {\bibfnamefont {P.}~\bibnamefont {Bosch}},
  \bibinfo {author} {\bibfnamefont {W.~E.}\ \bibnamefont {East}}, \bibinfo
  {author} {\bibfnamefont {S.~R.}\ \bibnamefont {Green}}, \ and\ \bibinfo
  {author} {\bibfnamefont {L.}~\bibnamefont {Lehner}},\ }\href {\doibase
  10.1103/PhysRevD.105.064046} {\bibfield  {journal} {\bibinfo  {journal}
  {Phys. Rev. D}\ }\textbf {\bibinfo {volume} {105}},\ \bibinfo {pages}
  {064046} (\bibinfo {year} {2022})},\ \Eprint
  {http://arxiv.org/abs/2112.11168} {arXiv:2112.11168 [gr-qc]} \BibitemShut
  {NoStop}%
\bibitem [{\citenamefont {Ma}\ \emph {et~al.}(2022)\citenamefont {Ma},
  \citenamefont {Mitman}, \citenamefont {Sun}, \citenamefont {Deppe},
  \citenamefont {H\'ebert}, \citenamefont {Kidder}, \citenamefont {Moxon},
  \citenamefont {Throwe}, \citenamefont {Vu},\ and\ \citenamefont
  {Chen}}]{Ma:2022wpv}%
  \BibitemOpen
  \bibfield  {author} {\bibinfo {author} {\bibfnamefont {S.}~\bibnamefont
  {Ma}}, \bibinfo {author} {\bibfnamefont {K.}~\bibnamefont {Mitman}}, \bibinfo
  {author} {\bibfnamefont {L.}~\bibnamefont {Sun}}, \bibinfo {author}
  {\bibfnamefont {N.}~\bibnamefont {Deppe}}, \bibinfo {author} {\bibfnamefont
  {F.}~\bibnamefont {H\'ebert}}, \bibinfo {author} {\bibfnamefont {L.~E.}\
  \bibnamefont {Kidder}}, \bibinfo {author} {\bibfnamefont {J.}~\bibnamefont
  {Moxon}}, \bibinfo {author} {\bibfnamefont {W.}~\bibnamefont {Throwe}},
  \bibinfo {author} {\bibfnamefont {N.~L.}\ \bibnamefont {Vu}}, \ and\ \bibinfo
  {author} {\bibfnamefont {Y.}~\bibnamefont {Chen}},\ }\href {\doibase
  10.1103/PhysRevD.106.084036} {\bibfield  {journal} {\bibinfo  {journal}
  {Phys. Rev. D}\ }\textbf {\bibinfo {volume} {106}},\ \bibinfo {pages}
  {084036} (\bibinfo {year} {2022})},\ \Eprint
  {http://arxiv.org/abs/2207.10870} {arXiv:2207.10870 [gr-qc]} \BibitemShut
  {NoStop}%
\bibitem [{\citenamefont {Mitman}\ \emph {et~al.}(2022)\citenamefont {Mitman}
  \emph {et~al.}}]{Mitman:2022qdl}%
  \BibitemOpen
  \bibfield  {author} {\bibinfo {author} {\bibfnamefont {K.}~\bibnamefont
  {Mitman}} \emph {et~al.},\ }\href@noop {} {\  (\bibinfo {year} {2022})},\
  \Eprint {http://arxiv.org/abs/2208.07380} {arXiv:2208.07380 [gr-qc]}
  \BibitemShut {NoStop}%
\bibitem [{\citenamefont {Cheung}\ \emph
  {et~al.}(2022{\natexlab{b}})\citenamefont {Cheung} \emph
  {et~al.}}]{Cheung:2022rbm}%
  \BibitemOpen
  \bibfield  {author} {\bibinfo {author} {\bibfnamefont {M.~H.-Y.}\
  \bibnamefont {Cheung}} \emph {et~al.},\ }\href@noop {} {\  (\bibinfo {year}
  {2022}{\natexlab{b}})},\ \Eprint {http://arxiv.org/abs/2208.07374}
  {arXiv:2208.07374 [gr-qc]} \BibitemShut {NoStop}%
\bibitem [{\citenamefont {Kehagias}\ \emph {et~al.}(2023)\citenamefont
  {Kehagias}, \citenamefont {Perrone}, \citenamefont {Riotto},\ and\
  \citenamefont {Riva}}]{Kehagias:2023ctr}%
  \BibitemOpen
  \bibfield  {author} {\bibinfo {author} {\bibfnamefont {A.}~\bibnamefont
  {Kehagias}}, \bibinfo {author} {\bibfnamefont {D.}~\bibnamefont {Perrone}},
  \bibinfo {author} {\bibfnamefont {A.}~\bibnamefont {Riotto}}, \ and\ \bibinfo
  {author} {\bibfnamefont {F.}~\bibnamefont {Riva}},\ }\href@noop {} {\
  (\bibinfo {year} {2023})},\ \Eprint {http://arxiv.org/abs/2301.09345}
  {arXiv:2301.09345 [gr-qc]} \BibitemShut {NoStop}%
\bibitem [{\citenamefont {Kehagias}\ and\ \citenamefont
  {Riotto}(2023)}]{Kehagias:2023mcl}%
  \BibitemOpen
  \bibfield  {author} {\bibinfo {author} {\bibfnamefont {A.}~\bibnamefont
  {Kehagias}}\ and\ \bibinfo {author} {\bibfnamefont {A.}~\bibnamefont
  {Riotto}},\ }\href@noop {} {\  (\bibinfo {year} {2023})},\ \Eprint
  {http://arxiv.org/abs/2302.01240} {arXiv:2302.01240 [gr-qc]} \BibitemShut
  {NoStop}%
\bibitem [{\citenamefont {Dorband}\ \emph {et~al.}(2006)\citenamefont
  {Dorband}, \citenamefont {Berti}, \citenamefont {Diener}, \citenamefont
  {Schnetter},\ and\ \citenamefont {Tiglio}}]{Dorband:2006gg}%
  \BibitemOpen
  \bibfield  {author} {\bibinfo {author} {\bibfnamefont {E.~N.}\ \bibnamefont
  {Dorband}}, \bibinfo {author} {\bibfnamefont {E.}~\bibnamefont {Berti}},
  \bibinfo {author} {\bibfnamefont {P.}~\bibnamefont {Diener}}, \bibinfo
  {author} {\bibfnamefont {E.}~\bibnamefont {Schnetter}}, \ and\ \bibinfo
  {author} {\bibfnamefont {M.}~\bibnamefont {Tiglio}},\ }\href {\doibase
  10.1103/PhysRevD.74.084028} {\bibfield  {journal} {\bibinfo  {journal} {Phys.
  Rev. D}\ }\textbf {\bibinfo {volume} {74}},\ \bibinfo {pages} {084028}
  (\bibinfo {year} {2006})},\ \Eprint {http://arxiv.org/abs/gr-qc/0608091}
  {arXiv:gr-qc/0608091} \BibitemShut {NoStop}%
\bibitem [{\citenamefont {Berti}\ \emph
  {et~al.}(2007{\natexlab{c}})\citenamefont {Berti}, \citenamefont {Cardoso},
  \citenamefont {Gonzalez},\ and\ \citenamefont {Sperhake}}]{Berti:2007dg}%
  \BibitemOpen
  \bibfield  {author} {\bibinfo {author} {\bibfnamefont {E.}~\bibnamefont
  {Berti}}, \bibinfo {author} {\bibfnamefont {V.}~\bibnamefont {Cardoso}},
  \bibinfo {author} {\bibfnamefont {J.~A.}\ \bibnamefont {Gonzalez}}, \ and\
  \bibinfo {author} {\bibfnamefont {U.}~\bibnamefont {Sperhake}},\ }\href
  {\doibase 10.1103/PhysRevD.75.124017} {\bibfield  {journal} {\bibinfo
  {journal} {Phys. Rev. D}\ }\textbf {\bibinfo {volume} {75}},\ \bibinfo
  {pages} {124017} (\bibinfo {year} {2007}{\natexlab{c}})},\ \Eprint
  {http://arxiv.org/abs/gr-qc/0701086} {arXiv:gr-qc/0701086} \BibitemShut
  {NoStop}%
\bibitem [{\citenamefont {Mroue}\ \emph {et~al.}(2013)\citenamefont {Mroue}
  \emph {et~al.}}]{Mroue:2013xna}%
  \BibitemOpen
  \bibfield  {author} {\bibinfo {author} {\bibfnamefont {A.~H.}\ \bibnamefont
  {Mroue}} \emph {et~al.},\ }\href {\doibase 10.1103/PhysRevLett.111.241104}
  {\bibfield  {journal} {\bibinfo  {journal} {Phys. Rev. Lett.}\ }\textbf
  {\bibinfo {volume} {111}},\ \bibinfo {pages} {241104} (\bibinfo {year}
  {2013})},\ \Eprint {http://arxiv.org/abs/1304.6077} {arXiv:1304.6077 [gr-qc]}
  \BibitemShut {NoStop}%
\bibitem [{\citenamefont {Boyle}\ \emph {et~al.}(2019)\citenamefont {Boyle}
  \emph {et~al.}}]{Boyle:2019kee}%
  \BibitemOpen
  \bibfield  {author} {\bibinfo {author} {\bibfnamefont {M.}~\bibnamefont
  {Boyle}} \emph {et~al.},\ }\href {\doibase 10.1088/1361-6382/ab34e2}
  {\bibfield  {journal} {\bibinfo  {journal} {Class. Quant. Grav.}\ }\textbf
  {\bibinfo {volume} {36}},\ \bibinfo {pages} {195006} (\bibinfo {year}
  {2019})},\ \Eprint {http://arxiv.org/abs/1904.04831} {arXiv:1904.04831
  [gr-qc]} \BibitemShut {NoStop}%
\bibitem [{\citenamefont {Ruiz}\ \emph {et~al.}(2008)\citenamefont {Ruiz},
  \citenamefont {Takahashi}, \citenamefont {Alcubierre},\ and\ \citenamefont
  {Nunez}}]{Ruiz:2007yx}%
  \BibitemOpen
  \bibfield  {author} {\bibinfo {author} {\bibfnamefont {M.}~\bibnamefont
  {Ruiz}}, \bibinfo {author} {\bibfnamefont {R.}~\bibnamefont {Takahashi}},
  \bibinfo {author} {\bibfnamefont {M.}~\bibnamefont {Alcubierre}}, \ and\
  \bibinfo {author} {\bibfnamefont {D.}~\bibnamefont {Nunez}},\ }\href
  {\doibase 10.1007/s10714-007-0570-8} {\bibfield  {journal} {\bibinfo
  {journal} {Gen. Rel. Grav.}\ }\textbf {\bibinfo {volume} {40}},\ \bibinfo
  {pages} {2467} (\bibinfo {year} {2008})},\ \Eprint
  {http://arxiv.org/abs/0707.4654} {arXiv:0707.4654 [gr-qc]} \BibitemShut
  {NoStop}%
\bibitem [{\citenamefont {Bhagwat}\ \emph {et~al.}(2018)\citenamefont
  {Bhagwat}, \citenamefont {Okounkova}, \citenamefont {Ballmer}, \citenamefont
  {Brown}, \citenamefont {Giesler}, \citenamefont {Scheel},\ and\ \citenamefont
  {Teukolsky}}]{Bhagwat:2017tkm}%
  \BibitemOpen
  \bibfield  {author} {\bibinfo {author} {\bibfnamefont {S.}~\bibnamefont
  {Bhagwat}}, \bibinfo {author} {\bibfnamefont {M.}~\bibnamefont {Okounkova}},
  \bibinfo {author} {\bibfnamefont {S.~W.}\ \bibnamefont {Ballmer}}, \bibinfo
  {author} {\bibfnamefont {D.~A.}\ \bibnamefont {Brown}}, \bibinfo {author}
  {\bibfnamefont {M.}~\bibnamefont {Giesler}}, \bibinfo {author} {\bibfnamefont
  {M.~A.}\ \bibnamefont {Scheel}}, \ and\ \bibinfo {author} {\bibfnamefont
  {S.~A.}\ \bibnamefont {Teukolsky}},\ }\href {\doibase
  10.1103/PhysRevD.97.104065} {\bibfield  {journal} {\bibinfo  {journal} {Phys.
  Rev. D}\ }\textbf {\bibinfo {volume} {97}},\ \bibinfo {pages} {104065}
  (\bibinfo {year} {2018})},\ \Eprint {http://arxiv.org/abs/1711.00926}
  {arXiv:1711.00926 [gr-qc]} \BibitemShut {NoStop}%
\bibitem [{\citenamefont {Bamber}\ \emph {et~al.}(2021)\citenamefont {Bamber},
  \citenamefont {Tattersall}, \citenamefont {Clough},\ and\ \citenamefont
  {Ferreira}}]{Bamber:2021knr}%
  \BibitemOpen
  \bibfield  {author} {\bibinfo {author} {\bibfnamefont {J.}~\bibnamefont
  {Bamber}}, \bibinfo {author} {\bibfnamefont {O.~J.}\ \bibnamefont
  {Tattersall}}, \bibinfo {author} {\bibfnamefont {K.}~\bibnamefont {Clough}},
  \ and\ \bibinfo {author} {\bibfnamefont {P.~G.}\ \bibnamefont {Ferreira}},\
  }\href {\doibase 10.1103/PhysRevD.103.124013} {\bibfield  {journal} {\bibinfo
   {journal} {Phys. Rev. D}\ }\textbf {\bibinfo {volume} {103}},\ \bibinfo
  {pages} {124013} (\bibinfo {year} {2021})},\ \Eprint
  {http://arxiv.org/abs/2103.00026} {arXiv:2103.00026 [gr-qc]} \BibitemShut
  {NoStop}%
\bibitem [{\citenamefont {{Dyson}}(2004)}]{2004Natur.427..297D}%
  \BibitemOpen
  \bibfield  {author} {\bibinfo {author} {\bibfnamefont {F.}~\bibnamefont
  {{Dyson}}},\ }\href {\doibase 10.1038/427297a} {\bibfield  {journal}
  {\bibinfo  {journal} {\nat}\ }\textbf {\bibinfo {volume} {427}},\ \bibinfo
  {pages} {297} (\bibinfo {year} {2004})}\BibitemShut {NoStop}%
\bibitem [{\citenamefont {Kelly}\ and\ \citenamefont
  {Baker}(2013)}]{Kelly:2012nd}%
  \BibitemOpen
  \bibfield  {author} {\bibinfo {author} {\bibfnamefont {B.~J.}\ \bibnamefont
  {Kelly}}\ and\ \bibinfo {author} {\bibfnamefont {J.~G.}\ \bibnamefont
  {Baker}},\ }\href {\doibase 10.1103/PhysRevD.87.084004} {\bibfield  {journal}
  {\bibinfo  {journal} {Phys. Rev. D}\ }\textbf {\bibinfo {volume} {87}},\
  \bibinfo {pages} {084004} (\bibinfo {year} {2013})},\ \Eprint
  {http://arxiv.org/abs/1212.5553} {arXiv:1212.5553 [gr-qc]} \BibitemShut
  {NoStop}%
\bibitem [{\citenamefont {Bernuzzi}\ and\ \citenamefont
  {Nagar}(2010)}]{Bernuzzi:2010ty}%
  \BibitemOpen
  \bibfield  {author} {\bibinfo {author} {\bibfnamefont {S.}~\bibnamefont
  {Bernuzzi}}\ and\ \bibinfo {author} {\bibfnamefont {A.}~\bibnamefont
  {Nagar}},\ }\href {\doibase 10.1103/PhysRevD.81.084056} {\bibfield  {journal}
  {\bibinfo  {journal} {Phys. Rev. D}\ }\textbf {\bibinfo {volume} {81}},\
  \bibinfo {pages} {084056} (\bibinfo {year} {2010})},\ \Eprint
  {http://arxiv.org/abs/1003.0597} {arXiv:1003.0597 [gr-qc]} \BibitemShut
  {NoStop}%
\bibitem [{\citenamefont {Barausse}\ \emph {et~al.}(2012)\citenamefont
  {Barausse}, \citenamefont {Buonanno}, \citenamefont {Hughes}, \citenamefont
  {Khanna}, \citenamefont {O'Sullivan},\ and\ \citenamefont
  {Pan}}]{Barausse:2011kb}%
  \BibitemOpen
  \bibfield  {author} {\bibinfo {author} {\bibfnamefont {E.}~\bibnamefont
  {Barausse}}, \bibinfo {author} {\bibfnamefont {A.}~\bibnamefont {Buonanno}},
  \bibinfo {author} {\bibfnamefont {S.~A.}\ \bibnamefont {Hughes}}, \bibinfo
  {author} {\bibfnamefont {G.}~\bibnamefont {Khanna}}, \bibinfo {author}
  {\bibfnamefont {S.}~\bibnamefont {O'Sullivan}}, \ and\ \bibinfo {author}
  {\bibfnamefont {Y.}~\bibnamefont {Pan}},\ }\href {\doibase
  10.1103/PhysRevD.85.024046} {\bibfield  {journal} {\bibinfo  {journal} {Phys.
  Rev. D}\ }\textbf {\bibinfo {volume} {85}},\ \bibinfo {pages} {024046}
  (\bibinfo {year} {2012})},\ \Eprint {http://arxiv.org/abs/1110.3081}
  {arXiv:1110.3081 [gr-qc]} \BibitemShut {NoStop}%
\bibitem [{\citenamefont {Abbott}\ \emph
  {et~al.}(2016{\natexlab{b}})\citenamefont {Abbott} \emph
  {et~al.}}]{LIGOScientific:2016vlm}%
  \BibitemOpen
  \bibfield  {author} {\bibinfo {author} {\bibfnamefont {B.~P.}\ \bibnamefont
  {Abbott}} \emph {et~al.} (\bibinfo {collaboration} {LIGO Scientific,
  Virgo}),\ }\href {\doibase 10.1103/PhysRevLett.116.241102} {\bibfield
  {journal} {\bibinfo  {journal} {Phys. Rev. Lett.}\ }\textbf {\bibinfo
  {volume} {116}},\ \bibinfo {pages} {241102} (\bibinfo {year}
  {2016}{\natexlab{b}})},\ \Eprint {http://arxiv.org/abs/1602.03840}
  {arXiv:1602.03840 [gr-qc]} \BibitemShut {NoStop}%
\bibitem [{\citenamefont {Abbott}\ \emph
  {et~al.}(2016{\natexlab{c}})\citenamefont {Abbott} \emph
  {et~al.}}]{LIGOScientific:2016lio}%
  \BibitemOpen
  \bibfield  {author} {\bibinfo {author} {\bibfnamefont {B.~P.}\ \bibnamefont
  {Abbott}} \emph {et~al.} (\bibinfo {collaboration} {LIGO Scientific,
  Virgo}),\ }\href {\doibase 10.1103/PhysRevLett.116.221101} {\bibfield
  {journal} {\bibinfo  {journal} {Phys. Rev. Lett.}\ }\textbf {\bibinfo
  {volume} {116}},\ \bibinfo {pages} {221101} (\bibinfo {year}
  {2016}{\natexlab{c}})},\ \bibinfo {note} {[Erratum: Phys.Rev.Lett. 121,
  129902 (2018)]},\ \Eprint {http://arxiv.org/abs/1602.03841} {arXiv:1602.03841
  [gr-qc]} \BibitemShut {NoStop}%
\bibitem [{\citenamefont {Maggio}\ \emph {et~al.}(2022)\citenamefont {Maggio},
  \citenamefont {Silva}, \citenamefont {Buonanno},\ and\ \citenamefont
  {Ghosh}}]{Maggio:2022hre}%
  \BibitemOpen
  \bibfield  {author} {\bibinfo {author} {\bibfnamefont {E.}~\bibnamefont
  {Maggio}}, \bibinfo {author} {\bibfnamefont {H.~O.}\ \bibnamefont {Silva}},
  \bibinfo {author} {\bibfnamefont {A.}~\bibnamefont {Buonanno}}, \ and\
  \bibinfo {author} {\bibfnamefont {A.}~\bibnamefont {Ghosh}},\ }\href@noop {}
  {\  (\bibinfo {year} {2022})},\ \Eprint {http://arxiv.org/abs/2212.09655}
  {arXiv:2212.09655 [gr-qc]} \BibitemShut {NoStop}%
\bibitem [{\citenamefont {Isi}\ \emph {et~al.}(2021)\citenamefont {Isi},
  \citenamefont {Farr}, \citenamefont {Giesler}, \citenamefont {Scheel},\ and\
  \citenamefont {Teukolsky}}]{Isi:2020tac}%
  \BibitemOpen
  \bibfield  {author} {\bibinfo {author} {\bibfnamefont {M.}~\bibnamefont
  {Isi}}, \bibinfo {author} {\bibfnamefont {W.~M.}\ \bibnamefont {Farr}},
  \bibinfo {author} {\bibfnamefont {M.}~\bibnamefont {Giesler}}, \bibinfo
  {author} {\bibfnamefont {M.~A.}\ \bibnamefont {Scheel}}, \ and\ \bibinfo
  {author} {\bibfnamefont {S.~A.}\ \bibnamefont {Teukolsky}},\ }\href {\doibase
  10.1103/PhysRevLett.127.011103} {\bibfield  {journal} {\bibinfo  {journal}
  {Phys. Rev. Lett.}\ }\textbf {\bibinfo {volume} {127}},\ \bibinfo {pages}
  {011103} (\bibinfo {year} {2021})},\ \Eprint
  {http://arxiv.org/abs/2012.04486} {arXiv:2012.04486 [gr-qc]} \BibitemShut
  {NoStop}%
\bibitem [{\citenamefont {Ghosh}\ \emph {et~al.}(2018)\citenamefont {Ghosh},
  \citenamefont {Johnson-Mcdaniel}, \citenamefont {Ghosh}, \citenamefont
  {Mishra}, \citenamefont {Ajith}, \citenamefont {Del~Pozzo}, \citenamefont
  {Berry}, \citenamefont {Nielsen},\ and\ \citenamefont
  {London}}]{Ghosh:2017gfp}%
  \BibitemOpen
  \bibfield  {author} {\bibinfo {author} {\bibfnamefont {A.}~\bibnamefont
  {Ghosh}}, \bibinfo {author} {\bibfnamefont {N.~K.}\ \bibnamefont
  {Johnson-Mcdaniel}}, \bibinfo {author} {\bibfnamefont {A.}~\bibnamefont
  {Ghosh}}, \bibinfo {author} {\bibfnamefont {C.~K.}\ \bibnamefont {Mishra}},
  \bibinfo {author} {\bibfnamefont {P.}~\bibnamefont {Ajith}}, \bibinfo
  {author} {\bibfnamefont {W.}~\bibnamefont {Del~Pozzo}}, \bibinfo {author}
  {\bibfnamefont {C.~P.~L.}\ \bibnamefont {Berry}}, \bibinfo {author}
  {\bibfnamefont {A.~B.}\ \bibnamefont {Nielsen}}, \ and\ \bibinfo {author}
  {\bibfnamefont {L.}~\bibnamefont {London}},\ }\href {\doibase
  10.1088/1361-6382/aa972e} {\bibfield  {journal} {\bibinfo  {journal} {Class.
  Quant. Grav.}\ }\textbf {\bibinfo {volume} {35}},\ \bibinfo {pages} {014002}
  (\bibinfo {year} {2018})},\ \Eprint {http://arxiv.org/abs/1704.06784}
  {arXiv:1704.06784 [gr-qc]} \BibitemShut {NoStop}%
\bibitem [{\citenamefont {Nollert}(1996)}]{Nollert:1996rf}%
  \BibitemOpen
  \bibfield  {author} {\bibinfo {author} {\bibfnamefont {H.-P.}\ \bibnamefont
  {Nollert}},\ }\href {\doibase 10.1103/PhysRevD.53.4397} {\bibfield  {journal}
  {\bibinfo  {journal} {Phys. Rev. D}\ }\textbf {\bibinfo {volume} {53}},\
  \bibinfo {pages} {4397} (\bibinfo {year} {1996})},\ \Eprint
  {http://arxiv.org/abs/gr-qc/9602032} {arXiv:gr-qc/9602032} \BibitemShut
  {NoStop}%
\bibitem [{\citenamefont {Barausse}\ \emph {et~al.}(2014)\citenamefont
  {Barausse}, \citenamefont {Cardoso},\ and\ \citenamefont
  {Pani}}]{Barausse:2014tra}%
  \BibitemOpen
  \bibfield  {author} {\bibinfo {author} {\bibfnamefont {E.}~\bibnamefont
  {Barausse}}, \bibinfo {author} {\bibfnamefont {V.}~\bibnamefont {Cardoso}}, \
  and\ \bibinfo {author} {\bibfnamefont {P.}~\bibnamefont {Pani}},\ }\href
  {\doibase 10.1103/PhysRevD.89.104059} {\bibfield  {journal} {\bibinfo
  {journal} {Phys. Rev. D}\ }\textbf {\bibinfo {volume} {89}},\ \bibinfo
  {pages} {104059} (\bibinfo {year} {2014})},\ \Eprint
  {http://arxiv.org/abs/1404.7149} {arXiv:1404.7149 [gr-qc]} \BibitemShut
  {NoStop}%
\bibitem [{\citenamefont {Gleiser}\ \emph {et~al.}(1996)\citenamefont
  {Gleiser}, \citenamefont {Nicasio}, \citenamefont {Price},\ and\
  \citenamefont {Pullin}}]{Gleiser:1995gx}%
  \BibitemOpen
  \bibfield  {author} {\bibinfo {author} {\bibfnamefont {R.~J.}\ \bibnamefont
  {Gleiser}}, \bibinfo {author} {\bibfnamefont {C.~O.}\ \bibnamefont
  {Nicasio}}, \bibinfo {author} {\bibfnamefont {R.~H.}\ \bibnamefont {Price}},
  \ and\ \bibinfo {author} {\bibfnamefont {J.}~\bibnamefont {Pullin}},\ }\href
  {\doibase 10.1088/0264-9381/13/10/001} {\bibfield  {journal} {\bibinfo
  {journal} {Class. Quant. Grav.}\ }\textbf {\bibinfo {volume} {13}},\ \bibinfo
  {pages} {L117} (\bibinfo {year} {1996})},\ \Eprint
  {http://arxiv.org/abs/gr-qc/9510049} {arXiv:gr-qc/9510049} \BibitemShut
  {NoStop}%
\bibitem [{\citenamefont {Campanelli}\ and\ \citenamefont
  {Lousto}(1999)}]{Campanelli:1998jv}%
  \BibitemOpen
  \bibfield  {author} {\bibinfo {author} {\bibfnamefont {M.}~\bibnamefont
  {Campanelli}}\ and\ \bibinfo {author} {\bibfnamefont {C.~O.}\ \bibnamefont
  {Lousto}},\ }\href {\doibase 10.1103/PhysRevD.59.124022} {\bibfield
  {journal} {\bibinfo  {journal} {Phys. Rev. D}\ }\textbf {\bibinfo {volume}
  {59}},\ \bibinfo {pages} {124022} (\bibinfo {year} {1999})},\ \Eprint
  {http://arxiv.org/abs/gr-qc/9811019} {arXiv:gr-qc/9811019} \BibitemShut
  {NoStop}%
\bibitem [{\citenamefont {Brizuela}\ \emph {et~al.}(2009)\citenamefont
  {Brizuela}, \citenamefont {Martin-Garcia},\ and\ \citenamefont
  {Tiglio}}]{Brizuela:2009qd}%
  \BibitemOpen
  \bibfield  {author} {\bibinfo {author} {\bibfnamefont {D.}~\bibnamefont
  {Brizuela}}, \bibinfo {author} {\bibfnamefont {J.~M.}\ \bibnamefont
  {Martin-Garcia}}, \ and\ \bibinfo {author} {\bibfnamefont {M.}~\bibnamefont
  {Tiglio}},\ }\href {\doibase 10.1103/PhysRevD.80.024021} {\bibfield
  {journal} {\bibinfo  {journal} {Phys. Rev. D}\ }\textbf {\bibinfo {volume}
  {80}},\ \bibinfo {pages} {024021} (\bibinfo {year} {2009})},\ \Eprint
  {http://arxiv.org/abs/0903.1134} {arXiv:0903.1134 [gr-qc]} \BibitemShut
  {NoStop}%
\bibitem [{\citenamefont {Ioka}\ and\ \citenamefont
  {Nakano}(2007)}]{Ioka:2007ak}%
  \BibitemOpen
  \bibfield  {author} {\bibinfo {author} {\bibfnamefont {K.}~\bibnamefont
  {Ioka}}\ and\ \bibinfo {author} {\bibfnamefont {H.}~\bibnamefont {Nakano}},\
  }\href {\doibase 10.1103/PhysRevD.76.061503} {\bibfield  {journal} {\bibinfo
  {journal} {Phys. Rev. D}\ }\textbf {\bibinfo {volume} {76}},\ \bibinfo
  {pages} {061503} (\bibinfo {year} {2007})},\ \Eprint
  {http://arxiv.org/abs/0704.3467} {arXiv:0704.3467 [astro-ph]} \BibitemShut
  {NoStop}%
\bibitem [{\citenamefont {Nakano}\ and\ \citenamefont
  {Ioka}(2007)}]{Nakano:2007cj}%
  \BibitemOpen
  \bibfield  {author} {\bibinfo {author} {\bibfnamefont {H.}~\bibnamefont
  {Nakano}}\ and\ \bibinfo {author} {\bibfnamefont {K.}~\bibnamefont {Ioka}},\
  }\href {\doibase 10.1103/PhysRevD.76.084007} {\bibfield  {journal} {\bibinfo
  {journal} {Phys. Rev. D}\ }\textbf {\bibinfo {volume} {76}},\ \bibinfo
  {pages} {084007} (\bibinfo {year} {2007})},\ \Eprint
  {http://arxiv.org/abs/0708.0450} {arXiv:0708.0450 [gr-qc]} \BibitemShut
  {NoStop}%
\bibitem [{\citenamefont {Pazos}\ \emph {et~al.}(2010)\citenamefont {Pazos},
  \citenamefont {Brizuela}, \citenamefont {Martin-Garcia},\ and\ \citenamefont
  {Tiglio}}]{Pazos:2010xf}%
  \BibitemOpen
  \bibfield  {author} {\bibinfo {author} {\bibfnamefont {E.}~\bibnamefont
  {Pazos}}, \bibinfo {author} {\bibfnamefont {D.}~\bibnamefont {Brizuela}},
  \bibinfo {author} {\bibfnamefont {J.~M.}\ \bibnamefont {Martin-Garcia}}, \
  and\ \bibinfo {author} {\bibfnamefont {M.}~\bibnamefont {Tiglio}},\ }\href
  {\doibase 10.1103/PhysRevD.82.104028} {\bibfield  {journal} {\bibinfo
  {journal} {Phys. Rev. D}\ }\textbf {\bibinfo {volume} {82}},\ \bibinfo
  {pages} {104028} (\bibinfo {year} {2010})},\ \Eprint
  {http://arxiv.org/abs/1009.4665} {arXiv:1009.4665 [gr-qc]} \BibitemShut
  {NoStop}%
\bibitem [{\citenamefont {Loutrel}\ \emph {et~al.}(2021)\citenamefont
  {Loutrel}, \citenamefont {Ripley}, \citenamefont {Giorgi},\ and\
  \citenamefont {Pretorius}}]{Loutrel:2020wbw}%
  \BibitemOpen
  \bibfield  {author} {\bibinfo {author} {\bibfnamefont {N.}~\bibnamefont
  {Loutrel}}, \bibinfo {author} {\bibfnamefont {J.~L.}\ \bibnamefont {Ripley}},
  \bibinfo {author} {\bibfnamefont {E.}~\bibnamefont {Giorgi}}, \ and\ \bibinfo
  {author} {\bibfnamefont {F.}~\bibnamefont {Pretorius}},\ }\href {\doibase
  10.1103/PhysRevD.103.104017} {\bibfield  {journal} {\bibinfo  {journal}
  {Phys. Rev. D}\ }\textbf {\bibinfo {volume} {103}},\ \bibinfo {pages}
  {104017} (\bibinfo {year} {2021})},\ \Eprint
  {http://arxiv.org/abs/2008.11770} {arXiv:2008.11770 [gr-qc]} \BibitemShut
  {NoStop}%
\bibitem [{\citenamefont {Ripley}\ \emph {et~al.}(2021)\citenamefont {Ripley},
  \citenamefont {Loutrel}, \citenamefont {Giorgi},\ and\ \citenamefont
  {Pretorius}}]{Ripley:2020xby}%
  \BibitemOpen
  \bibfield  {author} {\bibinfo {author} {\bibfnamefont {J.~L.}\ \bibnamefont
  {Ripley}}, \bibinfo {author} {\bibfnamefont {N.}~\bibnamefont {Loutrel}},
  \bibinfo {author} {\bibfnamefont {E.}~\bibnamefont {Giorgi}}, \ and\ \bibinfo
  {author} {\bibfnamefont {F.}~\bibnamefont {Pretorius}},\ }\href {\doibase
  10.1103/PhysRevD.103.104018} {\bibfield  {journal} {\bibinfo  {journal}
  {Phys. Rev. D}\ }\textbf {\bibinfo {volume} {103}},\ \bibinfo {pages}
  {104018} (\bibinfo {year} {2021})},\ \Eprint
  {http://arxiv.org/abs/2010.00162} {arXiv:2010.00162 [gr-qc]} \BibitemShut
  {NoStop}%
\bibitem [{\citenamefont {Bosch}\ \emph {et~al.}(2016)\citenamefont {Bosch},
  \citenamefont {Green},\ and\ \citenamefont {Lehner}}]{Bosch:2016vcp}%
  \BibitemOpen
  \bibfield  {author} {\bibinfo {author} {\bibfnamefont {P.}~\bibnamefont
  {Bosch}}, \bibinfo {author} {\bibfnamefont {S.~R.}\ \bibnamefont {Green}}, \
  and\ \bibinfo {author} {\bibfnamefont {L.}~\bibnamefont {Lehner}},\ }\href
  {\doibase 10.1103/PhysRevLett.116.141102} {\bibfield  {journal} {\bibinfo
  {journal} {Phys. Rev. Lett.}\ }\textbf {\bibinfo {volume} {116}},\ \bibinfo
  {pages} {141102} (\bibinfo {year} {2016})},\ \Eprint
  {http://arxiv.org/abs/1601.01384} {arXiv:1601.01384 [gr-qc]} \BibitemShut
  {NoStop}%
\bibitem [{\citenamefont {Bosch}\ \emph {et~al.}(2020)\citenamefont {Bosch},
  \citenamefont {Green}, \citenamefont {Lehner},\ and\ \citenamefont
  {Roussille}}]{Bosch:2019anc}%
  \BibitemOpen
  \bibfield  {author} {\bibinfo {author} {\bibfnamefont {P.}~\bibnamefont
  {Bosch}}, \bibinfo {author} {\bibfnamefont {S.~R.}\ \bibnamefont {Green}},
  \bibinfo {author} {\bibfnamefont {L.}~\bibnamefont {Lehner}}, \ and\ \bibinfo
  {author} {\bibfnamefont {H.}~\bibnamefont {Roussille}},\ }\href {\doibase
  10.1103/PhysRevD.102.044014} {\bibfield  {journal} {\bibinfo  {journal}
  {Phys. Rev. D}\ }\textbf {\bibinfo {volume} {102}},\ \bibinfo {pages}
  {044014} (\bibinfo {year} {2020})},\ \Eprint
  {http://arxiv.org/abs/1912.05598} {arXiv:1912.05598 [gr-qc]} \BibitemShut
  {NoStop}%
\bibitem [{\citenamefont {Capano}\ \emph {et~al.}(2021)\citenamefont {Capano},
  \citenamefont {Cabero}, \citenamefont {Westerweck}, \citenamefont {Abedi},
  \citenamefont {Kastha}, \citenamefont {Nitz}, \citenamefont {Wang},
  \citenamefont {Nielsen},\ and\ \citenamefont {Krishnan}}]{Capano:2021etf}%
  \BibitemOpen
  \bibfield  {author} {\bibinfo {author} {\bibfnamefont {C.~D.}\ \bibnamefont
  {Capano}}, \bibinfo {author} {\bibfnamefont {M.}~\bibnamefont {Cabero}},
  \bibinfo {author} {\bibfnamefont {J.}~\bibnamefont {Westerweck}}, \bibinfo
  {author} {\bibfnamefont {J.}~\bibnamefont {Abedi}}, \bibinfo {author}
  {\bibfnamefont {S.}~\bibnamefont {Kastha}}, \bibinfo {author} {\bibfnamefont
  {A.~H.}\ \bibnamefont {Nitz}}, \bibinfo {author} {\bibfnamefont {Y.-F.}\
  \bibnamefont {Wang}}, \bibinfo {author} {\bibfnamefont {A.~B.}\ \bibnamefont
  {Nielsen}}, \ and\ \bibinfo {author} {\bibfnamefont {B.}~\bibnamefont
  {Krishnan}},\ }\href@noop {} {\  (\bibinfo {year} {2021})},\ \Eprint
  {http://arxiv.org/abs/2105.05238} {arXiv:2105.05238 [gr-qc]} \BibitemShut
  {NoStop}%
\bibitem [{\citenamefont {Tattersall}\ \emph
  {et~al.}(2018{\natexlab{a}})\citenamefont {Tattersall}, \citenamefont
  {Ferreira},\ and\ \citenamefont {Lagos}}]{Tattersall:2017erk}%
  \BibitemOpen
  \bibfield  {author} {\bibinfo {author} {\bibfnamefont {O.~J.}\ \bibnamefont
  {Tattersall}}, \bibinfo {author} {\bibfnamefont {P.~G.}\ \bibnamefont
  {Ferreira}}, \ and\ \bibinfo {author} {\bibfnamefont {M.}~\bibnamefont
  {Lagos}},\ }\href {\doibase 10.1103/PhysRevD.97.044021} {\bibfield  {journal}
  {\bibinfo  {journal} {Phys. Rev. D}\ }\textbf {\bibinfo {volume} {97}},\
  \bibinfo {pages} {044021} (\bibinfo {year} {2018}{\natexlab{a}})},\ \Eprint
  {http://arxiv.org/abs/1711.01992} {arXiv:1711.01992 [gr-qc]} \BibitemShut
  {NoStop}%
\bibitem [{\citenamefont {Tattersall}\ \emph
  {et~al.}(2018{\natexlab{b}})\citenamefont {Tattersall}, \citenamefont
  {Ferreira},\ and\ \citenamefont {Lagos}}]{Tattersall:2018map}%
  \BibitemOpen
  \bibfield  {author} {\bibinfo {author} {\bibfnamefont {O.~J.}\ \bibnamefont
  {Tattersall}}, \bibinfo {author} {\bibfnamefont {P.~G.}\ \bibnamefont
  {Ferreira}}, \ and\ \bibinfo {author} {\bibfnamefont {M.}~\bibnamefont
  {Lagos}},\ }\href {\doibase 10.1103/PhysRevD.97.084005} {\bibfield  {journal}
  {\bibinfo  {journal} {Phys. Rev. D}\ }\textbf {\bibinfo {volume} {97}},\
  \bibinfo {pages} {084005} (\bibinfo {year} {2018}{\natexlab{b}})},\ \Eprint
  {http://arxiv.org/abs/1802.08606} {arXiv:1802.08606 [gr-qc]} \BibitemShut
  {NoStop}%
\bibitem [{\citenamefont {Cardoso}\ \emph {et~al.}(2019)\citenamefont
  {Cardoso}, \citenamefont {Kimura}, \citenamefont {Maselli}, \citenamefont
  {Berti}, \citenamefont {Macedo},\ and\ \citenamefont
  {McManus}}]{Cardoso:2019mqo}%
  \BibitemOpen
  \bibfield  {author} {\bibinfo {author} {\bibfnamefont {V.}~\bibnamefont
  {Cardoso}}, \bibinfo {author} {\bibfnamefont {M.}~\bibnamefont {Kimura}},
  \bibinfo {author} {\bibfnamefont {A.}~\bibnamefont {Maselli}}, \bibinfo
  {author} {\bibfnamefont {E.}~\bibnamefont {Berti}}, \bibinfo {author}
  {\bibfnamefont {C.~F.~B.}\ \bibnamefont {Macedo}}, \ and\ \bibinfo {author}
  {\bibfnamefont {R.}~\bibnamefont {McManus}},\ }\href {\doibase
  10.1103/PhysRevD.99.104077} {\bibfield  {journal} {\bibinfo  {journal} {Phys.
  Rev. D}\ }\textbf {\bibinfo {volume} {99}},\ \bibinfo {pages} {104077}
  (\bibinfo {year} {2019})},\ \Eprint {http://arxiv.org/abs/1901.01265}
  {arXiv:1901.01265 [gr-qc]} \BibitemShut {NoStop}%
\bibitem [{\citenamefont {McManus}\ \emph {et~al.}(2019)\citenamefont
  {McManus}, \citenamefont {Berti}, \citenamefont {Macedo}, \citenamefont
  {Kimura}, \citenamefont {Maselli},\ and\ \citenamefont
  {Cardoso}}]{McManus:2019ulj}%
  \BibitemOpen
  \bibfield  {author} {\bibinfo {author} {\bibfnamefont {R.}~\bibnamefont
  {McManus}}, \bibinfo {author} {\bibfnamefont {E.}~\bibnamefont {Berti}},
  \bibinfo {author} {\bibfnamefont {C.~F.~B.}\ \bibnamefont {Macedo}}, \bibinfo
  {author} {\bibfnamefont {M.}~\bibnamefont {Kimura}}, \bibinfo {author}
  {\bibfnamefont {A.}~\bibnamefont {Maselli}}, \ and\ \bibinfo {author}
  {\bibfnamefont {V.}~\bibnamefont {Cardoso}},\ }\href {\doibase
  10.1103/PhysRevD.100.044061} {\bibfield  {journal} {\bibinfo  {journal}
  {Phys. Rev. D}\ }\textbf {\bibinfo {volume} {100}},\ \bibinfo {pages}
  {044061} (\bibinfo {year} {2019})},\ \Eprint
  {http://arxiv.org/abs/1906.05155} {arXiv:1906.05155 [gr-qc]} \BibitemShut
  {NoStop}%
\bibitem [{\citenamefont {V\"olkel}\ \emph {et~al.}(2022)\citenamefont
  {V\"olkel}, \citenamefont {Franchini},\ and\ \citenamefont
  {Barausse}}]{Volkel:2022aca}%
  \BibitemOpen
  \bibfield  {author} {\bibinfo {author} {\bibfnamefont {S.~H.}\ \bibnamefont
  {V\"olkel}}, \bibinfo {author} {\bibfnamefont {N.}~\bibnamefont {Franchini}},
  \ and\ \bibinfo {author} {\bibfnamefont {E.}~\bibnamefont {Barausse}},\
  }\href {\doibase 10.1103/PhysRevD.105.084046} {\bibfield  {journal} {\bibinfo
   {journal} {Phys. Rev. D}\ }\textbf {\bibinfo {volume} {105}},\ \bibinfo
  {pages} {084046} (\bibinfo {year} {2022})},\ \Eprint
  {http://arxiv.org/abs/2202.08655} {arXiv:2202.08655 [gr-qc]} \BibitemShut
  {NoStop}%
\bibitem [{\citenamefont {Li}\ \emph {et~al.}(2023)\citenamefont {Li},
  \citenamefont {Wagle}, \citenamefont {Chen},\ and\ \citenamefont
  {Yunes}}]{Li:2022pcy}%
  \BibitemOpen
  \bibfield  {author} {\bibinfo {author} {\bibfnamefont {D.}~\bibnamefont
  {Li}}, \bibinfo {author} {\bibfnamefont {P.}~\bibnamefont {Wagle}}, \bibinfo
  {author} {\bibfnamefont {Y.}~\bibnamefont {Chen}}, \ and\ \bibinfo {author}
  {\bibfnamefont {N.}~\bibnamefont {Yunes}},\ }\href {\doibase
  10.1103/PhysRevX.13.021029} {\bibfield  {journal} {\bibinfo  {journal} {Phys.
  Rev. X}\ }\textbf {\bibinfo {volume} {13}},\ \bibinfo {pages} {021029}
  (\bibinfo {year} {2023})},\ \Eprint {http://arxiv.org/abs/2206.10652}
  {arXiv:2206.10652 [gr-qc]} \BibitemShut {NoStop}%
\bibitem [{\citenamefont {Hussain}\ and\ \citenamefont
  {Zimmerman}(2022)}]{Hussain:2022ins}%
  \BibitemOpen
  \bibfield  {author} {\bibinfo {author} {\bibfnamefont {A.}~\bibnamefont
  {Hussain}}\ and\ \bibinfo {author} {\bibfnamefont {A.}~\bibnamefont
  {Zimmerman}},\ }\href {\doibase 10.1103/PhysRevD.106.104018} {\bibfield
  {journal} {\bibinfo  {journal} {Phys. Rev. D}\ }\textbf {\bibinfo {volume}
  {106}},\ \bibinfo {pages} {104018} (\bibinfo {year} {2022})},\ \Eprint
  {http://arxiv.org/abs/2206.10653} {arXiv:2206.10653 [gr-qc]} \BibitemShut
  {NoStop}%
\bibitem [{\citenamefont {Gupta}\ \emph {et~al.}(2020)\citenamefont {Gupta},
  \citenamefont {Datta}, \citenamefont {Kastha}, \citenamefont {Borhanian},
  \citenamefont {Arun},\ and\ \citenamefont {Sathyaprakash}}]{Gupta:2020lxa}%
  \BibitemOpen
  \bibfield  {author} {\bibinfo {author} {\bibfnamefont {A.}~\bibnamefont
  {Gupta}}, \bibinfo {author} {\bibfnamefont {S.}~\bibnamefont {Datta}},
  \bibinfo {author} {\bibfnamefont {S.}~\bibnamefont {Kastha}}, \bibinfo
  {author} {\bibfnamefont {S.}~\bibnamefont {Borhanian}}, \bibinfo {author}
  {\bibfnamefont {K.~G.}\ \bibnamefont {Arun}}, \ and\ \bibinfo {author}
  {\bibfnamefont {B.~S.}\ \bibnamefont {Sathyaprakash}},\ }\href {\doibase
  10.1103/PhysRevLett.125.201101} {\bibfield  {journal} {\bibinfo  {journal}
  {Phys. Rev. Lett.}\ }\textbf {\bibinfo {volume} {125}},\ \bibinfo {pages}
  {201101} (\bibinfo {year} {2020})},\ \Eprint
  {http://arxiv.org/abs/2005.09607} {arXiv:2005.09607 [gr-qc]} \BibitemShut
  {NoStop}%
\bibitem [{\citenamefont {Datta}\ \emph {et~al.}(2022)\citenamefont {Datta},
  \citenamefont {Saleem}, \citenamefont {Arun},\ and\ \citenamefont
  {Sathyaprakash}}]{Datta:2022izc}%
  \BibitemOpen
  \bibfield  {author} {\bibinfo {author} {\bibfnamefont {S.}~\bibnamefont
  {Datta}}, \bibinfo {author} {\bibfnamefont {M.}~\bibnamefont {Saleem}},
  \bibinfo {author} {\bibfnamefont {K.~G.}\ \bibnamefont {Arun}}, \ and\
  \bibinfo {author} {\bibfnamefont {B.~S.}\ \bibnamefont {Sathyaprakash}},\
  }\href@noop {} {\  (\bibinfo {year} {2022})},\ \Eprint
  {http://arxiv.org/abs/2208.07757} {arXiv:2208.07757 [gr-qc]} \BibitemShut
  {NoStop}%
\bibitem [{\citenamefont {Johnson-McDaniel}\ \emph {et~al.}(2022)\citenamefont
  {Johnson-McDaniel}, \citenamefont {Ghosh}, \citenamefont {Ghonge},
  \citenamefont {Saleem}, \citenamefont {Krishnendu},\ and\ \citenamefont
  {Clark}}]{Johnson-McDaniel:2021yge}%
  \BibitemOpen
  \bibfield  {author} {\bibinfo {author} {\bibfnamefont {N.~K.}\ \bibnamefont
  {Johnson-McDaniel}}, \bibinfo {author} {\bibfnamefont {A.}~\bibnamefont
  {Ghosh}}, \bibinfo {author} {\bibfnamefont {S.}~\bibnamefont {Ghonge}},
  \bibinfo {author} {\bibfnamefont {M.}~\bibnamefont {Saleem}}, \bibinfo
  {author} {\bibfnamefont {N.~V.}\ \bibnamefont {Krishnendu}}, \ and\ \bibinfo
  {author} {\bibfnamefont {J.~A.}\ \bibnamefont {Clark}},\ }\href {\doibase
  10.1103/PhysRevD.105.044020} {\bibfield  {journal} {\bibinfo  {journal}
  {Phys. Rev. D}\ }\textbf {\bibinfo {volume} {105}},\ \bibinfo {pages}
  {044020} (\bibinfo {year} {2022})},\ \Eprint
  {http://arxiv.org/abs/2109.06988} {arXiv:2109.06988 [gr-qc]} \BibitemShut
  {NoStop}%
\bibitem [{\citenamefont {Silva}\ \emph {et~al.}(2022)\citenamefont {Silva},
  \citenamefont {Ghosh},\ and\ \citenamefont {Buonanno}}]{Silva:2022srr}%
  \BibitemOpen
  \bibfield  {author} {\bibinfo {author} {\bibfnamefont {H.~O.}\ \bibnamefont
  {Silva}}, \bibinfo {author} {\bibfnamefont {A.}~\bibnamefont {Ghosh}}, \ and\
  \bibinfo {author} {\bibfnamefont {A.}~\bibnamefont {Buonanno}},\ }\href@noop
  {} {\  (\bibinfo {year} {2022})},\ \Eprint {http://arxiv.org/abs/2205.05132}
  {arXiv:2205.05132 [gr-qc]} \BibitemShut {NoStop}%
\bibitem [{\citenamefont {Nee}\ \emph {et~al.}(2023)\citenamefont {Nee},
  \citenamefont {V\"olkel},\ and\ \citenamefont {Pfeiffer}}]{Nee:2023osy}%
  \BibitemOpen
  \bibfield  {author} {\bibinfo {author} {\bibfnamefont {P.~J.}\ \bibnamefont
  {Nee}}, \bibinfo {author} {\bibfnamefont {S.~H.}\ \bibnamefont {V\"olkel}}, \
  and\ \bibinfo {author} {\bibfnamefont {H.~P.}\ \bibnamefont {Pfeiffer}},\
  }\href@noop {} {\  (\bibinfo {year} {2023})},\ \Eprint
  {http://arxiv.org/abs/2302.06634} {arXiv:2302.06634 [gr-qc]} \BibitemShut
  {NoStop}%
\bibitem [{ICT()}]{ICTP_Cardoso}%
  \BibitemOpen
  \href@noop {} {}\bibinfo {note} {{Vitor Cardoso, ICTP Summer School Lectures
  2018: \\ \url{https://indico.ictp.it/event/8317/} }}\BibitemShut {NoStop}%
\bibitem [{\citenamefont {Alsing}\ \emph {et~al.}(2012)\citenamefont {Alsing},
  \citenamefont {Berti}, \citenamefont {Will},\ and\ \citenamefont
  {Zaglauer}}]{Alsing:2011er}%
  \BibitemOpen
  \bibfield  {author} {\bibinfo {author} {\bibfnamefont {J.}~\bibnamefont
  {Alsing}}, \bibinfo {author} {\bibfnamefont {E.}~\bibnamefont {Berti}},
  \bibinfo {author} {\bibfnamefont {C.~M.}\ \bibnamefont {Will}}, \ and\
  \bibinfo {author} {\bibfnamefont {H.}~\bibnamefont {Zaglauer}},\ }\href
  {\doibase 10.1103/PhysRevD.85.064041} {\bibfield  {journal} {\bibinfo
  {journal} {Phys. Rev. D}\ }\textbf {\bibinfo {volume} {85}},\ \bibinfo
  {pages} {064041} (\bibinfo {year} {2012})},\ \Eprint
  {http://arxiv.org/abs/1112.4903} {arXiv:1112.4903 [gr-qc]} \BibitemShut
  {NoStop}%
\bibitem [{\citenamefont {Cardoso}\ \emph {et~al.}(2003)\citenamefont
  {Cardoso}, \citenamefont {Dias},\ and\ \citenamefont
  {Lemos}}]{Cardoso:2002pa}%
  \BibitemOpen
  \bibfield  {author} {\bibinfo {author} {\bibfnamefont {V.}~\bibnamefont
  {Cardoso}}, \bibinfo {author} {\bibfnamefont {O.~J.~C.}\ \bibnamefont
  {Dias}}, \ and\ \bibinfo {author} {\bibfnamefont {J.~P.~S.}\ \bibnamefont
  {Lemos}},\ }\href {\doibase 10.1103/PhysRevD.67.064026} {\bibfield  {journal}
  {\bibinfo  {journal} {Phys. Rev. D}\ }\textbf {\bibinfo {volume} {67}},\
  \bibinfo {pages} {064026} (\bibinfo {year} {2003})},\ \Eprint
  {http://arxiv.org/abs/hep-th/0212168} {arXiv:hep-th/0212168} \BibitemShut
  {NoStop}%
\bibitem [{\citenamefont {Hadar}\ and\ \citenamefont
  {Kol}(2011)}]{Hadar:2009ip}%
  \BibitemOpen
  \bibfield  {author} {\bibinfo {author} {\bibfnamefont {S.}~\bibnamefont
  {Hadar}}\ and\ \bibinfo {author} {\bibfnamefont {B.}~\bibnamefont {Kol}},\
  }\href {\doibase 10.1103/PhysRevD.84.044019} {\bibfield  {journal} {\bibinfo
  {journal} {Phys. Rev. D}\ }\textbf {\bibinfo {volume} {84}},\ \bibinfo
  {pages} {044019} (\bibinfo {year} {2011})},\ \Eprint
  {http://arxiv.org/abs/0911.3899} {arXiv:0911.3899 [gr-qc]} \BibitemShut
  {NoStop}%
\bibitem [{\citenamefont {Hadar}\ \emph {et~al.}(2011)\citenamefont {Hadar},
  \citenamefont {Kol}, \citenamefont {Berti},\ and\ \citenamefont
  {Cardoso}}]{Hadar:2011vj}%
  \BibitemOpen
  \bibfield  {author} {\bibinfo {author} {\bibfnamefont {S.}~\bibnamefont
  {Hadar}}, \bibinfo {author} {\bibfnamefont {B.}~\bibnamefont {Kol}}, \bibinfo
  {author} {\bibfnamefont {E.}~\bibnamefont {Berti}}, \ and\ \bibinfo {author}
  {\bibfnamefont {V.}~\bibnamefont {Cardoso}},\ }\href {\doibase
  10.1103/PhysRevD.84.047501} {\bibfield  {journal} {\bibinfo  {journal} {Phys.
  Rev. D}\ }\textbf {\bibinfo {volume} {84}},\ \bibinfo {pages} {047501}
  (\bibinfo {year} {2011})},\ \Eprint {http://arxiv.org/abs/1105.3861}
  {arXiv:1105.3861 [gr-qc]} \BibitemShut {NoStop}%
\bibitem [{\citenamefont {Hadar}\ \emph {et~al.}(2014)\citenamefont {Hadar},
  \citenamefont {Porfyriadis},\ and\ \citenamefont
  {Strominger}}]{Hadar:2014dpa}%
  \BibitemOpen
  \bibfield  {author} {\bibinfo {author} {\bibfnamefont {S.}~\bibnamefont
  {Hadar}}, \bibinfo {author} {\bibfnamefont {A.~P.}\ \bibnamefont
  {Porfyriadis}}, \ and\ \bibinfo {author} {\bibfnamefont {A.}~\bibnamefont
  {Strominger}},\ }\href {\doibase 10.1103/PhysRevD.90.064045} {\bibfield
  {journal} {\bibinfo  {journal} {Phys. Rev. D}\ }\textbf {\bibinfo {volume}
  {90}},\ \bibinfo {pages} {064045} (\bibinfo {year} {2014})},\ \Eprint
  {http://arxiv.org/abs/1403.2797} {arXiv:1403.2797 [hep-th]} \BibitemShut
  {NoStop}%
\bibitem [{\citenamefont {Hadar}\ \emph {et~al.}(2015)\citenamefont {Hadar},
  \citenamefont {Porfyriadis},\ and\ \citenamefont
  {Strominger}}]{Hadar:2015xpa}%
  \BibitemOpen
  \bibfield  {author} {\bibinfo {author} {\bibfnamefont {S.}~\bibnamefont
  {Hadar}}, \bibinfo {author} {\bibfnamefont {A.~P.}\ \bibnamefont
  {Porfyriadis}}, \ and\ \bibinfo {author} {\bibfnamefont {A.}~\bibnamefont
  {Strominger}},\ }\href {\doibase 10.1007/JHEP07(2015)078} {\bibfield
  {journal} {\bibinfo  {journal} {JHEP}\ }\textbf {\bibinfo {volume} {07}},\
  \bibinfo {pages} {078} (\bibinfo {year} {2015})},\ \Eprint
  {http://arxiv.org/abs/1504.07650} {arXiv:1504.07650 [hep-th]} \BibitemShut
  {NoStop}%
\bibitem [{\citenamefont {Folacci}\ and\ \citenamefont {Ould
  El~Hadj}(2018)}]{Folacci:2018cic}%
  \BibitemOpen
  \bibfield  {author} {\bibinfo {author} {\bibfnamefont {A.}~\bibnamefont
  {Folacci}}\ and\ \bibinfo {author} {\bibfnamefont {M.}~\bibnamefont {Ould
  El~Hadj}},\ }\href {\doibase 10.1103/PhysRevD.98.084008} {\bibfield
  {journal} {\bibinfo  {journal} {Phys. Rev. D}\ }\textbf {\bibinfo {volume}
  {98}},\ \bibinfo {pages} {084008} (\bibinfo {year} {2018})},\ \Eprint
  {http://arxiv.org/abs/1806.01577} {arXiv:1806.01577 [gr-qc]} \BibitemShut
  {NoStop}%
\bibitem [{\citenamefont {Apte}\ and\ \citenamefont
  {Hughes}(2019)}]{Apte:2019txp}%
  \BibitemOpen
  \bibfield  {author} {\bibinfo {author} {\bibfnamefont {A.}~\bibnamefont
  {Apte}}\ and\ \bibinfo {author} {\bibfnamefont {S.~A.}\ \bibnamefont
  {Hughes}},\ }\href {\doibase 10.1103/PhysRevD.100.084031} {\bibfield
  {journal} {\bibinfo  {journal} {Phys. Rev. D}\ }\textbf {\bibinfo {volume}
  {100}},\ \bibinfo {pages} {084031} (\bibinfo {year} {2019})},\ \Eprint
  {http://arxiv.org/abs/1901.05901} {arXiv:1901.05901 [gr-qc]} \BibitemShut
  {NoStop}%
\bibitem [{\citenamefont {Lim}\ \emph {et~al.}(2019)\citenamefont {Lim},
  \citenamefont {Khanna}, \citenamefont {Apte},\ and\ \citenamefont
  {Hughes}}]{Lim:2019xrb}%
  \BibitemOpen
  \bibfield  {author} {\bibinfo {author} {\bibfnamefont {H.}~\bibnamefont
  {Lim}}, \bibinfo {author} {\bibfnamefont {G.}~\bibnamefont {Khanna}},
  \bibinfo {author} {\bibfnamefont {A.}~\bibnamefont {Apte}}, \ and\ \bibinfo
  {author} {\bibfnamefont {S.~A.}\ \bibnamefont {Hughes}},\ }\href {\doibase
  10.1103/PhysRevD.100.084032} {\bibfield  {journal} {\bibinfo  {journal}
  {Phys. Rev. D}\ }\textbf {\bibinfo {volume} {100}},\ \bibinfo {pages}
  {084032} (\bibinfo {year} {2019})},\ \Eprint
  {http://arxiv.org/abs/1901.05902} {arXiv:1901.05902 [gr-qc]} \BibitemShut
  {NoStop}%
\bibitem [{\citenamefont {Hughes}\ \emph {et~al.}(2019)\citenamefont {Hughes},
  \citenamefont {Apte}, \citenamefont {Khanna},\ and\ \citenamefont
  {Lim}}]{Hughes:2019zmt}%
  \BibitemOpen
  \bibfield  {author} {\bibinfo {author} {\bibfnamefont {S.~A.}\ \bibnamefont
  {Hughes}}, \bibinfo {author} {\bibfnamefont {A.}~\bibnamefont {Apte}},
  \bibinfo {author} {\bibfnamefont {G.}~\bibnamefont {Khanna}}, \ and\ \bibinfo
  {author} {\bibfnamefont {H.}~\bibnamefont {Lim}},\ }\href {\doibase
  10.1103/PhysRevLett.123.161101} {\bibfield  {journal} {\bibinfo  {journal}
  {Phys. Rev. Lett.}\ }\textbf {\bibinfo {volume} {123}},\ \bibinfo {pages}
  {161101} (\bibinfo {year} {2019})},\ \Eprint
  {http://arxiv.org/abs/1901.05900} {arXiv:1901.05900 [gr-qc]} \BibitemShut
  {NoStop}%
\bibitem [{\citenamefont {Lim}\ \emph {et~al.}(2022)\citenamefont {Lim},
  \citenamefont {Hughes},\ and\ \citenamefont {Khanna}}]{Lim:2022veo}%
  \BibitemOpen
  \bibfield  {author} {\bibinfo {author} {\bibfnamefont {H.}~\bibnamefont
  {Lim}}, \bibinfo {author} {\bibfnamefont {S.~A.}\ \bibnamefont {Hughes}}, \
  and\ \bibinfo {author} {\bibfnamefont {G.}~\bibnamefont {Khanna}},\ }\href
  {\doibase 10.1103/PhysRevD.105.124030} {\bibfield  {journal} {\bibinfo
  {journal} {Phys. Rev. D}\ }\textbf {\bibinfo {volume} {105}},\ \bibinfo
  {pages} {124030} (\bibinfo {year} {2022})},\ \Eprint
  {http://arxiv.org/abs/2204.06007} {arXiv:2204.06007 [gr-qc]} \BibitemShut
  {NoStop}%
\bibitem [{\citenamefont {Oshita}(2022)}]{Oshita:2022pkc}%
  \BibitemOpen
  \bibfield  {author} {\bibinfo {author} {\bibfnamefont {N.}~\bibnamefont
  {Oshita}},\ }\href@noop {} {\  (\bibinfo {year} {2022})},\ \Eprint
  {http://arxiv.org/abs/2208.02923} {arXiv:2208.02923 [gr-qc]} \BibitemShut
  {NoStop}%
\bibitem [{\citenamefont {Oshita}\ and\ \citenamefont
  {Tsuna}(2022)}]{Oshita:2022yry}%
  \BibitemOpen
  \bibfield  {author} {\bibinfo {author} {\bibfnamefont {N.}~\bibnamefont
  {Oshita}}\ and\ \bibinfo {author} {\bibfnamefont {D.}~\bibnamefont {Tsuna}},\
  }\href@noop {} {\  (\bibinfo {year} {2022})},\ \Eprint
  {http://arxiv.org/abs/2210.14049} {arXiv:2210.14049 [gr-qc]} \BibitemShut
  {NoStop}%
\bibitem [{\citenamefont {Rom}\ and\ \citenamefont {Sari}(2022)}]{Rom:2022uvv}%
  \BibitemOpen
  \bibfield  {author} {\bibinfo {author} {\bibfnamefont {B.}~\bibnamefont
  {Rom}}\ and\ \bibinfo {author} {\bibfnamefont {R.}~\bibnamefont {Sari}},\
  }\href {\doibase 10.1103/PhysRevD.106.104040} {\bibfield  {journal} {\bibinfo
   {journal} {Phys. Rev. D}\ }\textbf {\bibinfo {volume} {106}},\ \bibinfo
  {pages} {104040} (\bibinfo {year} {2022})},\ \Eprint
  {http://arxiv.org/abs/2204.11738} {arXiv:2204.11738 [gr-qc]} \BibitemShut
  {NoStop}%
\bibitem [{\citenamefont {Bishop}\ \emph {et~al.}(1996)\citenamefont {Bishop},
  \citenamefont {Gomez}, \citenamefont {Lehner},\ and\ \citenamefont
  {Winicour}}]{Bishop:1996gt}%
  \BibitemOpen
  \bibfield  {author} {\bibinfo {author} {\bibfnamefont {N.~T.}\ \bibnamefont
  {Bishop}}, \bibinfo {author} {\bibfnamefont {R.}~\bibnamefont {Gomez}},
  \bibinfo {author} {\bibfnamefont {L.}~\bibnamefont {Lehner}}, \ and\ \bibinfo
  {author} {\bibfnamefont {J.}~\bibnamefont {Winicour}},\ }\href {\doibase
  10.1103/PhysRevD.54.6153} {\bibfield  {journal} {\bibinfo  {journal} {Phys.
  Rev. D}\ }\textbf {\bibinfo {volume} {54}},\ \bibinfo {pages} {6153}
  (\bibinfo {year} {1996})},\ \Eprint {http://arxiv.org/abs/gr-qc/9705033}
  {arXiv:gr-qc/9705033} \BibitemShut {NoStop}%
\bibitem [{\citenamefont {Babiuc}\ \emph {et~al.}(2005)\citenamefont {Babiuc},
  \citenamefont {Szilagyi}, \citenamefont {Hawke},\ and\ \citenamefont
  {Zlochower}}]{Babiuc:2005pg}%
  \BibitemOpen
  \bibfield  {author} {\bibinfo {author} {\bibfnamefont {M.}~\bibnamefont
  {Babiuc}}, \bibinfo {author} {\bibfnamefont {B.}~\bibnamefont {Szilagyi}},
  \bibinfo {author} {\bibfnamefont {I.}~\bibnamefont {Hawke}}, \ and\ \bibinfo
  {author} {\bibfnamefont {Y.}~\bibnamefont {Zlochower}},\ }\href {\doibase
  10.1088/0264-9381/22/23/011} {\bibfield  {journal} {\bibinfo  {journal}
  {Class. Quant. Grav.}\ }\textbf {\bibinfo {volume} {22}},\ \bibinfo {pages}
  {5089} (\bibinfo {year} {2005})},\ \Eprint
  {http://arxiv.org/abs/gr-qc/0501008} {arXiv:gr-qc/0501008} \BibitemShut
  {NoStop}%
\bibitem [{\citenamefont {Winicour}(2009)}]{Winicour:2008vpn}%
  \BibitemOpen
  \bibfield  {author} {\bibinfo {author} {\bibfnamefont {J.}~\bibnamefont
  {Winicour}},\ }\href {\doibase 10.12942/lrr-2009-3} {\bibfield  {journal}
  {\bibinfo  {journal} {Living Rev. Rel.}\ }\textbf {\bibinfo {volume} {12}},\
  \bibinfo {pages} {3} (\bibinfo {year} {2009})},\ \Eprint
  {http://arxiv.org/abs/0810.1903} {arXiv:0810.1903 [gr-qc]} \BibitemShut
  {NoStop}%
\bibitem [{\citenamefont {Pereira}\ and\ \citenamefont
  {Sturani}(2022)}]{Pereira:2022kqn}%
  \BibitemOpen
  \bibfield  {author} {\bibinfo {author} {\bibfnamefont {T.}~\bibnamefont
  {Pereira}}\ and\ \bibinfo {author} {\bibfnamefont {R.}~\bibnamefont
  {Sturani}},\ }\href@noop {} {\  (\bibinfo {year} {2022})},\ \Eprint
  {http://arxiv.org/abs/2210.07299} {arXiv:2210.07299 [gr-qc]} \BibitemShut
  {NoStop}%
\end{thebibliography}%

\end{document}